\documentclass[12pt,prd, aps, showpacs, superscriptaddress,floatfix]{revtex4-1}
\usepackage{graphicx}
\usepackage{bm}
\usepackage{multirow}
\usepackage{axodraw}
\usepackage{epsfig}
\usepackage{psfrag}
\usepackage{soul}
\usepackage{multirow}
\usepackage{mathptmx}
\usepackage{mathrsfs}
\usepackage{amsmath, amssymb}
\usepackage{dcolumn}
\usepackage{epstopdf}
\usepackage{color}
\usepackage{hhline}
\usepackage{changebar}
\usepackage{subfigure}
\usepackage{placeins}
\usepackage{mathtools}
\usepackage{slashed}
\usepackage{longtable}
\usepackage{dsfont}

\usepackage{CJK}

\newcommand\riken{RIKEN-BNL Research Center, Brookhaven National
  Laboratory, Upton, NY 11973, USA}
\newcommand\bnl{Department of Physics, Brookhaven National Laboratory, Upton, NY 11973, USA}
\newcommand\edinb{SUPA, School of Physics, The University of
  Edinburgh, Edinburgh EH9 3JZ, UK}

\newcommand\cu{Physics Department, Columbia University, New York,
  NY 10027, USA}

\newcommand\uconn{Physics Department, University of Connecticut,
  Storrs, CT 06269-3046, USA}
\newcommand\soton{School of Physics and Astronomy, University of
  Southampton,  Southampton SO17 1BJ, UK}
\newcommand\kek{Institute of Particle and Nuclear Studies,
  KEK, Tsukuba, Ibaraki, 305-0801, Japan}
\newcommand\sokendai{Department of Particle and Nuclear Physics, Sokendai
Graduate University of Advanced Studies, Hayama, Kanagawa 240-0193, Japan}

\newcommand\cambridge{Department of Applied Mathematics \& Theoretical Physics, University of Cambridge, Cambridge CB3 0WA, UK}
\newcommand\trinityonleave{On leave from School of Mathematics, Trinity College Dublin, College Green, Dublin 2, Ireland}

\newcommand\yorkcanada{Department of Physics and Astronomy, York University, Toronto, Ontario, M3J 1P3, Canada}
\newcommand\cern{CERN, Physics Department, 1211 Geneva 23, Switzerland}

\newcounter{Outline}
\newcounter{Introduction}
\newcounter{Conclusions}
\newcounter{Acknowledgments}
\newcounter{Appendix}
\newcounter{Tables}
\newcounter{Figures}
\newcommand{\ba}{\begin{eqnarray}}
\newcommand{\ea}{\end{eqnarray}}
\newcommand{\bas}{\begin{eqnarray*}}
\newcommand{\eas}{\end{eqnarray*}}
\newcommand{\be}{\begin{equation}}
\newcommand{\ee}{\end{equation}}
\newcommand{\bes}{\begin{equation*}}
\newcommand{\ees}{\end{equation*}}
\newcommand{\bi}{\begin{itemize}}
\newcommand{\ei}{\end{itemize}}

\newcommand{\bcentre}{\begin{center}}
\newcommand{\ecentre}{\end{center}}

\newcommand{\Slash}[1]{\slashed{#1}}

\newcommand{\ident}{\mathds{1}}
\newcommand{\p}{{\cal P}}
\newcommand{\s}[1]{\Slash{#1}}
\newcommand{\la}{\langle}
\newcommand{\ra}{\rangle}

\newcommand{\bc}{\begin{center}}
\newcommand{\ec}{\end{center}}

\font\tenmsb=msbm10 scaled\magstep1
\font\sevenmsb=msbm7 scaled\magstep1
\font\fivemsb=msbm5 scaled\magstep1
\newfam\msbfam
\textfont\msbfam=\tenmsb
\scriptfont\msbfam=\sevenmsb
\scriptscriptfont\msbfam=\fivemsb

\newcommand{\mres}{m_{\rm res}}

\def\msbar{\overline{\mbox{\scriptsize MS}}}
\def\MSbar{\overline{\mbox{MS}}}
\def\rational#1#2{{\mathchoice{\textstyle{#1\over#2}}%
  {\scriptstyle{#1\over#2}}{\scriptscriptstyle{#1\over#2}}{#1/#2}}}
\def\half{\rational12}			    

\def\rmsub#1#2{#1_{\mbox{\tiny #2}}}	    

\def\ln{\mathop{\rm ln}}		    
\def\tr{\mathop{\rm tr}}		    
\def\det{\mathop{\rm det}}		    

\def\su3{SU(3)}

\def\tD{\mbox{D}\kern-0.65em\raise0.15ex\hbox{/}\kern0.15em} 
\def\sD{\mbox{\scriptsize D}\kern-0.5em\raise0.15ex\hbox{\scriptsize/}}
\def\ssD{\mbox{\tiny D}\kern-0.42em\raise0.15ex\hbox{\tiny/}}

\def\dslash{\hbox{\(\partial\)}\kern-0.5em\raise0.15ex\hbox{/}} 

\def\su3{SU(3)}

\def\mres{\rmsub{m}{res}}

\def\nicefrac#1#2{\leavevmode\kern.1em\raise.5ex\hbox{\the\scriptfont0 #1}\kern-
.1em/\kern-.15em\lower.25ex\hbox{\the\scriptfont0 #2}}

\newcommand{\bea}{\begin{eqnarray}}
\newcommand{\eea}{\end{eqnarray}}

\newcommand{\GeV}{\;\rm GeV}

\newcommand{\mobius}{M\"{o}bius }

\begin{document}
\bibliographystyle{apsrev}

\begin{CJK*}{UTF8}{}

\title{Domain wall QCD with physical quark masses}

\author{T.~Blum}\affiliation{\uconn}\affiliation{\riken}
\author{P.A.~Boyle}\affiliation{\edinb}
\author{N.H.~Christ}\affiliation{\cu}
\author{J.~Frison}\affiliation{\edinb}
\author{N.~Garron}\affiliation{\cambridge}\affiliation{\trinityonleave}
\author{R.J.~Hudspith}\affiliation{\yorkcanada}
\author{T.~Izubuchi}\affiliation{\riken}\affiliation{\bnl}
\author{T.~Janowski}\affiliation{\soton}
\author{C.~Jung}\affiliation{\bnl}

\author{A.~J\"{u}ttner}\affiliation{\soton}

\author{C.~Kelly}\affiliation{\riken}
\author{R.D.~Kenway}\affiliation{\edinb}
\author{C.~Lehner}\affiliation{\bnl}

\author{M.~Marinkovic}\affiliation{\soton}\affiliation{\cern}

\author{R.D.~Mawhinney}\affiliation{\cu}
\author{G.~McGlynn}\affiliation{\cu}
\author{D.J.~Murphy}\affiliation{\cu}
\CJKfamily{min}
\author{S.~Ohta (太田滋生)}\affiliation{\kek}\affiliation{\sokendai}\affiliation{\riken}
\author{A.~Portelli}\affiliation{\soton}
\author{C.T.~Sachrajda}\affiliation{\soton}
\author{A.~Soni}\affiliation{\bnl}

\collaboration{RBC and UKQCD Collaborations}
%
%
\mbox{}\hfill\noaffiliation{KEK-TH-1769, RBRC 1095, DAMTP-2014-86 }


\maketitle
\end{CJK*}


\newcommand{\ainvzero}{2.383(9)} 
\newcommand{\ainvone}{1.785(5)}  
\newcommand{\ainvtwo}{1.730(4)}  
\newcommand{\ainvthree}{2.359(7)} 
\newcommand{\ainvfour}{3.148(17)} 
\newcommand{\ainvfive}{1.378(7)} 

\newcommand{\Lfmtwo}{5.476(12)} 
\newcommand{\Lfmthree}{5.354(16)} 
\newcommand{\Lfmfour}{2.006(11)} 

\newcommand{\mpitwo}{139.2(4)} 
\newcommand{\mpithree}{139.2(5)} 
\newcommand{\mpifour}{371(5)} 

\newcommand{\mpiLtwo}{3.863(6)} 
\newcommand{\mpiLthree}{3.778(8)} 
\newcommand{\mpiLfour}{3.773(42)} 

\newcommand{\mpiLtwoshort}{3.86} 
\newcommand{\mpiLthreeshort}{3.78} 

\newcommand{\fpicont}{130.2(9)}
\newcommand{\fkcont}{155.5(8)}
\newcommand{\fkdfpicont}{1.195(5)}

\newcommand{\mudMSbar}{2.997(49)}
\newcommand{\msMSbar}{81.64(1.17)}

\newcommand{\BKRGI}{0.750(15)}
\newcommand{\BKMSbar}{0.530(11)}

\newcommand{\sqrttzero}{0.729(4)} 
\newcommand{\wzero}{0.874(5)}

\newcommand{\Lfmzero}{2.649(10)} 
\newcommand{\Lfmone}{2.653(7)} 
\newcommand{\Lfmfive}{4.581(23)} 

\newcommand{\mpiLzero}{3.122(12)} 
\newcommand{\mpiLone}{3.339(15)} 
\newcommand{\mpiLfive}{3.335(7)} 


\centerline{ABSTRACT}
We present results for several light hadronic quantities ($f_\pi$, $f_K$, $B_K$, $m_{ud}$, $m_s$, $t_0^{1/2}$, $w_0$) obtained from simulations of 2+1 flavor domain wall lattice QCD with large physical volumes and nearly-physical pion masses at two lattice spacings.  We perform a short, ${\cal O}(3)\%$, extrapolation in pion mass to the physical values by combining our new data in a simultaneous chiral/continuum `global fit' with a number of other ensembles with heavier pion masses.  We use the physical values of $m_\pi$, $m_K$ and $m_\Omega$ to determine the two quark masses and the scale - all other quantities are outputs from our simulations. We obtain results with sub-percent statistical errors and negligible chiral and finite-volume systematics for these light hadronic quantities, including: $f_\pi = \fpicont$ MeV; $f_K = \fkcont$ MeV; the average up/down quark mass and strange quark mass in the $\MSbar$ scheme at 3 GeV, $\mudMSbar$ and $\msMSbar$ MeV respectively; and the neutral kaon mixing parameter, $B_K$, in the RGI scheme, $\BKRGI$ and the $\MSbar$ scheme at 3 GeV, $\BKMSbar$.


\refstepcounter{section}
\setcounter{section}{0}


\newpage
\section{Introduction}
\label{sec:Introduction}

The low energy details of the strong interactions, encapsulated theoretically in the Lagrangian of QCD, are responsible for producing mesons and hadrons from quarks, creating most of the mass of the visible universe, and determining a vacuum state which exhibits symmetry breaking.  For many decades, the methods of numerical lattice QCD have been used to study these phenomena, both because of their intrinsic interest and because QCD effects are important for many precision tests of quark interactions in the Standard Model.  Many theoretical and computational advances have been made during this time and, in this paper, we report on the first simulations of 2+1 flavor QCD ({\em i.e.} QCD including the fermion determinant for $u$, $d$ and $s$ quarks with $m_u = m_d$) with essentially physical quark masses using a lattice fermion formulation which accurately preserves the continuum global symmetries of QCD at finite lattice spacing: domain wall fermions (DWF).

This isospin symmetric version of QCD requires three inputs to perform a simulation at a single lattice spacing: a bare coupling constant, a degenerate light quark mass ($m_u=m_d$), and a strange quark mass.  We fix these using the physical values for $m_\pi$, $m_K$, and $m_\Omega$.  In particular, for a fixed bare coupling, adjusting $m_u=m_d$ and $m_s$ until $m_\pi/m_\Omega$ and $m_K/m_\Omega$ take on their physical values leads to a determination of the lattice spacing, $a$, for this coupling.  All other low energy quantities, such as $f_\pi$ and $f_K$, are now predictions.  By repeating this for different lattice spacings, physical predictions in the continuum limit ($a \to 0$) for other low energy QCD observables are obtained.  In this work, we used results from our earlier simulations to estimate the input physical quark masses and then we make a modest correction in our results, using chiral perturbation theory and simple analytic ansatz, to adjust to the required quark mass values, a correction of less than 10\% in the quark mass. These physical quark mass simulations would not have been possible without IBM Blue Gene/Q resources~\cite{ibmbgqieee,ibmcodesign,originmass,Boyle:2012iy}.

For the past decade, the RBC and UKQCD collaborations have been steadily approaching the physical quark mass point with a series of 2+1 flavor domain wall fermion simulations. Recently~\cite{Arthur:2012opa} we reported on a combined analysis of three of our domain wall fermion ensembles with the Shamir kernel, namely our $32^3\times64$ and $24^3\times 64$ ensemble sets with the Iwasaki gauge action at $\beta=2.25$ and $\beta=2.13$ ($a^{-1}=\ainvzero$ GeV and $\ainvone$ GeV) and lightest unitary pion masses of $302(1)$ MeV and $337(2)$ MeV respectively, and our coarser $32^3\times 64$ Iwasaki+DSDR ensemble set with $\beta=1.75$ ($a^{-1}=\ainvfive$ GeV) but substantially lighter pion masses of $143(1)$ MeV partially-quenched and $171(1)$ MeV unitary. We refer to these as our 32I, 24I and 32ID ensembles, respectively. (The lattice spacings and other results for these ensembles quoted here come from global fits that include the new, physical quark mass ensembles, as well as new observable measurements on these older ensembles.  As such, central values have shifted from earlier published values, generally within the published errors. Also, the new errors are smaller, because of the increased data.)   For the latter 32ID ensembles, the use of a coarser lattice represented a compromise between the need to simulate with a large physical volume in order to keep finite-volume errors under control in the presence of such light pions and the prohibitive cost of increasing the lattice size. The DSDR term was used to suppress the dislocations in the gauge field that dominate the residual chiral symmetry breaking in the domain wall formulation at strong coupling. The addition of this ensemble set resulted in a factor of two reduction in the chiral extrapolation systematic error over our earlier analysis of the Iwasaki ensembles alone (24I and 32I)~\cite{Aoki:2010dy}, but the total errors on our physical predictions remained on the order of $2\%$. Now, combining algorithmic advances with the power of the latest generation of supercomputers, we are finally able to perform large volume simulations directly at the physical point without the need for such compromises.

In this paper we present an analysis of two 2+1 flavor domain wall ensembles simulated essentially at the physical point. The lattice sizes are $48^3\times 96$ and $64^3\times 128$ with physical volumes of $(\Lfmtwo\ {\rm fm})^3$ and $(\Lfmthree\ {\rm fm})^3$ ($m_\pi L = \mpiLtwoshort$ and $\mpiLthreeshort$). Throughout this document we refer to these ensembles with the labels 48I and 64I respectively. We utilize the \mobius domain wall action tuned such that the \mobius and Shamir kernels are identical up to a numerical factor, which allows us to simulate with a smaller fifth dimension, and hence a lower cost, for the same physics. This is discussed in more detail in Section~\ref{sec:SimulationDetails}. The values of $L_s$ are 24 and 12 for the 48I and 64I ensembles respectively. For the 48I ensemble, $L_s$ would have to be more than twice as large to achieve the same residual mass with the Shamir kernel.  The corresponding residual masses, $m_{\rm res}$, comprise $\sim 45\%$ of the physical light quark mass for the 48I ensemble, and $\sim 30\%$ for the 64I. We use the Iwasaki gauge action with $\beta=2.13$ and $2.25$, giving inverse lattice spacings of $a^{-1} = \ainvtwo$ GeV and $\ainvthree$ GeV, and the degenerate up/down quark masses were tuned to give (very nearly) physical pion masses of $\mpitwo$ MeV and $\mpithree$ MeV.

We also introduce a third ensemble generated with Shamir domain wall fermions and the Iwasaki gauge action at $\beta=2.37$, corresponding to an inverse lattice spacing of $\ainvfour$ GeV, with a lattice volume of $32^3\times 64$ and with $L_s=12$. The lightest unitary pion mass is $\mpifour$ MeV. Although these masses are unphysically heavy, this ensemble provides a third lattice spacing for each of the measured quantities, allowing us to bound the ${\cal O}(a^4)$ errors on our final results. We label this ensemble 32Ifine.

We have taken full advantage of each of our expensive 48I and 64I gauge configurations by developing a measurement package that uses EigCG to produce DWF eigenvectors in order to deflate subsequent quark mass solves, and that uses the all-mode-averaging (AMA) technique of Ref.~\cite{Blum:2012uh}.  In AMA, quark propagators are generated on every timeslice of the lattice but with reduced precision, and then corrected with a small number of precise measurements.  To reduce the fractional overhead of calculating eigenvectors and the large I/O demands of storing them, we share propagators between $m_\pi$, $m_K$, $f_\pi$, $f_K$, $B_K$, the $K_{l3}$ form factor $f_+^{K\pi}(q^2 = 0)$ and the $K \to (\pi \pi)_{{\rm I} = 2}$ amplitude.  (The last two quantities are not reported here.)  By putting so many measurements into a single job, the EigCG setup costs are only $\sim 20$\% of the total time, and we find this approach speeds up the measurement of these quantities by between 5 and 25 times, depending on the observable.  Here again the Blue Gene/Q has been invaluable, since it has a large enough memory to store the required eigenvectors and the reliability to run for sufficient time to use them in all of the above measurements.  In Section~\ref{sec:SimulationResults} we present the results of these measurements.

As mentioned already, in order to correct for the minor differences between the simulated and physical pion masses, we perform a short chiral extrapolation. As these new 48I and 64I ensembles have essentially the same quark masses, we must include data with other quark masses in order to determine the mass dependences. We achieve this by combining the 64I and 48I ensembles with the aforementioned $32^3\times64$ and $24^3\times 64$ Iwasaki gauge action ensemble sets (32I and 24I, respectively), and the $32^3\times64$ Iwasaki+DSDR ensemble set (32ID), in a simultaneous chiral/continuum `global fit'.  We also include the new 32Ifine ensemble, to give us a third lattice spacing with the same action, to improve the continuum extrapolation.  We note that these are the same kinds of fits we have used in our previous work with the 24I, 32I and 32ID ensembles - here we have the addition of very accurate data at physical quark masses.  In addition, we also have added Wilson flow measurements of the scale on all of our ensembles to the global fits.  While the Wilson flow scale in physical units is an output of our simulations, the relative values on the various ensembles provide additional accurate data that helps to constrain the lattice spacing determinations. In Section~\ref{sec:CombinedChiralFits} we discuss our fitting strategy in more detail and the fit results are presented in Section~\ref{sec:FitResults}.

Given the length of this paper and the many details discussed, we
present a summary of our physical results in Table~\ref{tab:summary_results} as the last part of this introduction.
These are continuum results for isospin symmetric 2+1 flavor QCD
without electromagnetic effects.  Our input values are $m_\pi$,
$m_K$, $m_\Omega$, and the results in Table \ref{tab:summary_results}
are outputs from our simulations. For results quoted in the $\MSbar$
scheme, the first error is statistical and the second is the error
from renormalization.  For other quantities, the error is the
statistical error.  The other usual sources of error (finite volume,
chiral extrapolation, continuum limit) have all been removed through
our measurements and any error estimates we can generate for these
possible systematic errors are dramatically smaller than the (already
small) statistical error quoted.  This is discussed at great length
in Section~\ref{sec:FitResults}.  The Conclusions section (Section~\ref{sec:Conclusions})
summarizes our results and gives comparisons of them
with experiment and/or the results of other lattice simulations.

\begin{table}[!ht] 
\centering
\begin{tabular}{cc}
Quantity & Value \\
\hline 
$f_\pi$ & $130.19 \pm 0.89$ MeV  \\
$f_K$ & $155.51 \pm 0.83$ MeV  \\
$f_K/f_\pi$ & $1.1945 \pm 0.0045 $  \\
$m_u = m_d(\msbar,3 \; {\rm GeV})$ & $2.997 \pm 0.036 \pm 0.033$ MeV \\
$m_s(\msbar,3 \; {\rm GeV})$ & $81.64 \pm 0.77 \pm 0.88$ MeV \\
$m_s/m_u = m_s/m_d$ & $27.34 \pm 0.21$ \\
$t_0^{1/2}$ & $0.7292 \pm 0.0041$ GeV$^{-1}$ \\
$w_0$ & $0.8742 \pm 0.0046$ GeV$^{-1}$ \\
$B_K({\rm SMOM}(\slashed q,\slashed q),3 \; {\rm GeV})$ & $0.5341 \pm 0.0018$  \\
$B_K(\msbar,3 \; {\rm GeV})$ & $0.5293 \pm 0.0017 \pm 0.0106$  \\
$\hat{B}_K$ & $0.7499 \pm 0.0024 \pm 0.0150$ \\
$L_4^{(2)}(\Lambda_{\chi PT} = 1 \; {\rm GeV}) $ & $ -0.000171 \pm 0.000064 $\\
$L_5^{(2)}(\Lambda_{\chi PT} = 1 \; {\rm GeV}) $ & $ 0.000513 \pm 0.000078 $\\
$L_6^{(2)}(\Lambda_{\chi PT} = 1 \; {\rm GeV}) $ & $ -0.000146 \pm 0.000036 $\\
$L_8^{(2)}(\Lambda_{\chi PT} = 1 \; {\rm GeV}) $ & $ 0.000631 \pm 0.000041 $\\
\end{tabular}
\label{tab:summary_results}
\caption{Summary of results from the simulations reported here.
The first error is the statistical error, which for most quantities
is much larger than any systematic error we can measure or estimate.
The exception is for the quantities in $\MSbar$ and $\hat{B}_K$.
For these quantities, the second error is the systematic error on the
renormalization, which is dominated by the perturbative matching between
the continuum RI-MOM scheme and the continuum $\MSbar$ scheme.}
\end{table}

The layout of this document is as follows: In Section~\ref{sec:SimulationDetails} we present the details of our new ensembles, including a more general discussion of the \mobius domain wall action. The associated simulated values of the pseudoscalar masses and decay constants, the $\Omega$-baryon mass, the vector and axial current renormalization factors, the neutral kaon mixing parameter, $B_K$, and the Wilson flow scales, $t_0^{1/2}$ and $w_0$, are given in Section~\ref{sec:SimulationResults}. In Section~\ref{sec:CombinedChiralFits} we provide an overview of our global fitting procedure for those quantities, the results of which are given in Section~\ref{sec:FitResults}. Finally, we present our conclusions in Section~\ref{sec:Conclusions}.

\section{Simulation details and ensemble properties}
\label{sec:SimulationDetails}
Substantial difficulties must be overcome in order to work with physical values of the light quark mass.  Common to all fermion formulations are the challenges of increasing the physical spacetime volume to avoid the large finite-volume errors that would result from decreasing the pion mass at fixed volume.  Similarly, the range of eigenvalues of the Dirac operator increases substantially, requiring many more iterations for the computation of its inverse and motivating the use of deflation and all-mode-averaging to reduce this computational cost.  For domain wall fermions it is also necessary to decrease the size of the residual chiral symmetry breaking to reduce the size of the residual mass to a level below that of the physical light quark masses.  While this could have been accomplished using the Shamir domain wall formulation~\cite{Shamir:1993zy,Furman:1994ky} used in previous RBC and UKQCD work, this would have required a doubling or tripling of the length of the fifth dimension, $L_s$, at substantial computational cost.  

Instead, our new, physical ensembles have been generated with a modified domain wall fermion action that suppresses residual chiral symmetry breaking, resulting in values for the residual mass that lie below that of the physical light quark, but without the substantial increase in $L_s$ that would have been required in the original domain wall framework.

We use the \mobius framework of Brower, Neff and Orginos~\cite{Brower:2004xi,Brower:2005qw,Brower:2012vk}.  Although the action has been changed, we remain within the subspace of the \mobius parametrization that preserves the $L_s\to\infty$ limit of domain wall fermions.  The changes to the Symanzik effective action resulting from this change in fermion formulation can be made arbitrarily small and are of the same size as the observed level of residual chiral symmetry breaking.  As discussed in Section~\ref{sec:mobius}, we are therefore able to combine our new ensembles in a continuum extrapolation with previous RBC and UKQCD ensembles.

\subsection{\mobius fermion formalism}
\label{sec:mobius}

In this section and in Appendix~\ref{appendix-mobiusconservedcurrents}  we describe the implementation of \mobius domain wall fermions, and provide a self-contained derivation of many of the properties of this formulation on which our calculation depends. 

Of central importance is the degree to which the present results from the \mobius version of the domain wall formalism can be combined with those from our earlier Shamir calculations when taking a continuum limit.  As reviewed below and in Appendix~\ref{appendix-mobiusconservedcurrents}, the Shamir and \mobius fermion formalisms result in very similar approximate sign functions, $\epsilon(H_M)$, having the form given in Eq.~\eqref{eq:sign_approx} below.  In fact, the only differences between the two functions $\epsilon(H_M)$ corresponding to Shamir and \mobius fermions is the choice of $L_s$ and an overall scale factor entering the definition of the kernel operator, $H_M$.  Thus, in the limit $L_s\to\infty$ both theories agree with the same, chirally symmetric, overlap theory.  The differences of both Shamir and \mobius fermions from that theory, and therefore from each other, vanish in this chiral limit.  Note, this equivalence in the chiral limit holds for both the fermion determinant that is used to generate the gauge ensembles (shown below) and for the 4-D propagators (shown in Appendix~\ref{appendix-mobiusconservedcurrents}) which determine all of the Green's functions which appear in our measurements and define our lattice approximation to QCD.

Thus, we expect that all details of the four dimensional approximation to QCD defined by the Shamir and \mobius actions must agree in the limit $L_s\to\infty$ and, in our case of finite $L_s$, will show differences on the order of the residual chiral symmetry breaking, the most accessible effect of finite $L_s$.  Since this constraint holds at finite lattice spacing, we conclude that the coefficients of the $O(a^2)$ corrections which appear in the four-dimensional, effective Symanzik Lagrangians for the Shamir and \mobius actions should agree at this same, sub-percent level, allowing a consistent continuum limit to be obtained from a combination of Shamir and \mobius results. 

To understand this argument in greater detail, it is useful to connect the Shamir and \mobius theories in two steps.  We might first discuss the relation between two Shamir theories: one with a smaller $L_s$ and larger residual chiral symmetry breaking, and a second with a larger value of $L_s$ and a value for $\mres$ below the physical light quark mass.  In the second step we can compare this large $L_s$ Shamir theory with a corresponding \mobius theory that has the same approximate degree of residual chiral symmetry breaking.  For example, when comparing our $\beta = 2.13$ Shamir and \mobius ensembles, we might begin with our $24^3\times 64$, $L_s=16$, 24I ensemble with $\mres a = 0.003154(15)$ which is larger than the physical light quark mass.  Next we consider a fictitious, $L_s=48$ ensemble which should have a value of $\mres$ very close to the 0.0006102(40) value of our 48I \mobius ensemble.  In this comparison we would work with the same Shamir formalism and simply approach the chiral limit more closely by increasing $L_s$ from 16 to 48.  Clearly the $5\times$ reduction in the light quark mass will produce a significant change in the theory, which to a large degree should be equivalent to reducing the input quark mass in a theory with a large fixed value of $L_s$.   Of course, there will be smaller changes as well.  In addition to reducing the size of $\mres$, we will also reduce the size of the dimension-five, $O(a)$ Sheikholeslami-Wohlert term (whose effects are expected to be at the $\mres a^2 \le 0.1\%$ level even for the smaller value of $L_s$).  There will be further small changes coming from approaching the $L_s\to\infty$ limit, for example the $3\%$ change in the lattice spacing discussed in Appendix~\ref{appendix-mresadep}.

The second comparison can be made between the fictitious $L_s=48$ Shamir ensemble and our actual 48I \mobius ensemble with $L_s=24$ and $b+c=2$.  Since the product of $L_s(b+c)$ is the same for these two examples, the approximate sign function will agree for eigenvalues of the kernel $H_M$ which are close to zero.  In fact, a study of the eigenvalues $\lambda$ of $H_M$ for the Shamir normalization shows that they lie in the range $0 \le \lambda \le 1.367(14)$ for $\beta=2.13$.  One can then examine the ratio of the two approximate sign functions, which determine the corresponding 4-D Dirac operators, over this entire eigenvalue range and show that the approximate Shamir and \mobius sign functions $\epsilon(H_M)$ agree at the 0.1\% level. Thus, in this second step we are comparing two extremely similar theories whose description of QCD is expected to differ in all aspects at the 0.1\% level.   We now turn to a detailed discussion of the Shamir and \mobius operators and their relation to the overlap theory.

Our conventions are as follows. The usual Wilson matrix is 
\begin{equation}D_W(M) = M+4 - \frac{1}{2} D_{\rm hop},\end{equation} 
where
\begin{equation}
D_{\rm hop} = (1-\gamma_\mu) U_\mu(x) \delta_{x+\mu,y} +
              (1+\gamma_\mu) U_\mu^\dagger(y) \delta_{x-\mu,y} \,.
\end{equation}
For our physical point ensembles we use a generalized form of the domain wall action~\cite{Brower:2004xi,Brower:2005qw,Brower:2012vk},
\begin{equation}
S^5 = \bar{\psi} D^5_{GDW} \psi\,,
\label{eq:Mobius_action}
\end{equation}
where
\begin{eqnarray}
D^5_{GDW}
&=&
\left(
\begin{array}{cccccc}
  \tilde{D} & - P_- & 0& \ldots & 0 &  m P_+ \\
-  P_+  & \ddots &  \ddots & 0      & \ldots &0 \\
0     & \ddots &  \ddots & \ddots & 0      &\vdots \\
\vdots& 0      &  \ddots & \ddots & \ddots & 0\\
0     & \ldots &    0    &  \ddots& \ddots & - P_- \\
m P_- & 0      & \ldots  &  0     &   - P_+   
& \tilde{D}
\end{array}
\right)\,,
\end{eqnarray}
and we define
\begin{eqnarray}
D_+ = (b D_W + 1) \quad;\quad
D_-= (1-c D_W )\quad;\quad
\tilde{D} = (D_-)^{-1} D_+\,.
\label{eq:Dtilde}
\end{eqnarray}

This generalized set of actions reduces to the 
standard Shamir action in the limit $b=1$, $c=0$, and it can also be
taken to give the polar approximation to the 
Neuberger overlap action as another limiting case~\cite{Borici:1999zw,Borici:1999da}. 
In all of our simulations we take the coefficients $b$ and $c$ as constant across the fifth dimension.
This setup is well known to yield a $\tanh$ approximation to the overlap sign function. 
Coefficients that vary across the fifth dimension can also be used to introduce other rational
approximations to the sign function, such as the Zolotarev approximation~\cite{Zolotarev1877,Edwards:1998yw,vandenEshof:2002ms}.

As in the Shamir domain wall fermion formulation we identify ``physical'', four-dimensional quark fields $q$ and $\bar{q}$ whose Green's functions define our domain wall fermion approximation to continuum QCD.   We choose to construct these as simple chiral projections of the five-dimensional fields $\psi$ and $\bar{\psi}$ which appear in the action given in Eq.~\eqref{eq:Mobius_action}:
\begin{equation}
\begin{array}{ccc}
q_R = P_+ \psi_{L_s} &\quad& q_L = P_- \psi_{1}\,,\\
\bar{q}_R = \bar{\psi}_{L_s} P_- & \quad & \bar{q}_L =  \bar{\psi}_{1} P_+\,.
\label{eq:physical_fields}
\end{array}
\end{equation}
While there is considerable freedom in this choice of the physical, four-dimensional quark fields, as is shown in Appendix~\ref{appendix-mobiusconservedcurrents}, this choice results in four-dimensional propagators which agree with those of the corresponding overlap theory up to a contact term in the $L_s\to\infty$ limit.  This choice is also dictated by the requirement that we be able to combine results from the present, physical point calculation with earlier results using Shamir fermions in taking a continuum limit.  With this choice both the \mobius and Shamir theories will yield 4-dimensional fermion propagators which differ only at the level of the residual chiral symmetry breaking.  The choice of physical quark fields given in Eq.~\eqref{eq:physical_fields} has the added benefits that the corresponding four-dimensional propagators satisfy a simple $\gamma^5$ hermiticity relation and a hermitian, partially-conserved axial current can be easily defined.

In practice, one solves for physical quark propagators using the linear system
\begin{equation} 
D_- D_{GDW}^5 \psi = D_- \eta\,.
\end{equation} 

To find the 4d effective action which corresponds to our choice of physical fields we must
first perform some changes to the field basis as follows. We write
\begin{eqnarray}
S^5 &=& \bar{\psi} D^5_{GDW} \psi 
= \bar\chi D^5_\chi \chi\,,
\end{eqnarray}
where, for now leaving a matrix $Q_-$ undefined,
$\chi = {\cal P}^{-1}\psi$, $
\bar\chi = \bar{\psi}\gamma_5 Q_-$, $D^5_\chi = Q_-^{-1} \gamma_5 D^5_{GDW} {\cal P}$, and
\begin{eqnarray}
{\cal P} &=& \left( 
\begin{array}{ccccc}
P_- & P_+     & 0      & \ldots & 0 \\
0   & \ddots & \ddots & 0 & \vdots \\
\vdots& 0 & \ddots & \ddots & 0 \\
0    & \ldots & 0 & \ddots & P_+ \\
P_+  & 0 & \ldots & 0 & P_-
\end{array}
\right)\,.
\end{eqnarray}
Then with 
\begin{equation}
\tilde{H} = \gamma_5 (D_-)^{-1} D_+ = \gamma_5 (H_-)^{-1}  H_+\,,
\end{equation}
and $H_- = \gamma_5 D_-$, $H_+=\gamma_5 D_+$ we may write
\begin{eqnarray}
D_\chi^5 &=& Q_-^{-1} \left[ 
\begin{array}{cccccc}
\tilde{H} &    P_-    & 0      & \ldots & 0 & m P_+   \\
- P_+         & \ddots  & \ddots & 0      & \ldots & 0 \\
0            & \ddots  & \ddots & \ddots & 0 & \vdots \\
\vdots       & 0       & \ddots & \ddots & \ddots & 0 \\
0            & \ldots  & 0      & \ddots & \ddots & P_-\\
-m P_-  &  0 & \ldots & 0 &- P_+ & \tilde{H}
\end{array}
\right]{\cal P }\,.
\end{eqnarray}
We may choose $Q_-$ to place the matrix $D_\chi^5$ in a particularly convenient form as follows,
\begin{equation}
\begin{array}{ccc}
Q_- &=& \tilde{H} P_- - P_+ = \gamma_5 [H_-]^{-1}[ H_+ P_- - H_-  P_+ ]\\
Q_+ &=& \tilde{H} P_+ + P_- = \gamma_5 [H_-]^{-1}[ H_+ P_+ - H_- P_- ]\,,
\end{array}
\end{equation}
and introduce the so-called transfer matrix as
\begin{equation}
\begin{array}{ccc}
T^{-1} &=& -(Q_-)^{-1} Q_+\nonumber\\
&=&
 - [ \gamma_5 \frac{(b+c) D_W}{2+(b-c) D_W}  -  1 ]^{-1}
  [ \gamma_5 \frac{(b+c) D_W}{2+(b-c) D_W}  +  1 ]\\
&=& -[ H_M - 1]^{-1} [ H_M + 1].
\end{array}
\end{equation}
Here the \mobius kernel is
\begin{equation}
H^M = \gamma_5\frac{(b+c) D_W}{2+(b-c)D_W}\,.
\end{equation}

We find $D^5_\chi$ takes the following form,
\begin{eqnarray}
D^5_\chi&=& 
\left[ 
\begin{array}{cccccc}
P_- - m P_+ & -T^{-1} & 0 & \ldots & \ldots & 0\\
0               & 1 & -T^{-1} & 0    & \ldots & \vdots\\
\vdots          & 0   & \ddots & \ddots & 0       & \vdots\\
\vdots          &\ldots& 0  & 1 & -T^{-1}     & 0  \\
0               &\ldots& \ldots&0 & 1     & -T^{-1}   \\
-T^{-1} (P_+ -m P_-) & 0    & \ldots&\ldots   & 0      & 1    
\end{array}
\right]\,,
\end{eqnarray}
for which we can perform a UDL decomposition around the top left block:
\begin{equation}
\left(
\begin{array}{cc}
D & C \\
B & A
\end{array}
\right)
= 
\left(
\begin{array}{cc}
1  &  C A^{-1} \\
0  & 1
\end{array}
\right)
\left(
\begin{array}{cc}
S_\chi & 0\\
0 & A
\end{array}
\right)
\left(
\begin{array}{cc}
1 &  0 \\
A^{-1} B  & 1
\end{array}
\right)\,.
\end{equation}
Here, the Schur complement is 
$
S_\chi = D - C A^{-1} B,
$
where
\begin{eqnarray}
A = \left(
\begin{array}{ccccc}
 1   & -T^{-1} & 0    & \ldots & \vdots\\
 0   & 1 & -T^{-1} & 0       & \vdots\\
 0   & 0  & 1 & -T^{-1}     & 0  \\
 0   & \ldots&0 & 1     & -T^{-1}   \\
 0    & \ldots&\ldots   & 0      & 1    
\end{array}
\right)& \quad\quad &
A^{-1} = \left(
\begin{array}{ccccc}
 1   & T^{-1} & T^{-2}   & \ldots & T^{-(L_s-2)}\\
 0   & 1 & T^{-1} & \ldots & T^{-(L_s-3)}\\
 0   & 0  & 1 & T^{-1}     & \vdots  \\
 0   & \ldots&0 & 1     & T^{-1}   \\
 0    & \ldots&\ldots   & 0      & 1    
\end{array}
\right),\\
D &=& P_- - m P_+,\\
C&=& (
\begin{array}{ccccc}
-T^{-1} & 0 & \ldots & \ldots & 0
\end{array}),\\
B^T&=& (
\begin{array}{ccccc}
0 &  & \ldots & 0 & -T^{-1} (P_+ - m P_-)
\end{array}),\\
C A^{-1} B
&=&   T^{-L_s} ( P_+-mP_- )\,.
\end{eqnarray}
Denoting the left and right factors as $U$ and $L(m)$ respectively, we 
write this factorization as $D_\chi^5 = U D_S(m) L(m) $.
The determinants of the $U$ and $L(m)$ are unity, and the determinant of the product is
simply 
\begin{equation}
\det D_\chi^5 = \det A \det S_\chi = \det S_\chi\,,
\end{equation}
 where
\begin{eqnarray}
S_\chi(m)&=& -(1 + T^{-L_s}) \gamma_5 
\left[ \frac{1+m}{2} + \frac{1-m}{2}\gamma_5 \frac{T^{-L_s}-1}{ T^{-L_s}+1} \right].
\end{eqnarray}
We can see that after the removal of the determinant of the Pauli Villars fields with $m=1$ in our ensembles we are left with the 
determinant of an effective overlap operator, which is the following rational function of the kernel:
\begin{equation}
\det D_{PV}^{-1} D(m)= \det D_{ov}  = \det\left(\frac{1+m}{2} + \frac{1-m}{2}\gamma_5 
\frac{ (1+ H_M)^{L_s}-(1 - H_M)^{L_s}}
     { (1+ H_M)^{L_s}+(1 - H_M)^{L_s}}\right)\,.
\end{equation}
We identify $D_{ov}$ as an approximation to the overlap operator with 
approximate sign function
\begin{equation}
\epsilon(H_M) = \frac{ (1+ H_M)^{L_s}-(1 - H_M)^{L_s}}
                     { (1+ H_M)^{L_s}+(1 - H_M)^{L_s}}\,,
\label{eq:sign_approx}
\end{equation}
with 
\begin{equation}
\lim_{L_s\to\infty} \epsilon(H_M)  = {\rm sgn} (H_M)\,.
\end{equation}
Note that since ${\rm sgn}(H_M) = {\rm sgn}(\alpha H_M)$ for all positive $\alpha$,
changing the \mobius parameters $b+c$ while keeping $b-c=1$ fixed leaves
our kernel $H_M$ proportional to the kernel for the Shamir formulation.
This therefore changes only the
approximation to the overlap sign function, but not the form of the $L_s\to\infty$
limit of the action. 

In this way, our new simulations with the \mobius action will 
differ from those with Shamir domain wall fermions only through terms proportional
to the residual chiral symmetry breaking. In particular the change of action
is not fundamentally different from simulating with a different $L_s$.

Other, equivalent views of this approximation to the sign function are useful.
Noting 
\begin{equation}
-\tanh \frac{1}{2} \log z = \frac{1-z}{1+z}\,,
\end{equation}
we see that since 
\begin{equation}
T^{-1} = \frac{1+H_M}{1-H_M} \iff H_M = \frac{1-T}{1+T}\,,
\end{equation}
we have
\begin{equation}
\frac{T^{-L_s} - 1}{T^{-L_s}+1} 
= \tanh \left( -\frac{L_s}{2} \log |T| \right)
= \tanh \left( L_s \tanh^{-1} H_M \right)\,,
\end{equation}
and for this reason our approximation to the sign function is often called the $\tanh$ approximation.

For eigenvalues of $H_M$ near zero, this $\tanh$ expression becomes a poor approximation to the sign function and it is for these small eigenvalues that the largest contributions to residual chiral symmetry breaking typically occur.  For small eigenvalues $\lambda$ of $H_M$, the $\tanh$ approximation is a steep, but not discontinuous, function at $\lambda=0$.  Examining Eq.~\eqref{eq:sign_approx} one can easily see that
\begin{equation}
\epsilon(\alpha \lambda ) \sim L_s \alpha \lambda\,,
\end{equation}
which approaches the discontinuity of the sign function only as $L_s\to\infty$.  The  quality of the sign function approximation for small eigenvalues can be improved by either increasing $L_s$ (at a linear cost) or by increasing the \mobius scale factor $\alpha=b+c$ while keeping $b-c = 1$ (close to cost-free), or both.  One concludes that the scale factor $b+c$ should be increased to the maximum extent consistent with keeping the upper edge of the spectrum of $H_M$ within the bounded region in which $\epsilon(H_M)$ is a good approximation to the sign function. In the limit of large $L_s$ a simulation with $(b+c)>1$ will have the same degree of chiral symmetry breaking as a simulation in which that scale factor has been set to one but with $L_s$ increased to $L_s(b+c)$.

In Appendix~\ref{appendix-mobiusconservedcurrents} we continue the above review of the relation between the DWF and overlap operators, demonstrating the equality of the Shamir and \mobius four-dimensional fermion propagators in the limit $L_s\to\infty$.   We also introduce a practical construction of the conserved vector and axial currents for \mobius fermions, appropriate for our choice of physical fermion fields.

\subsection{Simulation parameters and ensemble generation}

\begin{table}[t]
\begin{tabular}{c|ccc}
\hline\hline
	            & 48I                     & 64I                        & 32Ifine \\
\hline 
Size 	    & $48^3\times 96\times 24$        & $64^3\times 128\times 12$  & $32^3\times 64\times 12$     \\
$\beta$  	    & 2.13                    & 2.25                       & 2.37     \\
$am_l$    	    & 0.00078		      & 0.000678                   & 0.0047     \\
$am_h$    	    & 0.0362                  & 0.02661                    & 0.0186     \\
$\alpha$           & 2.0                     & 2.0                        & 1.0 \\
\hline
$a^{-1} ({\rm GeV})$      & \ainvtwo                & \ainvthree                 & \ainvfour \\
$L$ (fm)            & \Lfmtwo                 & \Lfmthree                  & \Lfmfour \\
$m_\pi L$           & \mpiLtwo                & \mpiLthree                 & \mpiLfour \\
$\langle P \rangle$ & 0.5871119(25)           & 0.6153342(21)              & 0.6388238(37)\\
$\langle\bar\psi\psi\rangle$ & 0.0006385(12)  & 0.0002928(9)               & 0.0006707(15)\\
$\langle\bar\psi\gamma^5\psi\rangle$ & -0.0000043(31) & -0.0000000(34)     & -0.0000013(26)\\
\end{tabular}
\caption{Input parameters and relevant quantities for the three new Iwasaki ensembles. Here $L$ is the spatial lattice extent in lattice units, and $\alpha = b+c$ is \mobius scaling factor (recall the 32Ifine is a Shamir DWF ensemble, and therefore has $\alpha=1.0$). The last three entries are the average plaquette, chiral condensate, and pseudoscalar condensate respectively. The lattice spacings are determined in Section~\ref{sec:FitResults} of this document.\label{tab-latinputparams} }
\end{table}

\begin{table}[t]
\begin{tabular}{c|c|c|c}
\hline\hline
	            & 48I                     & 64I                        & 32Ifine \\
\hline
Steps per traj.    & 15                      & 9                          & 6 \\
$\Delta\tau$       & 0.067                   & 0.111                      & 0.167 \\
Metropolis acc.    & 84\%                    & 87\%                       & 82\%\\
CG iters per traj. & $\sim 5.9 \times 10^5$ & $\sim 6.1 \times 10^5$      & $\sim 8.4\times 10^4$\\
\end{tabular}
\caption{The number of steps per HMC trajectory, the MD time-step $\Delta\tau$, the Metropolis acceptance and the total number of CG iterations for the three new ensembles.\label{tab-lathmcparams} }
\end{table}

We generated three domain wall ensembles with the Iwasaki gauge action. The 48I and 64I ensembles were generated with \mobius domain wall fermions and with (near-)physical pion masses, and the 32Ifine ensemble was generated with Shamir DWF and with a heavier mass but finer lattice spacing. The results from previous fits to our older ensembles were used to choose the input light and strange quark masses to the
simulations. The input parameters are listed in Table~\ref{tab-latinputparams}. As discussed above, the \mobius parameters for the 48I and 64I ensembles are chosen with $b-c=1$ such that the Shamir and \mobius kernels are identical. The values of $\alpha = b+c$, which to a first approximation gives the ratio of fifth-dimensional extents between the \mobius and the equivalent Shamir actions, are listed in the table.

We use an exact hybrid Monte Carlo algorithm for our ensemble generation, with five intermediate Hasenbusch masses, (0.005, 0.017, 0.07, 0.18, 0.45), for the two flavor part of the algorithm of both the 48I and 64I ensembles, and three intermediate masses, (0.005, 0.2, 0.6), for the 32Ifine. A rational approximation was used for the strange quark determinant. The integrator layout and parameters are given in Tables~\ref{tab-lathmcparams} and~\ref{tab-integrators}.

\begin{table}
\begin{tabular}{c|cccc}
\hline\hline
Level (i) & $S_i$                 & Integrator & $n_i$ & Step size (48I,64I,32Ifine)  \\
\hline
1         & $\sum S_Q + \sum S_R$ & FGI QPQPQ  & 1     & 1/15, 1/9, 1/6             \\
2         & $S_G$                 & FGI QPQPQ  & 4     &  -                     \\
\end{tabular}
\caption{The integrator layout for our three ensembles. Here $\sum S_Q$ and $\sum S_R$ are the sum of the quotient and rational quotient actions used for the light and strange quarks respectively. The sums are over the intermediate mass listed in the text. $S_G$ is the gauge action, FGI QPQPQ is a particular form of the force gradient integrator~\cite{Kennedy:2009fe}, and $n_i$ are the number of steps comprising a single update of the corresponding action. The coarsest time-steps are at level 1, and the step sizes are chosen such that the total trajectory length is 1 MD time unit. More detail regarding the notation and integrators can be found in Appendix A of Ref.~\cite{Arthur:2012opa}. \label{tab-integrators}}
\end{table}

Each trajectory of the 48I ensemble required 3.5 hours on 2 racks of Blue Gene/Q (BG/Q) ($2\times 1024$ nodes), and those of the 64I required 0.67 hours on 8 racks of BG/Q. We generated 2200 and 2850 trajectories for the 48I and 64I ensembles respectively. The first 1100 trajectories of the 64I ensemble were generated with $L_s = 10$ and produced a pion mass of about 170 MeV, due to the residual mass being larger than anticipated. Changing to $L_s=12$ reduced the residual mass, allowing us to simulate at essentially the physical pion mass. The 32Ifine ensemble required 5 minutes on 1 rack of BG/Q, and we generated 6940 trajectories for this ensemble.


\subsection{Ensemble properties}
%

\begin{figure}[tp]
\centering
\subfigure{\includegraphics[width=0.33\textwidth]{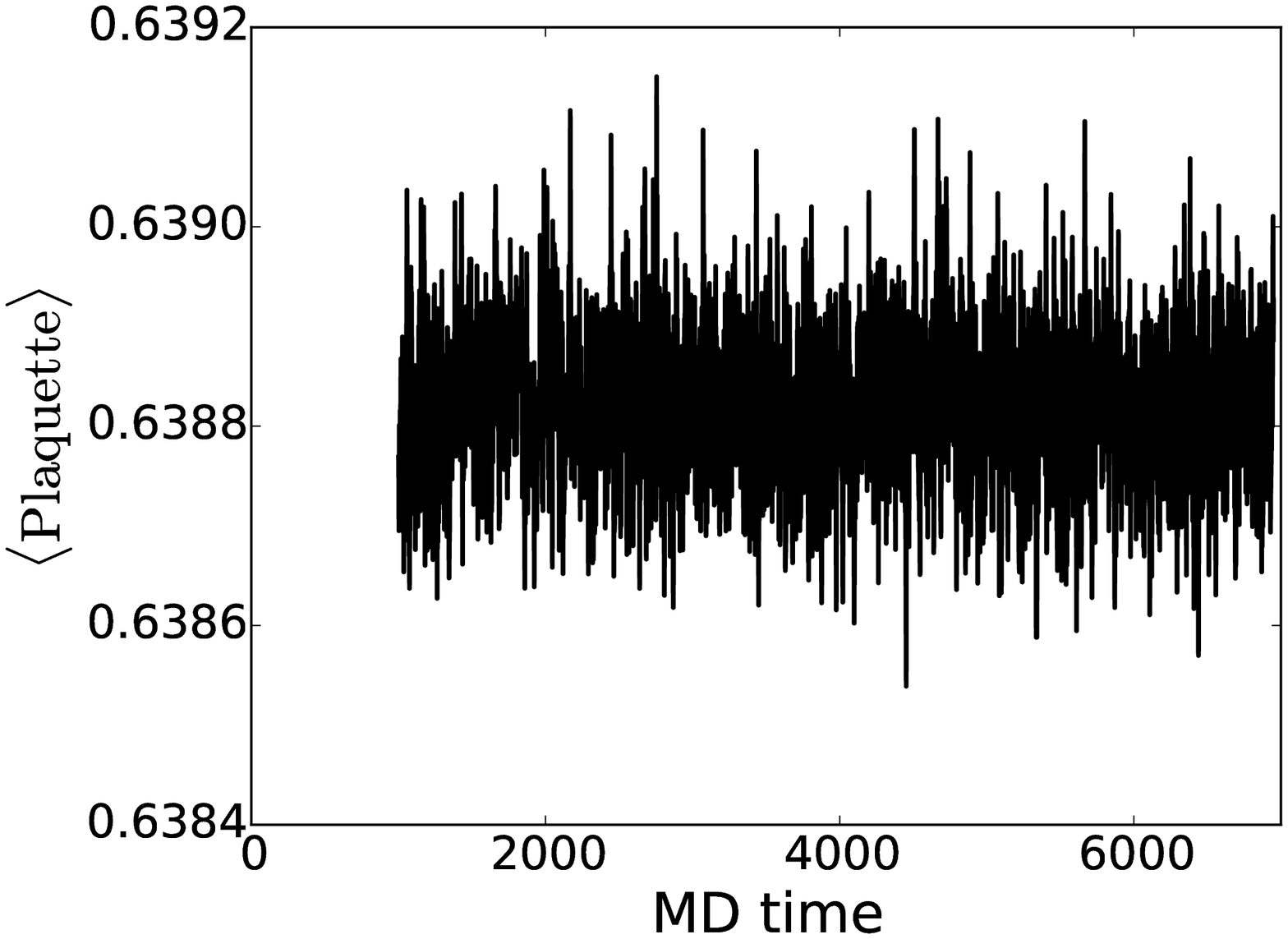} \includegraphics[width=0.33\textwidth]{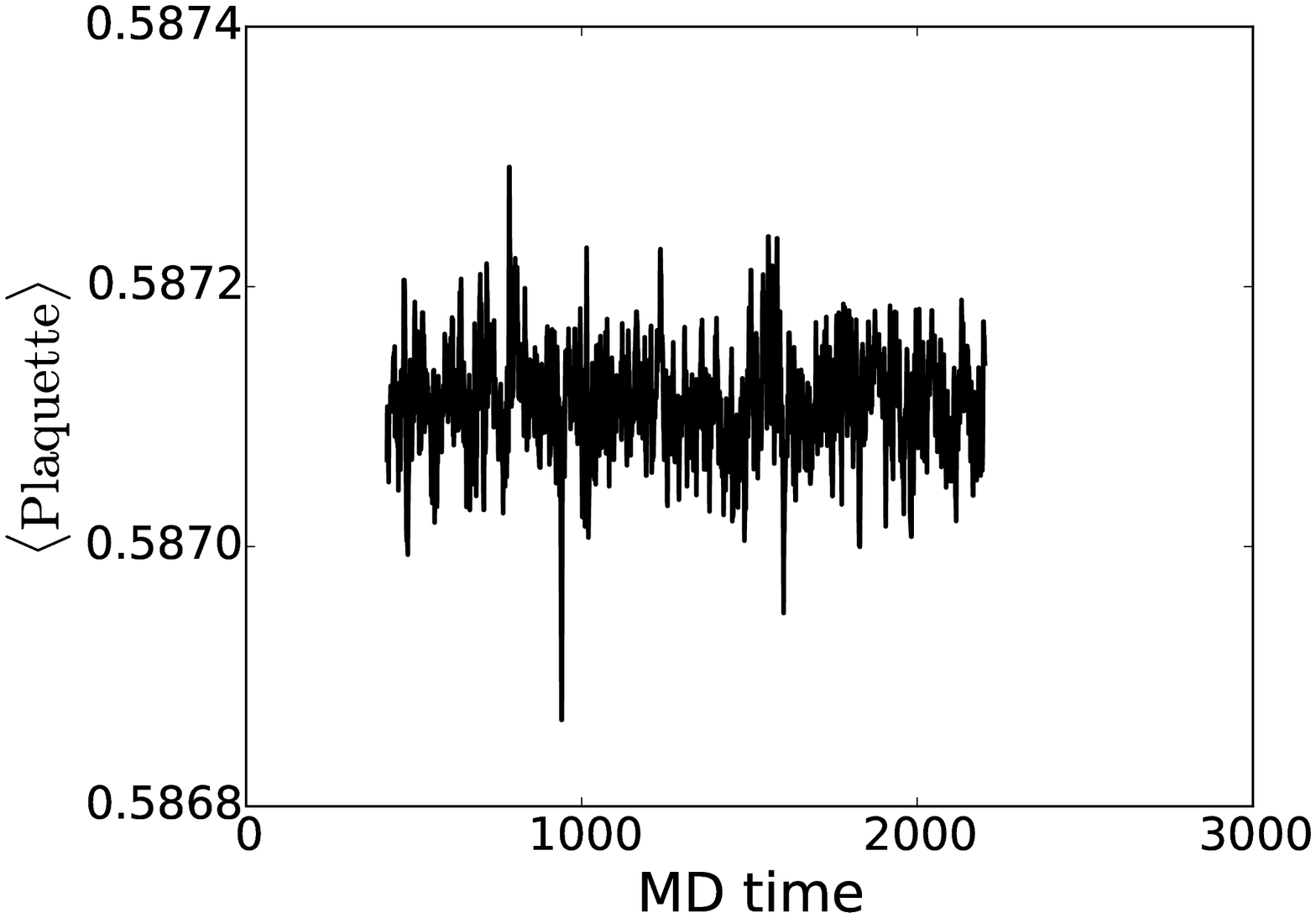} \includegraphics[width=0.33\textwidth]{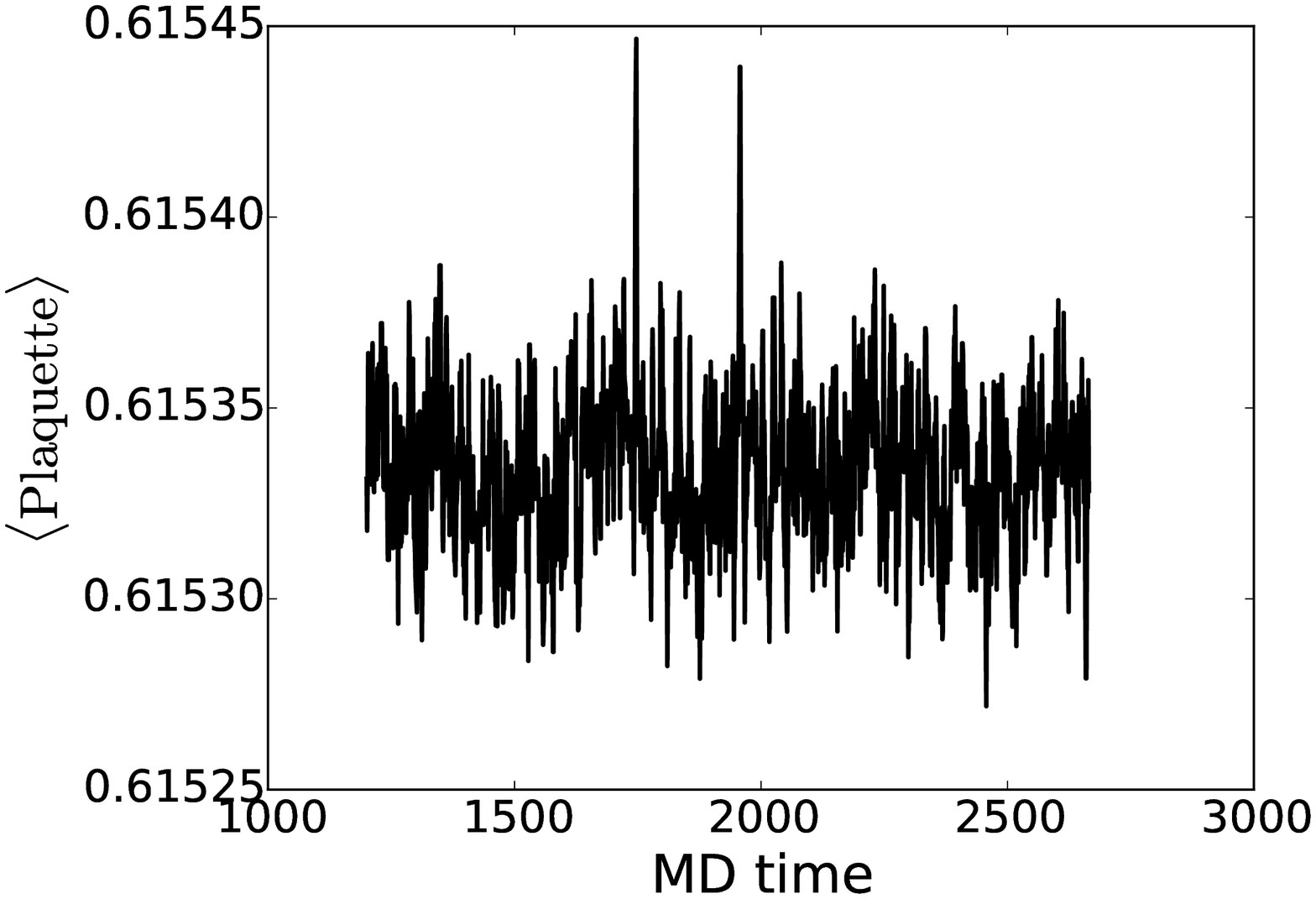}}
\subfigure{\includegraphics[width=0.33\textwidth]{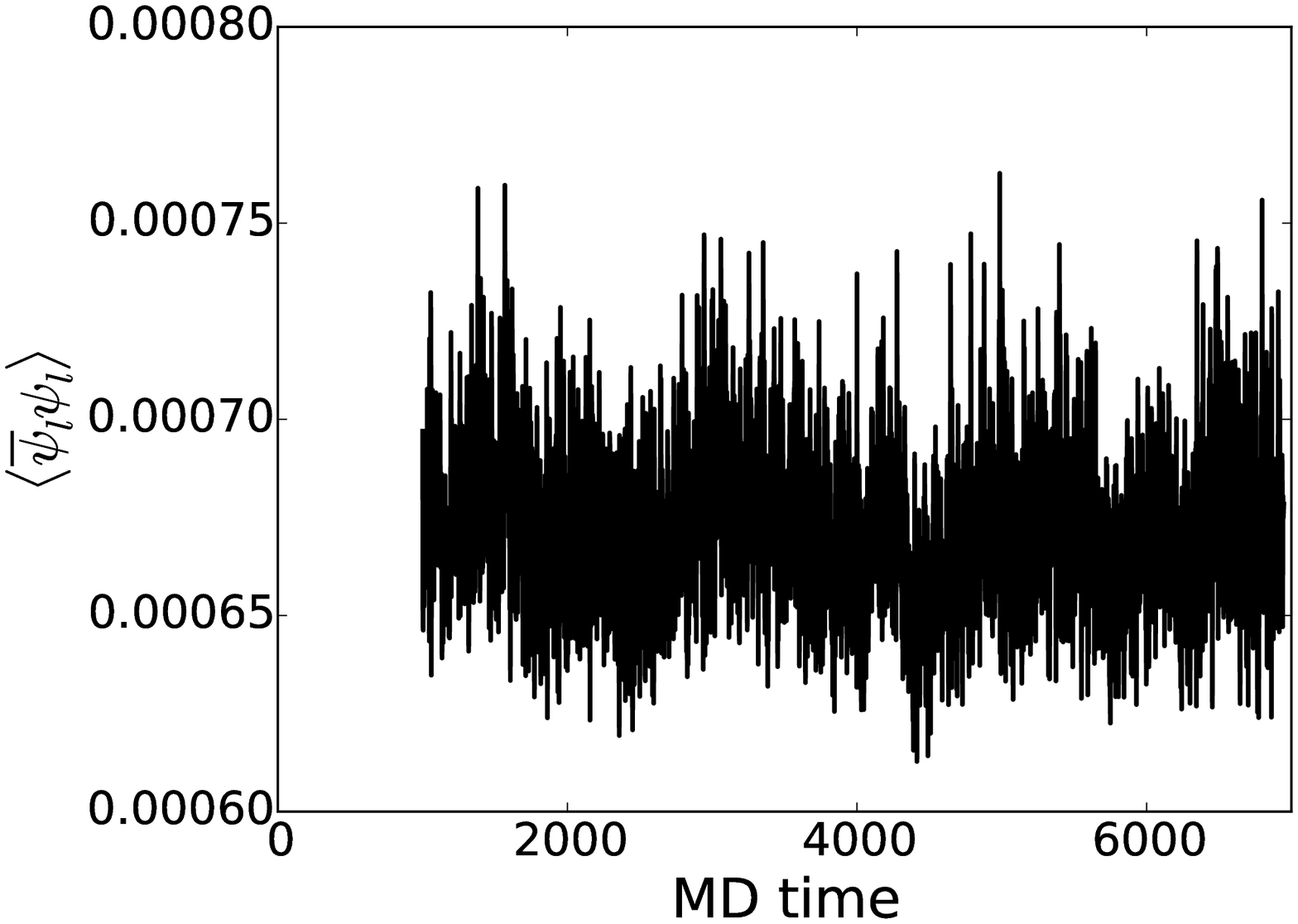} \includegraphics[width=0.33\textwidth]{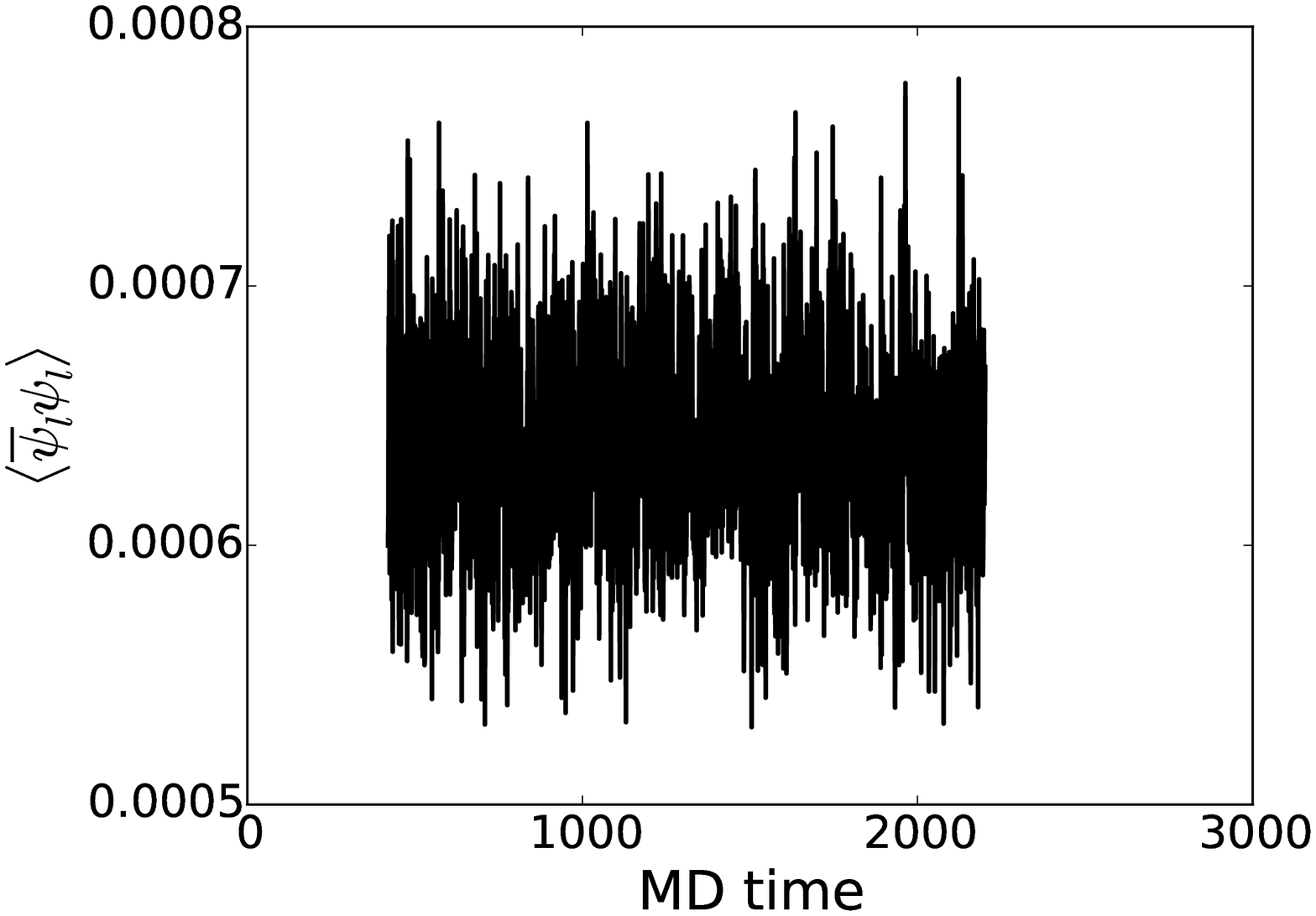} \includegraphics[width=0.33\textwidth]{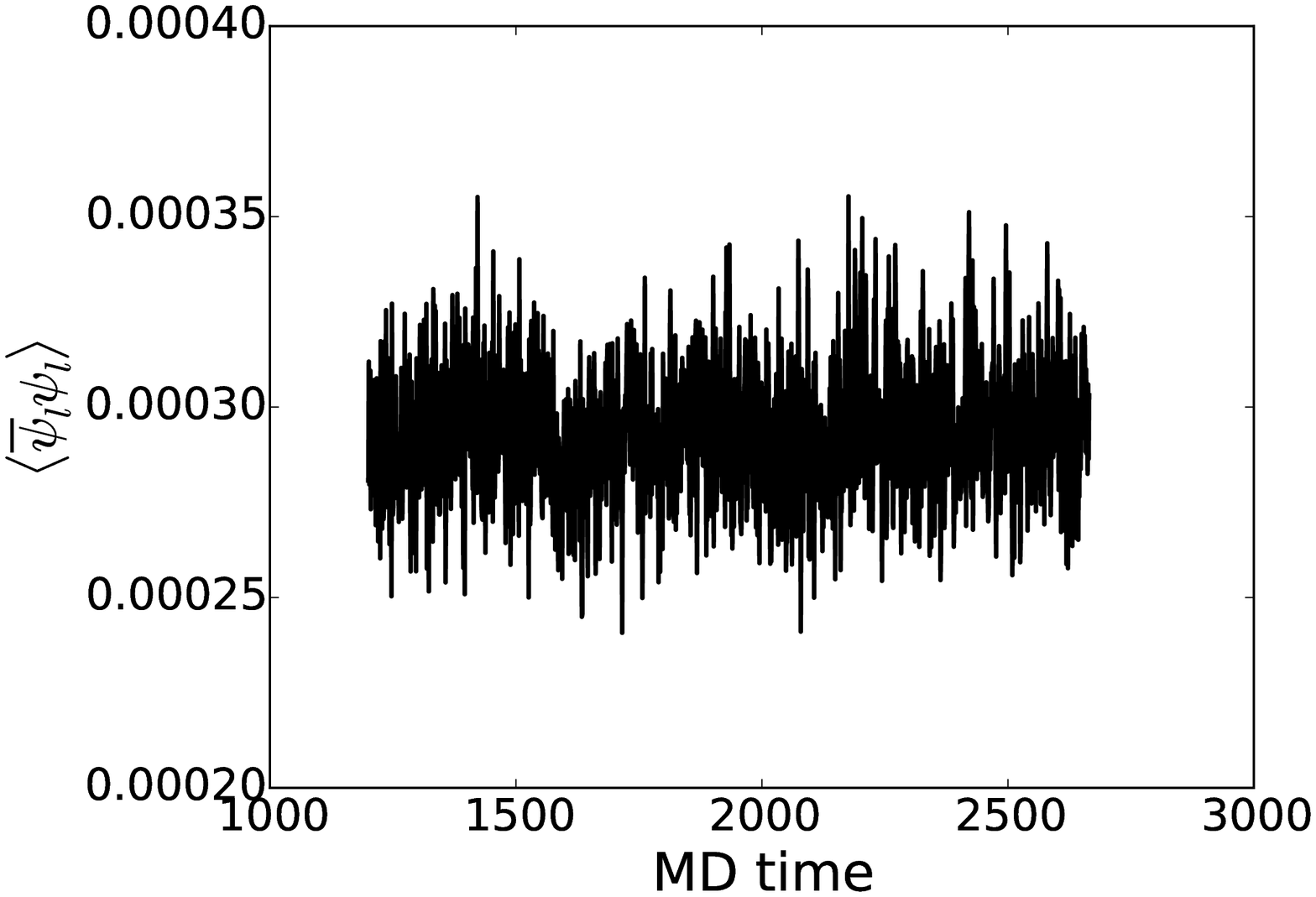}}
\subfigure{\includegraphics[width=0.33\textwidth]{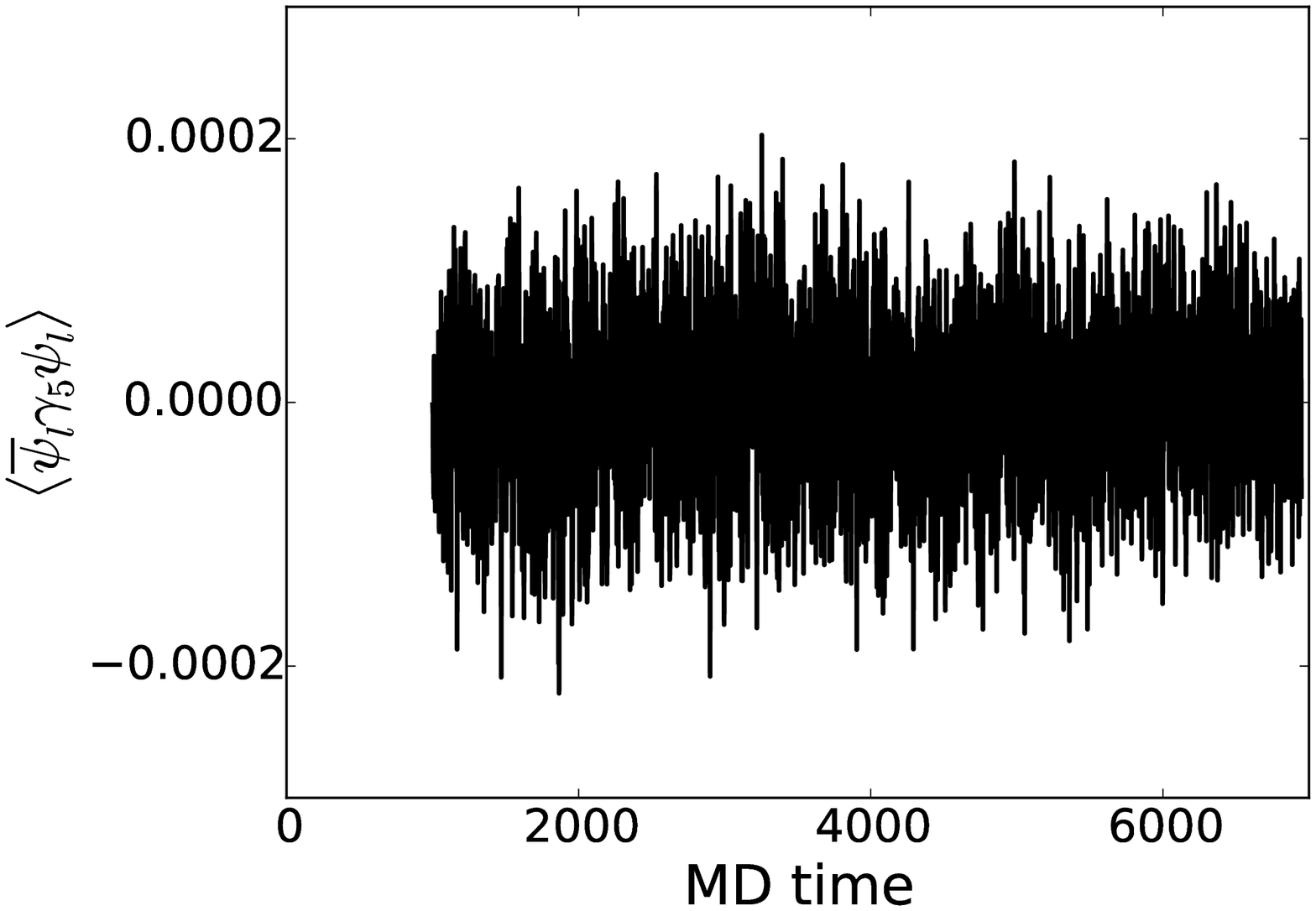} \includegraphics[width=0.33\textwidth]{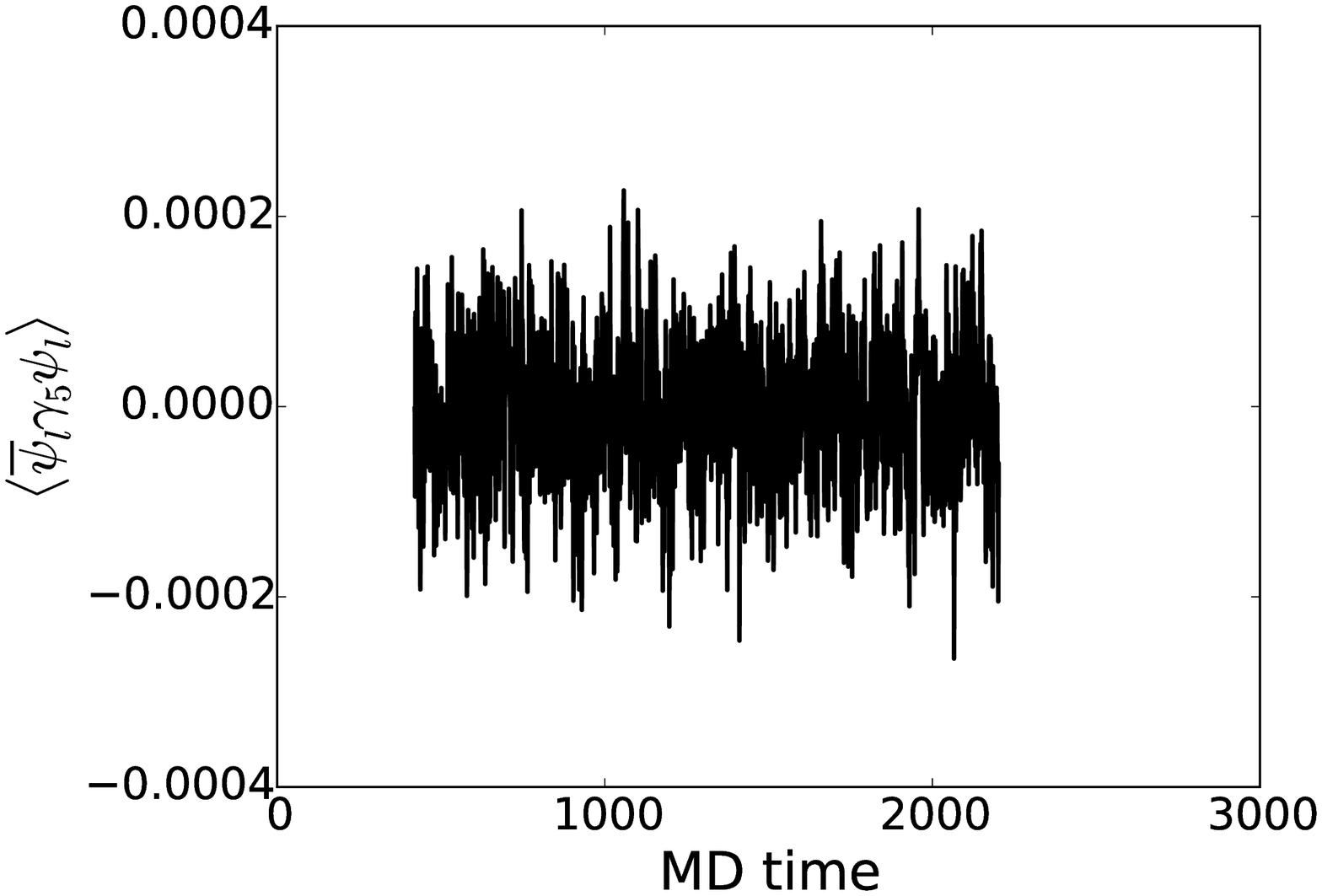} \includegraphics[width=0.33\textwidth]{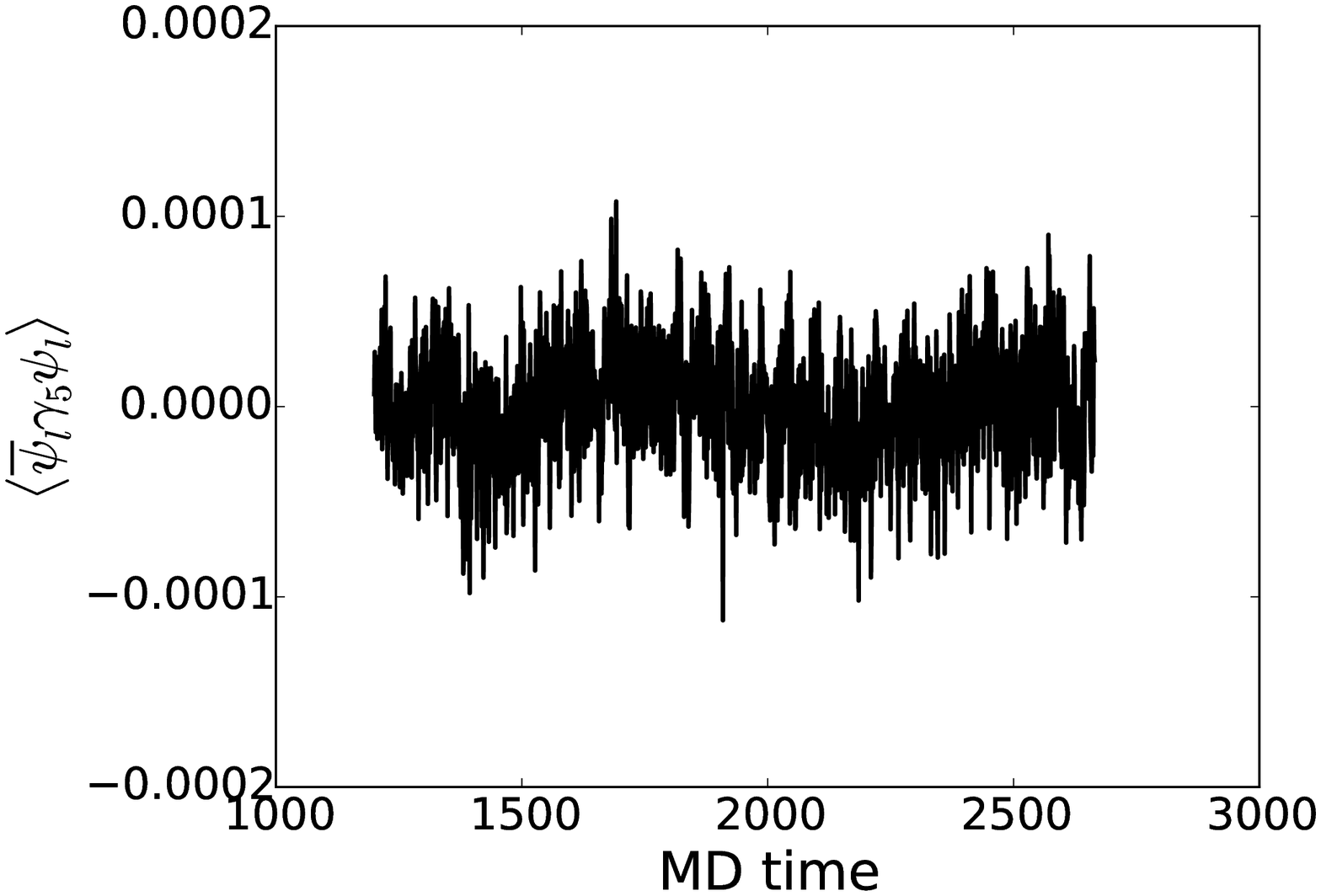}}
\subfigure{\includegraphics[width=0.33\textwidth]{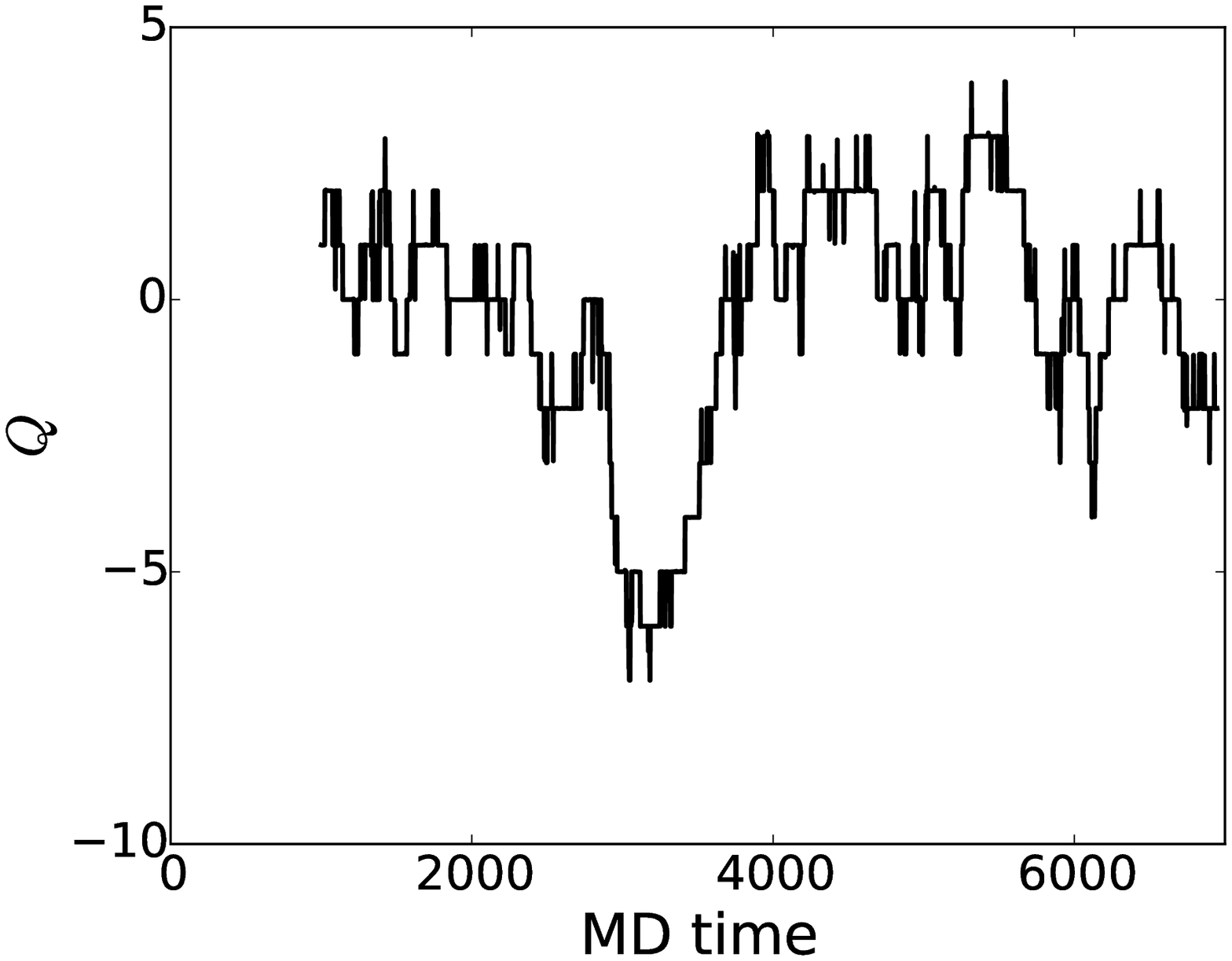} \includegraphics[width=0.33\textwidth]{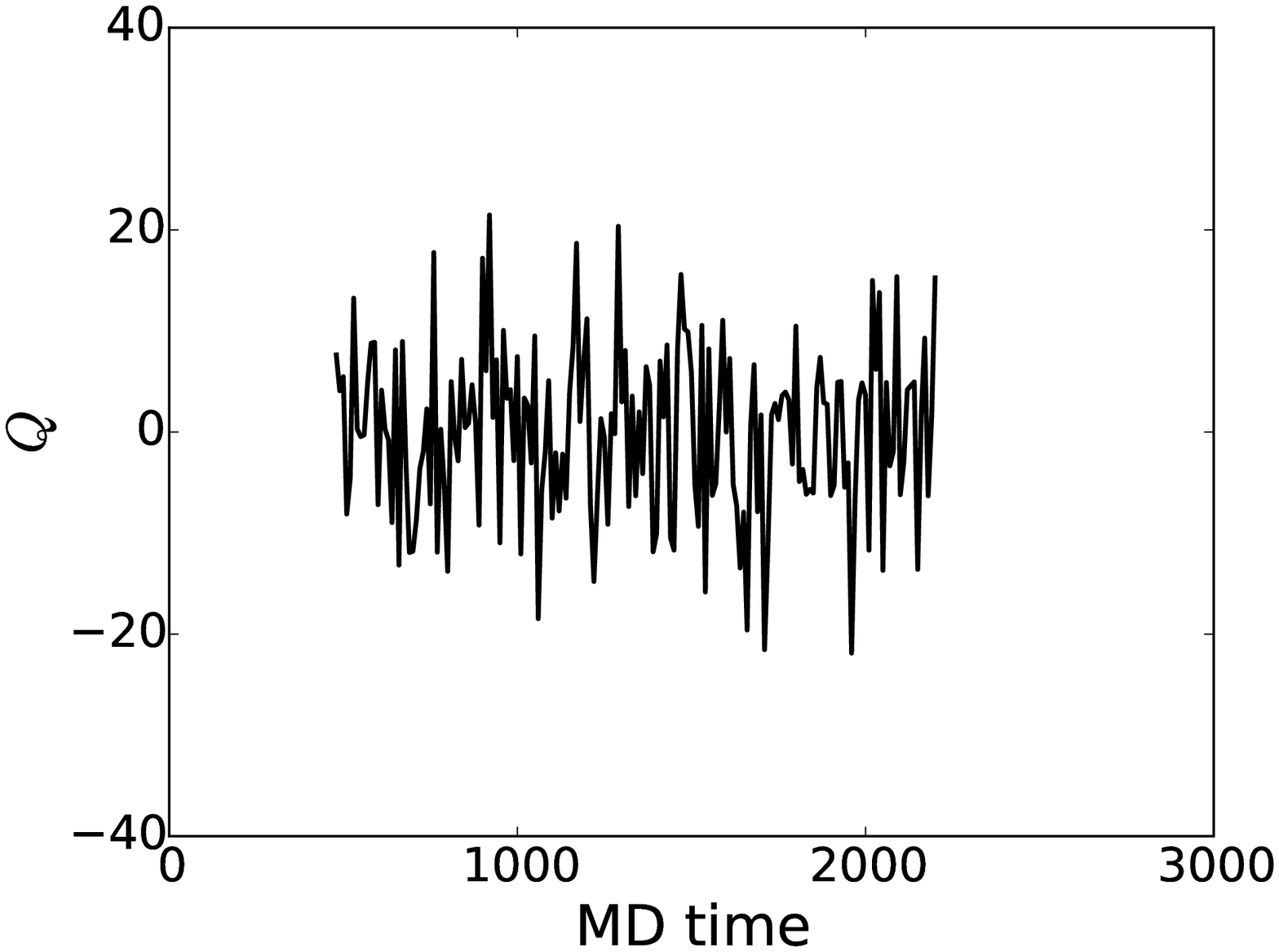} \includegraphics[width=0.33\textwidth]{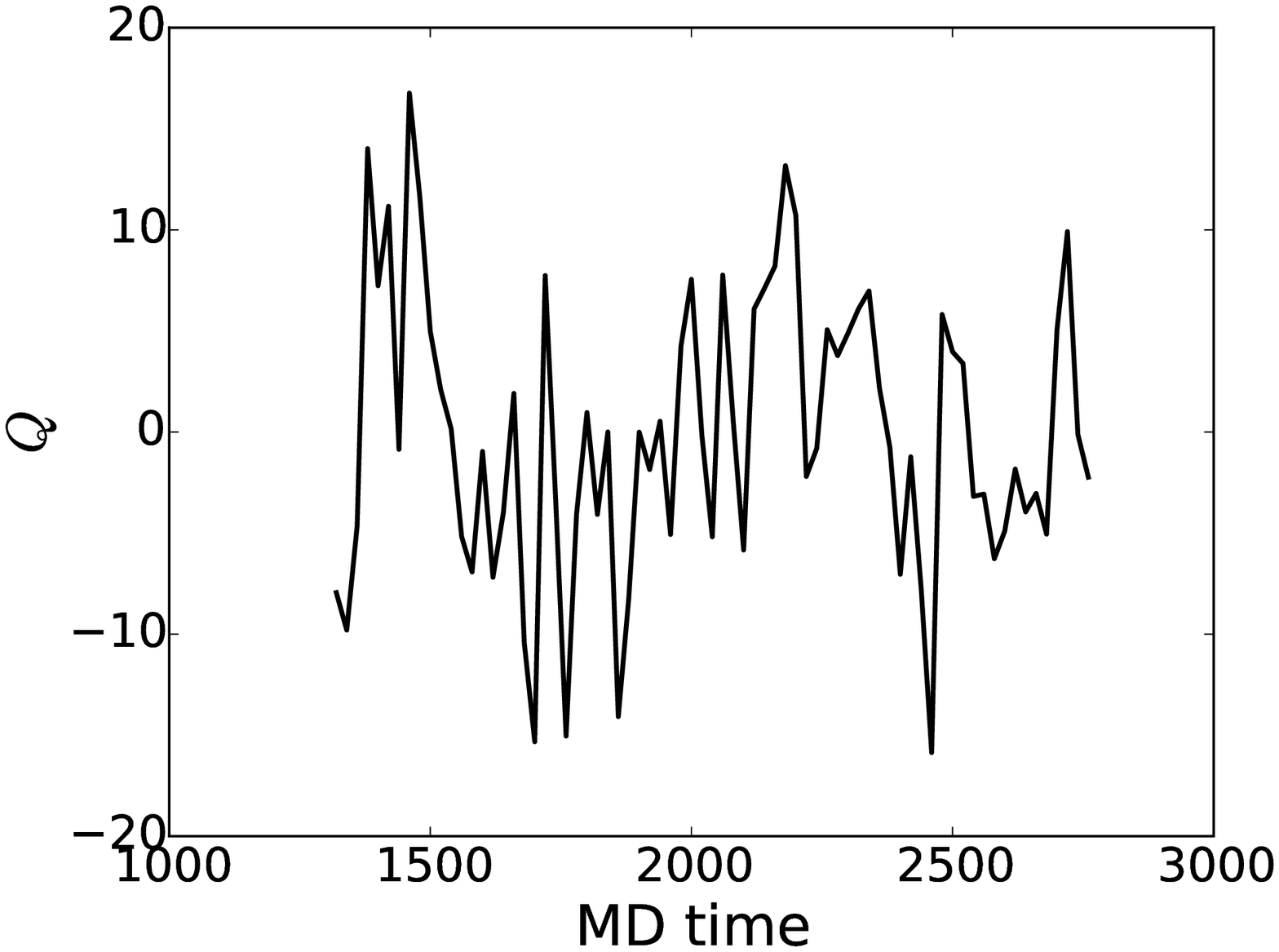}}
\caption{Monte Carlo evolution of the average plaquette (first row), light quark chiral condensate (second row), light quark pseudoscalar condensate (third row), and topological charge (fourth row) after thermalization on the 32Ifine (left column), 48I (middle column) and 64I (right column) ensembles.}
\label{fig:evolution_plots}
\end{figure}

\begin{figure}[tp]
\centering
\includegraphics*[width=0.3\textwidth]{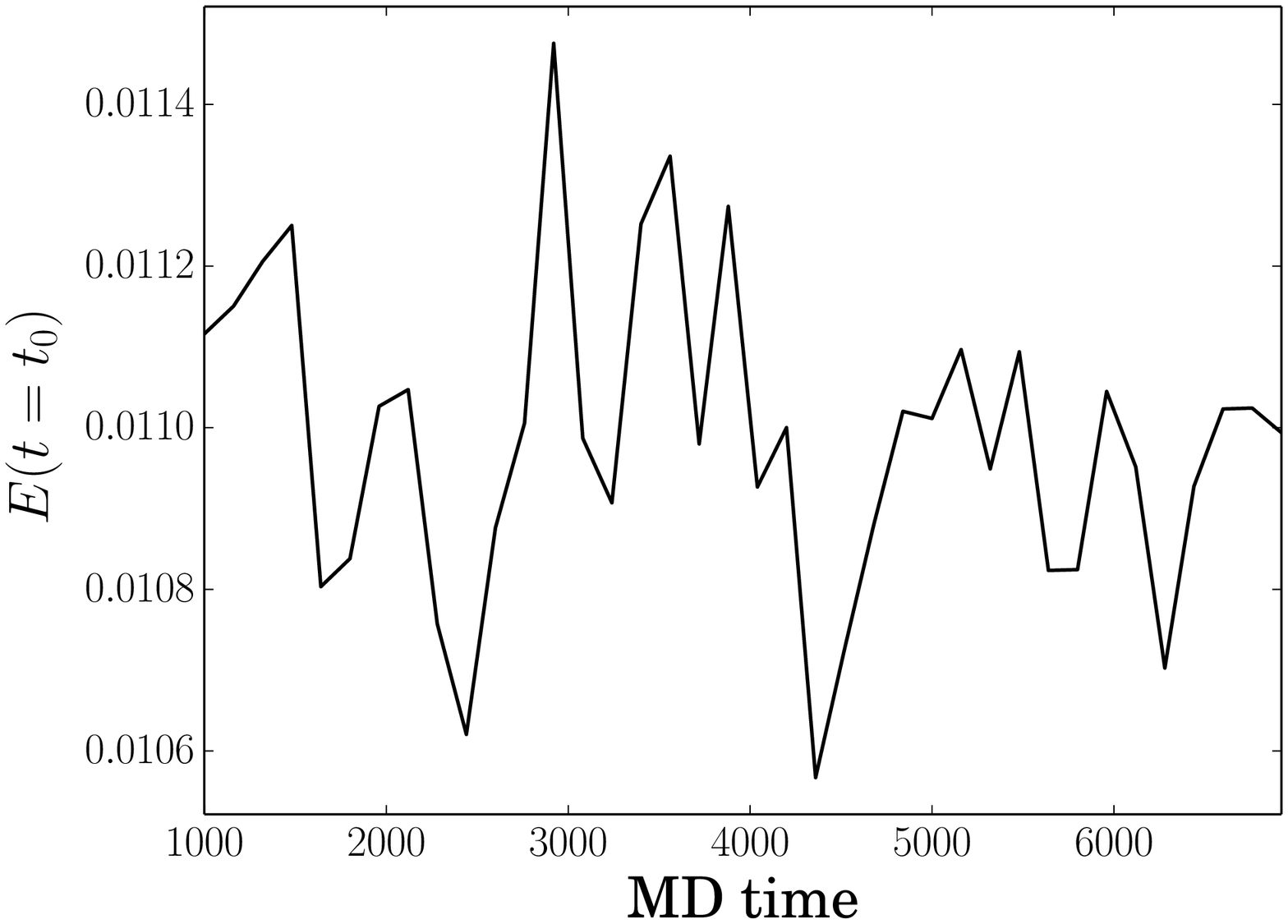}
\includegraphics*[width=0.3\textwidth]{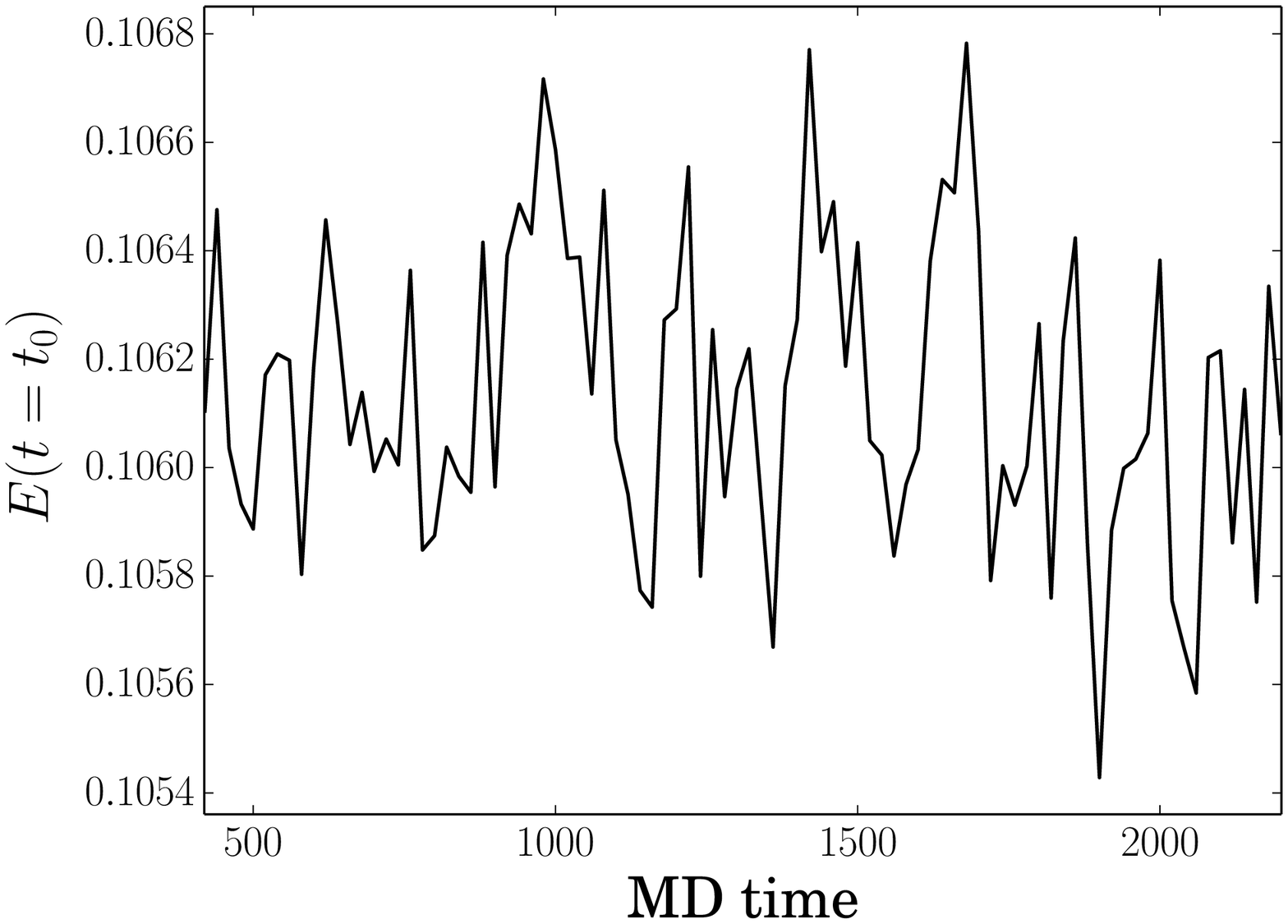}
\includegraphics*[width=0.3\textwidth]{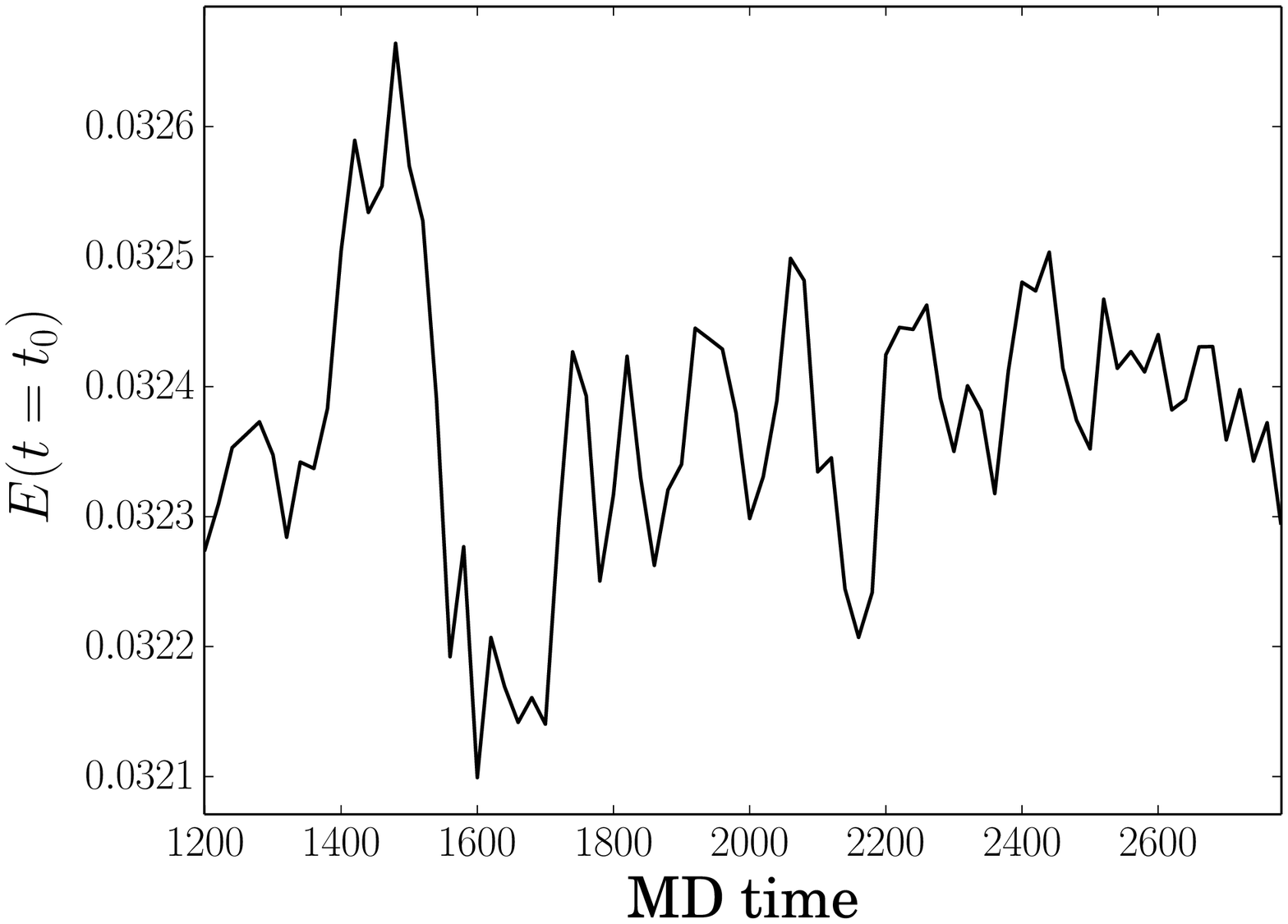}
\\
\includegraphics*[width=0.3\textwidth]{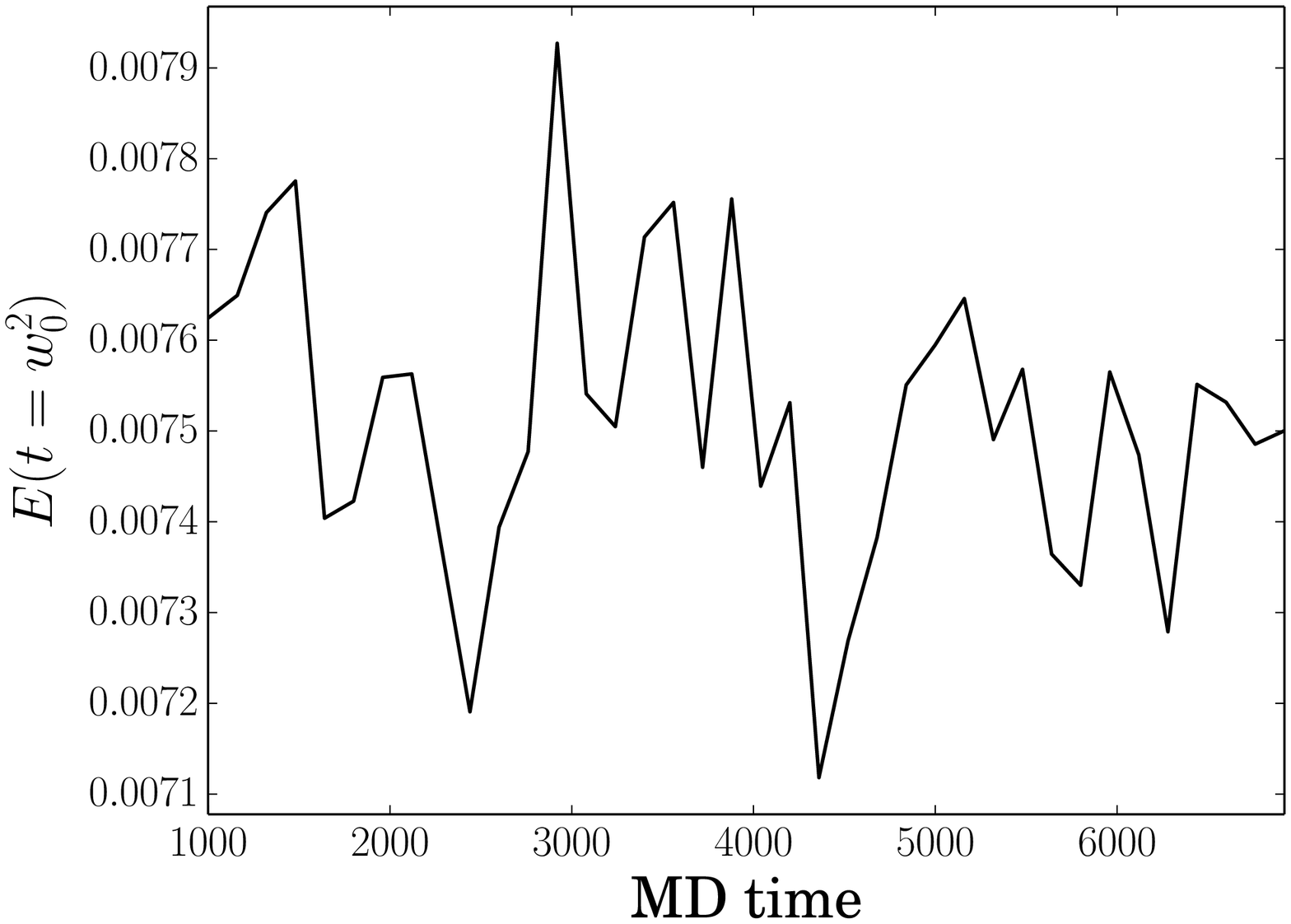}
\includegraphics*[width=0.3\textwidth]{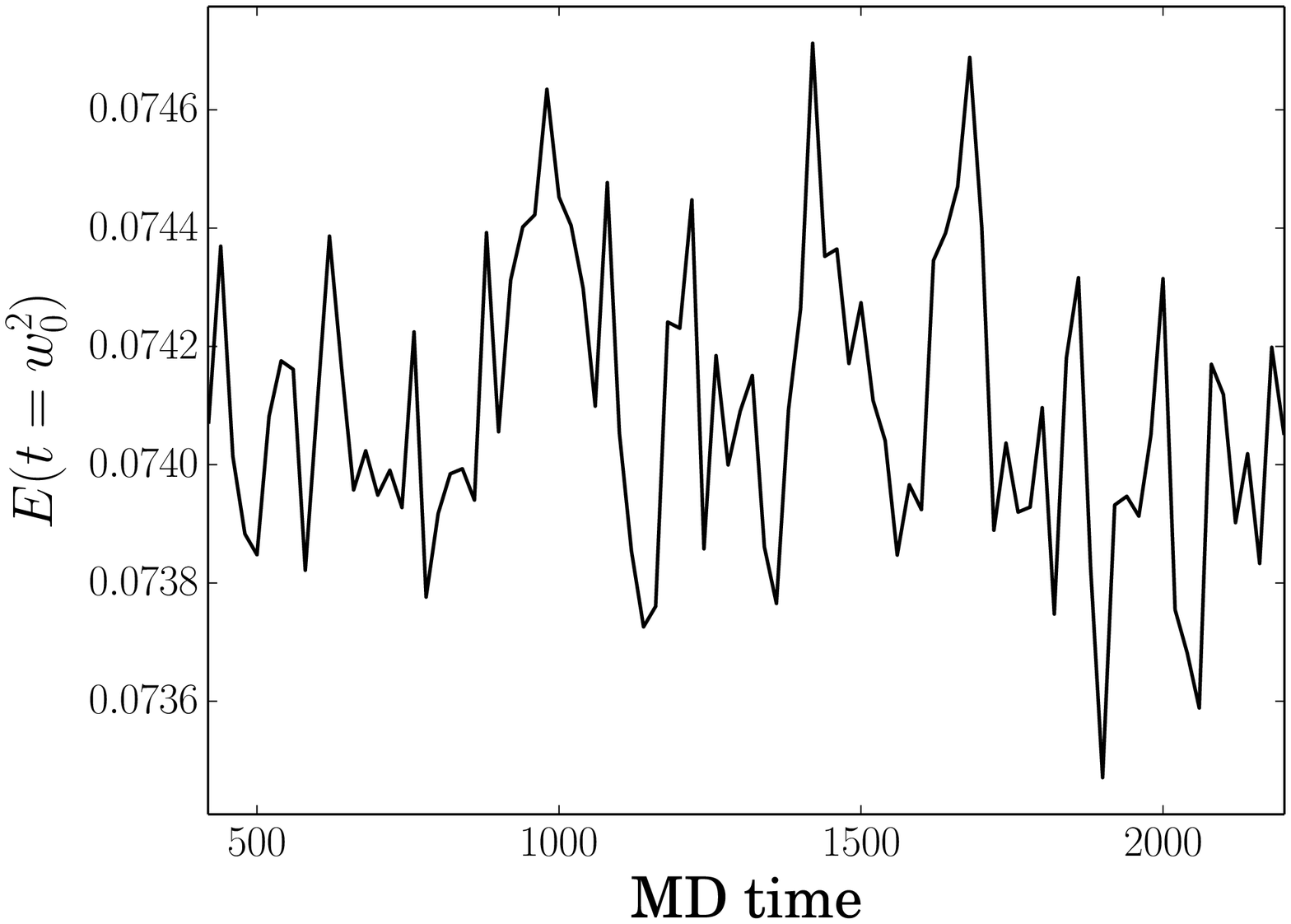}
\includegraphics*[width=0.3\textwidth]{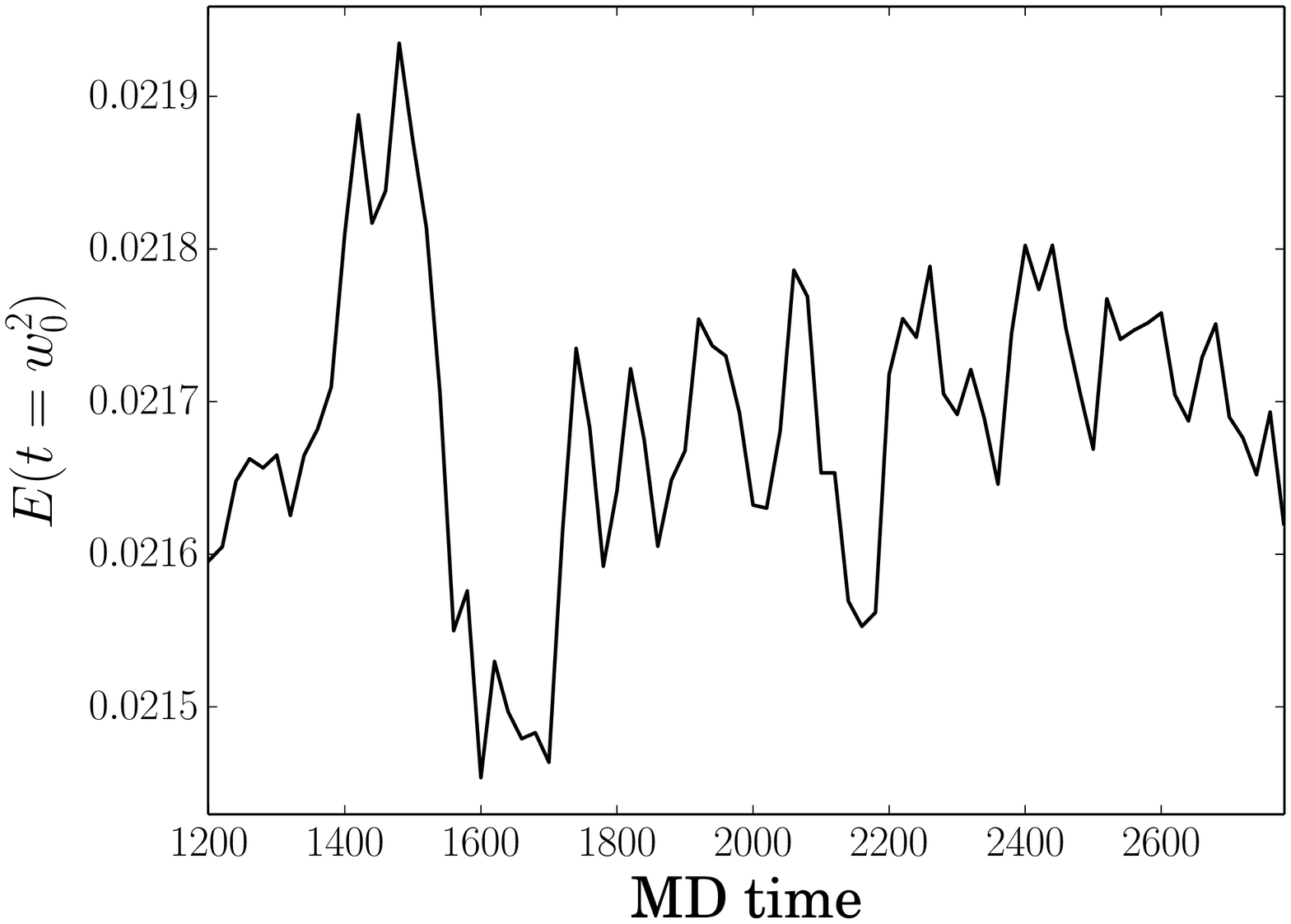}
\caption{Time history plots for the energy density evaluated at the Wilson flow times $t_0$ (top line) and $w_0^2$ (bottom line) on the 32Ifine (left column), 48I (middle column), 64I (right column). \label{fig:evolution_plots_t0w0} }
\end{figure}

In Figure~\ref{fig:evolution_plots} we plot the Monte Carlo evolution of the topological charge, plaquette, and the light quark scalar and pseudoscalar condensates, after thermalization. %
%
In addition we plot the time histories of the Clover-form energy density evaluated at the Wilson flow times $w_0^2$ and $t_0$ in Figure~\ref{fig:evolution_plots_t0w0}. %

\begin{figure}[tp]
\centering
\subfigure{\includegraphics[width=0.5\textwidth]{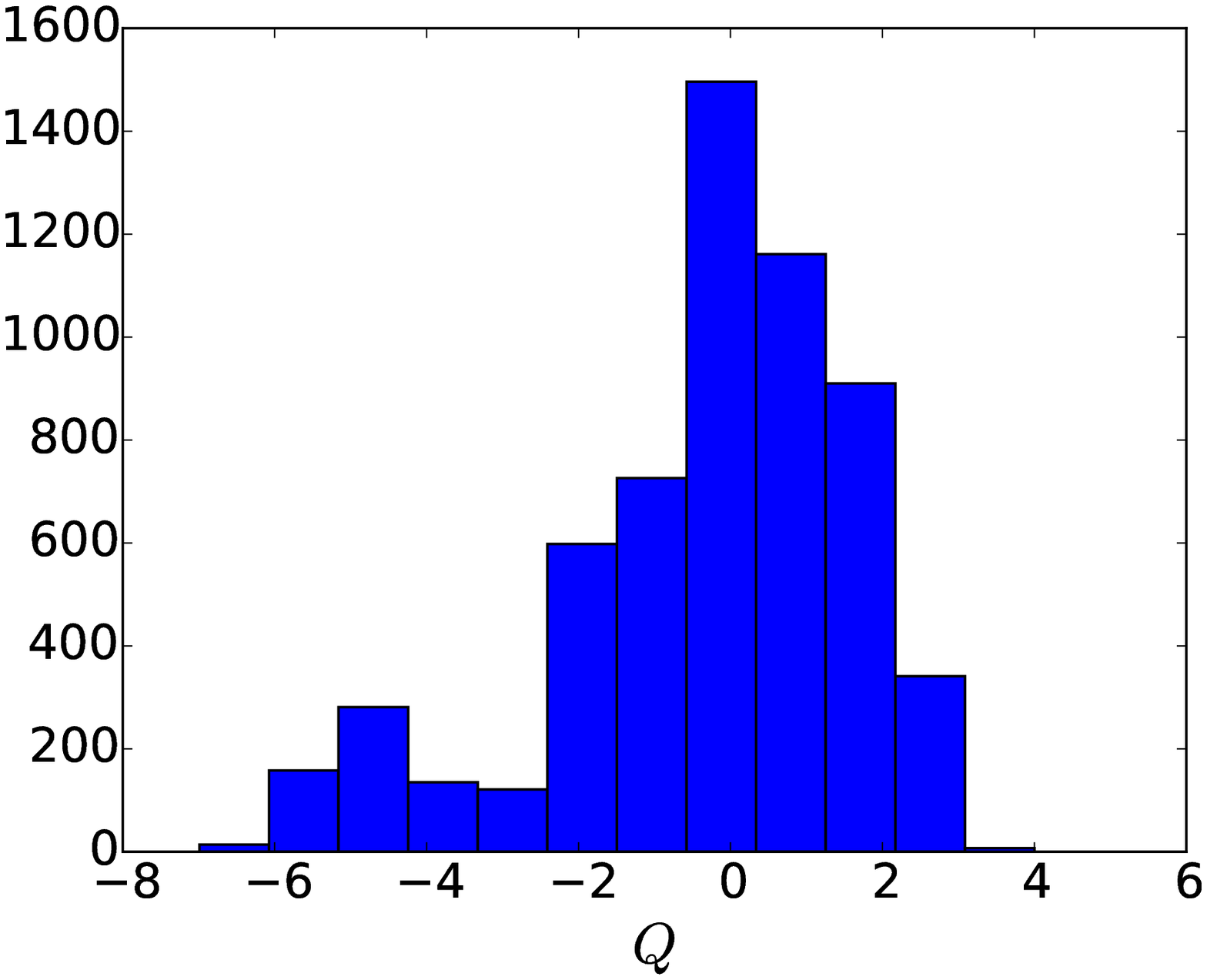} \includegraphics[width=0.5\textwidth]{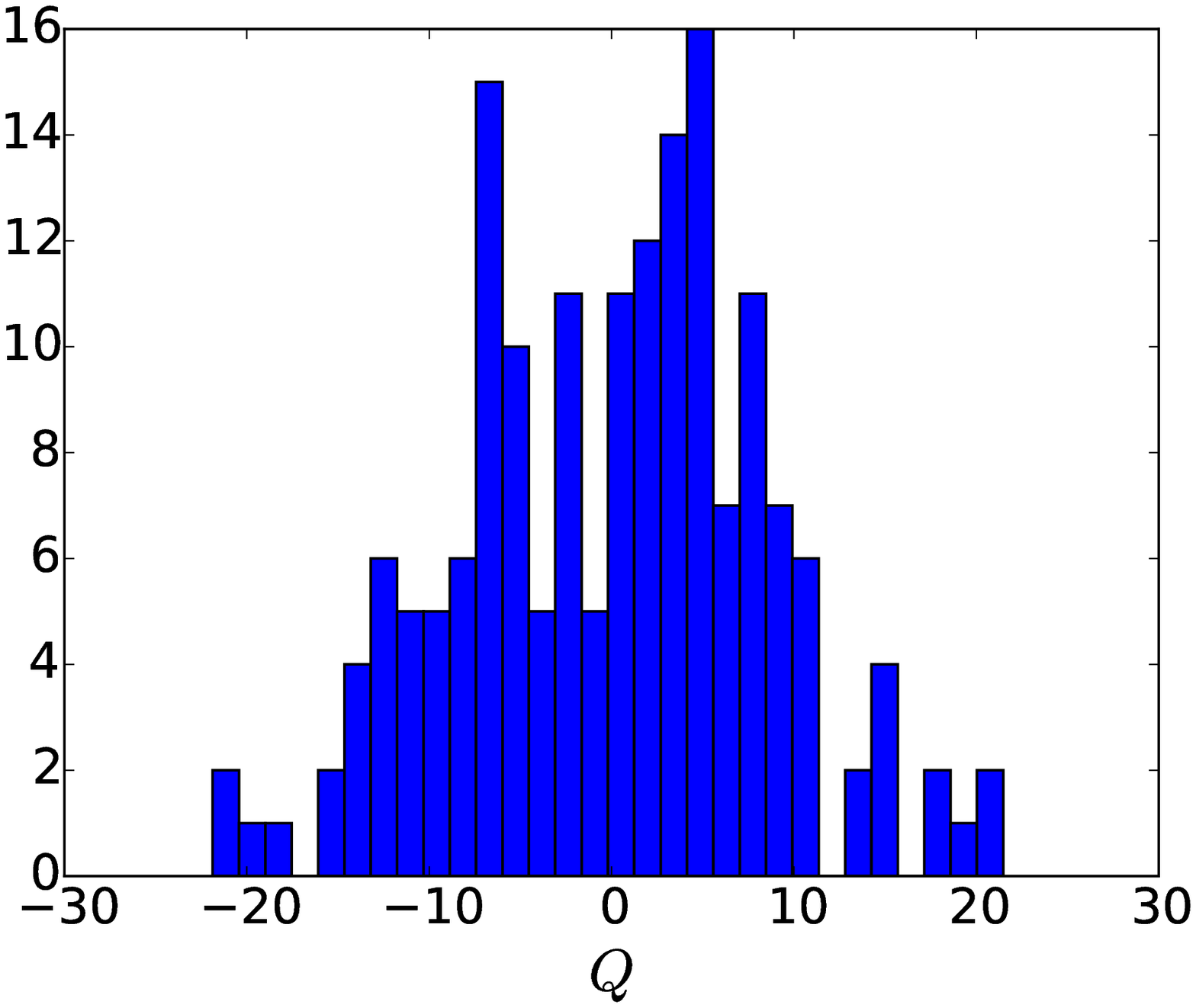}}
\subfigure{\includegraphics[width=0.5\textwidth]{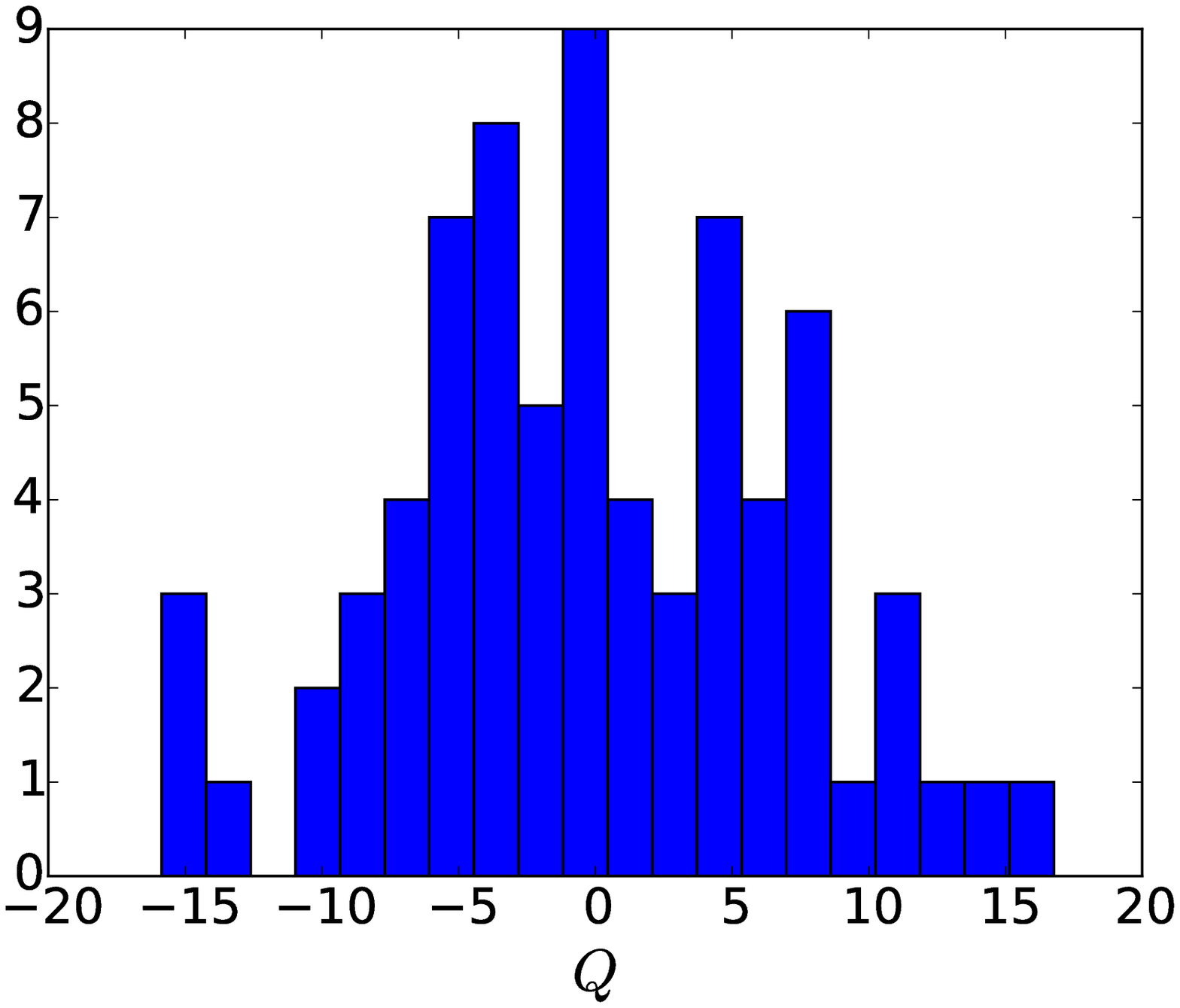}}
\caption{Topological charge distributions for the 32Ifine (top left), 48I (top right), and 64I (bottom) ensembles.}
\label{fig:tcharge_hist}
\end{figure}

We measured the topological charge by cooling the gauge fields with 60 rounds of APE~\cite{Albanese:1987ds} smearing (smearing coefficient 0.45), and then measured the field-theoretic topological charge density using the 5Li discretization of Ref.~\cite{deForcrand:1997sq}, which eliminates the $\mathcal{O}(a^{2})$ and $\mathcal{O}(a^{4})$ terms at tree level. In Figure~\ref{fig:tcharge_hist} we plot histograms of the topological charge distributions.

\begin{figure}[tp]
\centering
\subfigure{\includegraphics[width=0.5\textwidth]{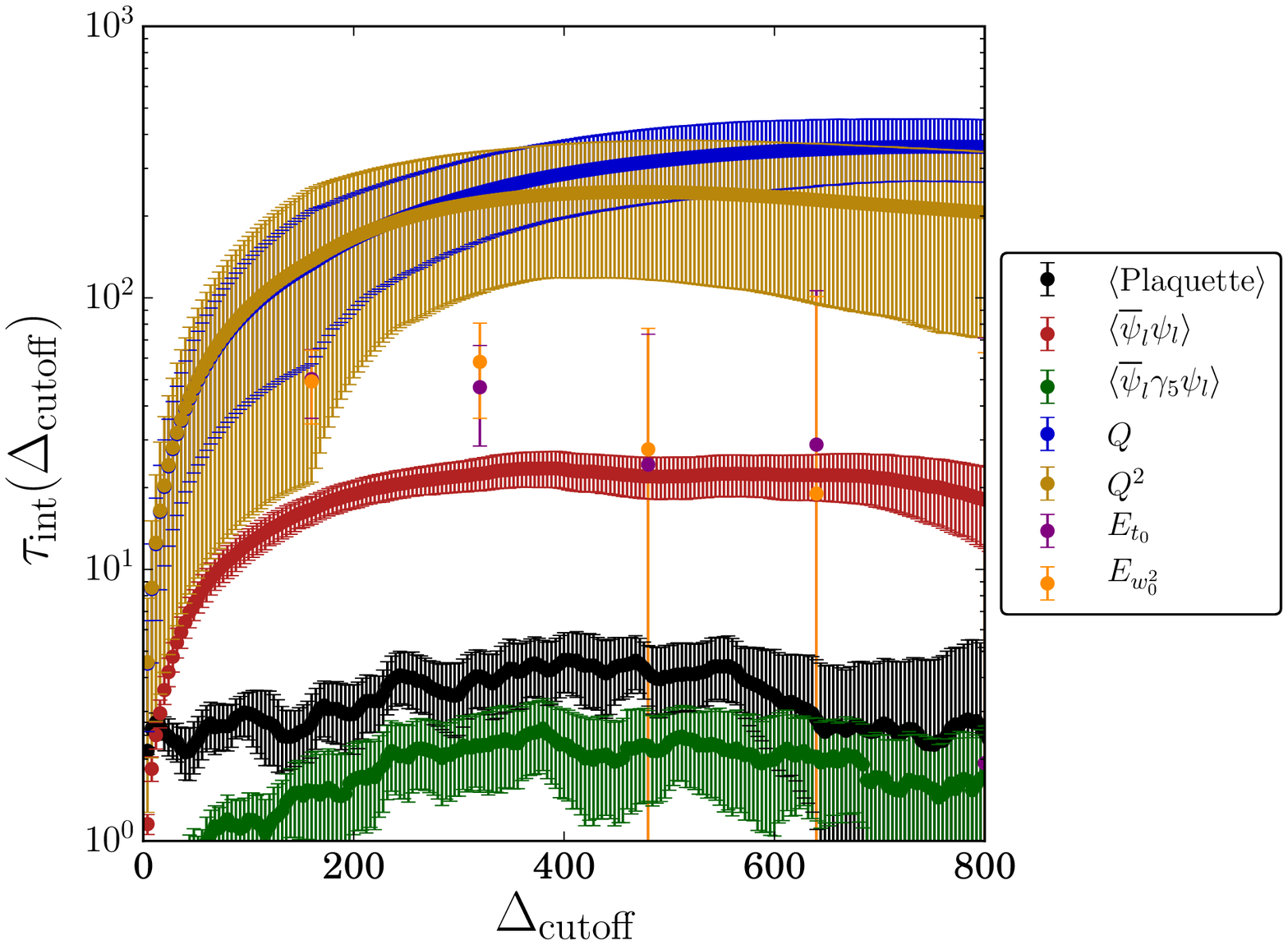} \includegraphics[width=0.5\textwidth]{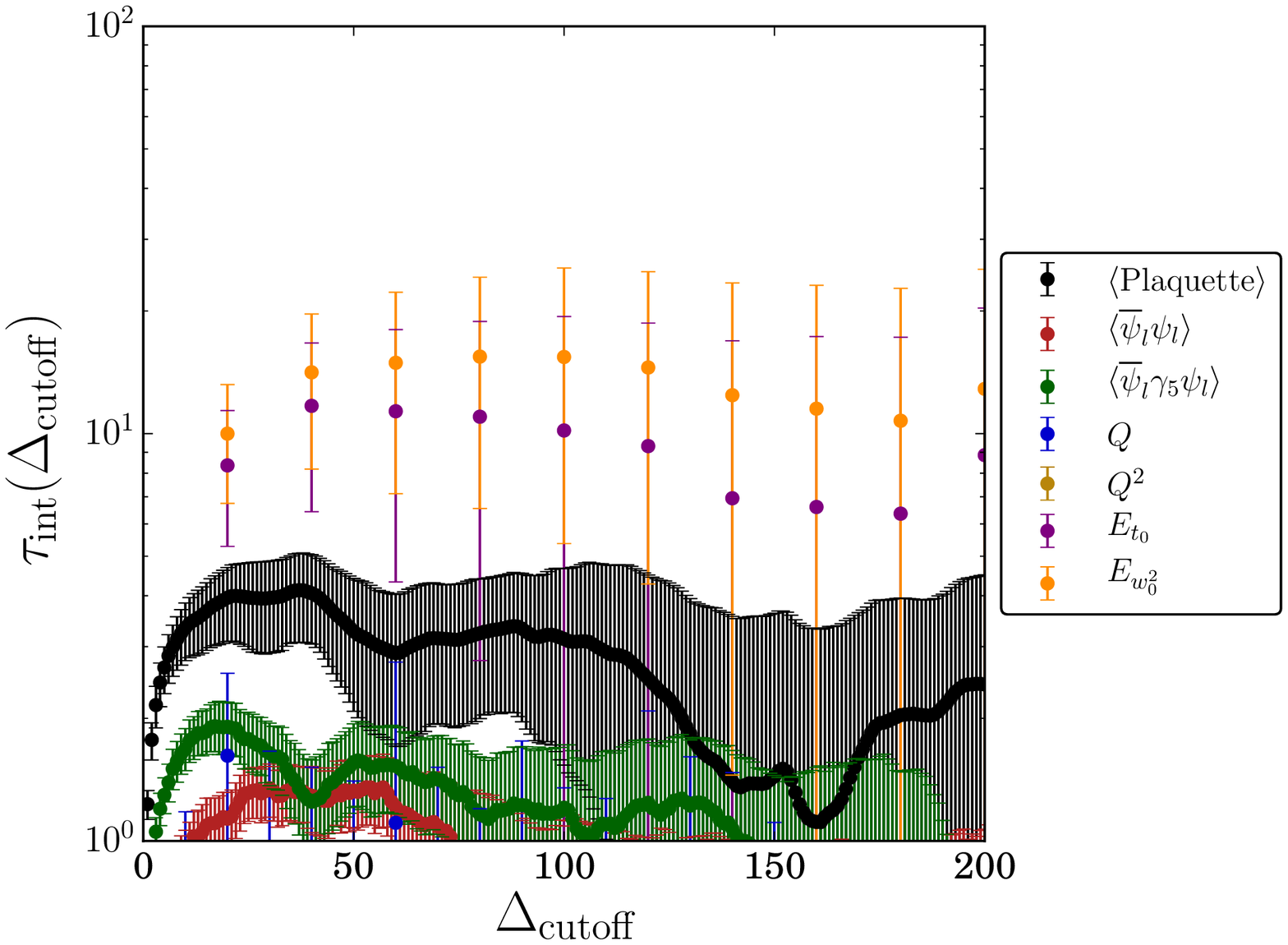}}
\subfigure{\includegraphics[width=0.5\textwidth]{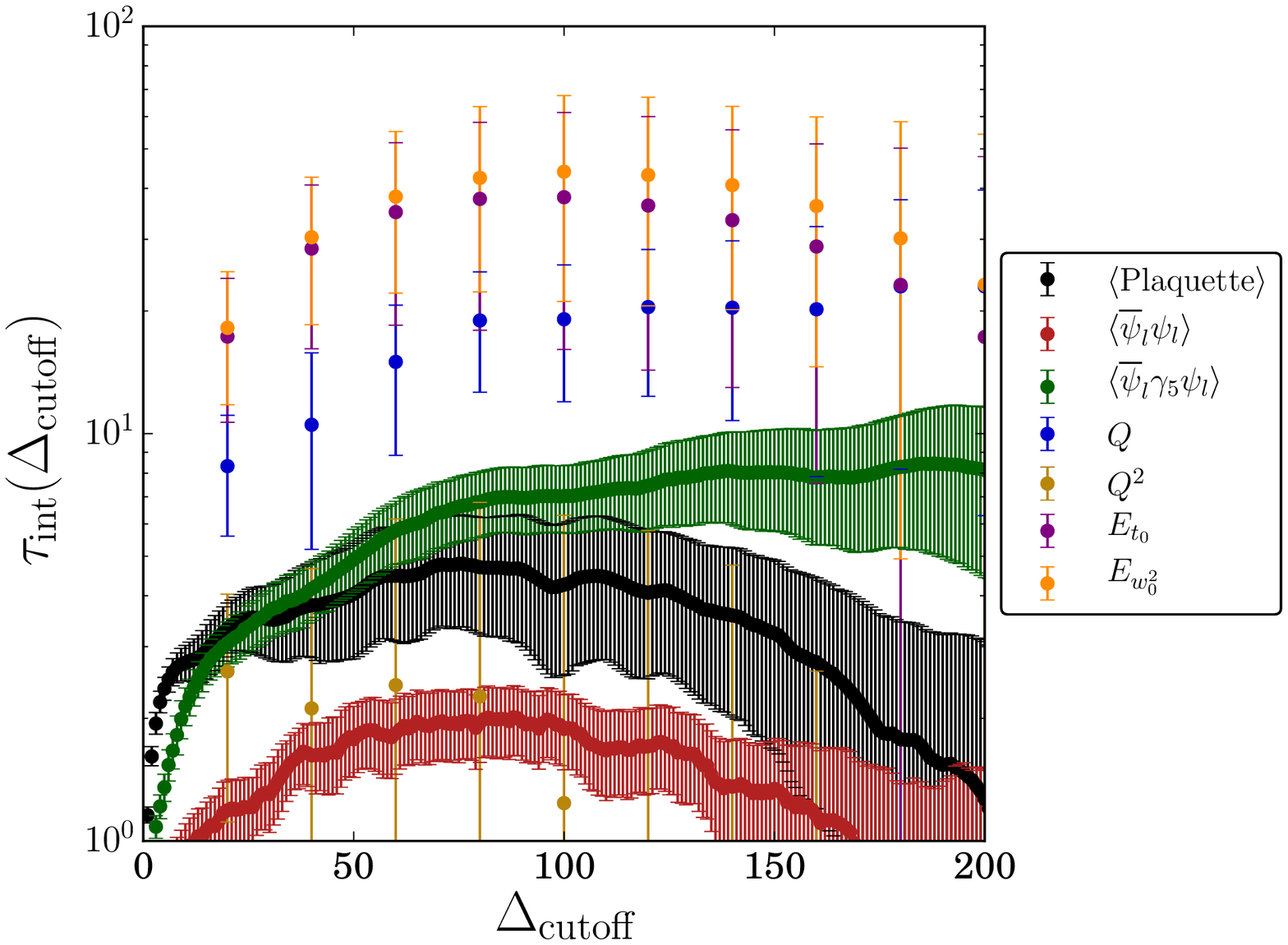}}

\caption{The integrated autocorrelation time as a function of the cutoff MD time separation, $\Delta_{\text{cutoff}}$, for the average plaquette; light quark scalar and pseudoscalar densities; topological charge $Q$ and its square; and the Clover-form energy densities evaluated at Wilson flow times $t_0$ and $w_0^2$, $E_{t_0}$ and $E_{w_0^2}$ respectively. These are plotted for the 32Ifine ensemble (top left), 48I (top right), and 64I (bottom) ensembles. The data has been binned over 960, 100, and 200 MD time units on the 32Ifine, 48I, and 64I ensemble, respectively.}
\label{fig:autocorrelation_time}
\end{figure}

In Figure~\ref{fig:autocorrelation_time} we plot the integrated autocorrelation time for the same observables on the 32Ifine, 48I, and 64I ensembles as a function of the cutoff in Molecular Dynamics (MD) time separation, $\Delta_{\text{cutoff}}$:
\begin{equation} \label{eqn:tint}
\tau_{\text{int}}(\Delta_{\text{cutoff}}) = 1 + 2 \sum\limits_{\Delta=1}^{\Delta_{\text{cutoff}}} C(\Delta)\,,
\end{equation}
where
\begin{equation} \label{eqn:autocorr}
C(\Delta) = \left< \frac{\left(Y(t)-\overline{Y}\right) \left(Y(t+\Delta)-\overline{Y}\right)}{\sigma^2} \right>_{t}
\end{equation}
is the autocorrelation function associated with the observable $Y(t)$. The mean and variance of $Y(t)$ are denoted $\overline{Y}$ and $\sigma^2$, and $\Delta$ is the lag measured in MD time units. The error on the integrated autocorrelation time is estimated using a method discussed in our earlier paper~\cite{Arthur:2012opa}: for each fixed $\Delta$ in Eq.~\eqref{eqn:autocorr} we bin the set of measurements $\left(Y(t)-\overline{Y}\right) \left(Y(t+\Delta)-\overline{Y}\right)$ over neighboring configurations and estimate the error on the mean $\left< \cdots \right>_{t}$ by bootstrap resampling. We then increase the bin size until the error bars stop growing, which we found to correspond to bin sizes of 960, 100, and 200 MD time units on the 32Ifine, 48I, and 64I ensemble, respectively. The error on $\tau_{\text{int}}$ is then computed from the bootstrap sum in Eq.~\eqref{eqn:tint}.

\begin{table}[t]
\begin{tabular}{c|ccccccc}
\hline\hline
Ensemble & $\langle P \rangle$ 	& $E_{t_0}$ 	& $E_{w_0^2}$ 	& $Q$ 		& $Q^2$ 	& $\langle \bar\psi\gamma^5\psi\rangle$ 	& $\langle \bar\psi\psi\rangle$ \\
\hline
32Ifine & 2.9(7)               	&  29(77)    	& 51(66) 	&  340(120)	&   240(140) 	&     2.6(8)    			&   24(4) \\    
48I     & 4.1(1.0) 		&  10(26) 	& 10(24) 	&  1.1(1.6) 	&   0.2(5) 	&   1.9(3) 				&    1.4(3) \\
64I     & 4.7(1.7) 		&  38(24) 	& 30(22) 	&  19(7)    	&   5(9)   	&   6(8) 				& 2.0(4)
\end{tabular}
\caption{Estimated integrated autocorrelation times for various quantities on the 32Ifine, 48I and 64I ensembles. \label{tab-autocorr} }
\end{table}

In Table~\ref{tab-autocorr} we tabulate estimates of the autocorrelation lengths for each of the various quantities included in the above figures. We can estimate $\tau_{\rm int}$ from the upper bound on the error for the slowest mode, which corresponds to the energy densities on the 64I and 48I ensembles, and the topological charge on the 32Ifine. This suggests $\tau_{\rm int}\sim 35$ MDTU for the 48I ensemble, $\tau_{\rm int}\sim 50$ MDTU for the 64I ensemble and $\tau_{\rm int}\sim 460$ MDTU for the 32Ifine ensemble.

For all quantities considered, we observe that the chosen bin sizes are sufficient to account for the autocorrelations suggested by Figure~\ref{fig:autocorrelation_time}. We also observe a significant decrease in the rate of tunneling between configurations with different topological charge as the lattice spacing becomes finer, as evidenced by the long autocorrelation time on the 32Ifine ensemble.

After generating our ensembles we discovered that there are spurious correlations between U(1) random numbers generated by the Columbia Physics System (CPS) random number generator (RNG) with a new seed. Fortunately, as discussed in Appendix~\ref{appendix-rng}, we determined that the correlation present in the freshly-seeded RNG state was lost during thermalization, and consequently that this had no measurable effect on our thermalized gauge configurations or measurements.

\section{Simulation measurement results}
\label{sec:SimulationResults}
In this section we present the results of fitting to a number of observables on the 48I and 64I ensembles. On the 48I ensemble we used data from 80 configurations in the range 420--2000 with a separation of 20 MD time units. The 64I measurements were performed on 40 configurations in the range 1200--2760 and separated by 40 MD time units. The data on both ensembles were binned over 5 successive configurations, corresponding to 100 MD time units and 200 MD time units respectively. On the 64I ensemble, we measured the cheaper Wilson flow scales every 20 configurations (as opposed to every 40 for the other measurements) in the range 1200--2780 and binned over 10 successive configurations. We also present similar results computed on 36 configurations of the 32Ifine ensemble in the range 1000--6600, measuring every 160 MD time units and using a bin size of 6 configurations (960 MD time units). 


\begin{figure}[tp]
\centering
\includegraphics[width=0.5\textwidth]{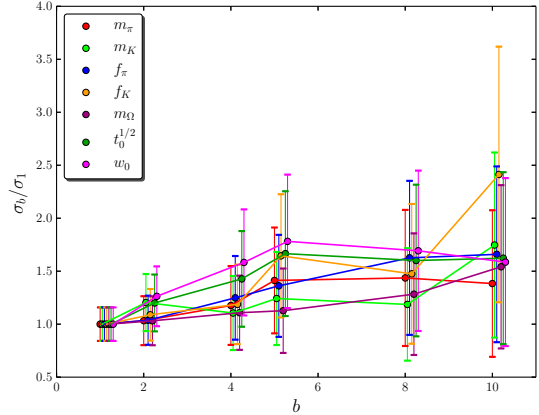}
\caption{The dependence of the error for the simulated data on the 64I ensemble. The vertical axis plots the ratio $\sigma_{b}/\sigma_1$ for bin size $b$ along the horizontal axis, where $\sigma$ is the statistical error and the subscript indicates the bin size for which that error was computed. The upper and lower bounds were obtained by varying $\sigma_b$ by $1/\sqrt{N}$, where $N$ is the number of samples.\label{fig-binerr64I} }
\end{figure}

With the bin sizes given above, the number of binned samples on the 48I, 64I and 32Ifine ensembles are 16, 8 and 6 respectively. We emphasize however that each measurement on the 64I ensemble is obtained from an average over 128 timeslices, and those on the 48I and 32Ifine over 96 and 64 timeslices, respectively. Nevertheless, the numbers of binned samples on the 64I and 32Ifine ensembles are considerably smaller than those typically encountered in lattice simulations and we therefore provide evidence that our use of this small number of large bins does to not lead to an inaccurate assignment of errors.

 
First, based on the integrated autocorrelation times determined in the previous section, the expected effective time separation between uncorrelated measurements is $\sim 100$ MDTU on the 64I ensemble, half of the actual bin size chosen. (Recall this is estimated as $2\times \tau_{\rm int}$). Our choice is therefore quite conservative. For the 48I and 32Ifine ensembles the time separation between uncorrelated measurements is $\sim 70$ and $\sim 920$ MDTU, respectively, which are comparable to our bin sizes of 100 and 960. However, these estimates are obtained from the energy densities and topological charge respectively, and the latter may be misleadingly large for the following reason. In a study by the ALPHA collaboration~\cite{Schaefer:2010hu} the authors point out that for an HMC algorithm which is invariant under parity, such as ours, the correlations seen in parity-even observables, which we study, will correspond to modes in the HMC evolution which are determined by parity-even quantities such as $Q^2$. We have included this quantity also in Figure~\ref{fig:autocorrelation_time} and Table~\ref{tab-autocorr}, for which we observe substantially smaller autocorrelation lengths, suggesting that our 48I and 32Ifine bin sizes are also quite conservative.

 
Of the 32Ifine and 64I data sets, the latter is the most important to our analysis. In Figure~\ref{fig-binerr64I} we plot the error on the 64I simulated data as a function of increasing bin size, where we estimate the error on the error as $\sim 1/\sqrt{N}$ were $N$ is the number of binned samples. In Figure~\ref{fig-binerrglobal} of Section~\ref{sec:FitResults} we show a similar plot but for the physical predictions of our global fits, again as a function of the 64I bin size. From these figures we observe no statistically significant dependence on the 64I bin size, suggesting that we are not underestimating our errors by making our choice of 100 MDTU bins for this ensemble.


The ability to generate physical mass ensembles forced us to seek dramatic improvements in our measurement strategy, since the statistical error for kaon observables increases with decreasing light quark mass (holding the strange quark mass fixed). For an example of this behavior, consider the kaon two-point function,
\begin{equation}
C(t) = \sum_{\vec y,\vec x} [\bar s u](\vec y,t)[\bar u s](\vec x,0)\,,
\end{equation}
which in the limit of large $t$ goes as
\begin{equation}
\langle C(t) \rangle = Ae^{-m_K t} + \ldots\,,
\end{equation}
where $\langle .. \rangle$ is the average over the gauge field ensemble. The standard deviation on this quantity, i.e. its statistical error, goes as $\sqrt{\langle C^2(t) \rangle}$, which contains two strange quark propagators and two light quark propagators. This quantity can also be represented as a linear combination of exponentially decaying terms:
\begin{equation}
\langle C^2(t) \rangle = B e^{-(m_{s\bar s}+m_\pi) t} + \ldots\,,
\end{equation}
where $m_{s\bar s}$ is the mass of the $s\bar s$ state. The signal-to-noise ratio goes as $\exp\left(-[m_K - (m_{s\bar s}+m_\pi)/2] t\right)$ in the large time limit, and therefore decays faster with lighter pions.

The first component of our measurement strategy involves maximally reusing propagators for all of our measurements, which include $m_\pi$, $m_K$, $m_\Omega$, $f_\pi$, $f_K$, $B_K$, $f^{K\pi}_+(0)$ and $K \to (\pi \pi)_{I=2}$. (Note that the latter two quantities are not reported on in this document.) Reusing propagators requires choosing a common source for our propagators that remains satisfactory across the entire range of measurements. Also, since we measure both two- and three-point functions, we need to be able to control the spatial momentum of the sources in order to project out unwanted momenta. We performed numerous studies of Coulomb gauge fixed wall sources and Coulomb gauge fixed box sources for many of these observables.  (The box sources were generically chosen so that an integer multiple of their linear dimension would fit in the lattice volume, allowing us to obtain zero momentum projections by using all possible box sources.)  While the box sources showed faster projection onto the desired ground state, the 
statistical errors on the wall sources were much smaller, such that the errors on the measured quantities per unit of computer time were essentially the same.  From these studies, we chose to use the simple Coulomb gauge fixed wall sources.

In previous work on the $\eta-\eta^\prime$ mass, which involves disconnected quark diagrams, we found that translating $n$-point functions over all possible temporal source locations reduced the error essentially as the square-root of the number of translations~\cite{Christ:2010dd}.  The calculation of such a large number of quark propagators on a single configuration can be accomplished much more quickly by a deflation algorithm. The EigCG algorithm~\cite{Stathopoulos:2007zi} was used for $K \to (\pi \pi)_{I=0,2}$ measurements at unphysical kinematics in Ref.~\cite{Qi_Liu_thesis}, and was adopted for this calculation. Measurements were again performed for all temporal translations of the $n$-point functions, and a factor of 7 speed-up was achieved. The major drawback of EigCG is the considerable memory footprint. However, BG/Q partitions have large memory and therefore this issue can be managed. In practice we found that only a fraction of the vectors generated by the EigCG method were good representations of the true eigenmodes, and in future we may be able to reduce the CG time further by pre-calculating exact low-modes using the implicitly restarted Lanczos algorithm with Chebyshev acceleration~\cite{Calvetti94animplicitly}. 

An alternative approach to generating a large number of quark propagators is to use inexact deflation.~\cite{Luscher:2007se}. This approach had not been optimally formulated for the domain wall operator when the measurements on our new ensembles were begun. However, a new formulation of inexact deflation appropriate to DWF, known as HDCG~\cite{Boyle:2014rwa}, has since been developed, and has been shown to be more efficient than EigCG; this technique is now being used for our valence measurements.

The final component of our measurement package is the use of the all-mode averaging (AMA)~\cite{Blum:2012uh} method to further reduce the cost of translating the propagator sources along the temporal direction. AMA is a generalization of low-mode averaging, in which one constructs an approximate propagator using exact low eigenmodes and a polynomial approximation to the high modes obtained by applying deflated conjugate gradient (CG) to a source vector on each temporal slice and averaging over the solutions. The stopping condition on the deflated conjugate gradient can generally be relaxed, reducing the iteration
count. The remaining bias in the observable is corrected using a small number of exact solves obtained using the low modes and a precise deflated CG solve from a single timeslice for the high-mode contribution. The benefit of this procedure is that the CG solves used for the polynomial approximation can be performed very cheaply using inexact `sloppy' stopping conditions of $10^{-4}$ or $10^{-5}$ as many of the low modes are already projected out exactly. The net result of combining the sloppy translated 
solution with the (typically small) bias correction is an exact result calculated many times more cheaply than if we were to perform precise deflated solves on every timeslice. 

In order to avoid any bias due to the even-odd decomposed Dirac operator used in the CG, we calculate the eigenvectors using EigCG on a volume source spanning the entire four-dimensional volume, and the temporal slices where we perform the exact solves are chosen randomly for each configuration. We calculate low modes in single precision using EigCG in order to reduce the memory footprint, and also perform the sloppy solves in single precision. For the exact solves, we achieve double precision accuracy through multiple restarts of single precision solves, restarting the solve by correcting the defect as calculated in double precision. For the zero-momentum strange quark propagators required, we do a standard, accurate CG solve for sources on every timeslice. On the 48I, we performed our measurements using single rack BG/Q partitions (1024 nodes), calculating 600 low modes with EigCG (filling the memory) and running continuously for 5.5 days. (Note that this timing includes non-zero momentum light quark solves for measurements of $f_+^K$ and additional light quark solves for $K\rightarrow(\pi\pi)_{I=2}$, which are not reported in this document.) For the 64I ensemble, the measurements were performed 
on between 8 and 32 rack BG/Q partitions at the ALCF and 1500 low modes were calculated by EigCG.  On a 32 rack partition, the latter took 5.3 hours and the solver sustained 1 PFlops. (The EigCG setup time is efficiently amortized in these calculations by using the EigCG eigenvectors to deflate a large number of solves.) 

The Coulomb gauge-fixing matrices for the 64I ensemble were not computed on the BG/Q and were instead determined separately (and more quickly) on a cluster, using the timeslice-by-timeslice Coulomb gauge FASD algorithm~\cite{Hudspith:2014oja}.

We simultaneously fit the residual mass, pseudoscalar masses and decay constants, axial and vector current renormalization coefficients ($Z_{A}$ and $Z_{V}$, respectively), and kaon bag parameter ($B_{lh}$). A separate fit was performed for the $\Omega$-baryon mass. The values for these observables obtained on each lattice, as well as the statistical errors computed by jackknife resampling, are summarized in Table~\ref{tab:results}. The corresponding fit ranges are summarized in Tables~\ref{tab:2pt_fit_ranges} and~\ref{tab:3pt_fit_ranges}. In the following sections we discuss the fit procedures and plot effective masses and amplitudes for each observable. 

\begin{table}[!ht] 
\centering
\begin{tabular}{c||c|c|c}
 & 32Ifine & 48I & 64I \\ 
\hline \hline
$m_{ll}$ & 0.11790(131) & 0.08049(13) & 0.05903(13) \\
$m_{lh}$ & 0.17720(118) & 0.28853(14) & 0.21531(17) \\
$f_{ll}$ & 0.04846(32) & 0.07580(8) & 0.05550(10) \\
$f_{lh}$ & 0.05358(22) & 0.09040(9) & 0.06653(10) \\
$Z_{A}$ & 0.77779(29) & 0.71191(5) & 0.74341(5) \\
$Z_{V}$ & 0.77700(8) & 0.71076(25) & 0.74293(14) \\
$B_{lh}$ & 0.5437(85) & 0.5841(6) & 0.5620(6) \\
$m_{hhh}$ & 0.5522(29) & 0.9702(10) & 0.7181(7) \\
$m_{hhh}'$ & 0.811(49) & 1.273(10) & 0.937(7) \\
$m_{\text{res}}$ & 0.0006296(58) & 0.0006102(40) & 0.0003116(23) \\
\hline
$w_{0}$ & 2.664(16) & 1.50125(94) & 2.0495(15) \\
$t_{0}^{1/2}$ & 2.2860(63) & 1.29659(28) & 1.74496(62) \\
\hline
$m_{ll}/m_{hhh}$ & 0.2135(26) & 0.08296(17) & 0.08220(20) \\
$m_{lh}/m_{hhh}$ & 0.3209(25) & 0.29740(32) & 0.29983(37) \\
\end{tabular}
\caption{Summary of fit results in lattice units.\label{tab:results}}
\end{table}

\begin{table}[!ht]
\centering
\begin{tabular}{c||c|c|c}
 Correlator & 32Ifine & 48I & 64I \\ 
\hline \hline
$PP^{WL} (ll)$ & 10:31 & 15:48 & 12:60 \\
$PP^{WW} (ll)$ & 10:31 & 10:35 & 10:61 \\
$AP^{WL} (ll)$ & 10:31 & 10:46 & 10:60 \\
$PP^{WL} (lh)$ & 10:31 & 14:40 & 17:49 \\
$PP^{WW} (lh)$ & 10:31 & 14:33 & 14:45 \\
$AP^{WL} (lh)$ & 10:31 & 12:40 & 20:49 \\
$Z_{A}$ & 11:52 & 6:89 & 10:117 \\
$\Omega$ & 6:20 & 5:17 & 5:19 \\
$m_{\text{res}}$ & 6:57 & 9:86 & 10:117 \\
\end{tabular}
\caption{Summary of fit ranges $t_{\text{min}}/a \le t/a \le t_{\text{max}}/a$ used for each two-point correlator and ensemble.}
\label{tab:2pt_fit_ranges}
\end{table}

\begin{table}[!ht]
\centering
\begin{tabular}{c||c|c|c}
 Ensemble & Quantity & $|t_{\text{source}} - t_{\text{sink}}|/a$ & $t_{\text{skip}}/a$ \\ 
\hline \hline
\multirow{2}{*}{32Ifine} & $Z_{V}$ & 16:4:32 & 8 \\
 & $B_{lh}$ & 52:4:56 & 10 \\
\hline
 \multirow{2}{*}{48I} & $Z_{V}$ & 12:4:24 & 6 \\
 & $B_{lh}$ & 20:4:40 & 10 \\
\hline
 \multirow{2}{*}{64I} & $Z_{V}$ & 15:5:40 & 6 \\
 & $B_{lh}$ & 25:5:40 & 10 \\
\end{tabular}
\caption{Summary of fit ranges used for each three-point correlator and ensemble. We simultaneously fit to all source-sink separations in the given range, where the operator insertion is evaluated at times which are at least $t_{\text{skip}}/a$ time slices away from the sources and sinks.}
\label{tab:3pt_fit_ranges}
\end{table}

\subsection{Residual mass}

\begin{figure}[h]
\centering
\subfigure{\includegraphics[width=0.5\textwidth]{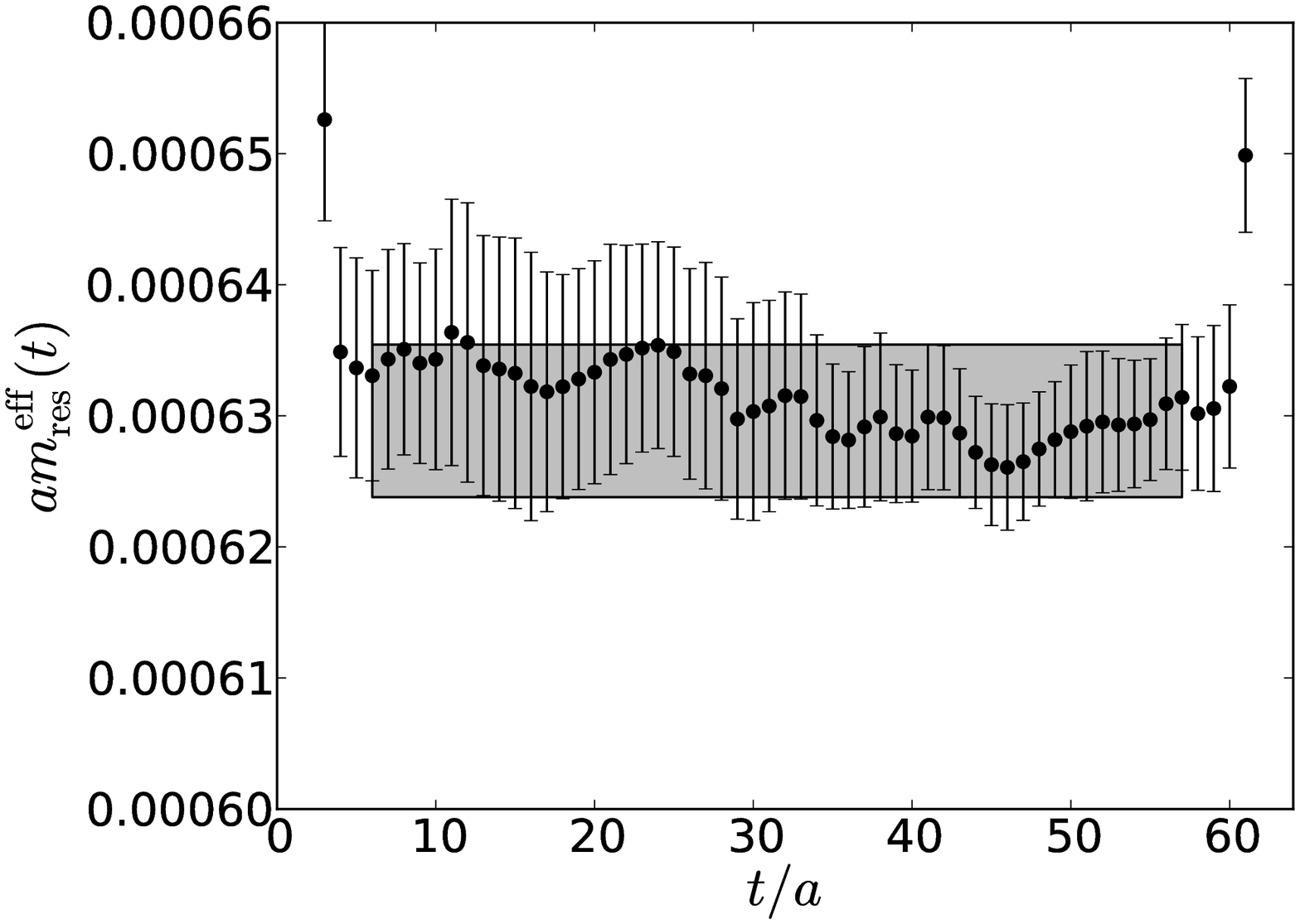} \includegraphics[width=0.5\textwidth]{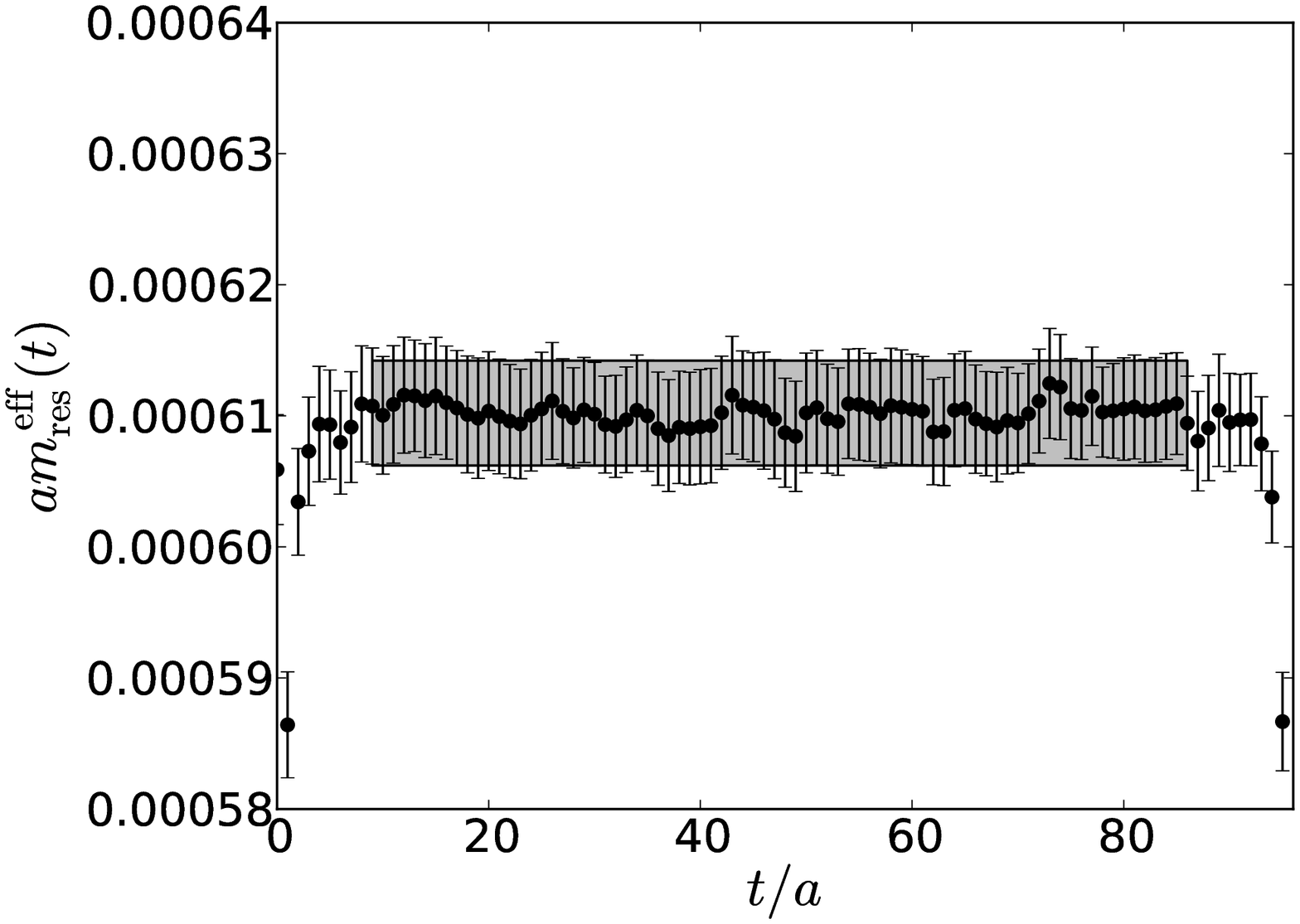}}
\subfigure{\includegraphics[width=0.5\textwidth]{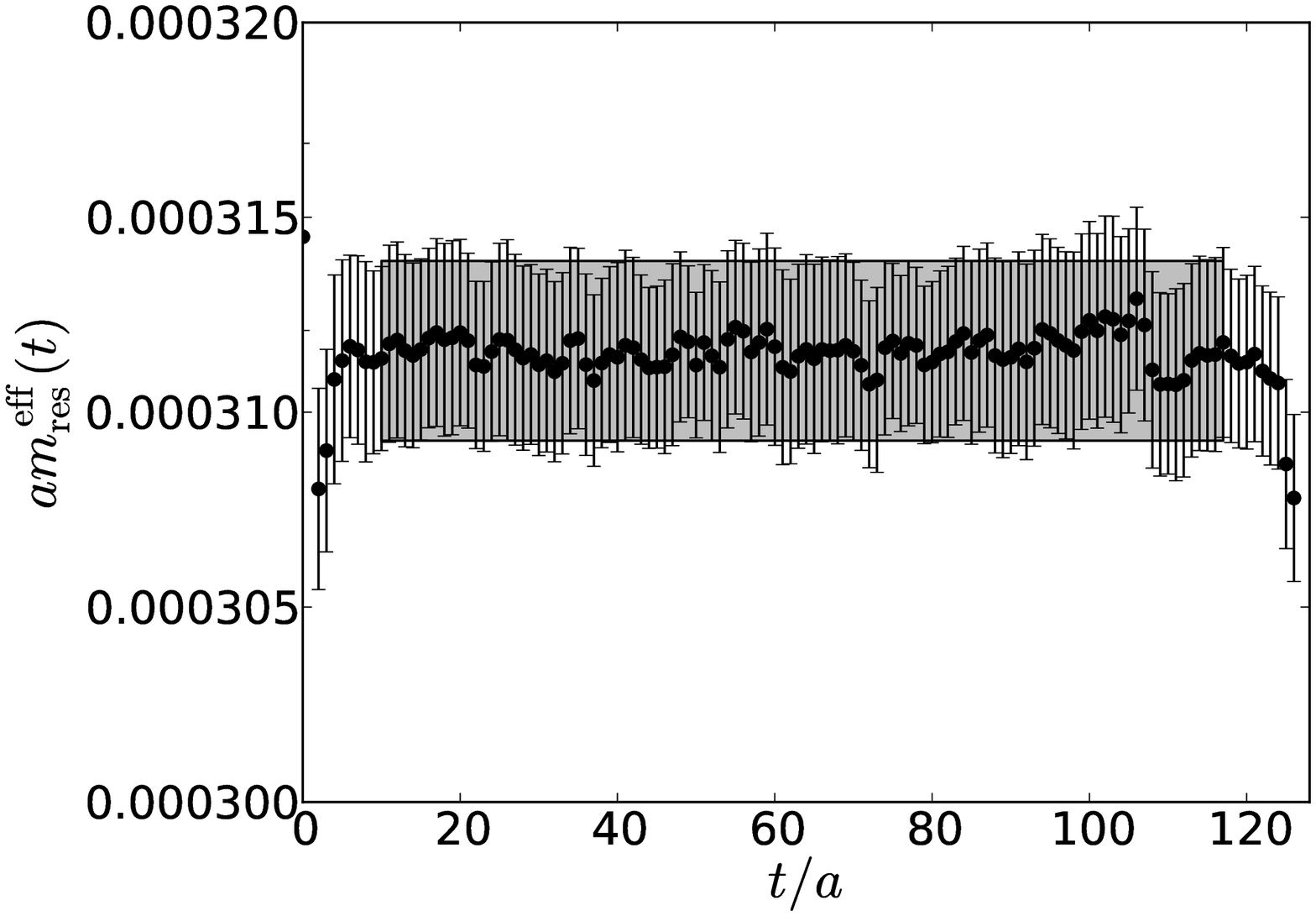}}
\caption{Effective $m_{\text{res}}$ on the 32Ifine (top left), 48I (top right), and 64I (bottom) ensembles.}
\label{fig:mres_eff_amplitude}
\end{figure}

For domain wall fermions, the leading effect of having a finite fifth dimension is an additive renormalization to the bare quark masses known as the residual mass, $m_{\rm res}$. We extract the residual mass from the ratio
\begin{equation}
\mathcal{C}_{m_{\text{res}}}(t) = \frac{\langle 0 | \sum_{\vec{x}} j^{a}_{5 q}(\vec{x},t) | \pi \rangle}{\langle 0 | \sum_{\vec{x}} j^{a}_{5}(\vec{x},t) | \pi \rangle},
\end{equation}
where $j^{a}_{5 q}$ is the pseudoscalar density evaluated at the midpoint of the fifth dimension, and $j^{a}_{5}$ is the physical pseudoscalar density constructed from the surface fields (cf. Ref.~\cite{Blum:2000kn}, Eqs.~(8) and~(9) ). In Figure~\ref{fig:mres_eff_amplitude} we plot the effective residual mass, as well as the fit, on each ensemble.

\subsection{Pseudoscalar Masses}
\begin{figure}[h]
\centering
\subfigure{\includegraphics[width=0.5\textwidth]{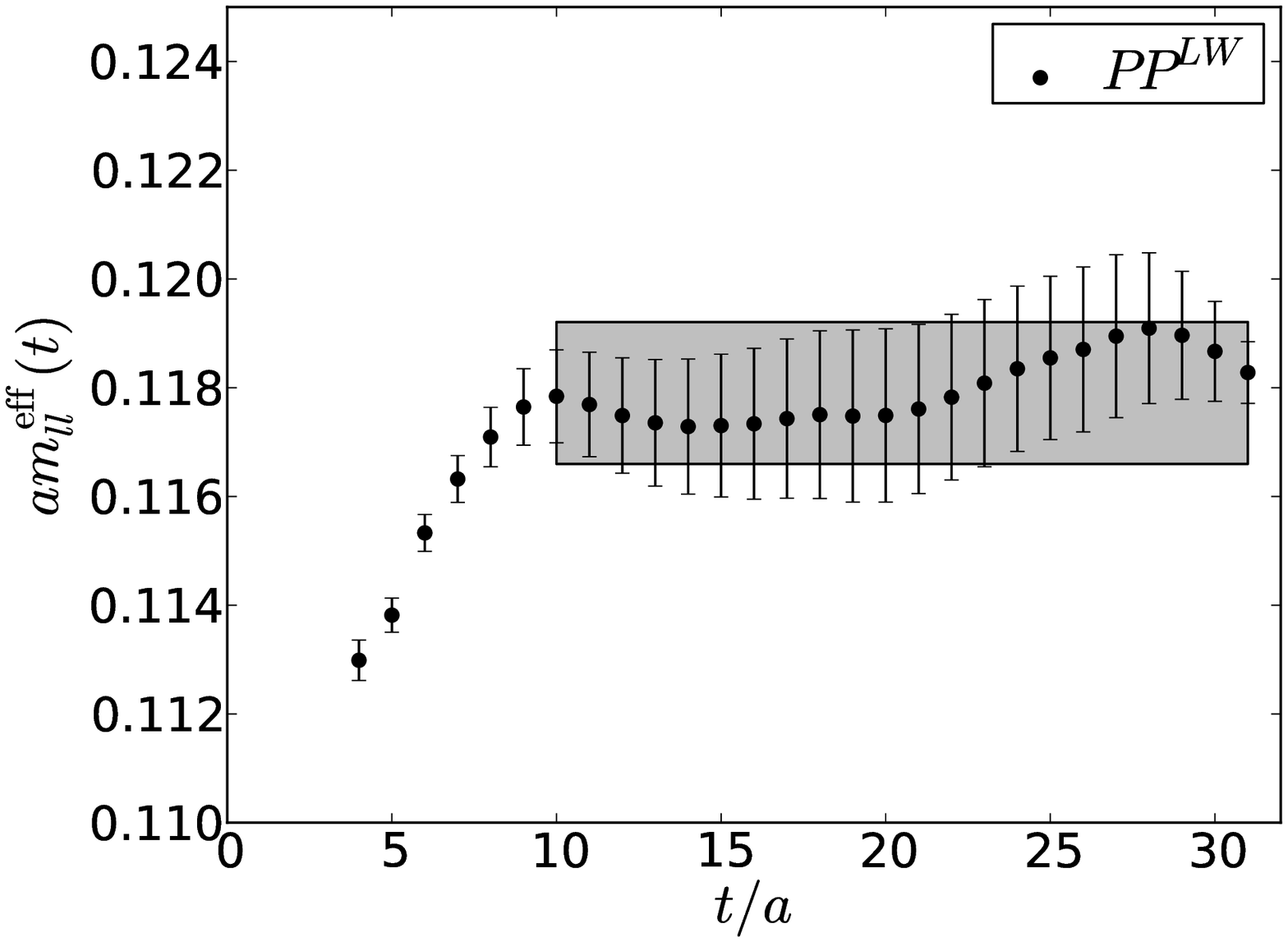} \includegraphics[width=0.5\textwidth]{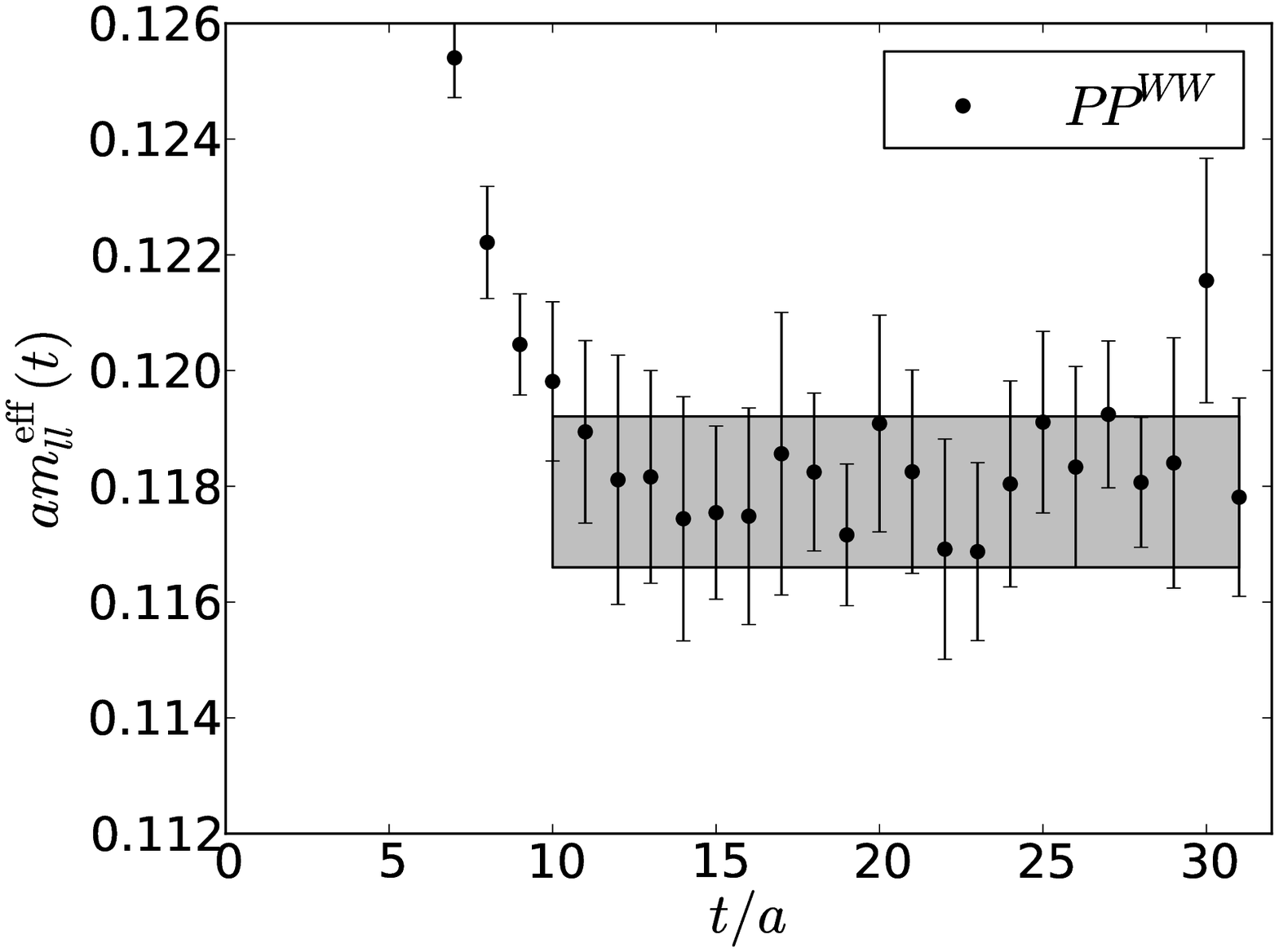}}
\subfigure{\includegraphics[width=0.5\textwidth]{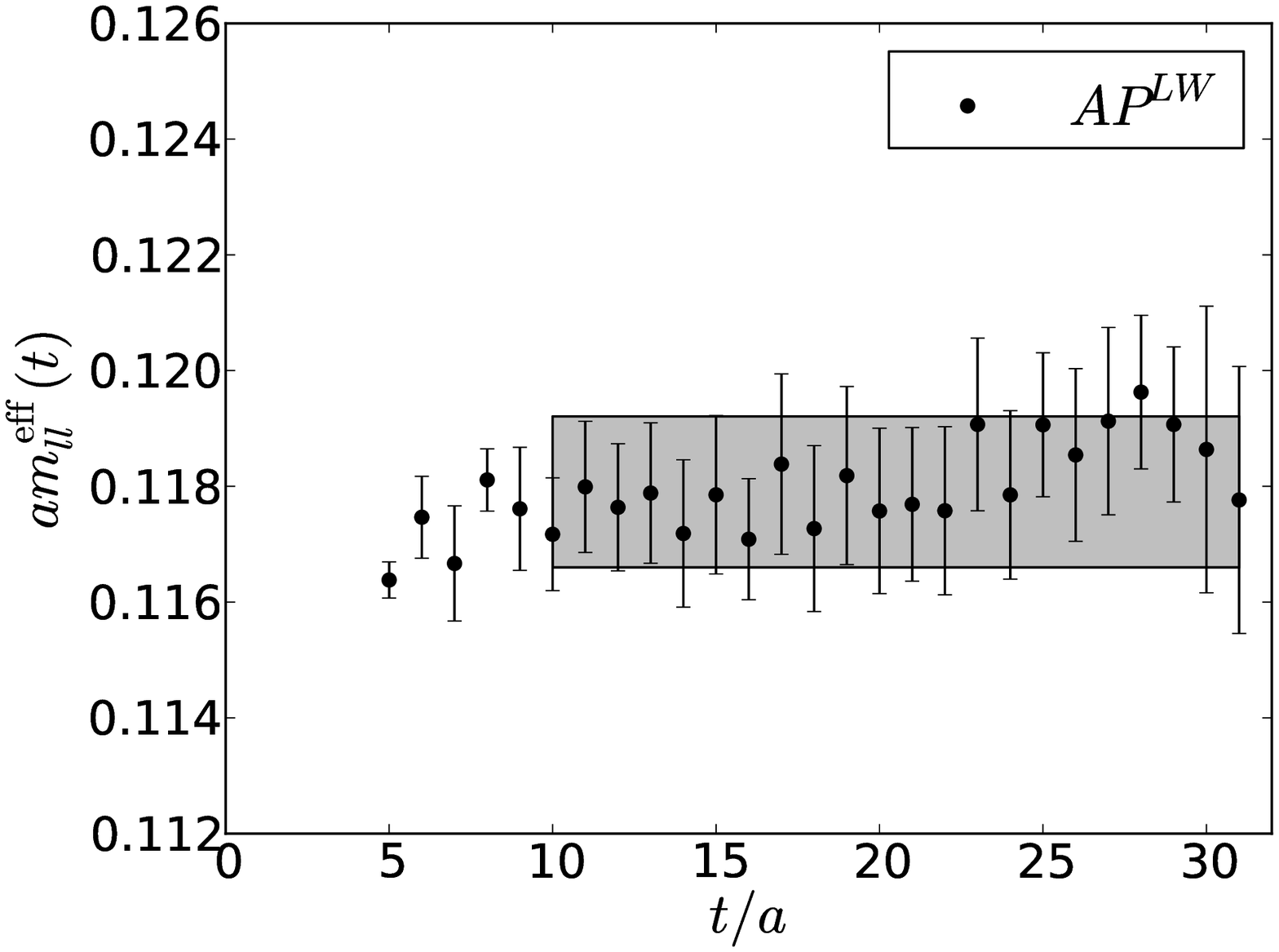}}
\caption{Effective $m_{ll}$ on the 32Ifine ensemble. We fit a common value of the mass to all three correlators.}
\label{fig:pion_eff_mass_32Ifine}
\end{figure}

\begin{figure}[h]
\centering
\subfigure{\includegraphics[width=0.5\textwidth]{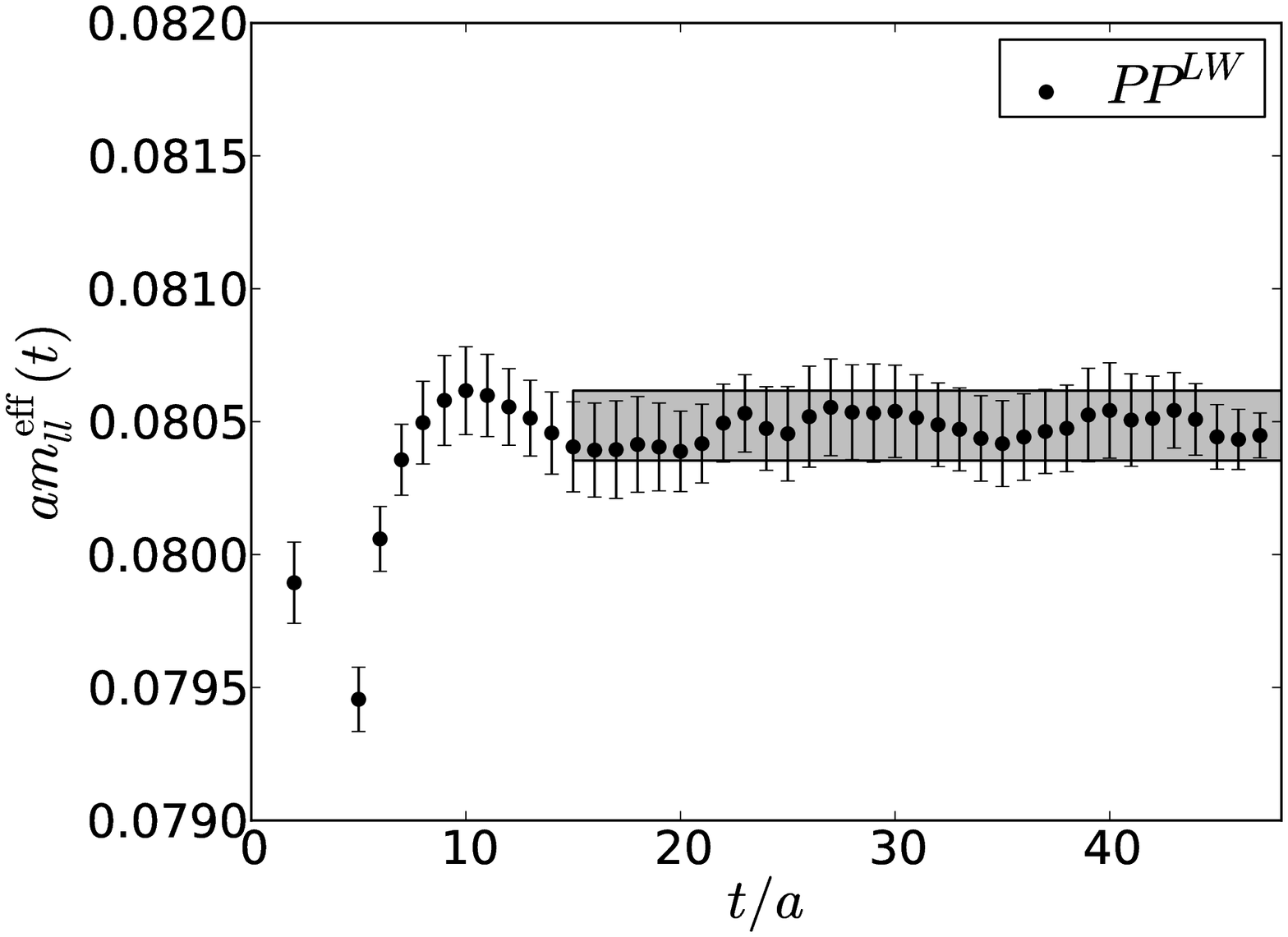} \includegraphics[width=0.5\textwidth]{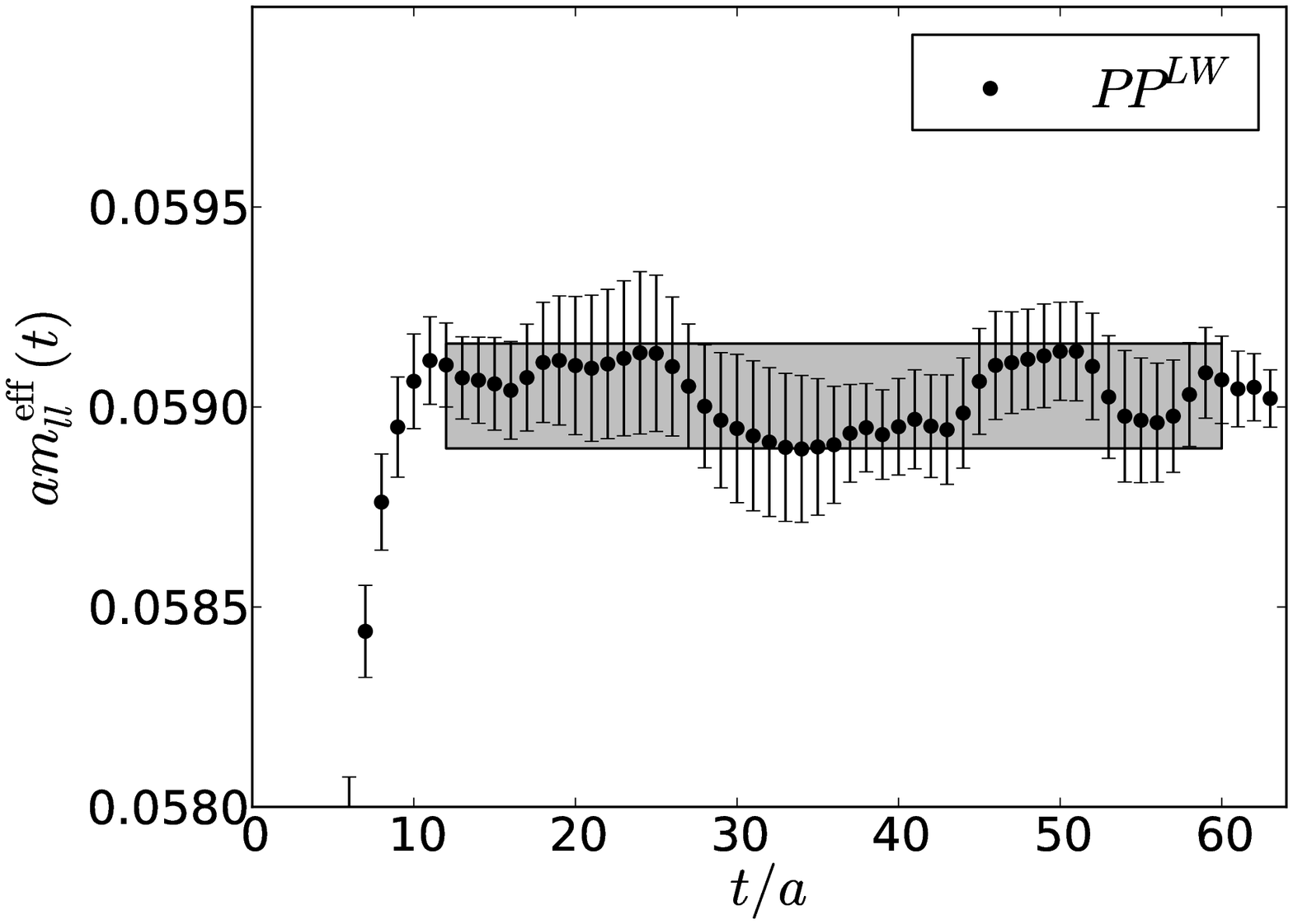}}
\subfigure{\includegraphics[width=0.5\textwidth]{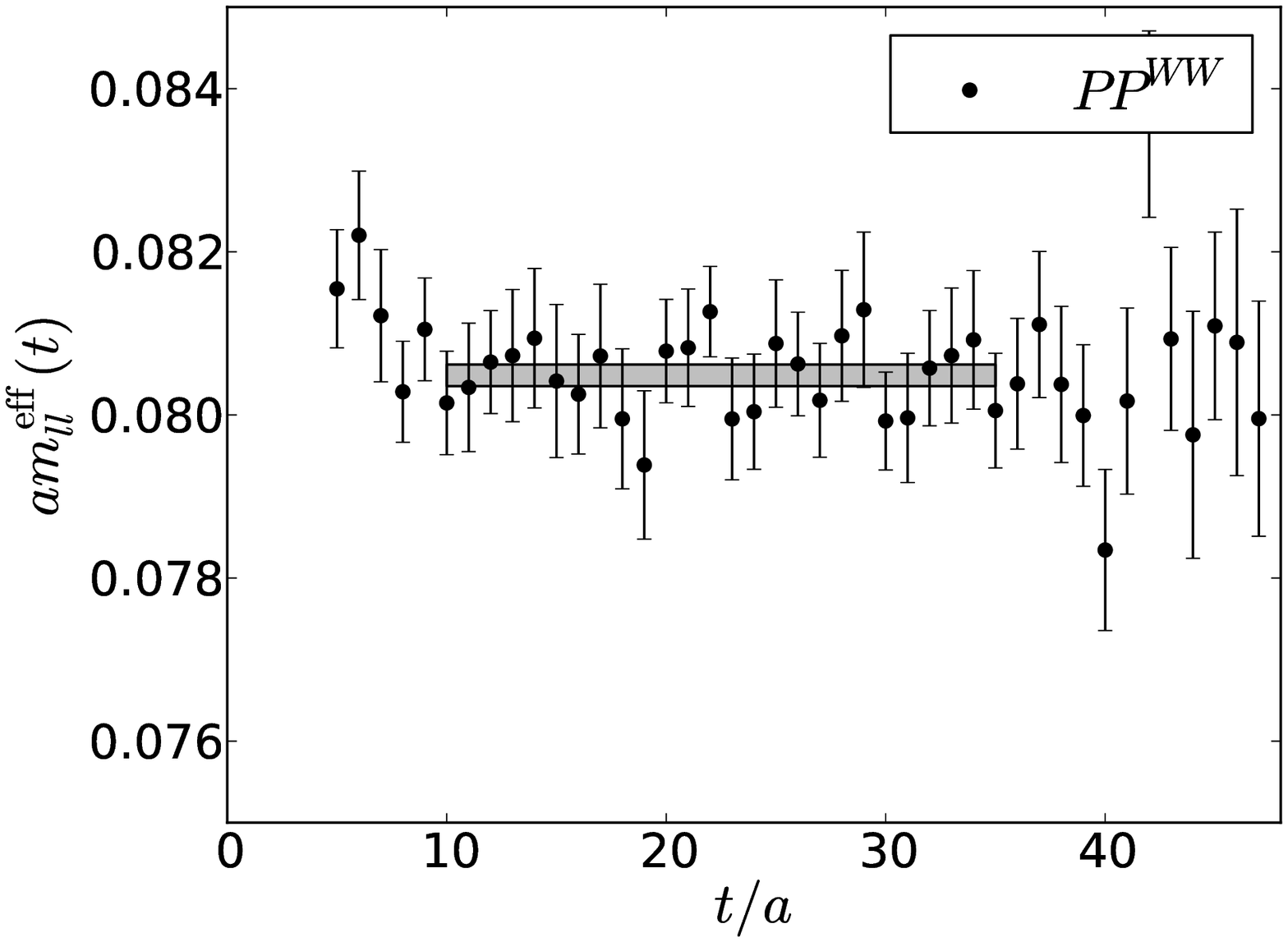} \includegraphics[width=0.5\textwidth]{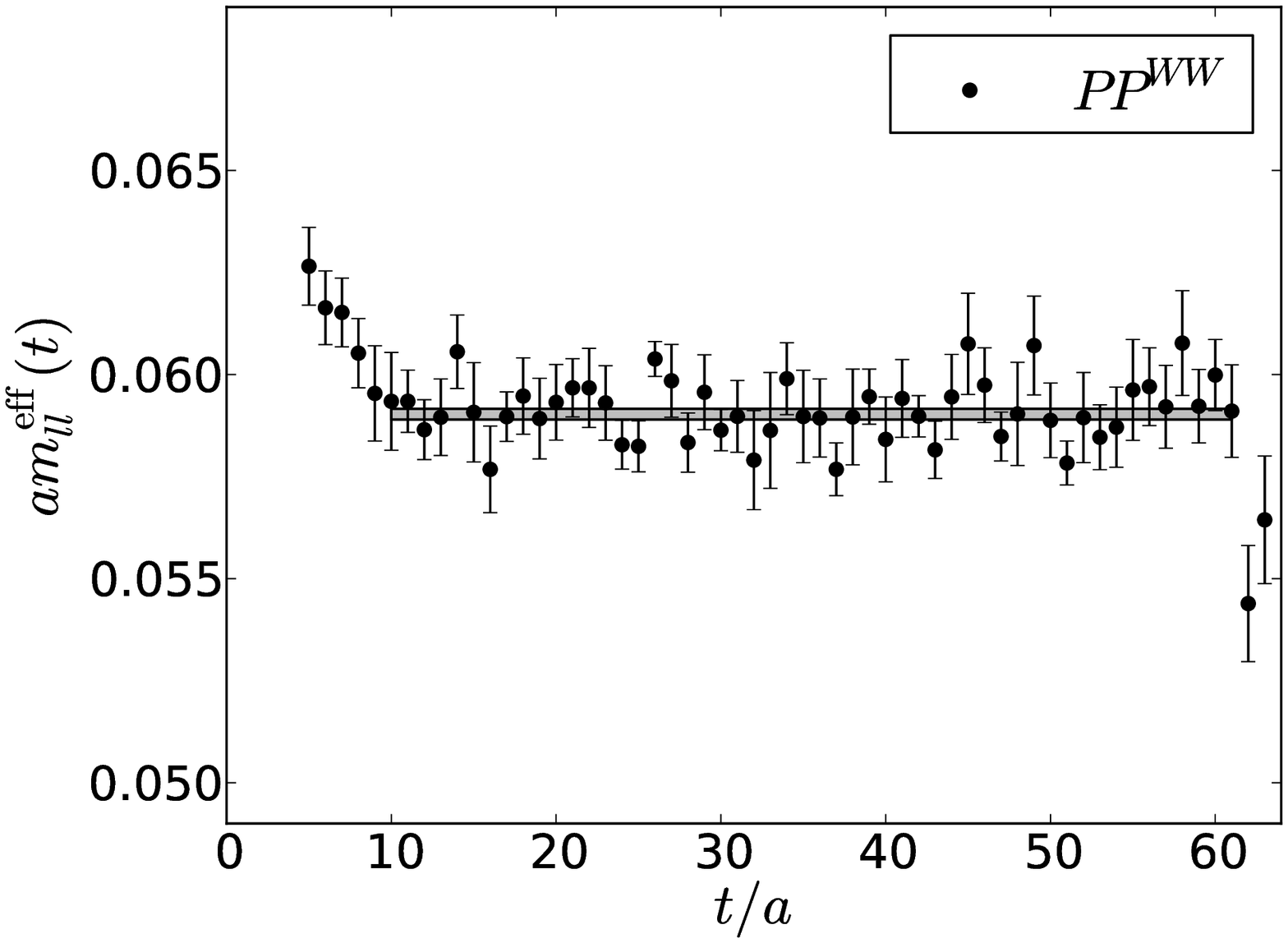}}
\subfigure{\includegraphics[width=0.5\textwidth]{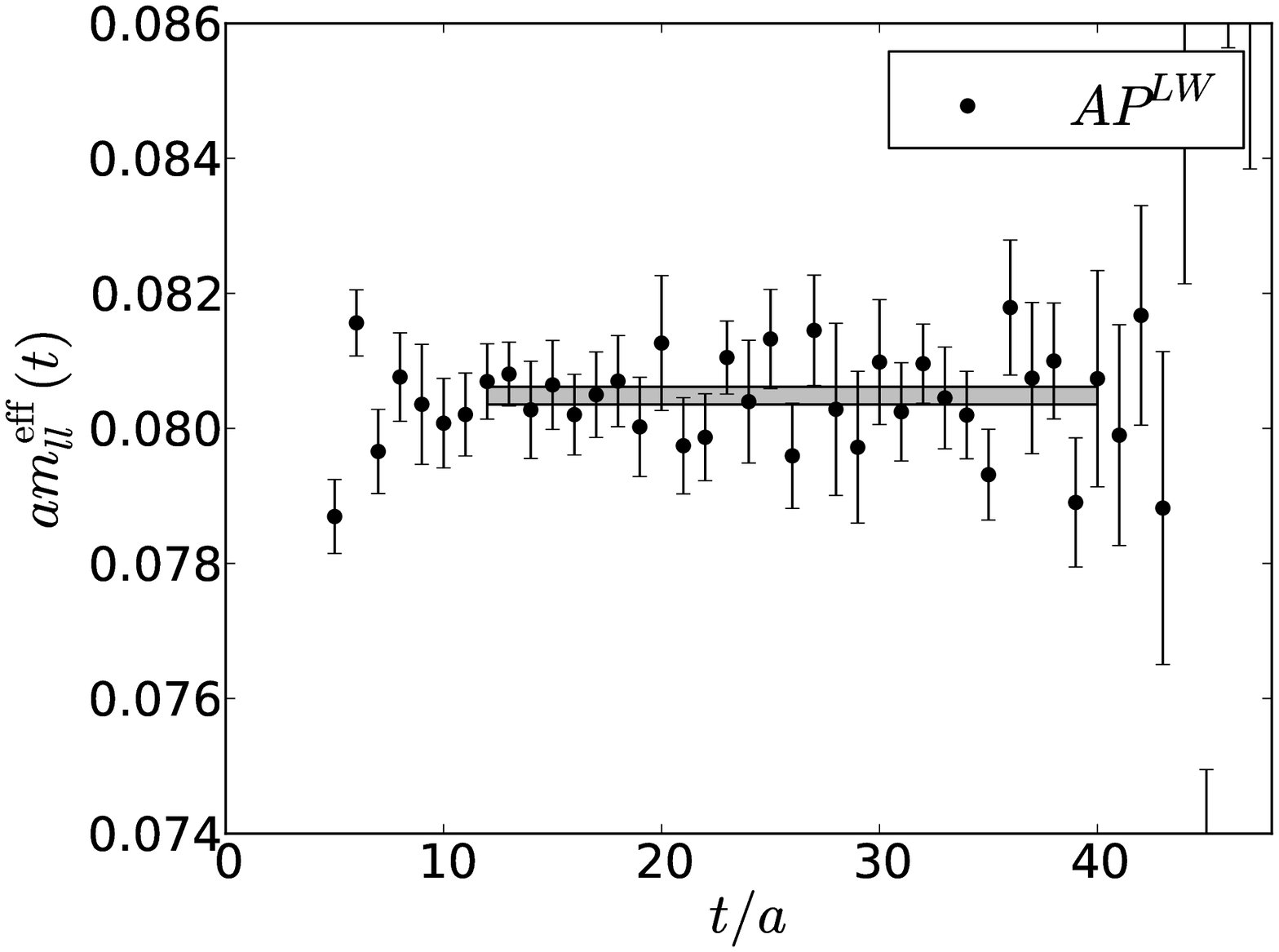} \includegraphics[width=0.5\textwidth]{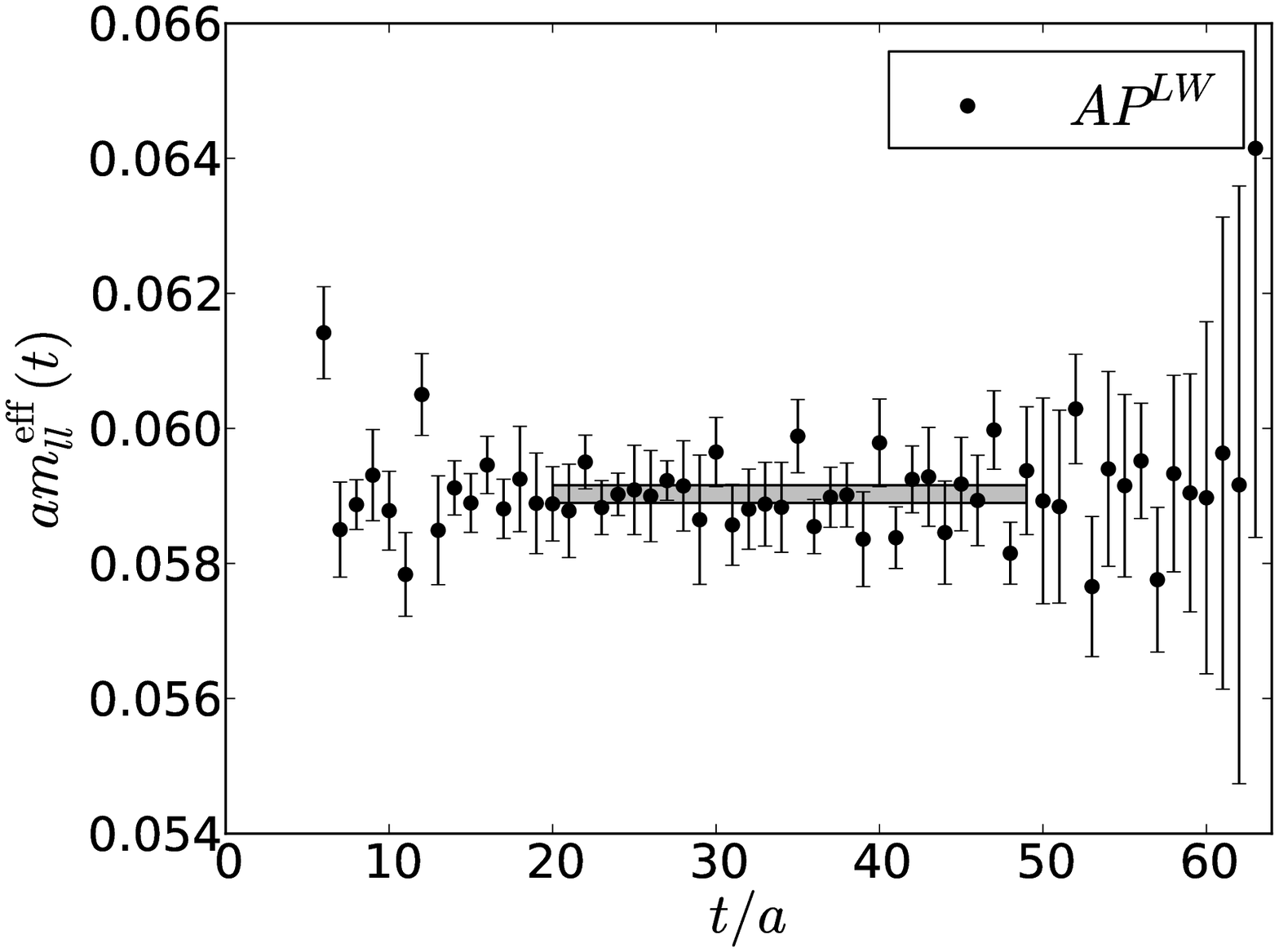}}
\caption{Effective $m_{ll}$ on the 48I (left column) and 64I (right column) ensembles. We fit a common value of the mass to all three correlators on a given ensemble.}
\label{fig:pion_eff_mass_48I_64I}
\end{figure}

\begin{figure}[h]
\centering
\subfigure{\includegraphics[width=0.5\textwidth]{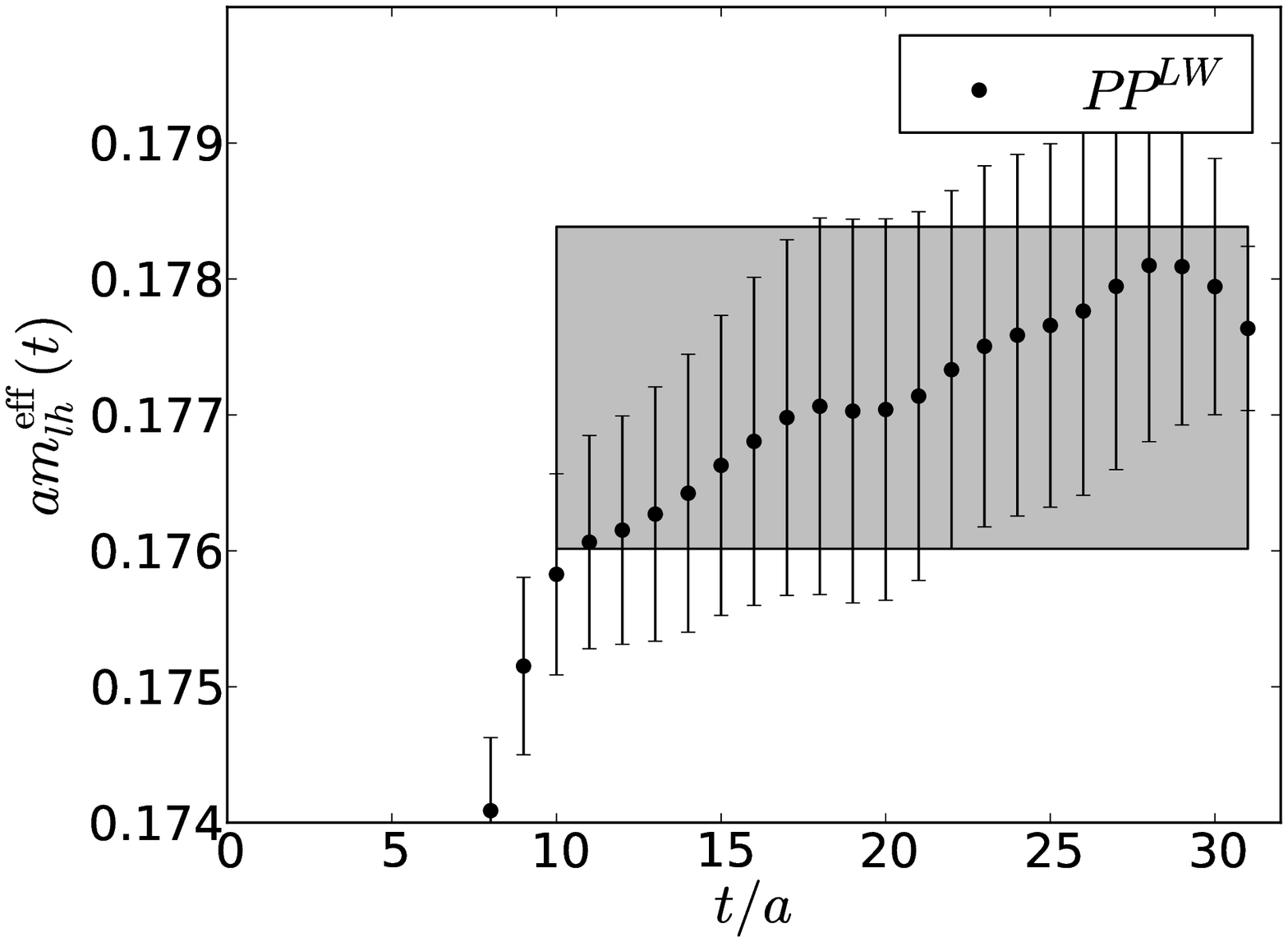} \includegraphics[width=0.5\textwidth]{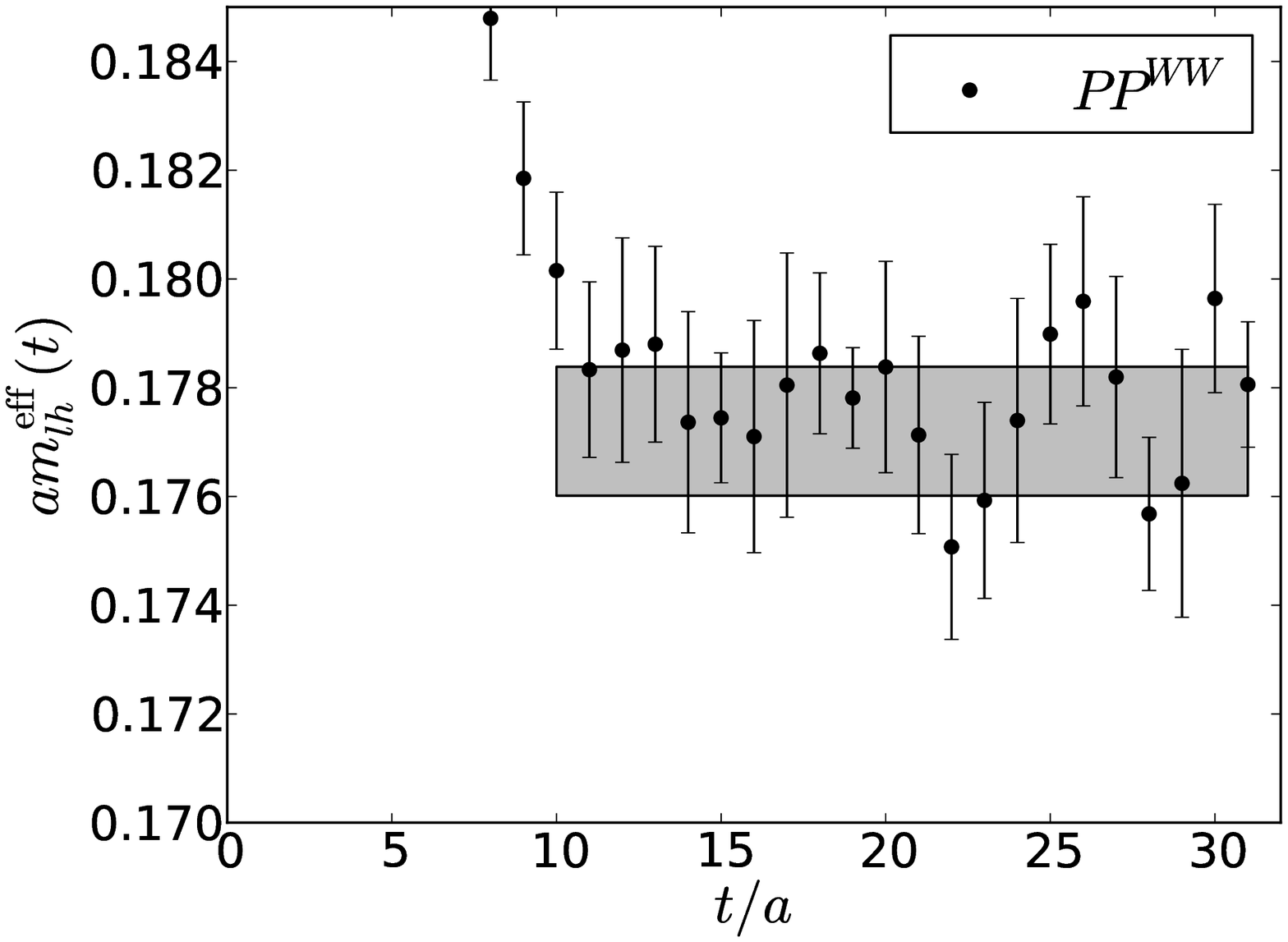}}
\subfigure{\includegraphics[width=0.5\textwidth]{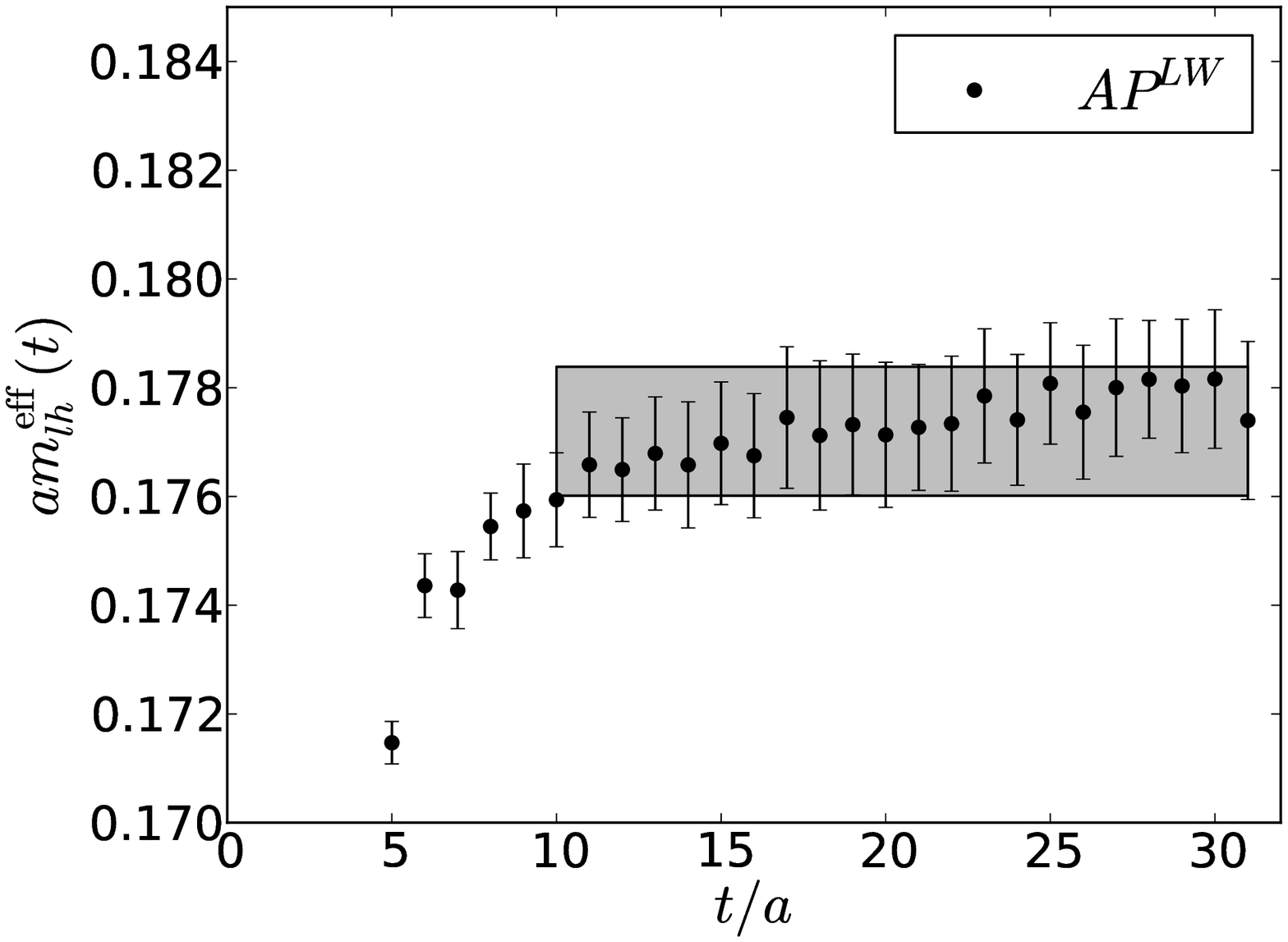}}
\caption{Effective $m_{lh}$ on the 32Ifine ensemble. We fit a common value of the mass to all three correlators.}
\label{fig:kaon_eff_mass_32Ifine}
\end{figure}

\begin{figure}[h]
\centering
\subfigure{\includegraphics[width=0.5\textwidth]{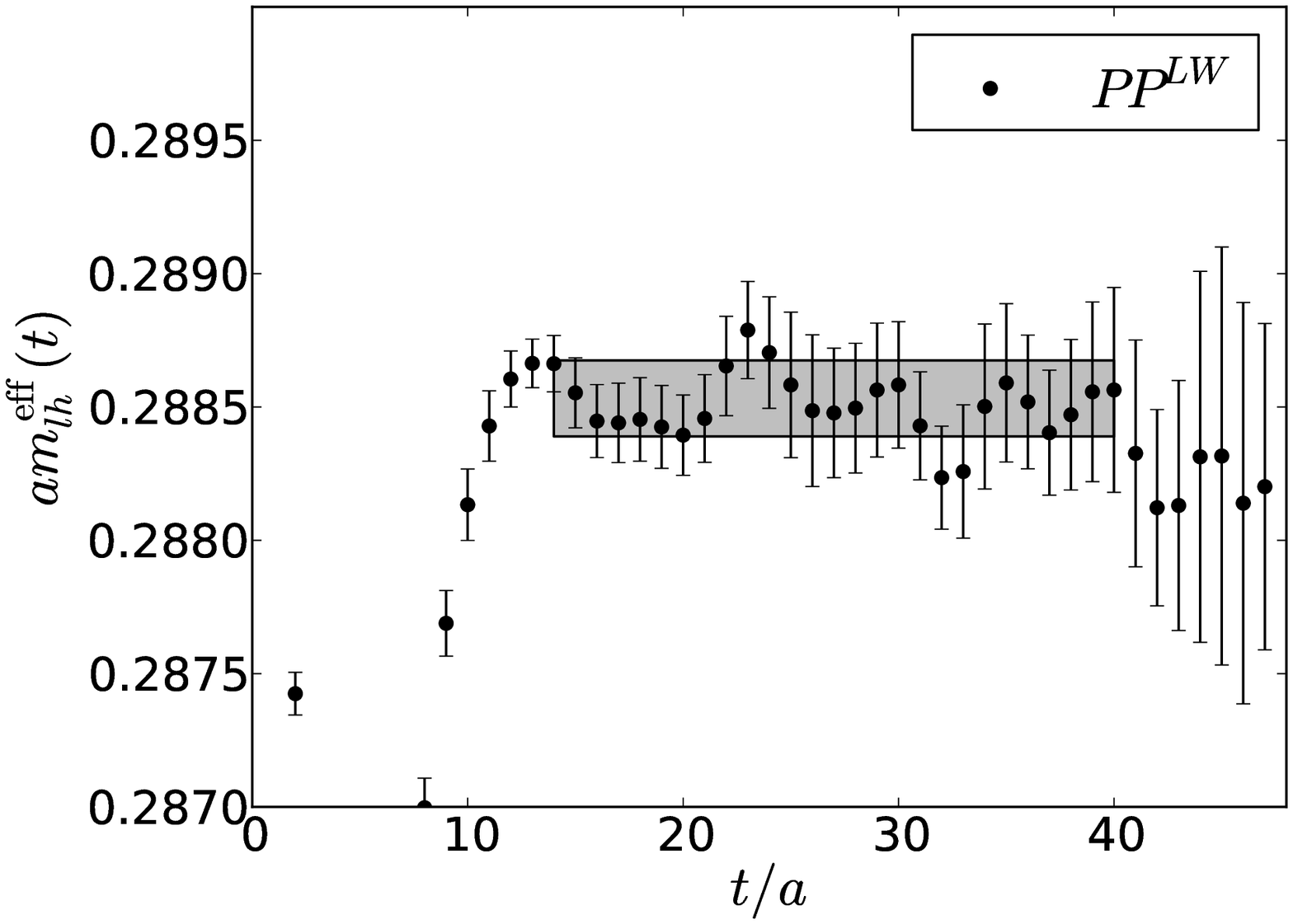} \includegraphics[width=0.5\textwidth]{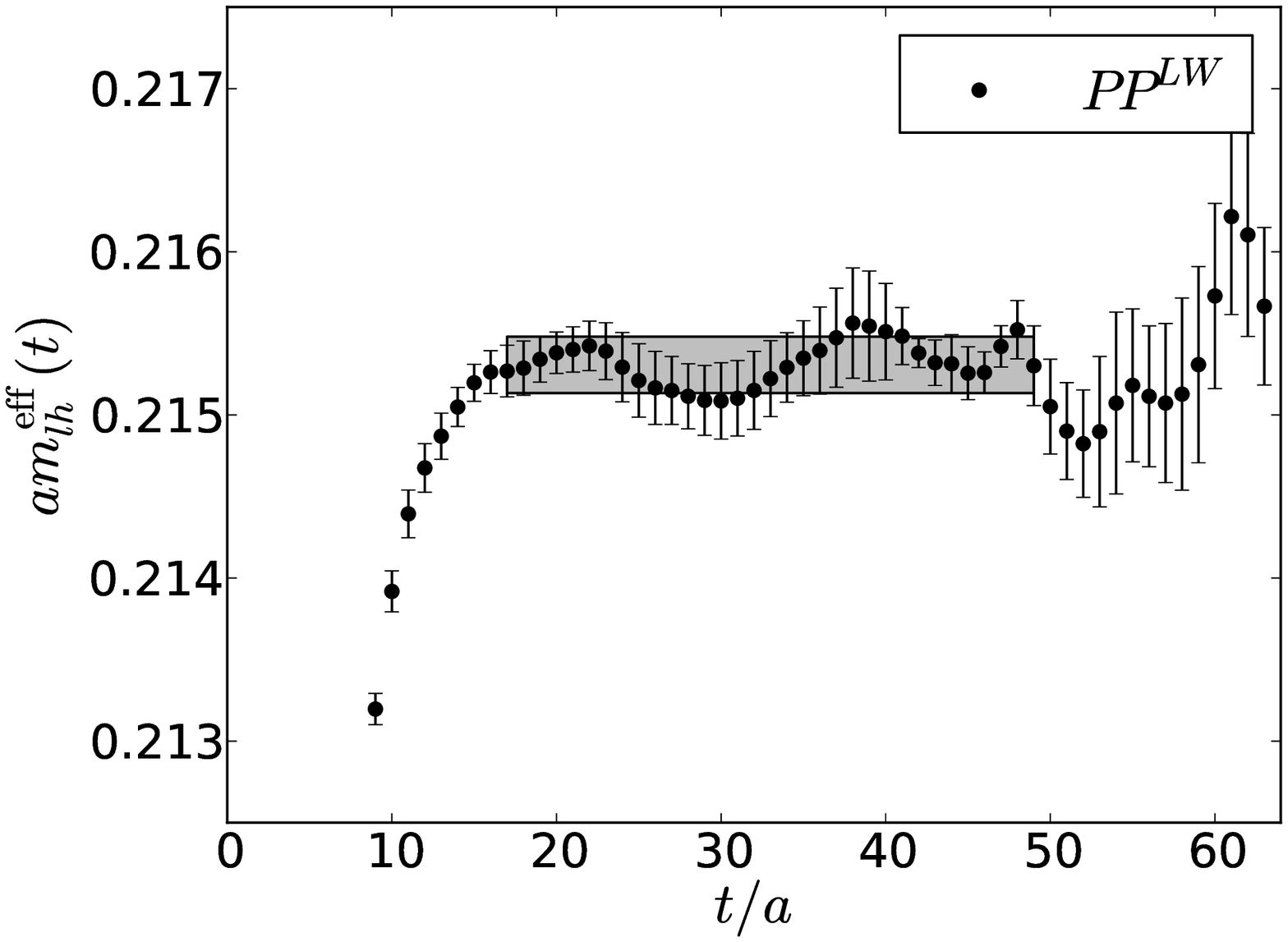}}
\subfigure{\includegraphics[width=0.5\textwidth]{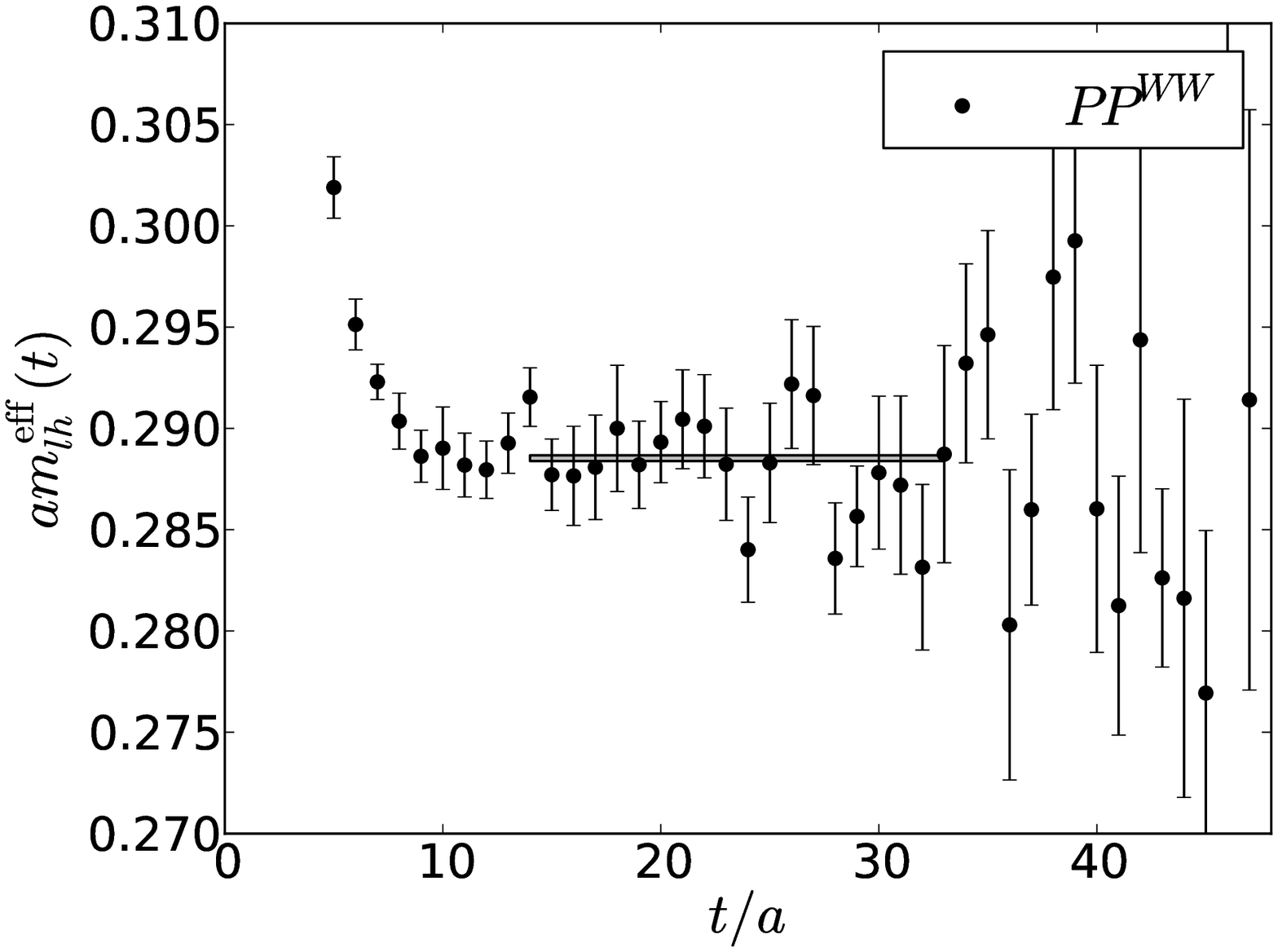} \includegraphics[width=0.5\textwidth]{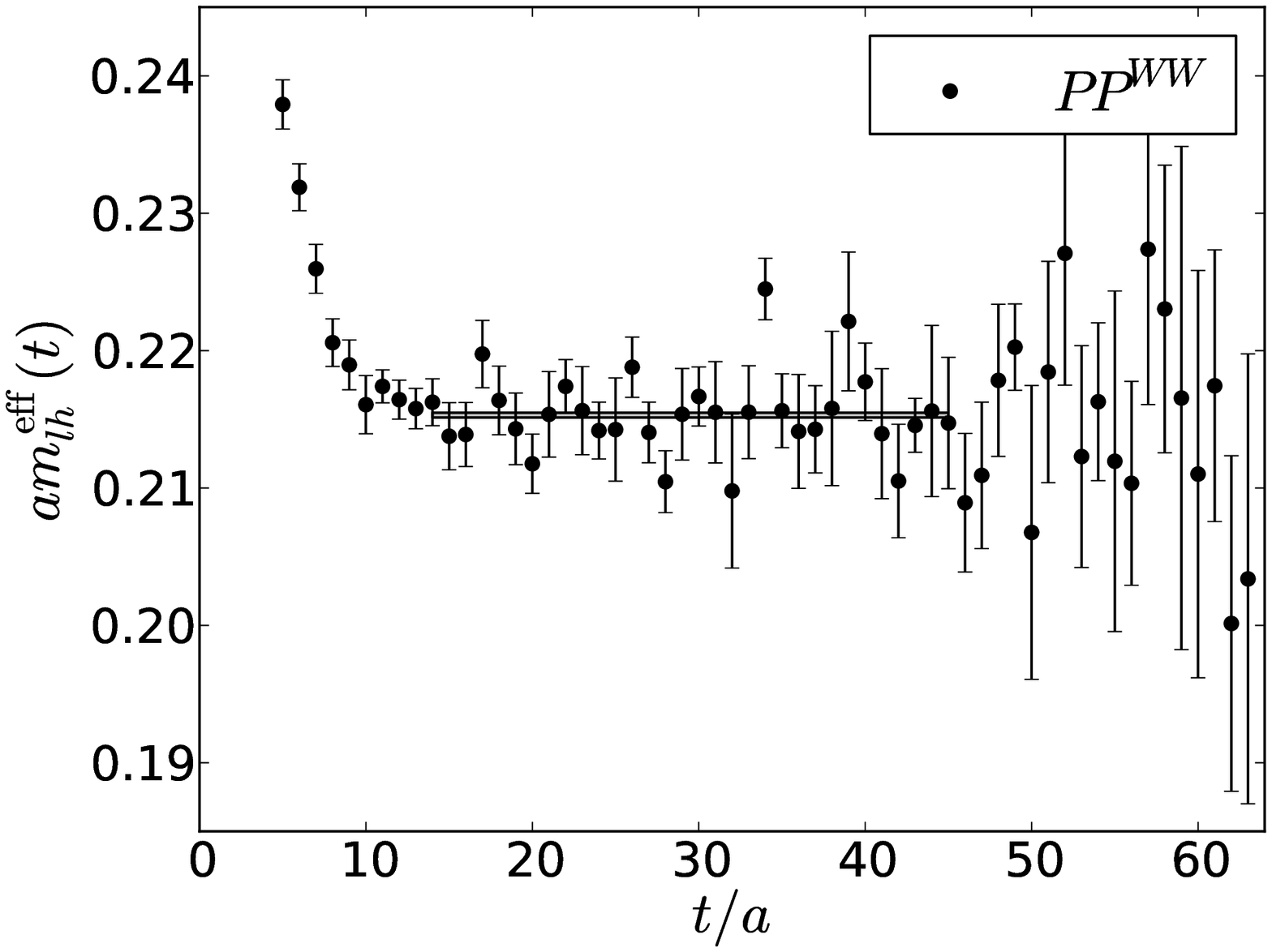}}
\subfigure{\includegraphics[width=0.5\textwidth]{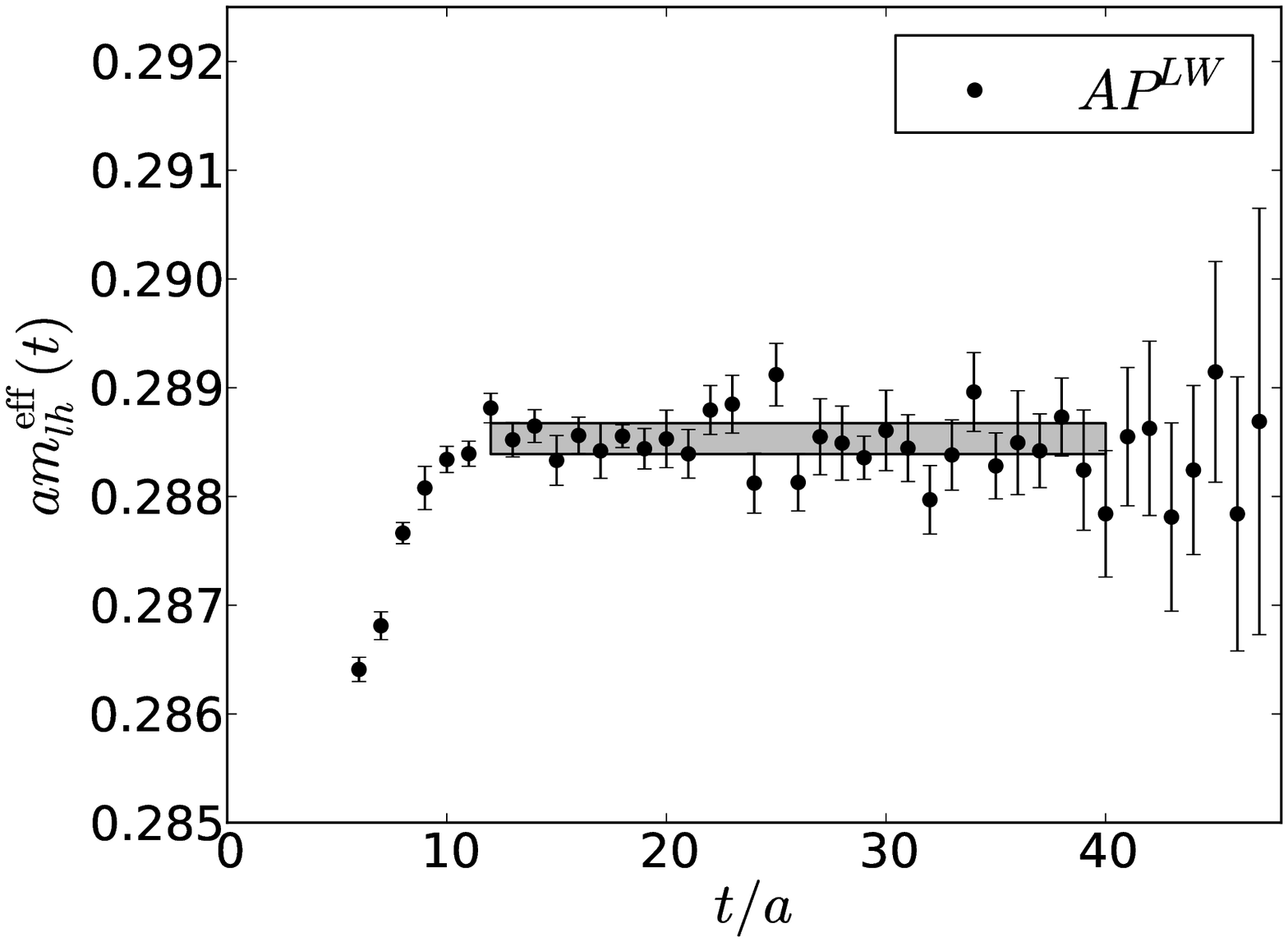} \includegraphics[width=0.5\textwidth]{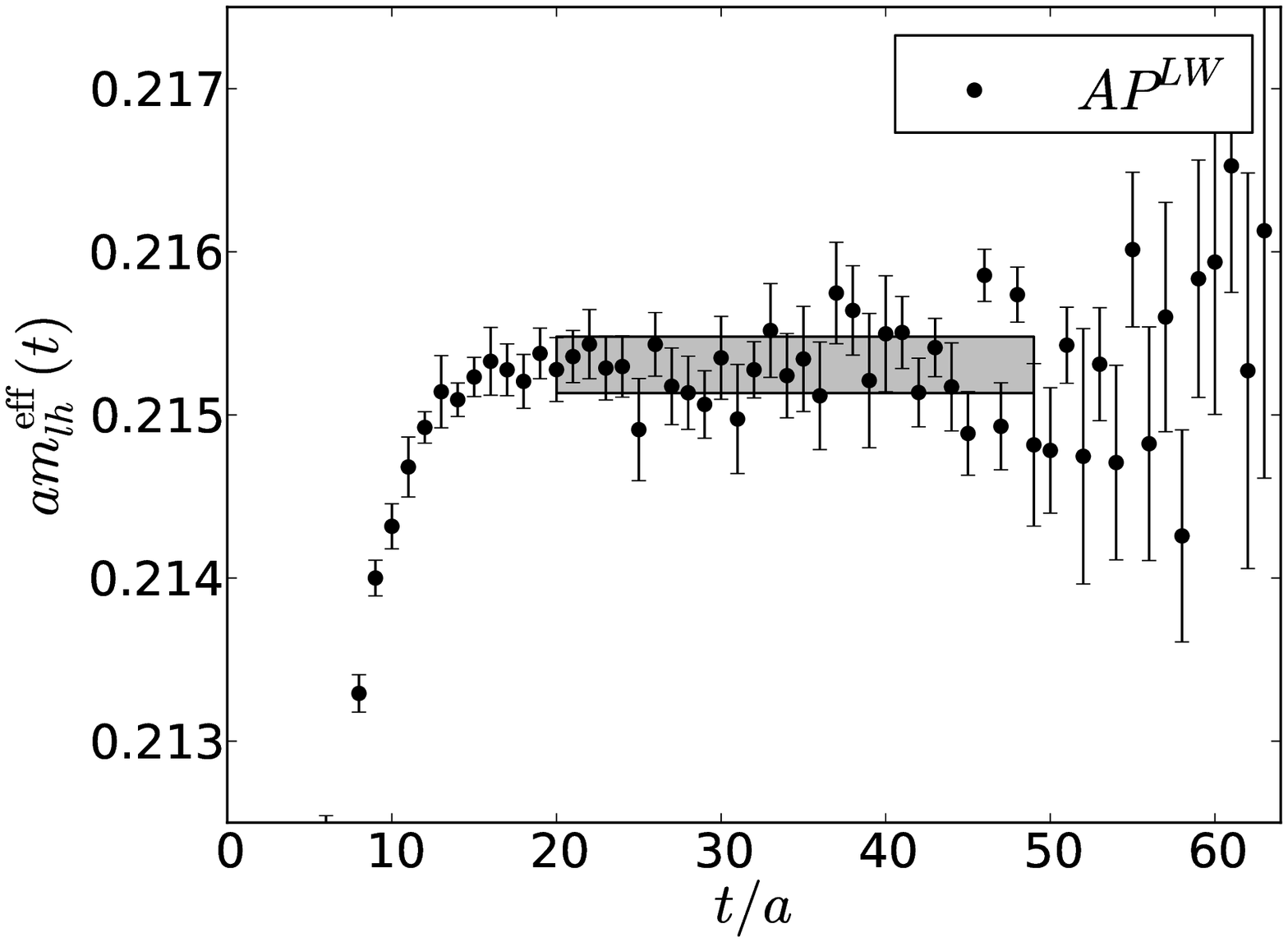}}
\caption{Effective $m_{lh}$ on the 48I (left column) and 64I (right column) ensembles. We fit a common value of the mass to all three correlators on a given ensemble.}
\label{fig:kaon_eff_mass_48I_64I}
\end{figure}

The masses of the pion and kaon at the simulated quark masses, denoted $m_{ll}$ and $m_{lh}$ respectively, were extracted by fitting to two-point functions of the form
\begin{equation}
\mathcal{C}^{s_{1} s_{2}}_{\mathcal{O}_{1} \mathcal{O}_{2}} (t) = \langle 0 | \mathcal{O}_{1}^{s_{1}}(t) \mathcal{O}_{2}^{s_{2}}(0) | 0 \rangle,.
\end{equation}
Here the subscripts indicate the interpolating operators and the superscripts denote the operator smearing used for the sink and source, respectively. In the following we have used Coulomb gauge-fixed wall ($W$) sources, and both local ($L$) and Coulomb gauge-fixed wall sinks. We extract the pseudoscalar meson masses by fitting three correlators simultaneously: $PP^{LW}$, $PP^{WW}$, and $AP^{LW}$, where $P$ is the pseudoscalar operator and $A$ is the temporal component of the axial current. These are fit to the following analytic form for the ground state of a Euclidean two-point correlation function:
\begin{equation} \label{eqn:ps_mass_fit_form}
\mathcal{C}_{\mathcal{O}_{1} \mathcal{O}_{2}}^{s_{1} s_{2}}(t) = \frac{\langle 0 | \mathcal{O}_{1}^{s_{1}} | X \rangle \langle X | \mathcal{O}_{2}^{s_{2}} | 0 \rangle}{2 m_{X} V} \left( e^{-m_{X} t} \pm e^{- m_{X} ( N_{t} - t ) } \right)\,,
\end{equation}
where the + (-) sign corresponds to the $PP$ ($AP$) correlators, and $X$ denotes the physical state to which the operators couple. In the following sections we use
\begin{equation}
\mathcal{N}_{\mathcal{O}_{1} \mathcal{O}_{2}}^{s_{1} s_{2}} \equiv \frac{\langle 0 | \mathcal{O}_{1}^{s_{1}} | X \rangle \langle X | \mathcal{O}_{2}^{s_{2}} | 0 \rangle}{2 m_{X} V}
\end{equation}
to denote the amplitude for a given correlator. The effective mass plots associated with these correlators, as well as the fitted masses, are shown in Figures~\ref{fig:pion_eff_mass_32Ifine},~\ref{fig:pion_eff_mass_48I_64I},~\ref{fig:kaon_eff_mass_32Ifine}, and~\ref{fig:kaon_eff_mass_48I_64I}.

\subsection{Pseudoscalar Decay Constants and Axial Current Renormalization}
\label{sec-decayconstmeas}

The pseudoscalar decay constants, $f_{\pi}$ and $f_{K}$, are defined in terms of the coupling of the pseudoscalar meson fields to the local four-dimensional axial current $A^{a}_{\mu}$:
\begin{equation} \label{eqn:pseudoscalar_decay} \begin{dcases}
\langle 0 | A^{a}_{\mu}(x) | \pi^{b}(p) \rangle = -i \delta^{a b} f_{\pi} p_{\mu} e^{i p \cdot x} \\
\langle 0 | A^{a}_{\mu}(x) | K^{b}(p) \rangle = -i \delta^{a b} f_{K} p_{\mu} e^{i p \cdot x}
\end{dcases}\,, \end{equation} 
where
\begin{equation}
A_{\mu}^{a}(x) = \overline{q}(x) \gamma_{\mu} \gamma_{5} \lambda^{a} q(x)
\end{equation}
is formed from the surface fields $q(x)$. In order to match this operator to the physically normalized Symanzik-improved axial operator $A_\mu^{Sa}$, we must derive the appropriate renormalization factor, $Z_A$. In the domain wall fermion formalism it is also possible to define a five-dimensional current $\mathcal{A}_{\mu}^{a}$ which satisfies the discretized partially-conserved axial current (PCAC) relation,
%
\begin{equation}\label{eqn:PCAC}
\Delta^-_\mu \langle \pi(x)| {\cal A}^a_\mu(y) \rangle = \langle \pi(x)|2m j^a_5(y) +2j^a_{5q}(y) \rangle,
\end{equation}
where $\Delta^-_\mu$ is the backwards discretized derivative. The factor relating this to the Symanzik current is denoted $Z_{\cal A}$. 

In the past, we took advantage of the fact that $Z_{\mathcal{A}} = 1 + {\cal O}(m_{\text{res}})$ to approximate $Z_A$ as $Z_A/Z_{\cal A}$, which can be computed directly via the following ratio:
\begin{equation} \label{eqn:za_ratio}
Z_{A} \approx \frac{Z_{A}}{Z_{\cal A}} = \frac{\langle 0 | \sum_{\vec{x}} \mathcal{A}_{\mu}^{a}(\vec{x},t) | \pi \rangle}{\langle 0 | \sum_{\vec{x}} A_{\mu}^{a}(\vec{x},t) | \pi \rangle}\,.
\end{equation}
The 5-D current ${\cal A}_\mu^a(x)$ is properly defined as the current carried by the link {\it between} $x$ and $x+\mu$, whereas the 4-D current $A_\mu^a(x)$ is defined on the lattice site $x$. The correlation functions $C(t+\half)=\sum_{\vec x}\langle {\cal A}_0^a(\vec x,t)\pi^a(\vec 0,0)\rangle$ and $L(t)=\sum_{\vec x}\langle A_0^a(\vec x,t)\pi^a(\vec 0,0)\rangle$, that one would use to compute the above ratio, are therefore not defined at the same temporal coordinate. By taking appropriate combinations of these correlators one can remove the associated ${\cal O}(a)$ error and reduce the ${\cal O}(a^2)$ error. $Z_{A}/Z_{\cal A}$ is then computed via the following ratio:~\cite{Blum:2000kn} 
\begin{equation}
R(t) = \frac{1}{2}\left[
\frac{C(t-\frac{1}{2}) +C(t+\frac{1}{2}) }{2 L(t)}
+
\frac{2 C(t+\frac{1}{2}) }{L(t-1)+L(t+1)}
\right]\,.
\end{equation}

While the 1--2\% $m_{\rm res}$ errors associated with the above determination of $Z_A$ could be neglected in our earlier work, where we were far from the chiral limit and the statistical errors were larger than in the current work, in Refs.~\cite{Aoki:2010dy} and~\cite{Sharpe:2007yd} it was shown that a better approximation could be obtained via the {\it vector} current. The local vector current operator formed from the domain wall surface fields is
\begin{equation}
V_{\mu}^{a}(x) = \overline{q}(x) \gamma_{\mu} \lambda^{a} q(x)\,,
\end{equation}
which is related to the Symanzik vector current $V_\mu^{Sa}$ by a renormalization coefficient $Z_V$ which was shown to be equal to $Z_A$ up to terms ${\cal O}(m_{\rm res}^2)$~\cite{Aoki:2010dy}. There is also a five-dimensional conserved vector current ${\cal V}_\mu^a$ for which the renormalization factor, $Z_{\cal V}$, is unity, and we can obtain a significantly better approximation to $Z_A$ by computing $Z_V/Z_{\cal V}$ on the lattice:
\begin{equation}
Z_{A} \approx \frac{Z_V}{Z_{\cal V}} = \frac{\langle 0 | \sum_{\vec{x},i} {\cal V}_i^a(\vec{x},t) V_i^a(\vec 0,0)| 0 \rangle}{\langle 0 | \sum_{\vec{x},i} V_i^{a}(\vec{x},t) V_i^a(\vec 0,0) | 0 \rangle}\,.
\end{equation}

Below we determine both $Z_A/Z_{\cal A}$ and $Z_V/Z_{\cal V}$, but use only the latter to renormalize our decay constants.

\subsubsection{Determination of $Z_A/Z_{\cal A}$}

\begin{figure}[h]
\centering
\subfigure{\includegraphics[width=0.5\textwidth]{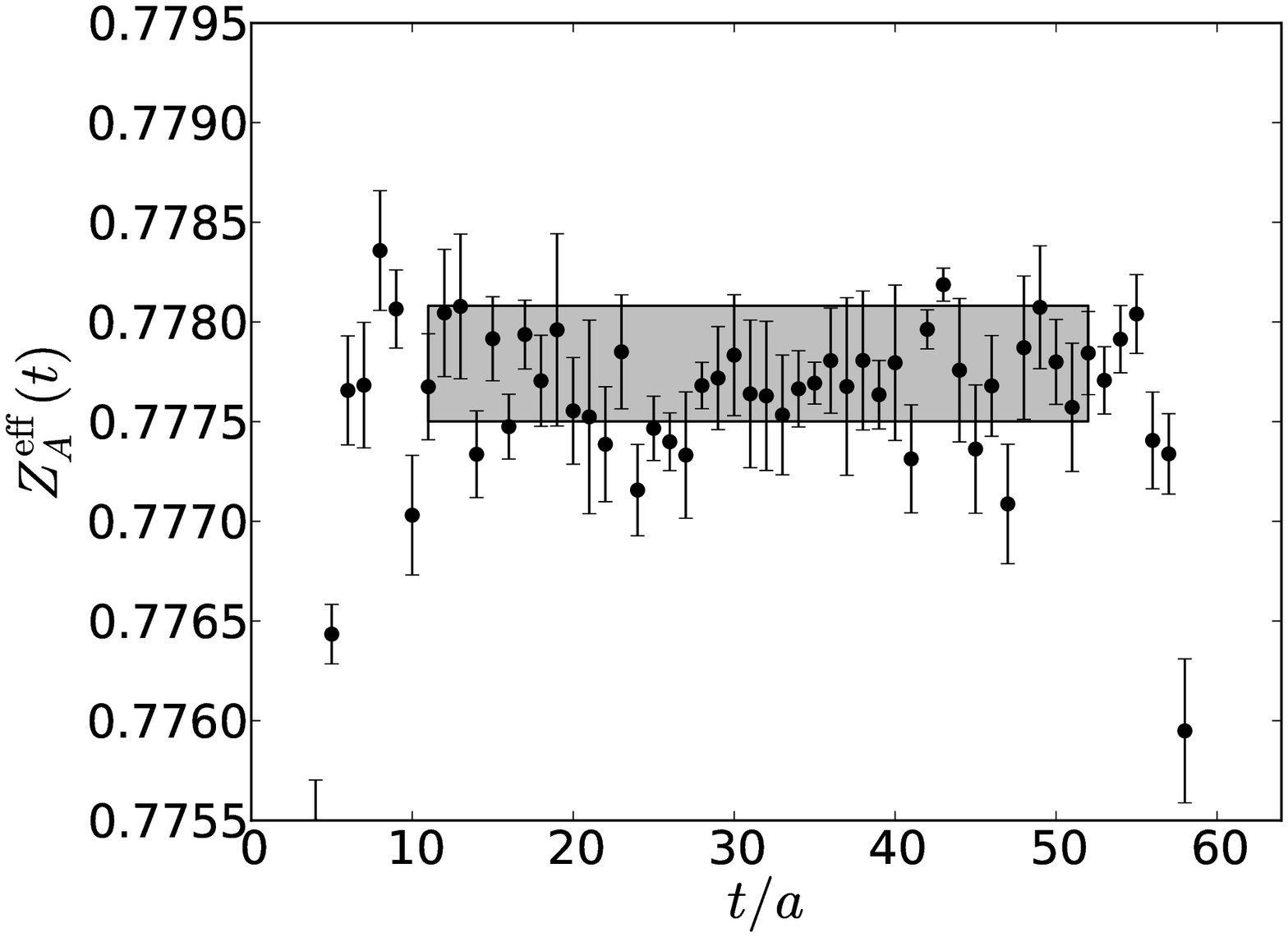} \includegraphics[width=0.5\textwidth]{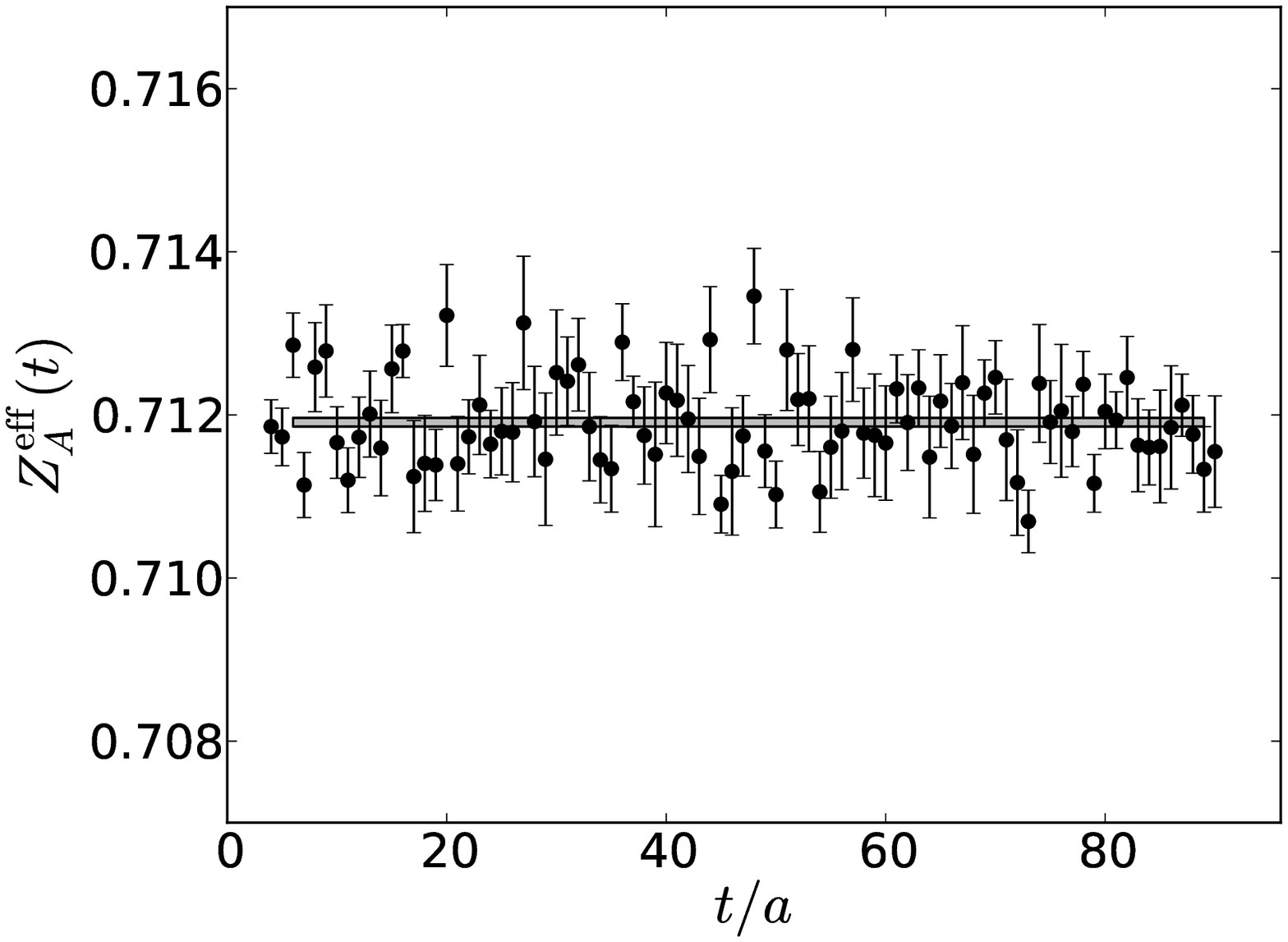}}
\subfigure{\includegraphics[width=0.5\textwidth]{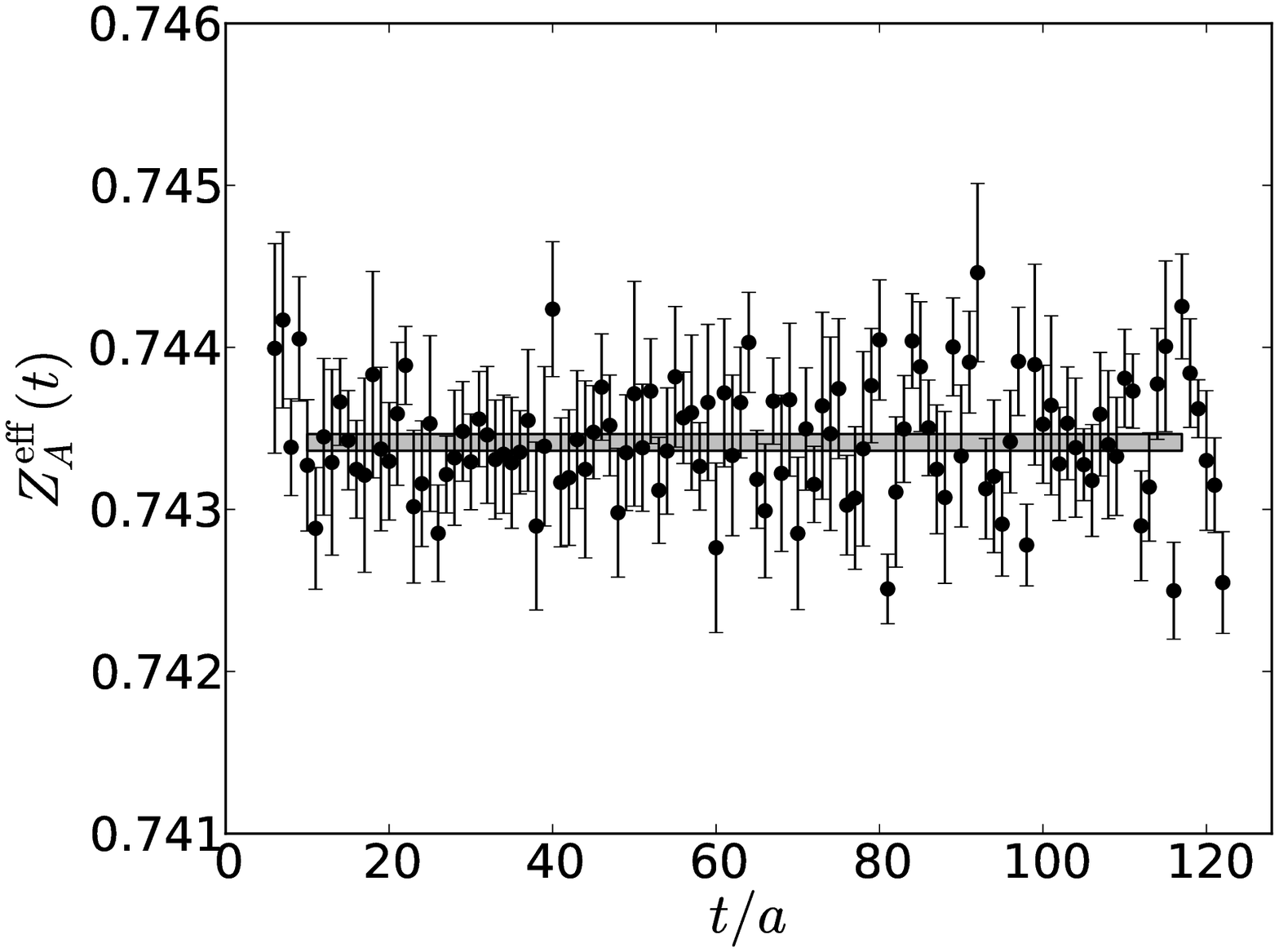}}
\caption{Effective $Z_{A}$ on the 32Ifine (top left), 48I (top right), and 64I (bottom) ensembles.}
\label{fig:za}
\end{figure}

We introduce a practical approach to the conserved axial current for \mobius fermions in Appendix~\ref{appendix-mobiusconservedcurrents} and Ref.~\cite{Boyle:2014hxa}. For the numerical determination of $Z_A$, the explicit construction of the current, used in Eq.~\eqref{eqn:za_ratio}, can be avoided with an alternate determination that utilizes the ratio of the divergences of the four-dimensional and five-dimensional axial currents:
\begin{equation} \label{eqn:za_fit_ratio}
Z_{A} \approx \frac{Z_{A}}{Z_{\cal A}} = \frac{\langle 0 | \sum_{\vec{x}} \partial_{\mu} \mathcal{A}_{\mu}^{a}(\vec{x},t) | \pi \rangle}{\langle 0 | \sum_{\vec{x}} \partial_{\mu} A_{\mu}^{a}(\vec{x},t) | \pi \rangle} = \frac{2 m \langle 0 | \sum_{\vec{x}} j^{a}_{5}(\vec{x},t) | \pi \rangle + 2 \langle 0 | \sum_{\vec{x}} j^{a}_{5 q}(\vec{x},t) | \pi \rangle}{\langle 0 | \sum_{\vec{x}} \partial_{\mu} A_{\mu}^{a}(\vec{x},t) | \pi \rangle}\,,
\end{equation}
where the last equality follows from the PCAC relation, Eq.~\eqref{eqn:PCAC}. We extract $Z_{A}$ from our lattice data using the improved ratio
\begin{equation} \begin{dcases}
C_{\cal A}(t) \equiv \langle 0 | \sum_{\vec{x}} \partial_{\mu} \mathcal{A}_{\mu}^{a}(\vec{x},t) | \pi \rangle \\
C_{A} \left( t-\frac{1}{2} \right) \equiv \langle 0 | \sum_{\vec{x}} \partial_{\mu} A_{\mu}^{a}(\vec{x},t) | \pi \rangle \\
Z_{A}^{\text{eff}}(t) = \frac{1}{2} \left[ \frac{C_{\cal A}(t-1) + C_{\cal A}(t)}{2 C_{A}(t-\frac{1}{2})} + \frac{2 C_{\cal A}(t)}{C_{A}(t+\frac{1}{2}) + C_{A}(t-\frac{1}{2})} \right]
\end{dcases}\,, \end{equation}
which is also constructed to minimize errors at $\mathcal{O}(a^{2})$~\cite{Blum:2000kn}. The translation by $\frac{1}{2}$ in the argument of the correlation function associated with $A_{\mu}^{a}$ arises from the divergence. The five-dimensional current $\mathcal{A}_{\mu}^{a}$, by contrast, is defined on the links between lattice sites, so its divergence is centered on the lattice. In Figure~\ref{fig:za} we plot the effective $Z_{A}$ and fit on each ensemble. 

\subsubsection{Determination of $Z_V/Z_{\cal V}$}
\label{sec-zvdeterm}

\begin{figure}[h]
\centering
\subfigure{\includegraphics[width=0.5\textwidth]{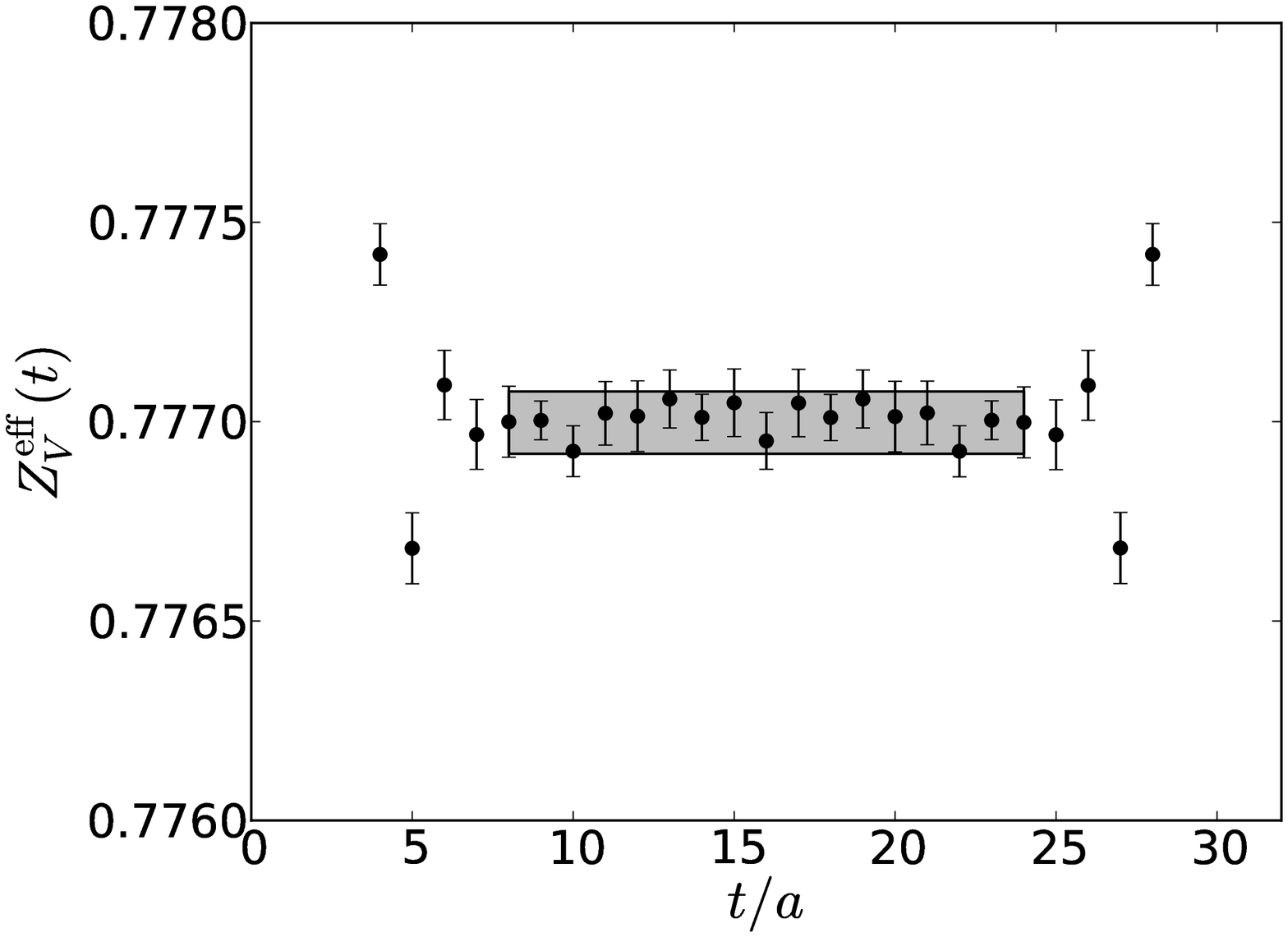} \includegraphics[width=0.5\textwidth]{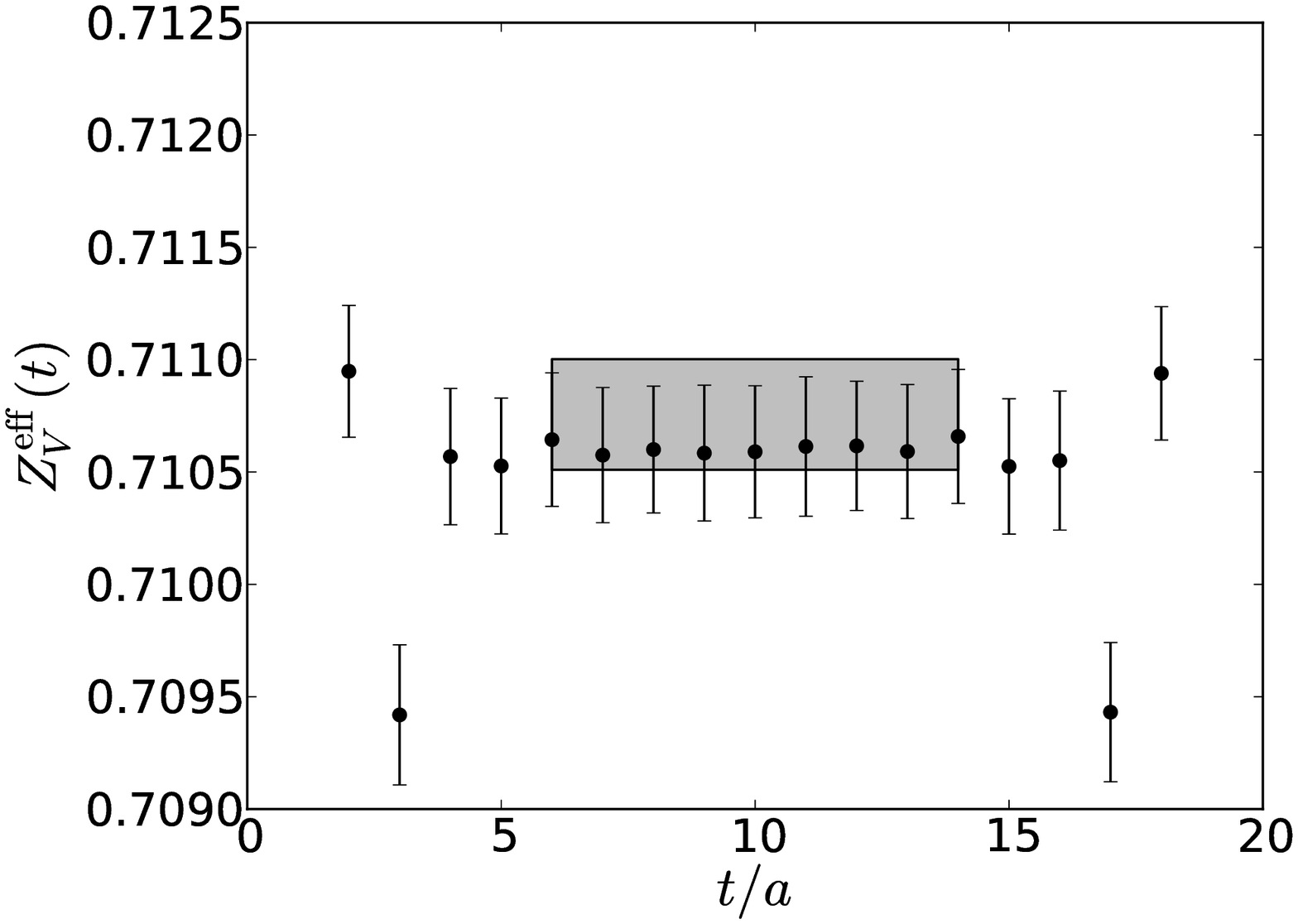}}
\subfigure{\includegraphics[width=0.5\textwidth]{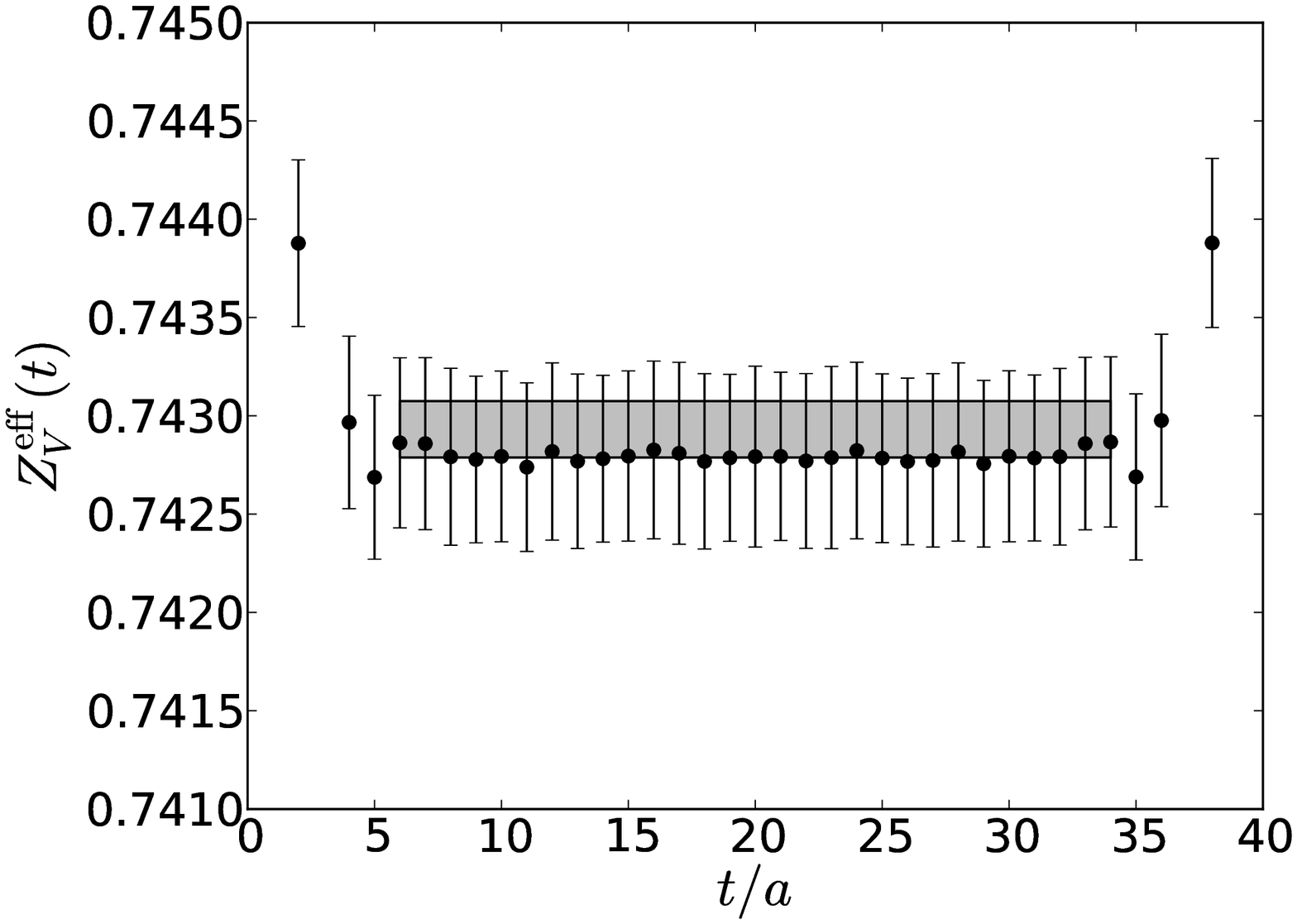}}
\caption{Effective $Z_{V}$ on the 32Ifine (top left), 48I (top right), and 64I (bottom) ensembles, for $\pi-\pi$ separations of 32 time units, 20 time units, and 40 time units, respectively. Note that in each case the fit is performed using several $\pi-\pi$ separations, not just the separation plotted here.}
\label{fig:zv}
\end{figure}

Since the relatively noisy $\rho$ meson is the lightest state to which the vector current couples, computing $Z_{V}$ accurately requires a different approach from that used for $Z_A$ (Eq.~\eqref{eqn:za_fit_ratio}). Instead, we calculate the pion electromagnetic form factors $f_{ll}^{+}(q^{2})$ and $f_{ll}^{-}(q^{2})$, defined by the matrix element
\begin{equation} \label{eqn:pion_em_ff}
\langle \pi(p_{1}) | V_{\mu} | \pi(p_{2}) \rangle = f_{ll}^{+}(q^{2}) \left( p_{2} + p_{1} \right)_{\mu} + f_{ll}^{-}(q^{2}) \left( p_{2} - p_{1} \right)_{\mu}\,,
\end{equation} 
where $q = p_{2} - p_{1}$ is the momentum transfer. Current conservation implies $f_{ll}^{-}(q^2) = 0$ for all $q^2$, leaving only the vector form factor, $f_{ll}^{+}$. For two pions at rest, $f_{ll}^{+}(0) = 1$, and we can fit $Z_{V}$ from the temporal component of Eq.~\eqref{eqn:pion_em_ff}. We fit to the ratio
\begin{equation} \label{eqn:zv_ratio}
\frac{\tilde{\mathcal{C}}^{WW}_{PP}(t_{\text{snk}})}{\mathcal{C}_{PVP}(t_{\text{src}},t,t_{\text{snk}})} \stackrel{t,|t_{\text{src}}-t_{\text{snk}}| \gg 1}{\cong} Z_{V}\,,
\end{equation}
where 
\begin{equation}
\tilde{\mathcal{C}}^{WW}_{PP}(t) = \mathcal{C}^{WW}_{PP}(t) - \frac{1}{2} \mathcal{C}^{WW}_{PP} \left( \frac{N_{t}}{2} \right) e^{-m_{ll}(N_{t}/2-t)}
\end{equation}
is the pion two-point function, Eq.~\eqref{eqn:ps_mass_fit_form}, with the around-the-world state removed using the fitted pion mass, and $\mathcal{C}^{WW}_{PVP}(t_{\text{src}},t,t_{\text{snk}})$ is the three-point function defined by the matrix element, Eq.~\eqref{eqn:pion_em_ff}. On the 32Ifine and 48I ensembles, this matrix element was computed for all $\pi-\pi$ separations, $t_{\rm sink}-t_{\rm src}$, that are a multiple of 4. For the 64I ensemble we computed on separations that are multiples of 5. We determine the ranges of $\pi-\pi$ separations to use in the fit by plotting the midpoint of Eq.~\eqref{eqn:zv_ratio} as a function of the $\pi-\pi$ separation on each ensemble and looking for a plateau: based on this analysis we chose to include $\pi-\pi$ separations in the range 16--32 on the 32Ifine ensemble, 12--24 on the 48I ensemble, and 15--40 on the 64I ensemble. In Figure~\ref{fig:zv} we illustrate this method by plotting Eq.~\eqref{eqn:zv_ratio} for a single $\pi-\pi$ separation included in the fit, as well as the fitted value for $Z_{V}$, on each ensemble.

\subsubsection{Determination of the Decay Constants}

\begin{figure}[h]
\centering
\subfigure{\includegraphics[width=0.5\textwidth]{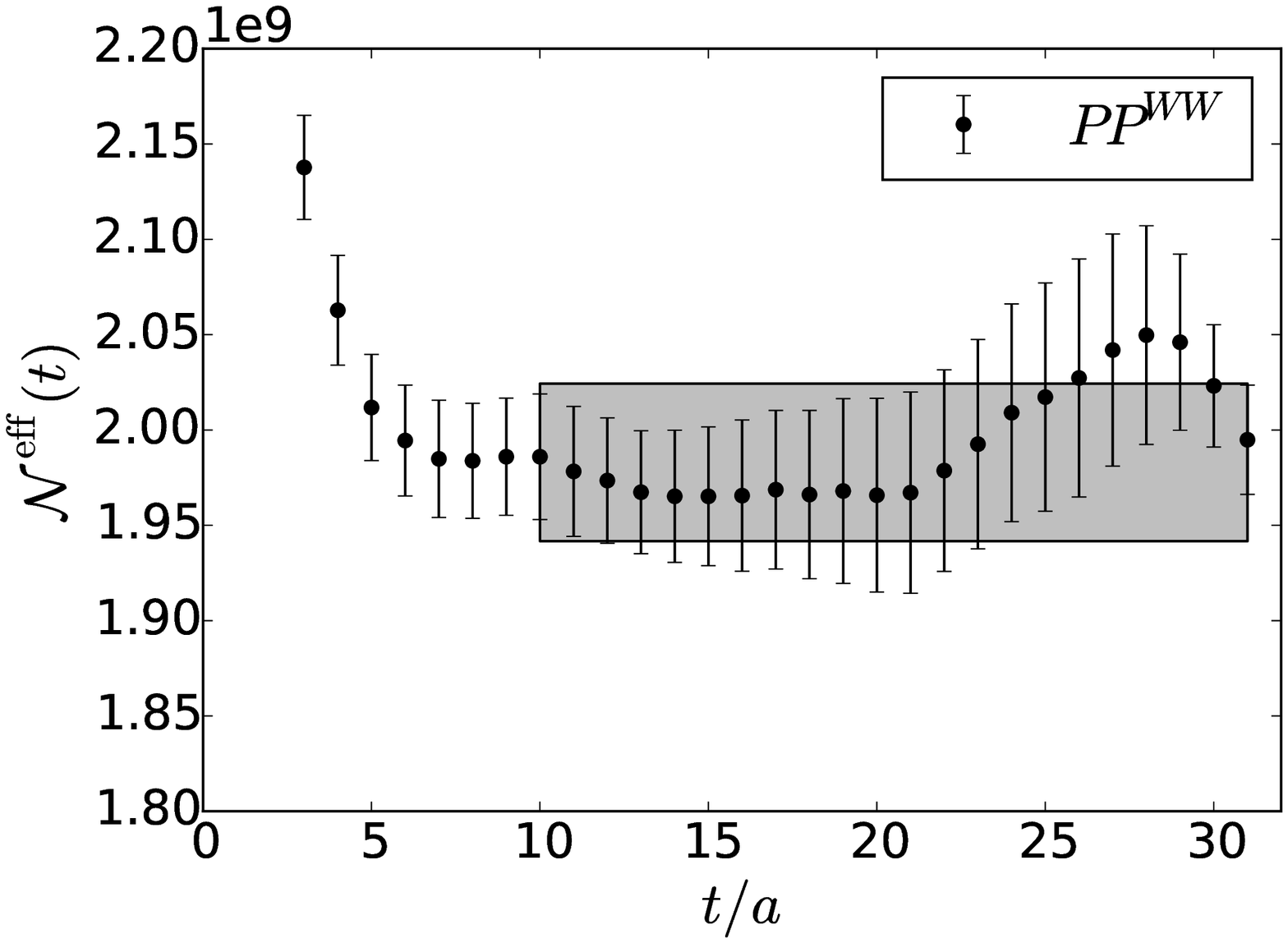} \includegraphics[width=0.5\textwidth]{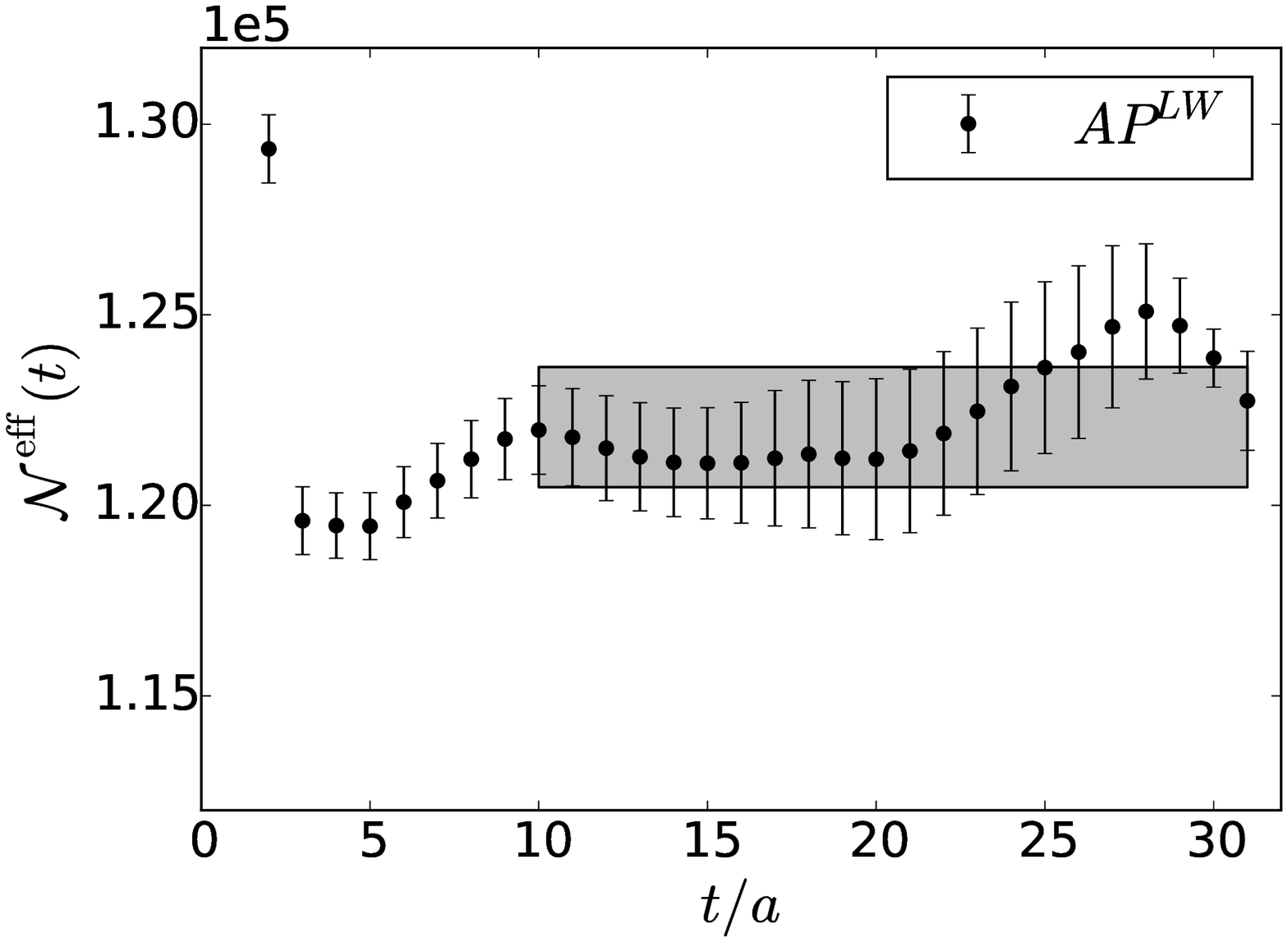}}
\subfigure{\includegraphics[width=0.5\textwidth]{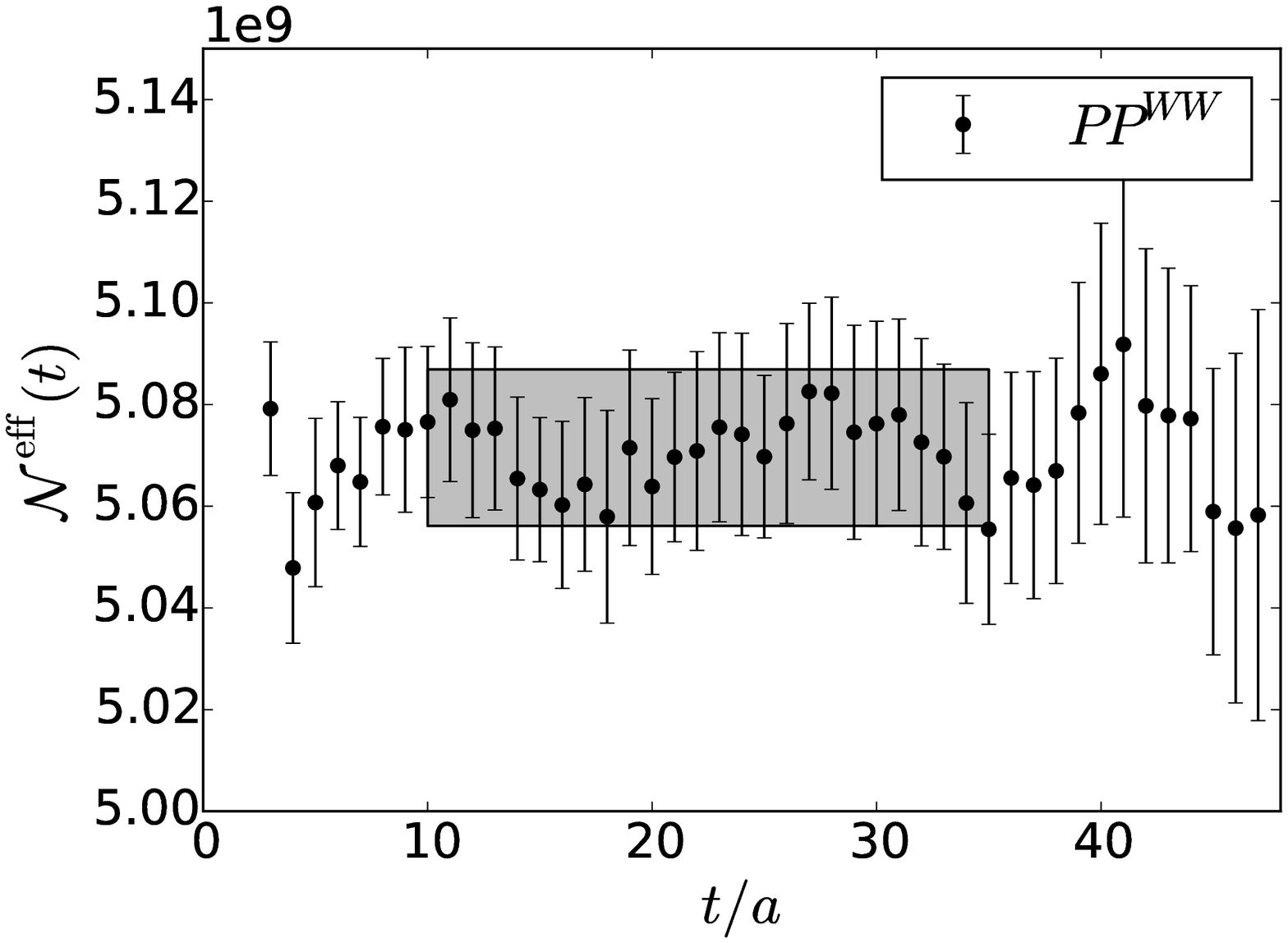} \includegraphics[width=0.5\textwidth]{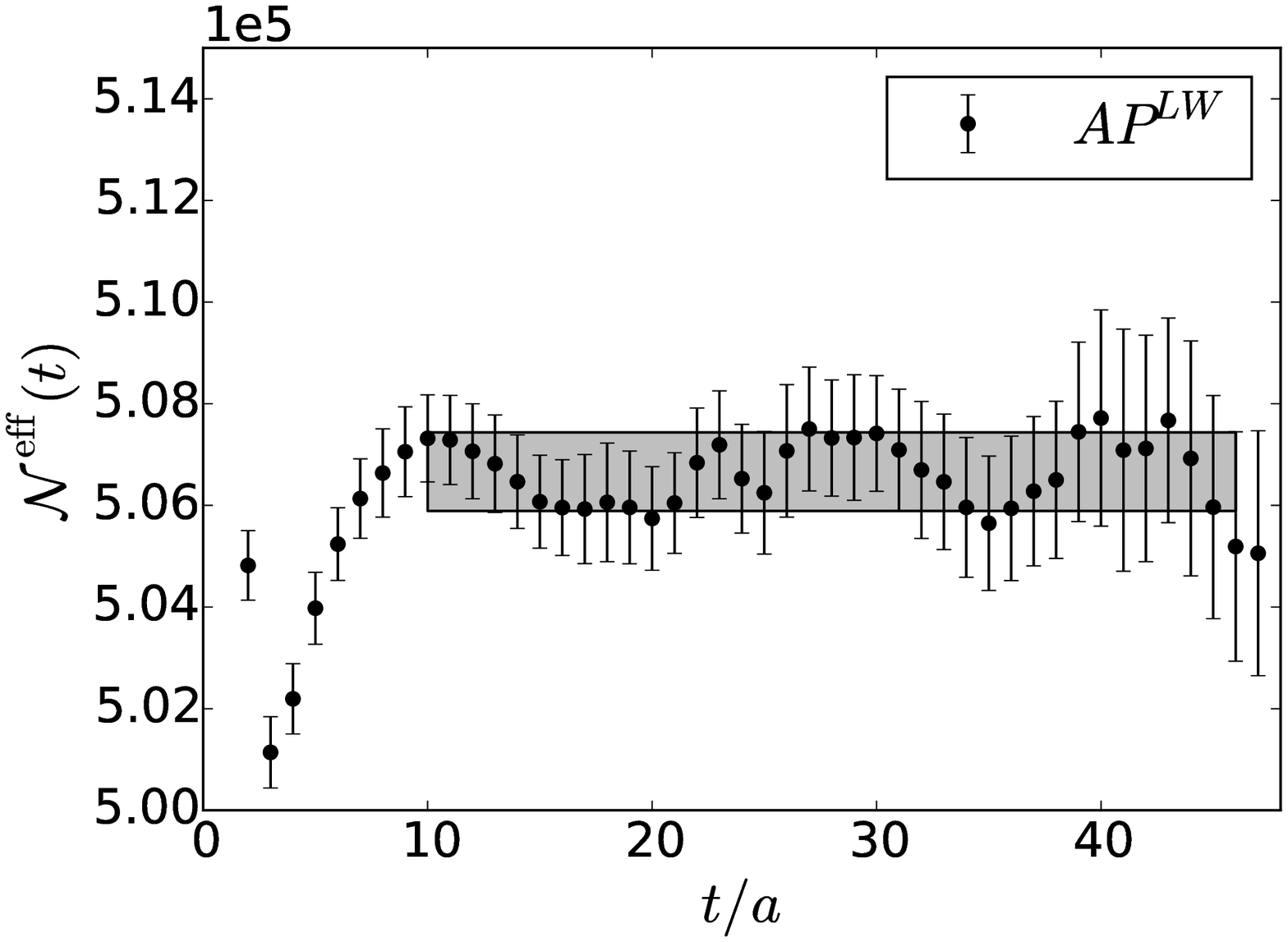}}
\subfigure{\includegraphics[width=0.5\textwidth]{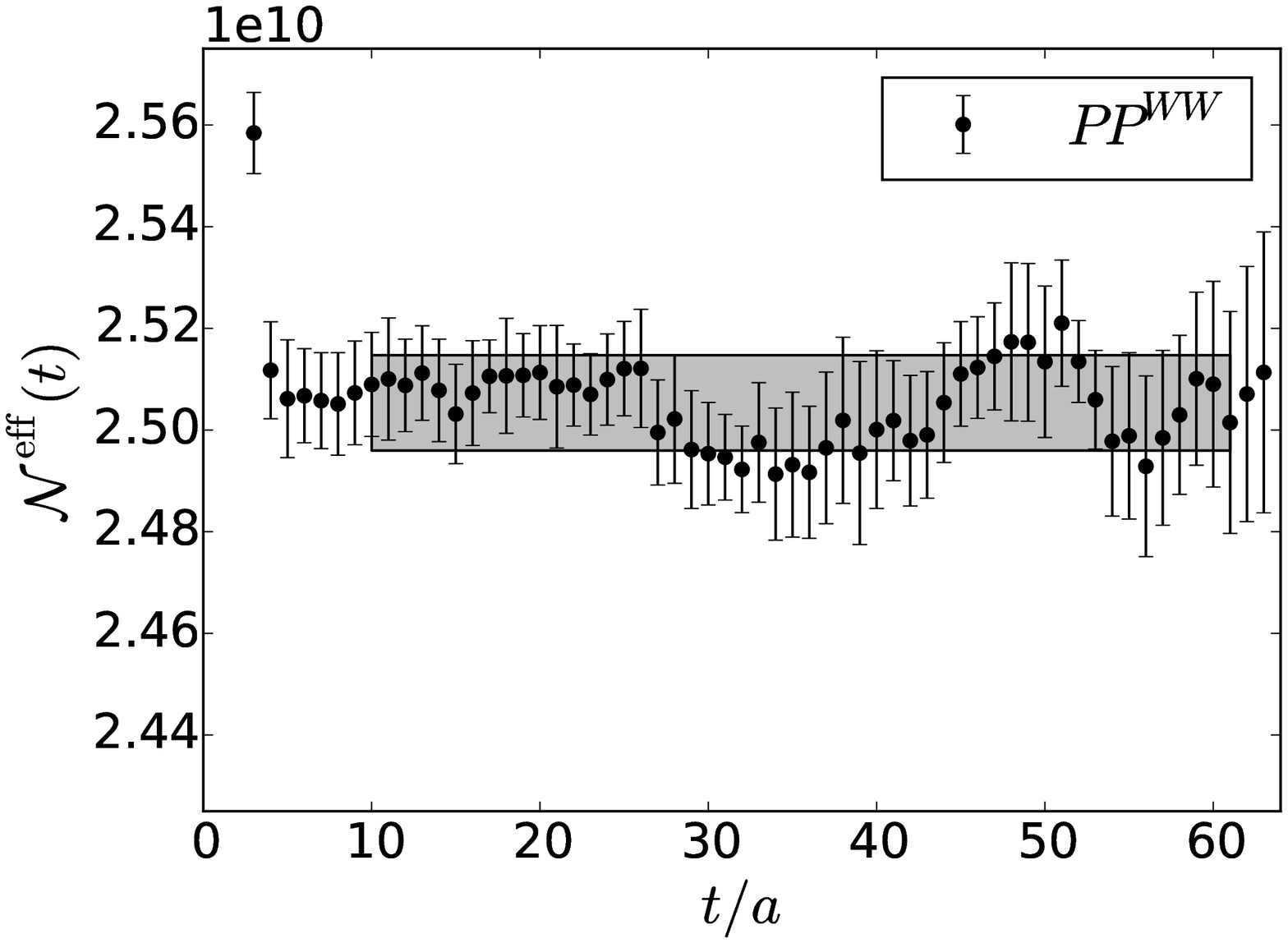} \includegraphics[width=0.5\textwidth]{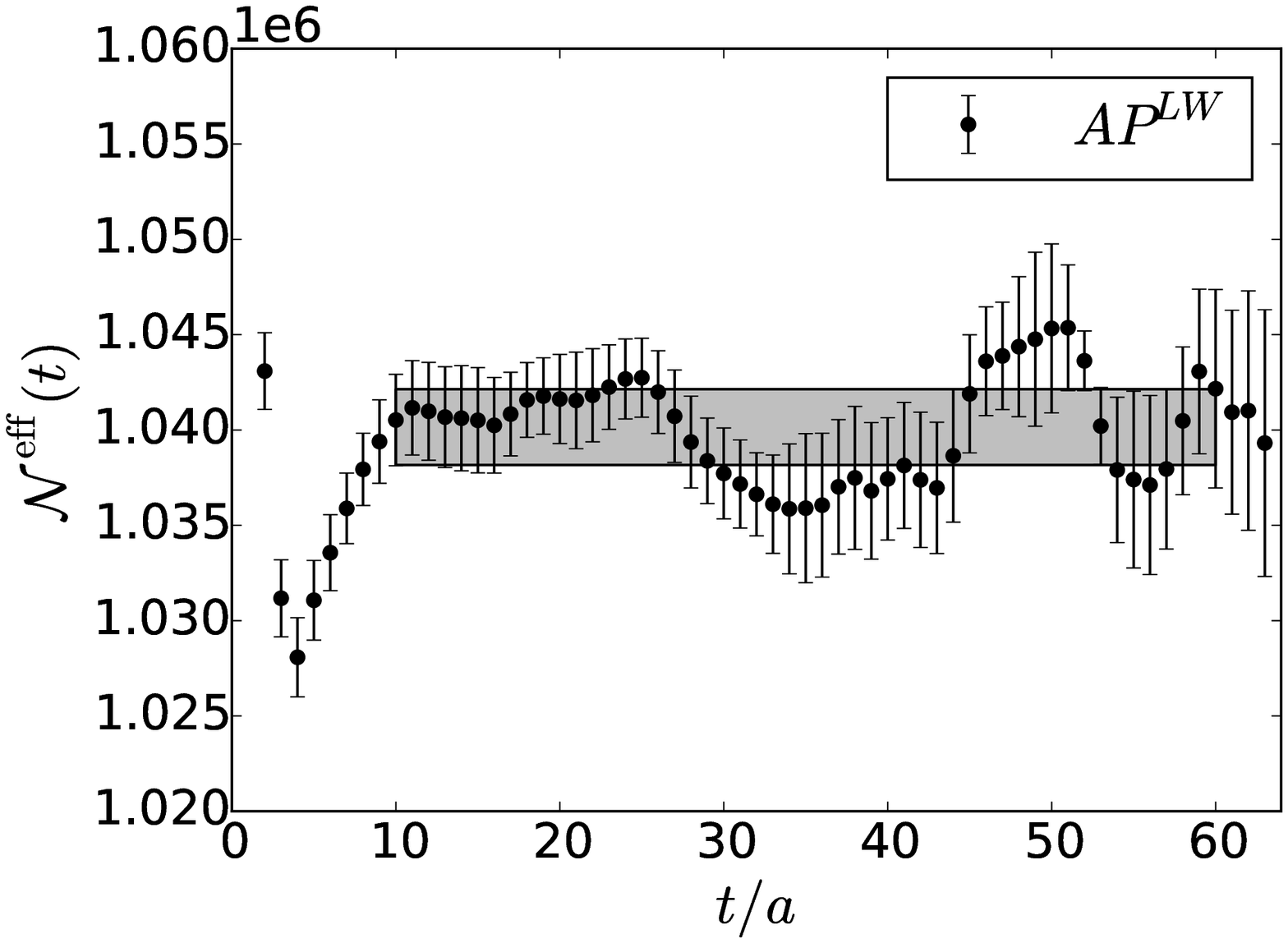}}
\caption{Effective amplitudes, defined by Eq.~\eqref{eqn:eff_amp},  associated with $f_{ll}$ on the 32Ifine (top), 48I (middle), and 64I (bottom) ensembles.}
\label{fig:pion_decay_constant}
\end{figure}

\begin{figure}[h]
\centering
\subfigure{\includegraphics[width=0.5\textwidth]{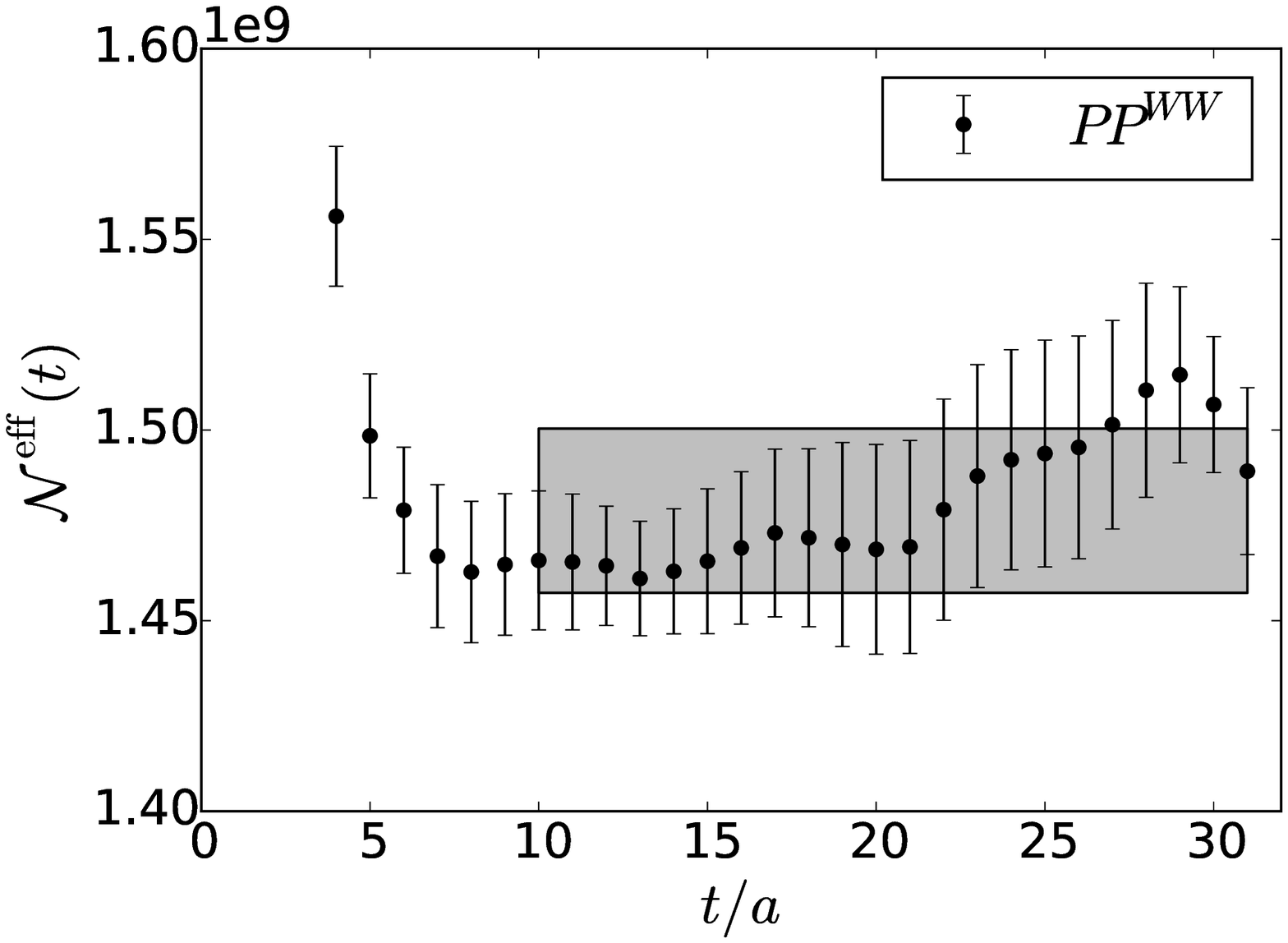} \includegraphics[width=0.5\textwidth]{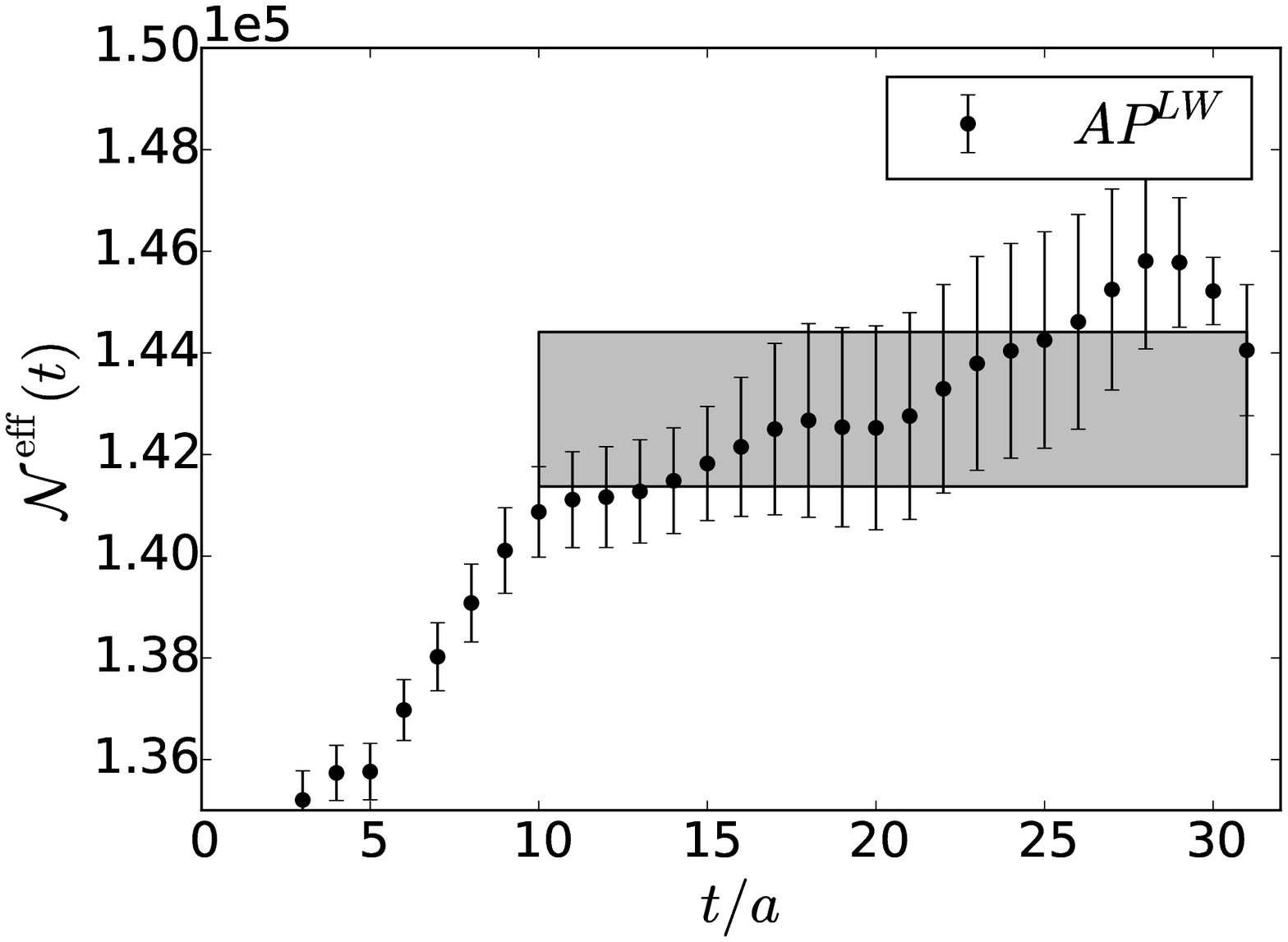}}
\subfigure{\includegraphics[width=0.5\textwidth]{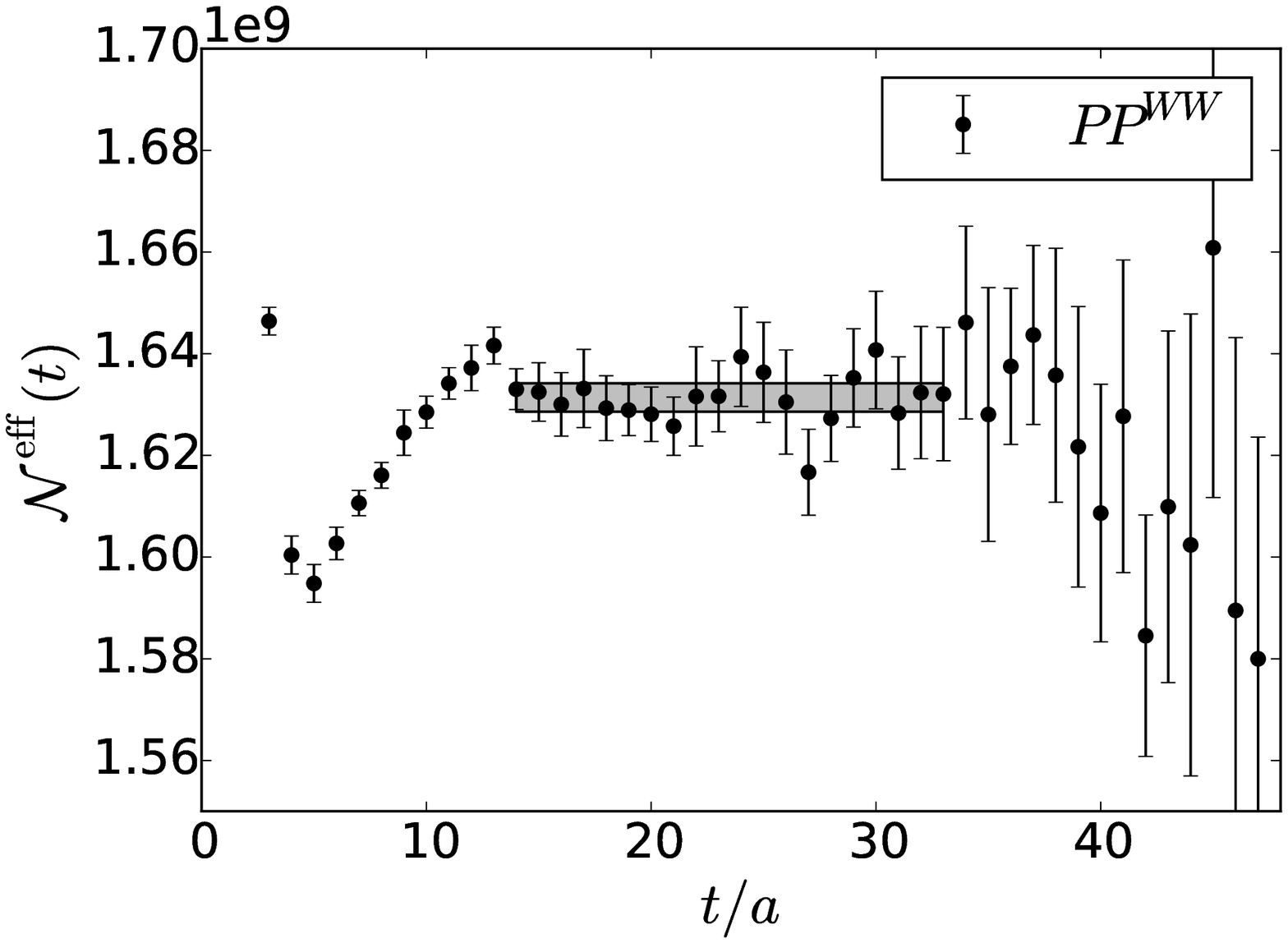} \includegraphics[width=0.5\textwidth]{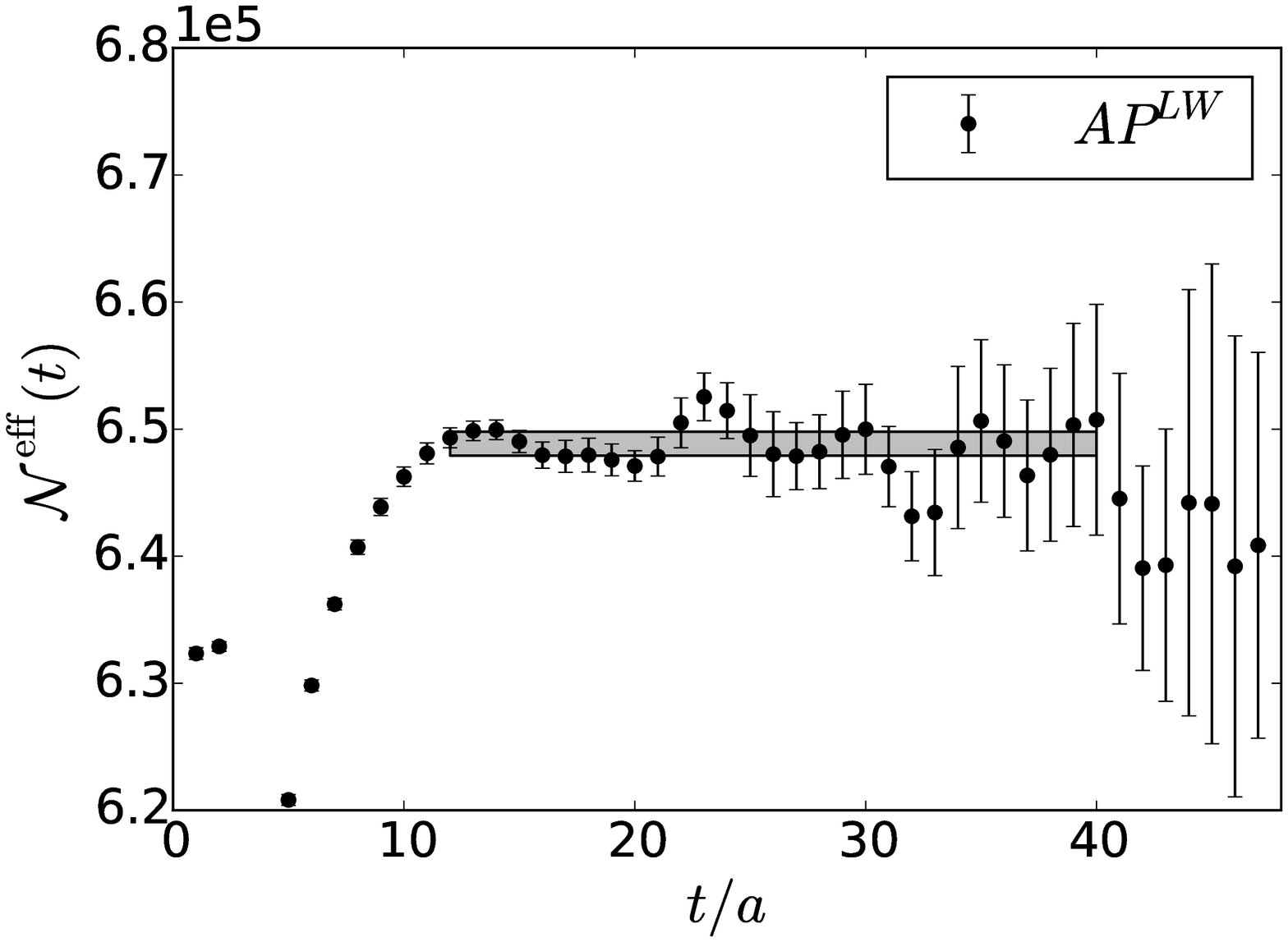}}
\subfigure{\includegraphics[width=0.5\textwidth]{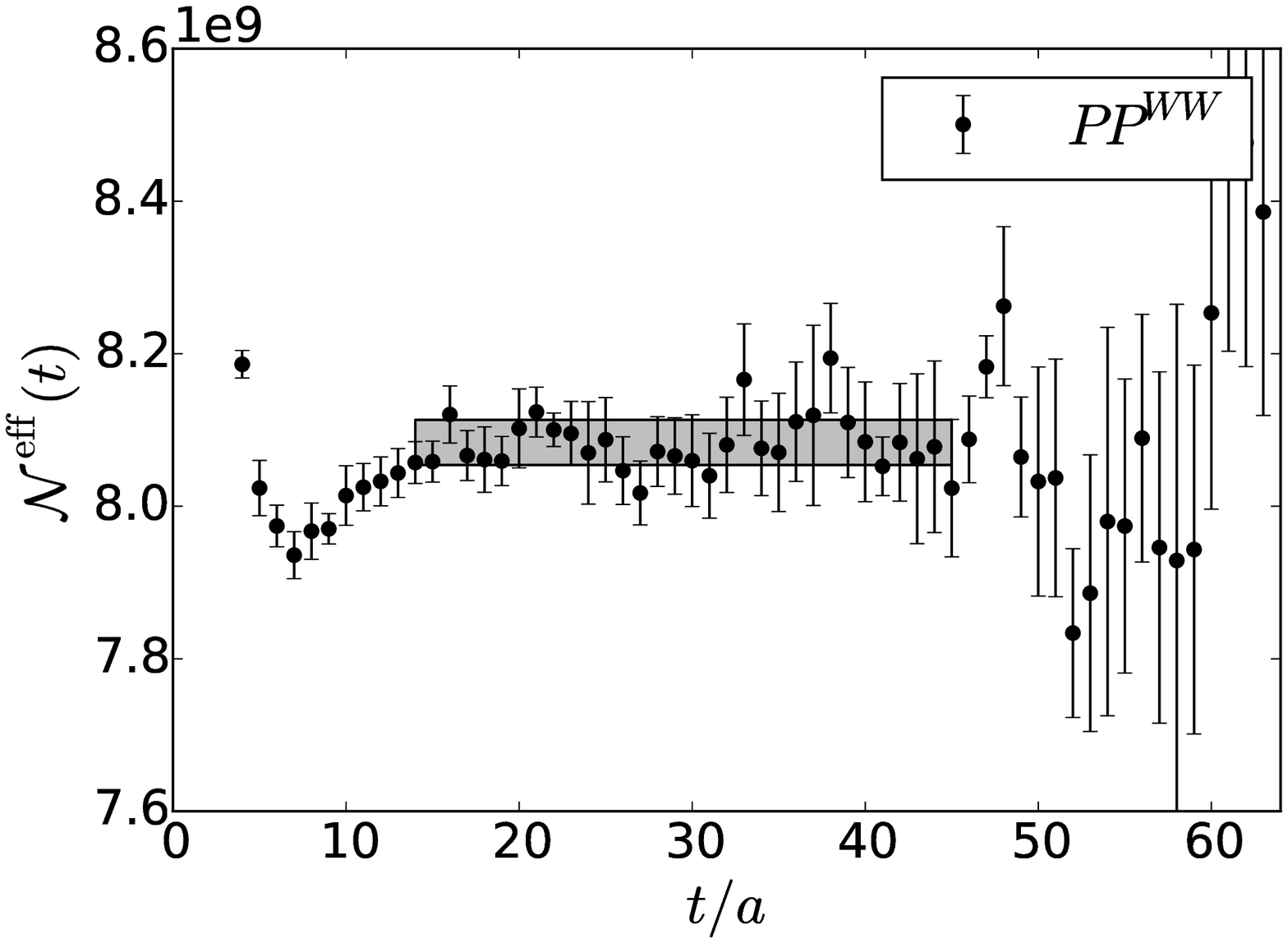} \includegraphics[width=0.5\textwidth]{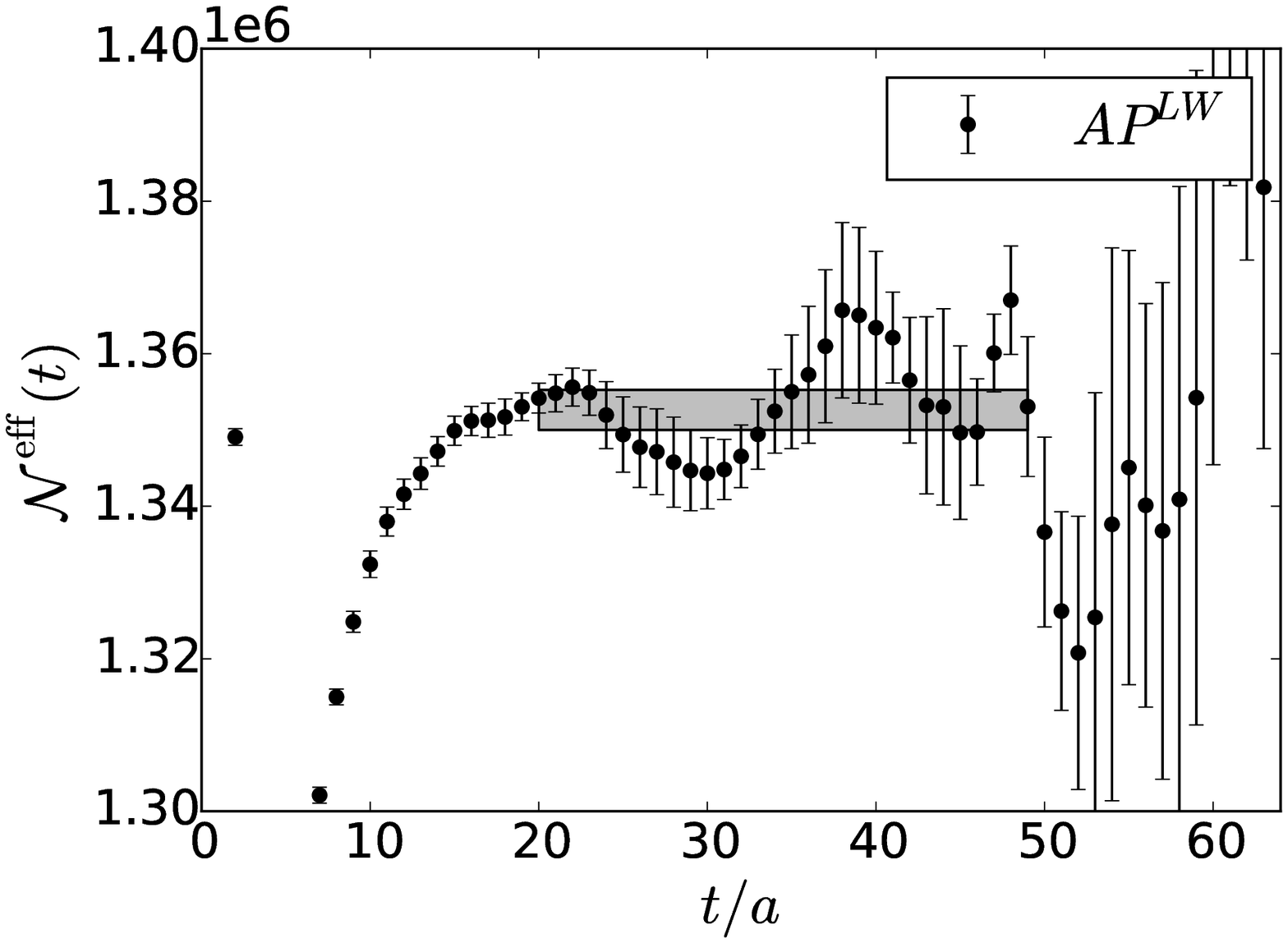}}
\caption{Effective amplitudes, defined by Eq.~\eqref{eqn:eff_amp},  associated with $f_{lh}$ on the 32Ifine (top), 48I (middle), and 64I (bottom) ensembles.}
\label{fig:kaon_decay_constant}
\end{figure}

The light-light pseudoscalar decay constant can be computed from $Z_{V}$ and the amplitudes of the $PP$ and $AP$ correlators as
\begin{equation}
f_{ll} = Z_{V} \sqrt{ \frac{2}{m_{ll} V} \frac{{\mathcal{N}_{AP}^{LW}}^{2}}{\mathcal{N}_{PP}^{WW}} },
\end{equation}
and likewise for the heavy-light pseudoscalar. In Figures~\ref{fig:pion_decay_constant} and~\ref{fig:kaon_decay_constant} we plot the effective amplitudes,
\begin{equation} \label{eqn:eff_amp} \begin{dcases}
\mathcal{N}_{PP}^{\text{eff}}(t) = \frac{\mathcal{C}_{PP}(t)}{\exp\left(-m t\right) + \exp\left(-m (N_{t}-t)\right)} \\
\mathcal{N}_{AP}^{\text{eff}}(t) = \frac{\mathcal{C}_{AP}(t)}{\exp\left(-m t\right) - \exp\left(-m (N_{t}-t)\right)} \\
m = m^{\text{eff}}(t)\,,
\end{dcases} \end{equation}
associated with $f_{ll}$ and $f_{lh}$. 

\subsection{Neutral Kaon Mixing Parameter}

\begin{figure}[h]
\centering
\subfigure{\includegraphics[width=0.5\textwidth]{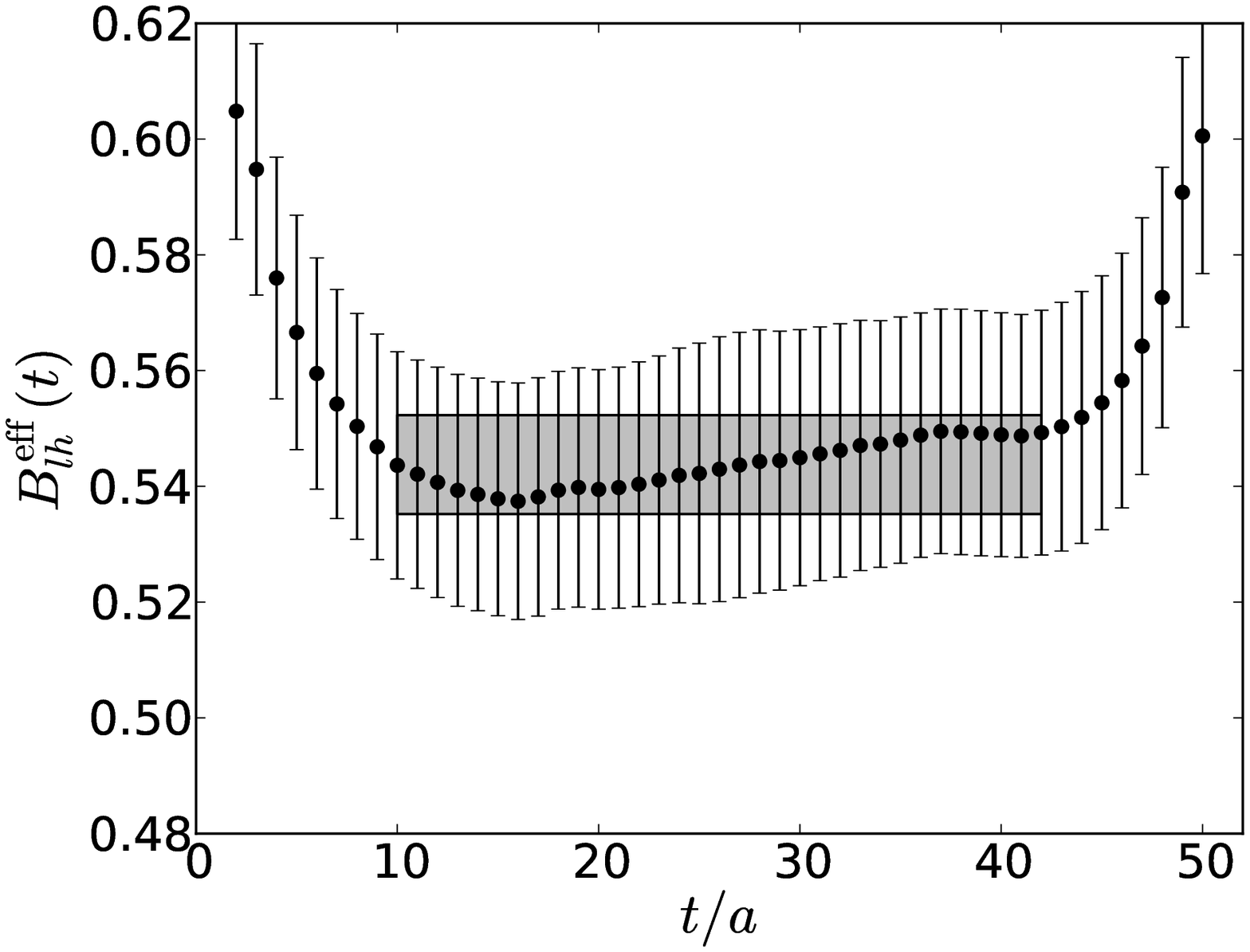} \includegraphics[width=0.5\textwidth]{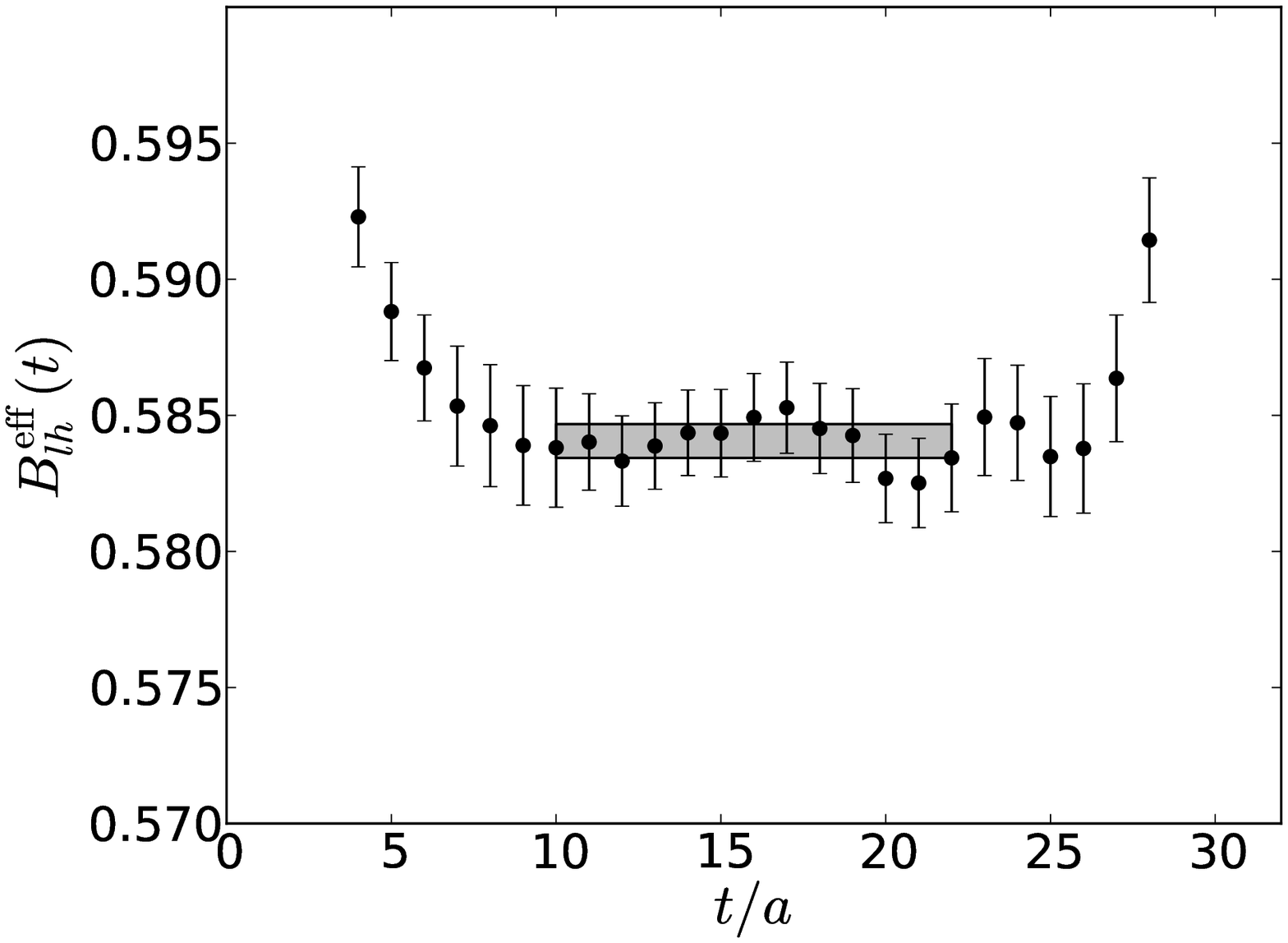}}
\subfigure{\includegraphics[width=0.5\textwidth]{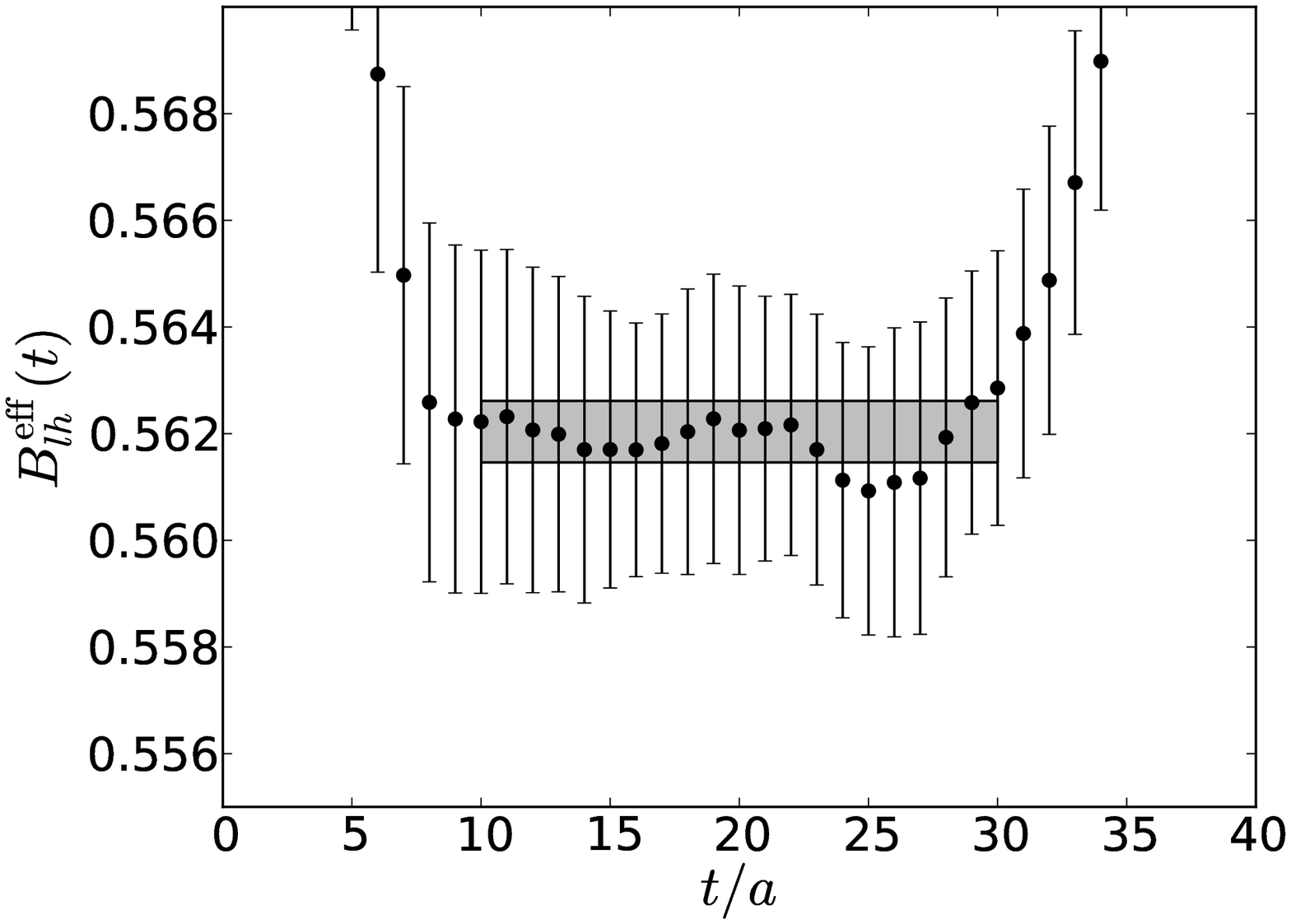}}
\caption{Effective $B_{lh}$ on the 32Ifine (top left), 48I (top right), and 64I (bottom) ensembles, for $K-K$ separations of 52 time units, 32 time units, and 40 time units, respectively. Note that in each case the fit is performed using several $K-\bar K$ separations, not just the separation plotted here.}
\label{fig:bk}
\end{figure}

We compute the neutral kaon mixing parameter, $B_{lh}$, from the ratio
\begin{equation} \label{eqn:bk_correlator}
\frac{ \langle \overline{K^{0}} | \mathcal{O}_{VV+AA} | K^{0} \rangle }{ \frac{8}{3} \langle K^{0} | A_{0} | 0 \rangle \langle 0 | A_{0} | \overline{K^{0}} \rangle } \cong B_{l h}\,,
\end{equation}
where $\mathcal{O}_{VV+AA}$ is the $\Delta S=2$ four-quark operator responsible for the mixing:
\begin{equation}
\mathcal{O}_{VV+AA} = \overline{s} \gamma_{\mu} \left( \mathbf{1} - \gamma_{5} \right) d \cdot \overline{s} \gamma^{\mu} \left( \mathbf{1} - \gamma_{5} \right) d\,.
\end{equation}
The matrix element in the numerator of Eq.~\eqref{eqn:bk_correlator} was computed for $K-\bar K$ separations which are a multiple of 4 (5) on the 32Ifine/48I (64I) ensemble. On the 32Ifine ensemble we use linear combinations of propagators with periodic and antiperiodic boundary conditions in the temporal direction to effectively double the time extent of the lattice for the $B_{lh}$ correlators, a technique we have also employed in previous calculations~\cite{Arthur:2012opa}. We determine appropriate ranges of $K-\bar K$ separations to include in the fit using the same procedure as described in the previous section for $Z_{V}$. We chose separations of 52 and 56 time units on the 32Ifine ensemble, $20,24\ldots 40$ on the 48I ensemble, and $25,30\ldots 40$ on the 64I ensemble. In Figure~\ref{fig:bk} we plot the $B_{lh}$ effective amplitude for a single $K-\bar K$ separation included in the fit, as well as the fitted value for $B_{lh}$, on each ensemble. 

\subsection{Omega Baryon Mass}
\label{section:omega}
\begin{figure}[h]
\centering
\subfigure{\includegraphics[width=0.5\textwidth]{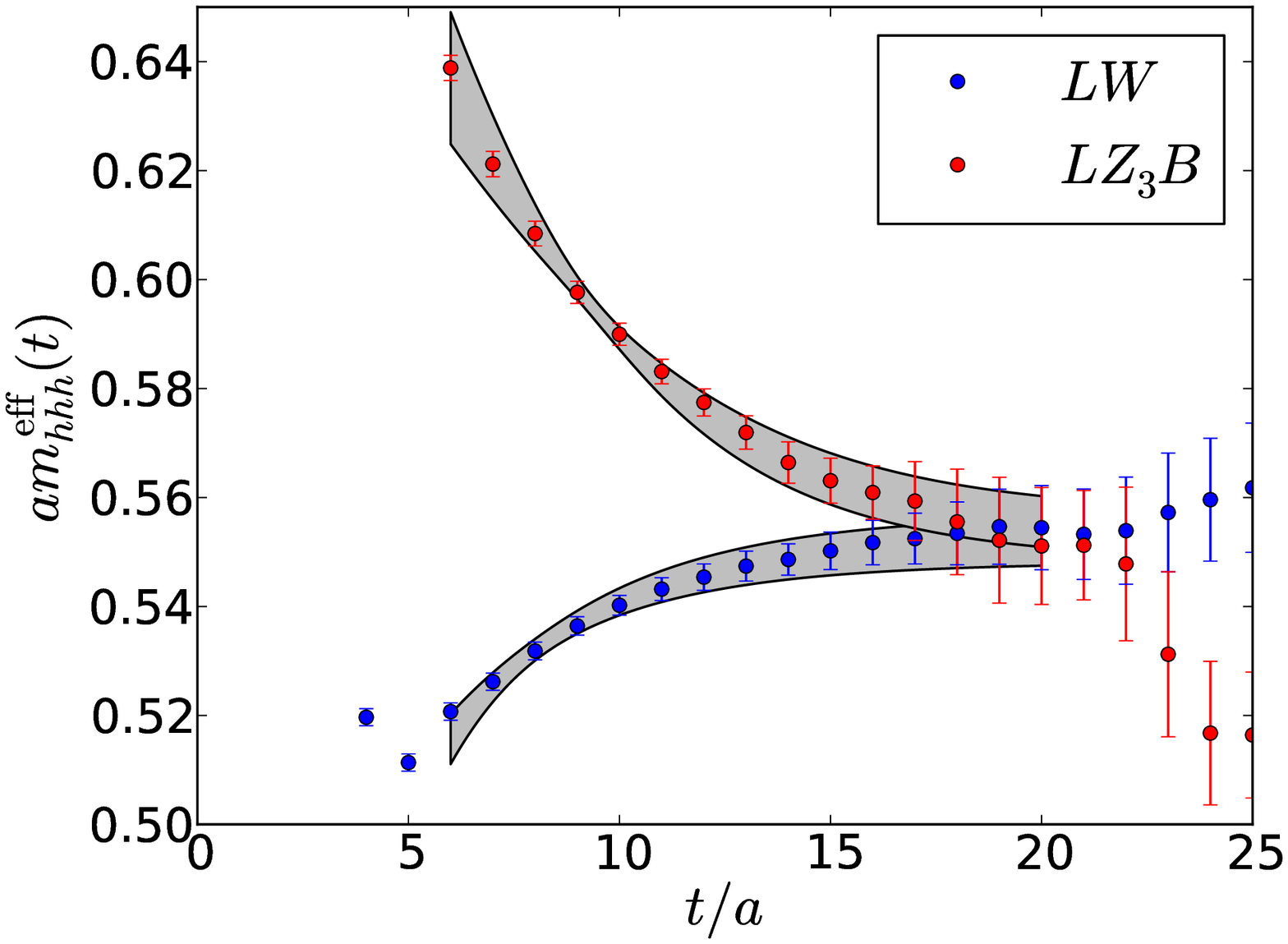} \includegraphics[width=0.5\textwidth]{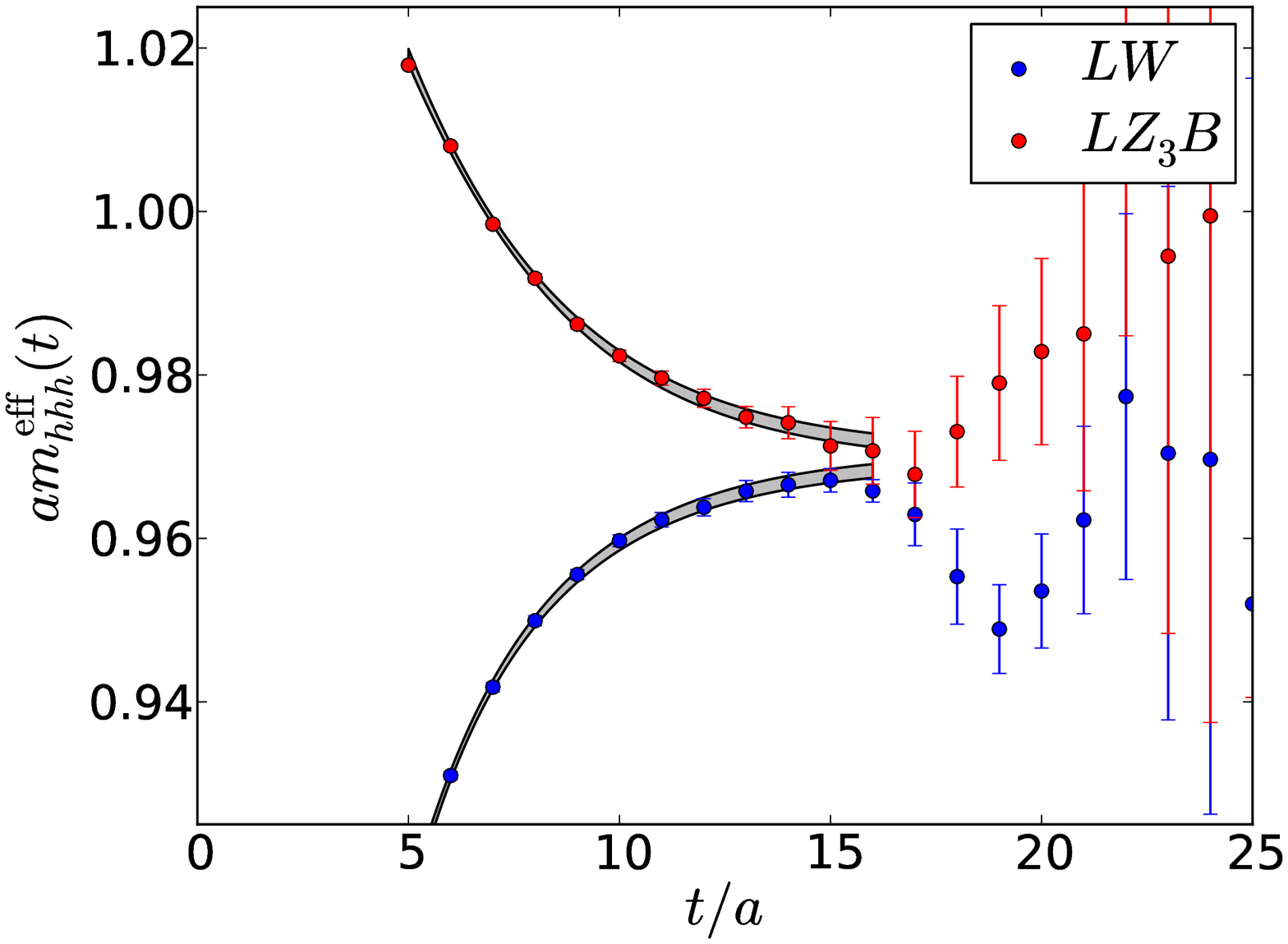}}
\subfigure{\includegraphics[width=0.5\textwidth]{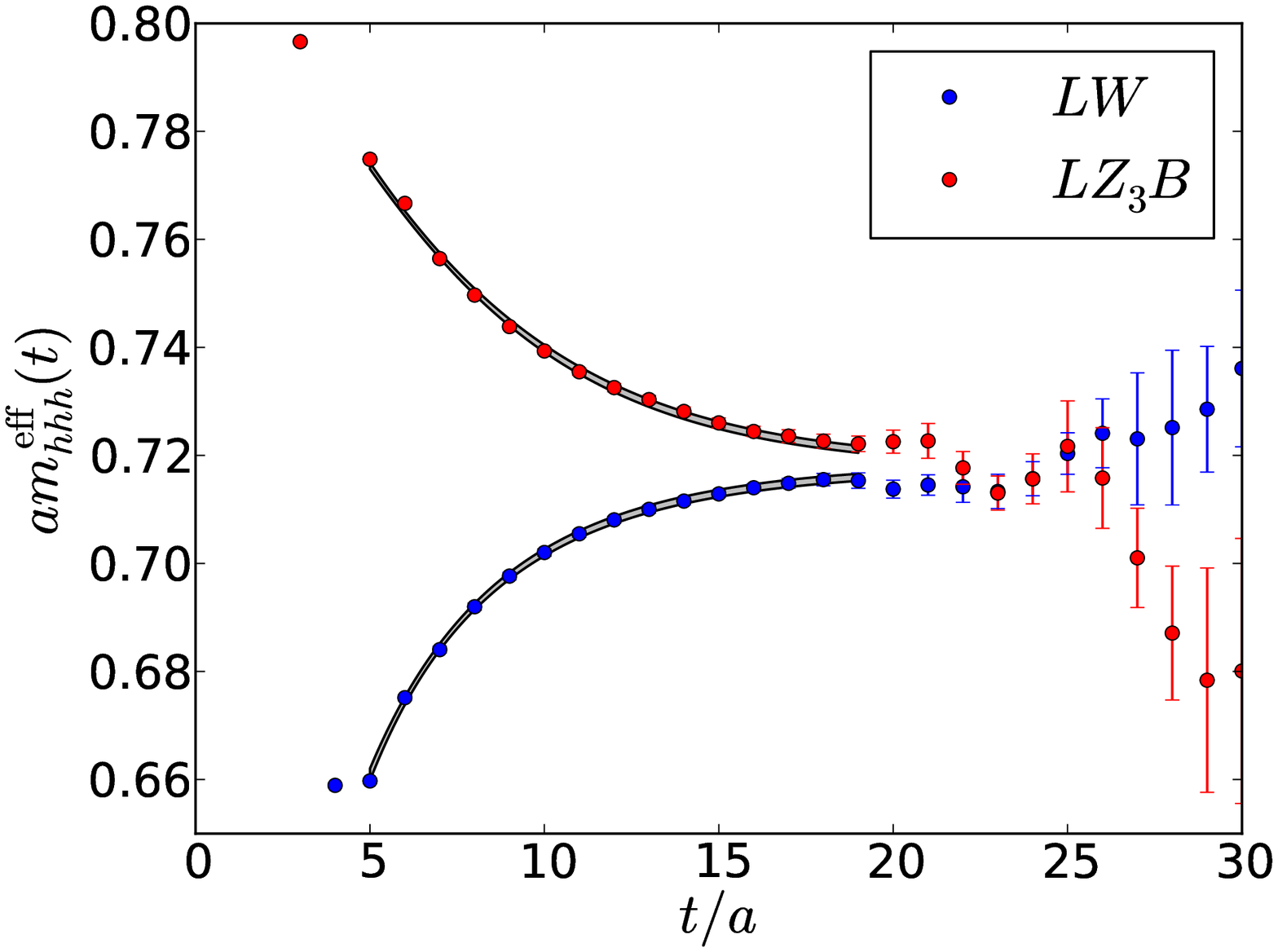}}
\caption{
The effective mass of the Omega baryon obtained using both our wall (LW) and Z$_3$ box source (LZ$_3$B) on the 32Ifine (top left), 48I (top right), and 64I (bottom) ensembles. The correlation functions are simultaneously fit to a two-exponential fit form, and the effective mass determined from the fit function (obtained by applying the same technique as used to extract the effective mass from the raw data) is overlayed with the data.
}
\label{fig:omega_eff_mass}
\end{figure}

We measured the $\Omega$-baryon mass $m_{hhh}$ from the two-point correlator
\begin{equation} \label{eqn:omega_correlator}
\mathcal{C}_{\Omega \Omega}^{s_{1} s_{2}}(t) = \sum\limits_{i=1}^{3} \langle 0 | \mathcal{O}_{\Omega}^{s_{1}}(\vec{x},t)_{i} \overline{\mathcal{O}}_{\Omega}^{s_{2}}(0)_{i} | 0 \rangle\,,
\end{equation}
using an interpolating operator
\begin{equation}
\mathcal{O}_{\Omega}(x)_{i} = \varepsilon_{a b c} \left( s_{a}^{\mathsf{T}}(x) C \gamma_{i} s_{b}(x) \right) s_{c}(x),
\end{equation}
where $C$ denotes the charge conjugation matrix. We performed measurements using both Coulomb gauge-fixed wall sources and $Z_{3}$ box ($Z_{3}B$) sources, and, in both cases, a local (point) sink. The correlator, Eq.~\eqref{eqn:omega_correlator}, is a $4 \times 4$ matrix in spin space which couples to both positive ($+$) and negative ($-$) parity states, and has the asymptotic form
\begin{equation}
\mathcal{C}^{s_{1} s_{2}}_{\Omega}(t) \stackrel{t \gg 1}{\cong} \sum\limits_{\vec{p}} \left( \frac{1}{2} \left( \mathbf{1} + \gamma_{4} \right) \mathcal{A}_{+}^{s_{1} s_{2}}(\vec{p}) e^{-E^{+}_{\vec{p}} t} - \frac{1}{2} \left( \mathbf{1} - \gamma_{4} \right) \mathcal{A}_{-}^{s_{1} s_{2}}(\vec{p}) e^{-E^{-}_{\vec{p}} t} \right)
\end{equation}
for large $t$. The fit to extract $m_{hhh}$ is performed by first projecting onto the positive parity component,
\begin{equation}
\mathcal{P}_{+} \mathcal{C}_{\Omega}^{s_{1} s_{2}} = \frac{1}{4} \tr \left\{ \frac{1}{2} \left( \mathbf{1} + \gamma_{4} \right) \mathcal{C}_{\Omega}^{s_{1} s_{2}} \right\}\,,
\end{equation}
for each source type, and then performing a simultaneous fit of both correlators to a sum of two exponential functions with common mass terms :
\begin{equation} \begin{dcases}
\mathcal{C}_{\Omega \Omega}^{LW}(t) = \mathcal{N}_{\Omega \Omega}^{LW} e^{-m_{hhh} t} + \tilde{\mathcal{N}}_{\Omega \Omega}^{LW} e^{-m_{hhh}' t} \\
\mathcal{C}_{\Omega \Omega}^{L Z_{3}B}(t) = \mathcal{N}_{\Omega \Omega}^{L Z_{3}B} e^{-m_{hhh} t} + \tilde{\mathcal{N}}_{\Omega \Omega}^{L Z_{3}B} e^{-m_{hhh}' t}
\end{dcases} . \end{equation}
One can also include terms proportional to $e^{-m_{-} \left(N_{t} - t\right)}$, where $m_{-}$ is the mass of the ground state in the negative parity channel, to account for around-the-world contamination effects, but we find that our lattices are sufficiently large and the masses of these states sufficiently heavy that including these terms has no statistically significant influence on the fitted $\Omega$ mass. Using multiple source types and double-exponential fits to common masses allows us to reduce the statistical error on the $\Omega$ baryon mass $m_{hhh}$, as well as to also fit the mass of the first excited state in the positive parity channel $m_{hhh}'$. Figure~\ref{fig:omega_eff_mass} plots the effective $\Omega$-baryon mass on each ensemble.

\begin{figure}[h]
\includegraphics[width=0.32\textwidth]{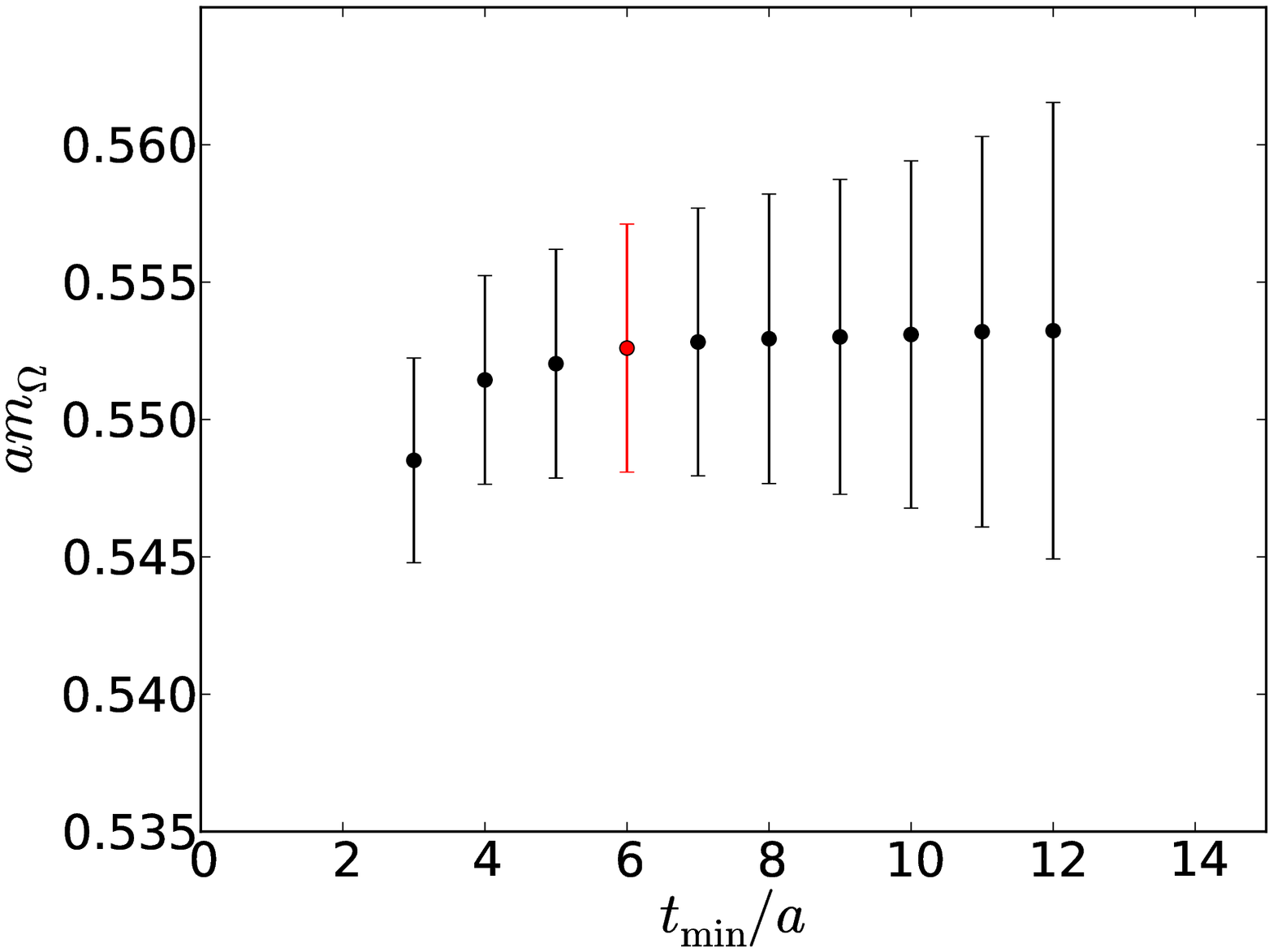}
\includegraphics[width=0.32\textwidth]{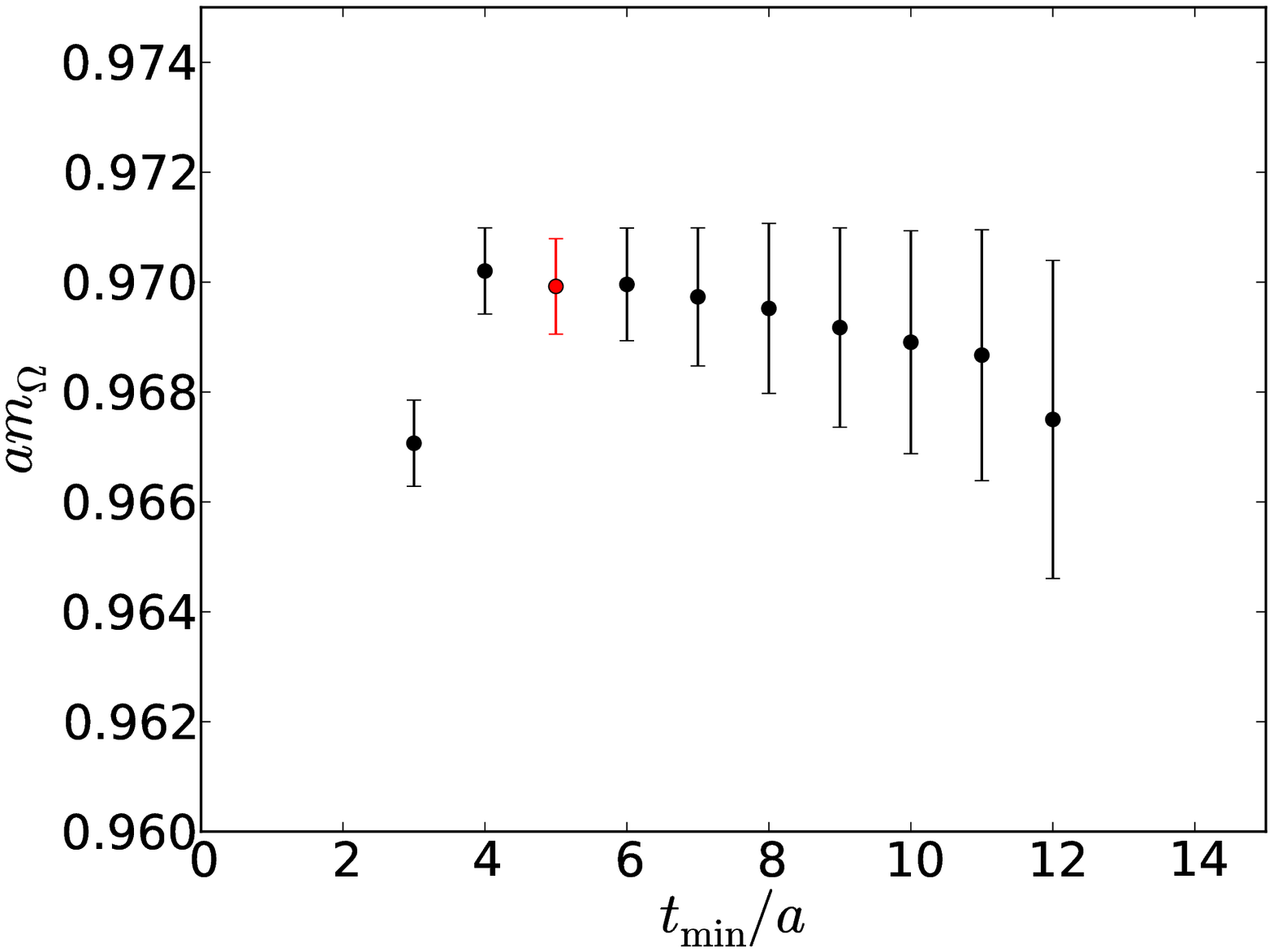}
\includegraphics[width=0.32\textwidth]{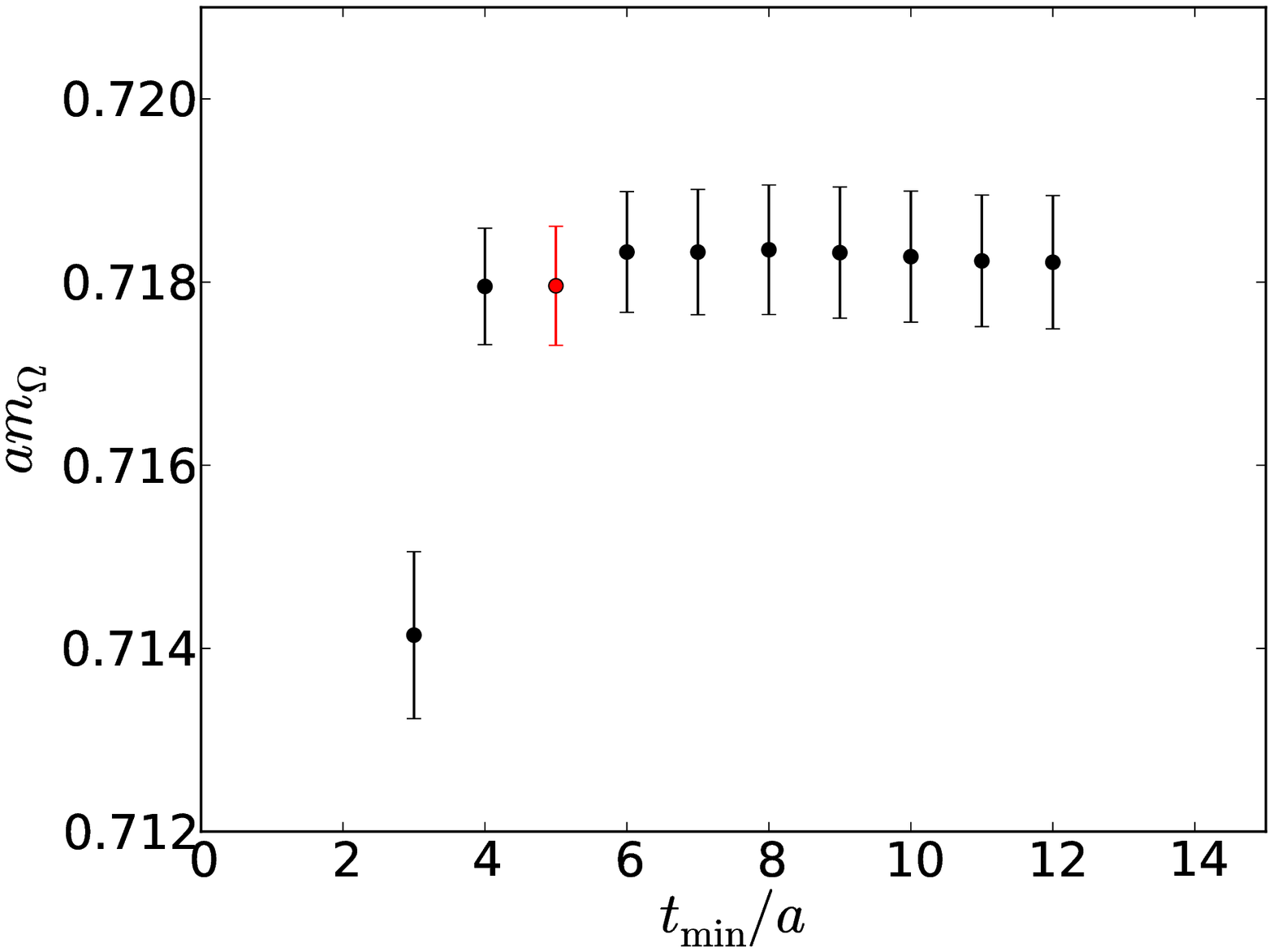}\\
\includegraphics[width=0.32\textwidth]{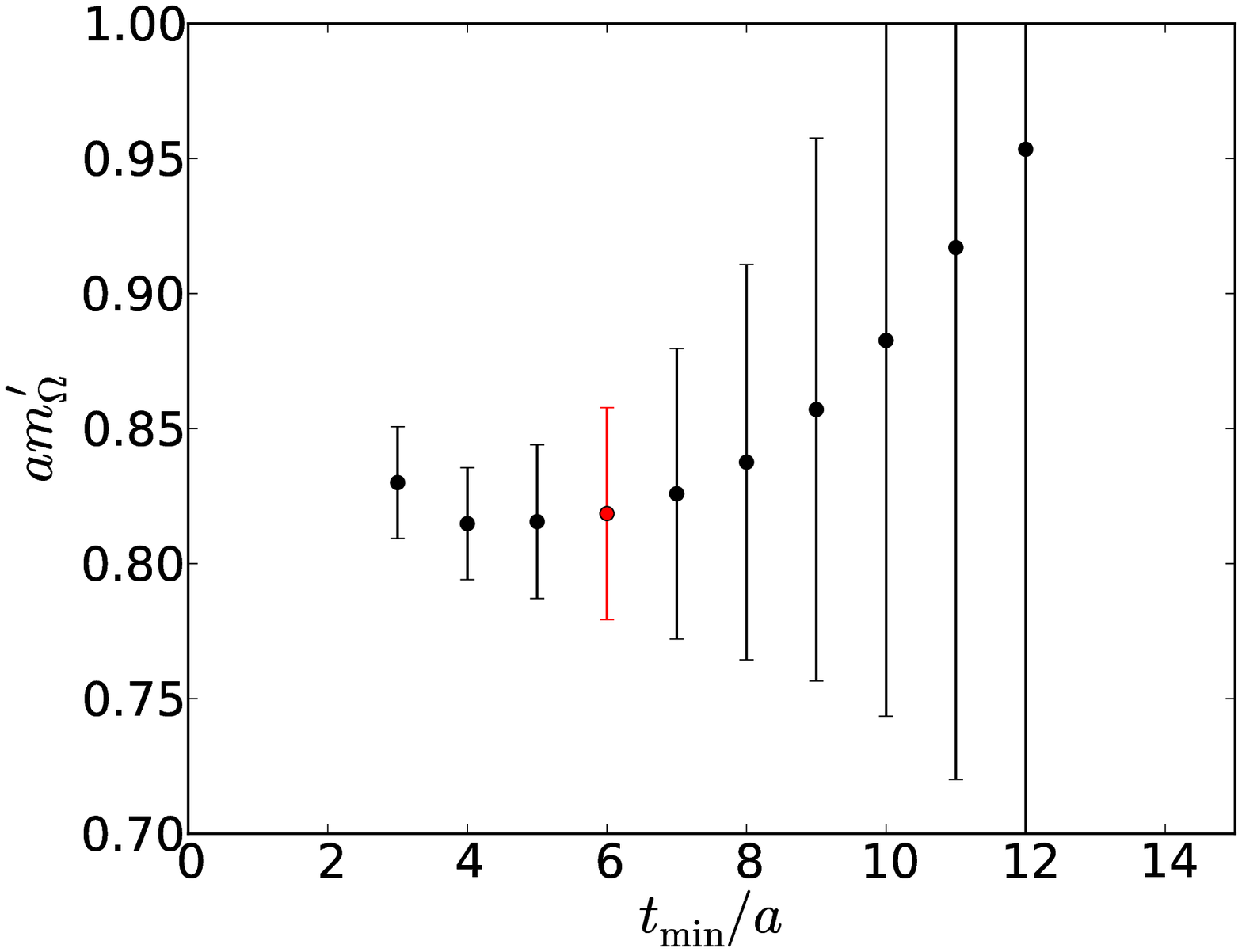}
\includegraphics[width=0.32\textwidth]{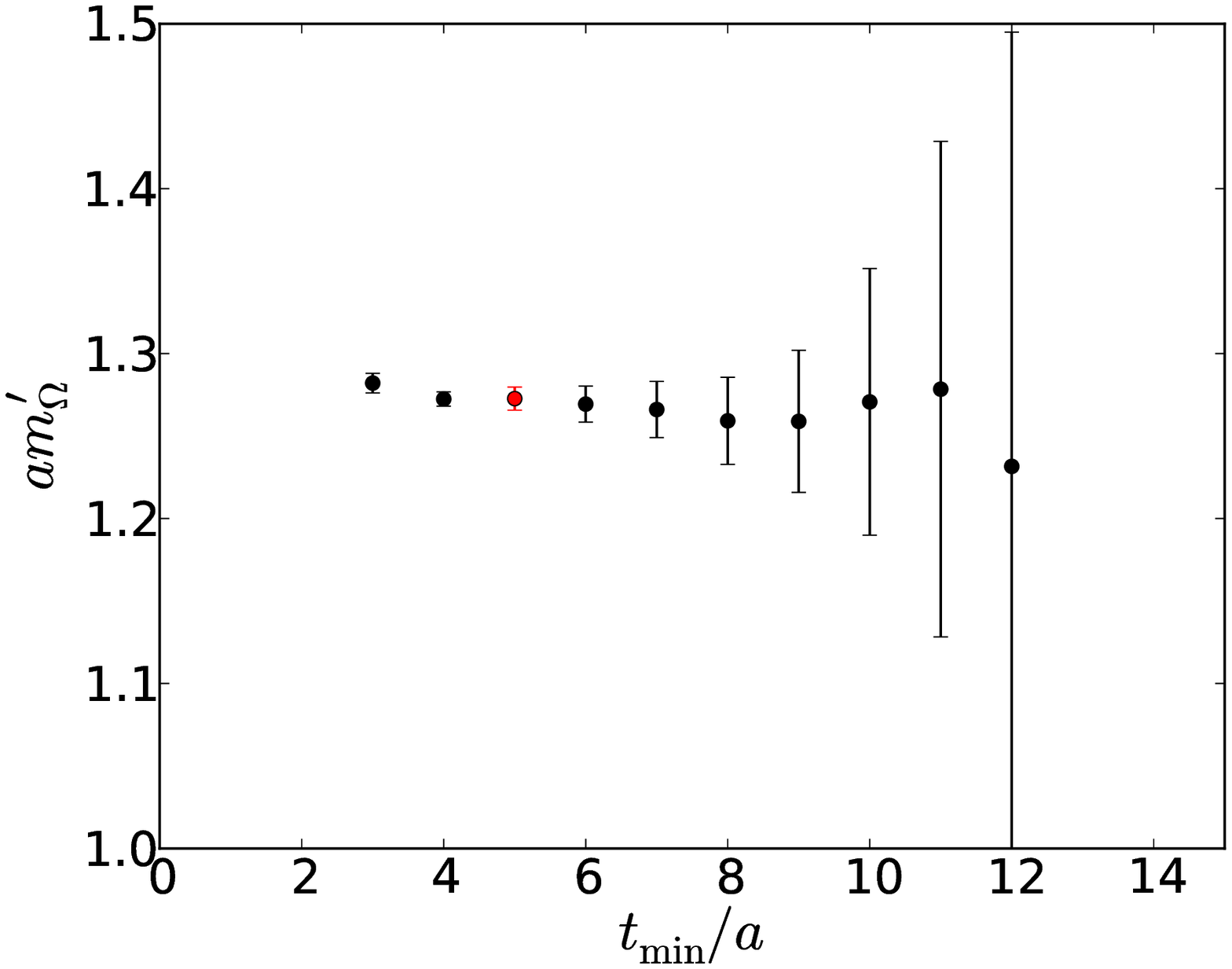}
\includegraphics[width=0.32\textwidth]{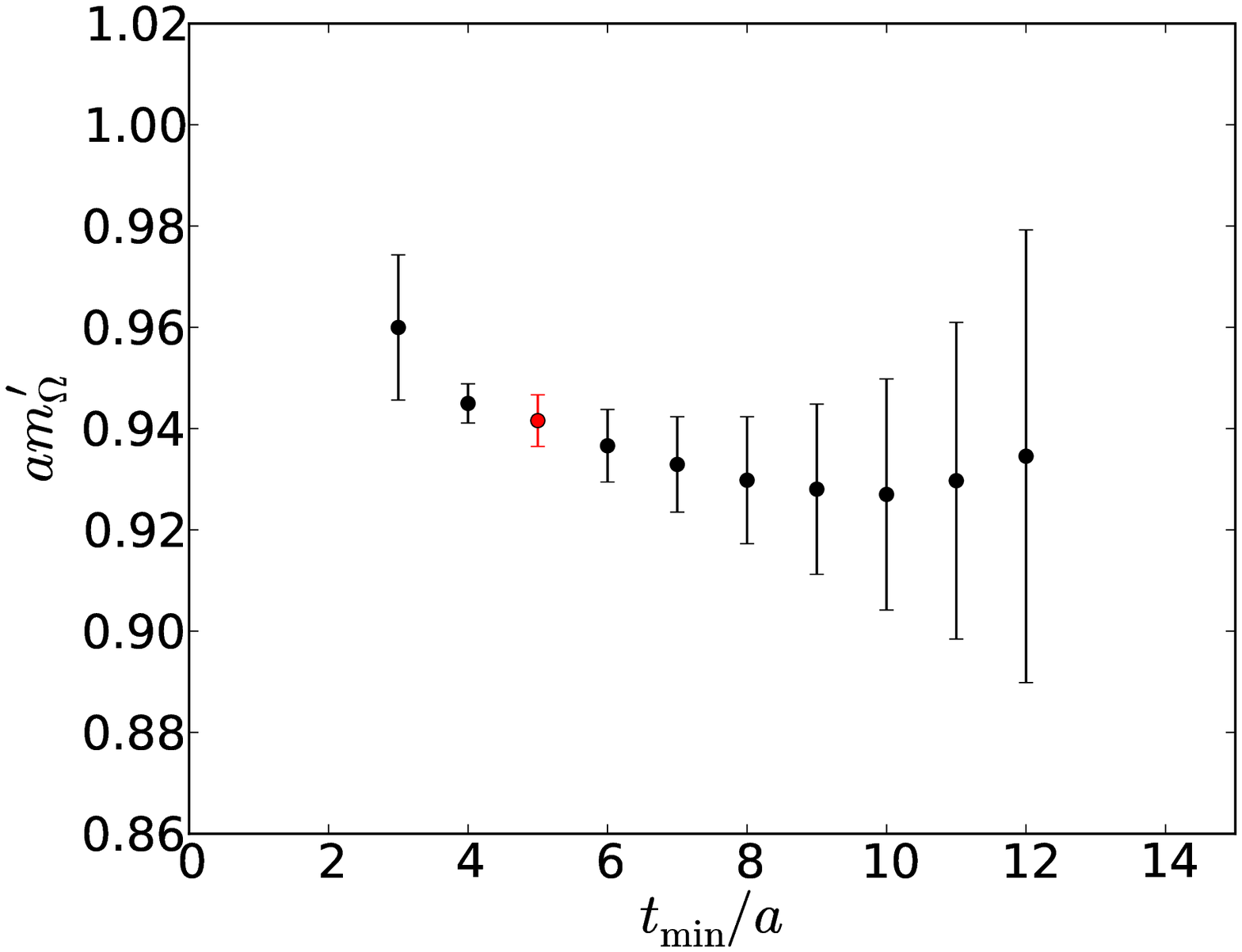}
\caption{The stability, as a function of the lower bound on the fit range $t_{\rm min}/a$, of our fitted Omega baryon ground state (upper row) and first-excited state (lower row) for the 32Ifine (left), 48I (middle) and 64I (right) ensembles. The point in red indicates our final value.\label{fig-omegatmindep} }
\end{figure}

In Figure~\ref{fig-omegatmindep} we plot the dependence of our fitted ground and excited state energies on the lower temporal bound of the fit. The upper bound of the fit window is fixed at 20, 16, and 19 on the 32Ifine, 48I, and 64I ensembles, respectively. We observe excellent stability for bounds above $t_{\rm min}=4$, suggesting that we have good resolution on both the ground and excited states, and that contamination of our results by higher-energy excited states can be discounted. In practice we use $t_{\rm min}=5$ for both the 48I and 64I ensembles, and $t_{\rm min}=6$ for the 32Ifine ensemble.

\subsection{Wilson flow scales}
\label{sec-wflowmeas}

\begin{figure}[h]
\centering
\includegraphics[width=0.48\textwidth]{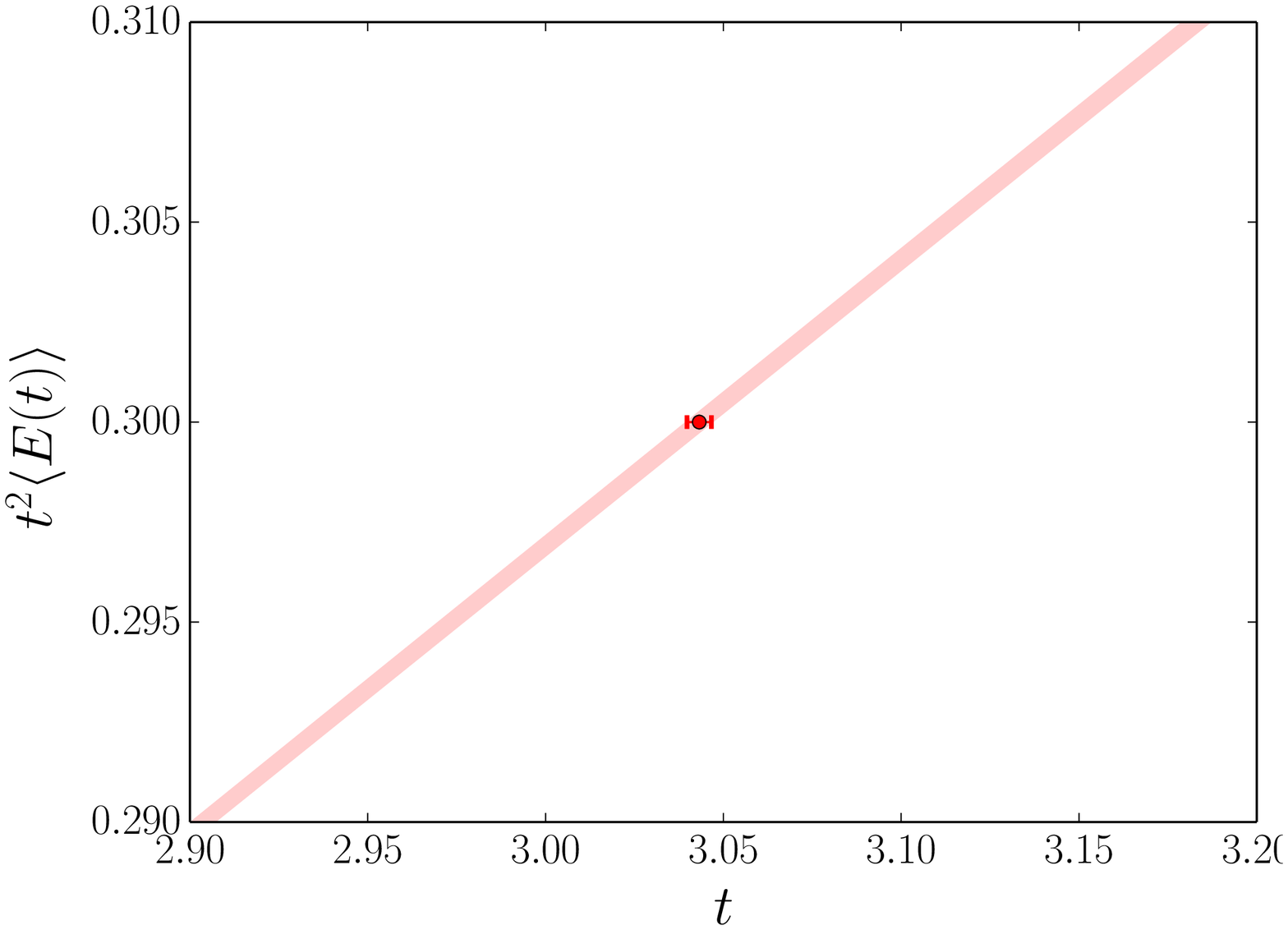}
\includegraphics[width=0.48\textwidth]{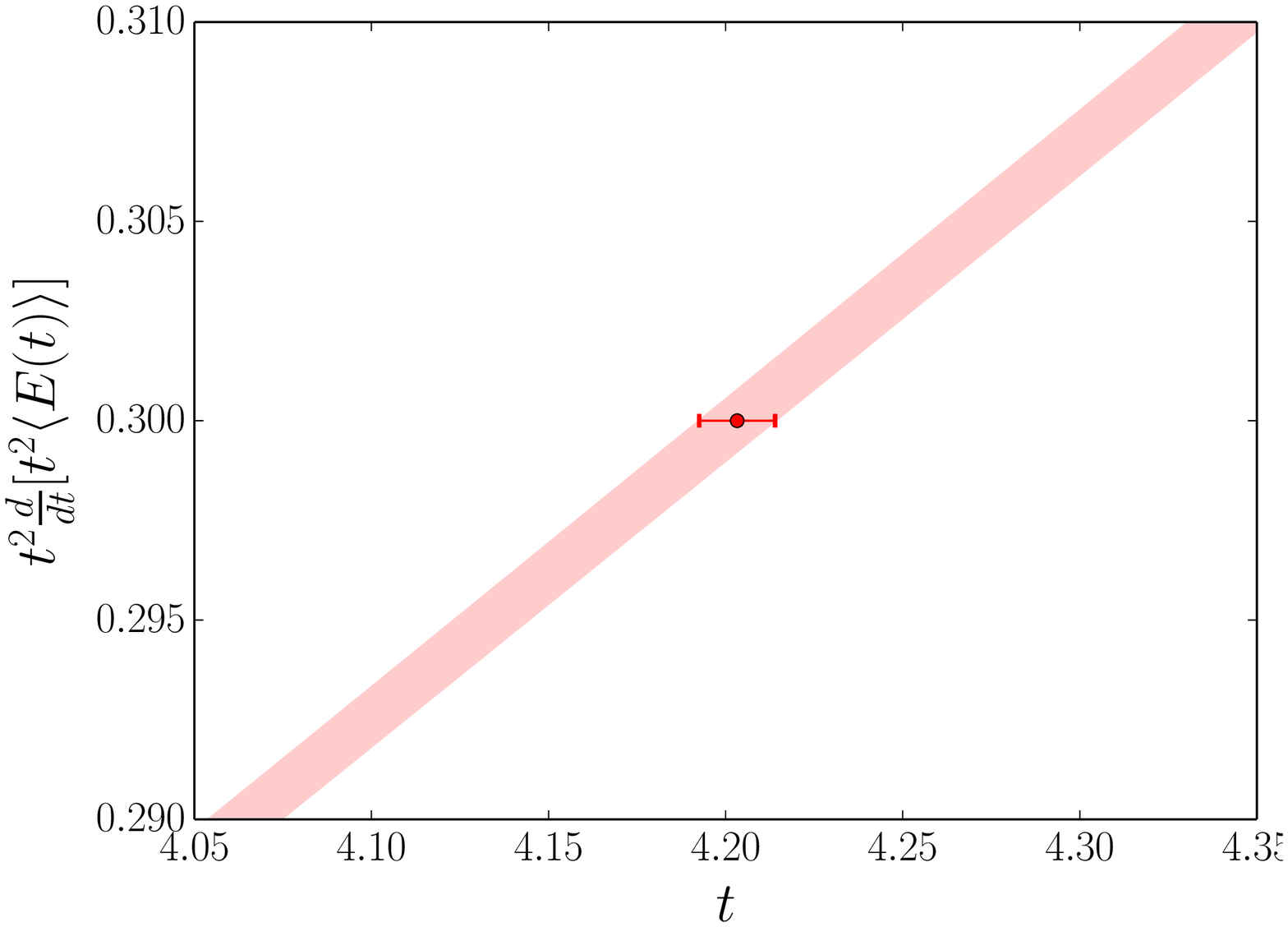}
\caption{The interpolation in Wilson flow time $t$ on the 64I ensemble of the functions of the action density used to define $t_0$ and $w_0$ respectively. The red point is the interpolated value.\label{fig-wflow64Iplots} }
\end{figure}

The Wilson flow scales, $t_0^{1/2}$ and $w_0$, are quantities with the dimension of length defined via the following equations:~\cite{Luscher:2010iy}
\begin{equation}
t^2 \langle E(t) \rangle |_{t = t_0} = 0.3\,,
\end{equation}
and~\cite{Borsanyi:2012zs}
\begin{equation}
t \frac{d}{dt} (t^2 \langle E(t) \rangle) |_{t = w_0^2} = 0.3\,,
\end{equation}
where $E$ is the discretized Yang-Mills action density,
\begin{equation}
E=\frac{1}{2}{\rm tr}(F_{\mu\nu}F_{\mu\nu})\,.
\end{equation}
We determine the action density using the clover discretization, for which $F_{\mu\nu}$ is estimated at each lattice site from the clover of four $1 \times 1$ plaquettes in the $\mu-\nu$ plane. We find that this leads to smaller discretization errors (especially for $t_0$) than estimating $F_{\mu\nu}$ directly from the plaquette via
\begin{equation}
\langle P \rangle = 1 - \frac{a^4}{36} \langle E \rangle + O(a^6)
\end{equation}
which is in agreement with some previous experience~\cite{Luscher:2010iy}. In Figure~\ref{fig-wflow64Iplots} we show an example of the interpolation of the two scales on the 64I ensemble. The final results for all ensembles are listed in Table~\ref{tab:results}.

\section{Simultaneous chiral/continuum fitting procedure}
\label{sec:CombinedChiralFits}
The bare quark masses for the 48I and 64I ensembles were chosen based on the results for the physical quark masses at equivalent bare couplings obtained in Ref.~\cite{Arthur:2012opa}. The simulated values for the dimensionless ratios $m_\pi/m_\Omega$ and $m_K/m_\Omega$ are shown in Table~\ref{tab:results}. Since we are not simulating electromagnetism, we compare to the following physical values:  $m_\pi = 135.0$ MeV, $m_K = 495.7$ MeV and $m_\Omega = 1.67225$ GeV. Clearly our simulations are very close to the physical point, yet we must perform the very modest extrapolation in order to obtain precise physical results.

\subsection{Summary of global fit procedure}

In Refs.~\cite{Aoki:2010dy,Arthur:2012opa} we have detailed a strategy for performing simultaneous chiral and continuum `global' fits to our lattice data. In this document we perform such fits to the following quantities: $m_\pi$, $m_K$, $f_\pi$, $f_K$, $m_\Omega$ and the Wilson flow scales $t_0^{1/2}$ and $w_0$. We parametrize the mass dependence of each quantity using three ans\"{a}tze (where applicable): NLO partially-quenched chiral perturbation theory with and without finite-volume corrections (i.e. infinite volume $\chi$PT), which we henceforth refer to as the `ChPTFV' and `ChPT' ans\"{a}tze respectively; and a linear `analytic' ansatz. For the ChPT and ChPTFV ans\"{a}tze we use heavy-meson $\chi$PT~\cite{Roessl:1999iu,Allton:2008pn} to describe quantities with valence strange quarks. %
%
For the convenience of the reader, we have collected the various ChPT and analytic fit forms in Appendix~\ref{appendix-fitforms}. In this appendix we also specify the new fit functions that we use to describe the Wilson flow scales, $t_0^{1/2}$ and $w_0$.


We use the difference between the results obtained for each ansatz to estimate our systematic errors. In order to account for discretization effects, we include in each fit form an $a^2$ term. As discussed in Ref.~\cite{Arthur:2012opa}, we neglect higher order effects including terms in $a^4$ and $a^2 \ln(a^2)$. The fits are performed to dimensionless data, with the parameters determined in the bare normalization of a reference ensemble $r$. The bare lattice quark masses and data on other ensemble sets are `renormalized' into this scheme via additional fit parameters: For an ensemble $e$, these are $Z_l^e$, $Z_h^e$ for normalizing the light and heavy quark masses respectively, and $R_a^e$ for the scale. These are defined as follows:
\begin{equation}\begin{array}{lcl}
\displaystyle Z_{l/h}^e = \frac{1}{R_a^e} \frac{(a \tilde m_{l/h} )^r }{ (a \tilde m_{l/h})^e } & {\rm and} & R_a^e = a^r/a^e\,,
\end{array}\end{equation}
where $a$ is the lattice spacing and $\tilde m = m+m_{\rm res}$. %
%
Note that the scheme used for the quark masses is implicitly mass dependent, hence we allow for different parameters to renormalize the heavy ($Z_h$) and light ($Z_l$) quarks. In practice this dependence is very weak and $Z_l$ and $Z_h$ differ only at the percent level even on our coarsest lattices (cf. Table~\ref{tab-globalfitparams}) despite the order of magnitude difference in the mass scales. Within a large range of light quark masses we previously observed no measurable dependence~\cite{Aoki:2010dy}, which motivated our choice to obtain these quantities as free parameters in the global fit (`generic scaling') rather than by matching at a single mass (`fixed trajectory').%

The procedure for obtaining the general dimensionless fit form for a quantity $Q$ is described in Appendix~\ref{appendix-dimlessforms}. 

We choose a continuum scaling trajectory along which $m_\pi/m_\Omega$ and $m_K/m_\Omega$ match their physical values. Here we include the $\Omega$ baryon mass due to the ease of obtaining an precise lattice measurement and its simple quark mass dependence. This procedure defines $m_\pi$, $m_K$ and $m_\Omega$ as having no lattice spacing dependence. After performing the fit, we obtain the lattice spacing for the reference ensemble by comparing the value of any of the aforementioned quantities to the corresponding physical value after extrapolating to the physical quark masses. The lattice spacings for the other ensembles are then obtained by dividing this value by $R_a^e$. An alternate choice of scaling trajectory, for example using $f_\pi$ in place of $m_\Omega$, would reintroduce the scale dependence on $m_\Omega$ and remove it from $f_\pi$; the values of each $a^2$ coefficient are therefore dependent on the choice of scaling trajectory, but the continuum limit is guaranteed to be the same (up to our ability to measure and extrapolate the quantities in question). Note that the inclusion of the Wilson flow data results in significant improvements in the statistical error on the lattice spacings compared to our previous determinations due to its influence on the shared ratios $R_a$. 

While the data on a given ensemble can be expected to be highly correlated, the estimated correlation matrices tend to suffer from having a poor condition number preventing their use in correlated fits. As a result, our global fits are performed assuming a diagonal correlation matrix. This approach can result in larger jackknife statistical errors than for correlated fits, however in the past~\cite{Allton:2008pn} we have experimented with performing partially-correlated fits where increasingly large numbers of leading eigenvectors were included in the estimate, and found little difference between the uncorrelated and correlated results. With uncorrelated fits the $\chi^2/{\rm d.o.f}$ may not be a reliable indicator of the goodness of fit, and to assess their quality we instead generate histograms of the deviation between the data and the fit.


\subsection{Details specific to this calculation}
\label{sec-chiraldescrspeccal}

Using our simultaneous fit strategy, we combine our 64I and 48I physical point ensembles with a number of existing domain wall ensembles: the `24I' and `32I' ensembles with lattice volumes $24^3\times 64\times 16$ and $32^3\times 64\times 16$ and Shamir domain wall fermions with the Iwasaki gauge action at bare couplings $\beta=2.13$ and $2.25$ respectively (equal to the 48I and 64I bare couplings respectively) described in Refs.~\cite{Allton:2008pn} and~\cite{Aoki:2010dy}; the `32ID' ensembles with volume $32^3\times 64\times 32$ and Shamir domain wall fermions with the Iwasaki+DSDR gauge action at $\beta = 1.75$ described in Ref.~\cite{Arthur:2012opa}; and finally the `32Ifine' ensemble with volume $32^3\times 64\times 12$ and Shamir domain wall fermions with the Iwasaki gauge action at $\beta = 2.37$ described in this document. %
%
%
For the convenience of the reader, we summarize the input parameters of the 24I, 32I and 32ID ensembles along with a number of relevant quantities including the range of pion masses, the lattice spacing and physical lattice size, in Table~\ref{tab-latinputparamsoldens}.

\begin{table}[t]
\begin{tabular}{c|c|c|c}
\hline\hline
	            & 32I                     & 24I                        & 32ID \\
\hline 
Size 	    & $32^3\times 64\times 16$        & $24^3\times 64\times 16$  & $32^3\times 64\times 32$     \\
Action      & Shamir DWF + I            & Shamir DWF + I      & Shamir DWF + ID \\
$\beta$  	    & 2.13                    & 2.25                       & 1.75     \\
\hline
$a^{-1} ({\rm GeV})$      & \ainvzero                & \ainvone                 & \ainvfive \\
$L$ (fm)            & \Lfmzero                 & \Lfmone                  & \Lfmfive \\
$m_\pi L$           & \mpiLzero                & \mpiLone                 & \mpiLfive \\
\hline
$m_\pi$ unitary (MeV)             & 302.4(1.2)--360.1(1.4)  &  339.7(1.3)--339.7(1.3) & 172.4(0.9)--315.5(1.6) \\
$m_\pi$ lightest PQ (MeV)  & 232.4(1.1)  &  248.3(1.2) & 143.8(0.8) \\

\end{tabular}
\caption{Input parameters and relevant quantities for the 32I, 24I and 32ID ensembles. For the action, I stands for the Iwasaki gauge action, and ID the Iwasaki action with the DSDR term. Here $L$ is the spatial lattice extent and $m_\pi L$ is given for the lightest partially-quenched pion at the simulated strange quark mass. The last two rows list the range of unitary pion masses and the lightest partially-quenched pion mass (PQ) mass, respectively. The full set of corresponding bare quark masses are given in Table~\ref{tab-mllist_oldens}. The lattice spacings used here are determined in Section~\ref{sec-physicalpred} \label{tab-latinputparamsoldens} }  
\end{table}


Following our earlier analyses, we use the 32I ensemble set as the reference ensemble against which the `scaling parameters', $Z_{l/h}$ and $R_a$, are defined.

\subsubsection{Ensemble-specific parameters}

As discussed in Section~\ref{sec:SimulationDetails}, the M\"{o}bius parameters of the 48I and 64I ensembles are chosen to ensure the equivalence of the M\"{o}bius and Shamir kernels; as a result, the ensembles with the Iwasaki gauge action can all be described by the same continuum scaling trajectory, i.e. with the same $a^2$ scaling coefficients. As described in Ref.~\cite{Arthur:2012opa}, additional parameters must be introduced to describe the lattice spacing dependence of the 32ID ensembles, which use the Iwasaki+DSDR gauge action to suppress the dislocations that enhance the domain wall residual chiral symmetry breaking on this coarse lattice.

Note that while the 32ID ensemble is the only data set with the Iwasaki+DSDR gauge action, the five additional $a^2$ terms for $f_\pi$, $f_K$, $w_0$, $t_0^{1/2}$ and $B_K$, are completely determined from the overall relative normalization of these data under the $\chi^2$ minimization. This leaves more than sufficient data to determine $Z_l$, $Z_h$ and $R_a$ on this ensemble set and to help constrain the coefficients of the mass terms that are common to all ensembles in these fits. Although the 32ID ensemble set is coarse ($a^{-1}= 1.38(1)$ GeV), we observe discretization effects only at the 5\% level suggesting a discretization systematic error arising from higher order (${\cal O}(a^4)$) terms at the 0.25\% scale, small enough to be neglected. The inclusion of these ensembles in the global fits is discussed at length in Ref.~\cite{Arthur:2012opa}.

The 48I and 64I ensembles have identical bare couplings to the 24I and 32I ensembles respectively, yet differ in their values of the total quark mass, $L_s$ and \mobius scale parameter $\alpha$.  The change in residual chiral symmetry breaking resulting from the changes in $L_s$ and $\alpha$ gives rise to a shift in the bare mass parameter of the low-energy effective Lagrangian, which we account for at leading order in our fits by renormalizing the quark masses as $\tilde m = m+m_{\rm res}$. Higher order effects such as those of order $\mres a^2$ are small enough to be ignored.  After performing this correction we might assume that the scaling parameters $Z_l$, $Z_h$ and $R_a$ (or equivalently the lattice spacing) for the 48I and 24I ensembles should be identical, and likewise for the 64I and 32I ensembles. However when we performed our global fits we found that the 48I lattice spacing is $3.2(2)\%$ larger than that of the 24I ensemble, and the 64I lattice spacing is $1.1(2)\%$ larger than the 32I value. We saw no statistically discernible differences in $Z_l$ and $Z_h$. 

As we mentioned in Section~\ref{sec:mobius} and discuss in detail in Appendix~\ref{appendix-mresadep}, the observed change in the lattice spacings can be expected to originate from the changes in the effective extent of the fifth dimension, $L_s' = \alpha L_s$, which differs by a factor of 3 between the 48I/24I ensembles, and a factor of 1.5 between the 64I/32I.  At finite $L_s'$ the Symanzik effective Lagrangian contains the leading-order operator $\beta_{\mathrm{eff}}{\rm tr}(F_{\mu\nu}F_{\mu\nu})$. A change in $L_s'$ which causes a 0.0025 change in $\mres$ should also be expected to cause a $\approx 0.0025$ change $\beta_{\mathrm{eff}}$, a change which results in an exponentially-enhanced change in the resulting lattice spacing. Recall that the $5.6\%$ change in the coupling between a $\beta = 2.13$, $a^{-1} = 1.75$ GeV ensemble and a $\beta=2.25$, $a^{-1} = 2.38$ GeV ensemble, gives rise to a 36\% change in the inverse lattice spacing.  Thus, we might expect a 3\% change in $a^{-1}$ to result from a 0.5\% change in the effective coupling, not far from the change we observe. We discuss in Appendix~\ref{appendix-mresadep} how changes of this size are not unreasonable, and provide additional numerical evidence for the observed change in lattice scale.
 
Finite $L_s'$ effects will also give rise to other higher order effects of a similar size. For example, we might expect ${\cal O}(0.5\%)$ shifts in the $a^2$ scaling coefficients of the various quantities included in our global fits. However, in Section~\ref{sec:FitResults} we find that even on the coarser 48I ensemble, the discretization effects are only at the 2-3\% level (cf. Table~\ref{tab-48I64I-datacorrections}), suggesting negligible, $0.02\%$ finite $L_s'$ effects. We again emphasize that for our large values of $L_s'$, it is only the exponentially enhanced dependence of the lattice spacing upon the Symanzik coefficients that gives rise to observable finite-$L_s'$ dependence in this quantity. We do not expect any other observable effects.


\begin{table}[t]
\begin{tabular}{c|ccc|ccc}
\hline\hline
Scheme & 48I & 24I & \% diff. & 64I & 32I & \% diff.\\
\hline
$\s{q}$      & 1.43613(80) & 1.4386(12)  & 0.17\%  &  1.43998(80) & 1.4396(37) & 0.03\% \\
$\gamma_\mu$ & 1.52070(89) & 1.5235(13) & 0.18\%   & 1.51764(98) &  1.5192(39) & 0.1\% \\
\end{tabular}
\caption{A comparison of the quark mass renormalization factors $Z_m$ between the 48I/24I and 64I/32I pairs of ensembles, giving the values and their percentage difference. The renormalization scale is 3 GeV and the definitions of the schemes are given in Appendix~\ref{appendix-npr} alongside details of the computation of the 24I and 32I values. Those for the 48I and 64I are not used later in the analysis and are presented here only for comparison. Note that unlike the 24I and 32I values, those for the 48I and 64I ensembles are not extrapolated to the chiral limit as they are computed at only a single mass but for other ensembles we have observed no significant mass dependence for these non-exceptional schemes. \label{tab-zmcomparison} }
\end{table}

Additional evidence for the closeness of our \mobius and Shamir ensembles can be obtained by comparing the renormalization factors for the quark masses, $Z_m$, and the kaon bag parameter, $Z_{B_K}$. The former are computed for the 32I and 24I ensembles in Appendix F, for use in obtaining renormalized physical quark masses later in this document. There we do not present the computation of the corresponding factors for the 48I and 64I ensembles as they are not needed in our later analysis. Nevertheless, we have computed these values, and we list them alongside the 24I and 32I numbers in Table~\ref{tab-zmcomparison}. We observe only tiny, 0.2\% scale differences between the 48I/24I values and even smaller $<0.1\%$ differences for the 64I/32I ensembles. Comparing the values for $Z_{B_K}$ in Table~\ref{tab:finalZbk} we again see differences only at the 0.25\% scale. This strongly suggests that finite-$L_s$ effects have no significant impact upon the UV physics other than through the exponentially enhanced dependence of the lattice spacing upon a shift in the bare coupling at the 0.5\% scale. In addition, these observations justify our fixing both $Z_l$ and $Z_h$ to be the same for the 24I and 48I ensembles, and also for the 32I and 64I ensembles.


\subsubsection{Weighted global fits}

The fits are performed independently for each superjackknife sample $J$ by minimizing $\chi^2_J$ under changes in the set of fit parameters ${\bf c}_J$ of the function $f$. $\chi^2_J$ is defined as
\begin{equation}
\chi^2_J = \sum_i \frac{ [y_{iJ} - f({\bf x}_{iJ}, {\bf c}_J)]^2 }{\sigma^2_i}\,
\end{equation}
were $y_{iJ}$ is the $J^{\rm th}$ superjackknife sample of a measurement $i$ and ${\bf x}_{iJ}$ are the associated input parameters (quark masses, etc). $\sigma_i$ is the error on the measurement, and provides the weight of each data point in the fit.

The na\"{i}ve $\chi^2$-minimization procedure weights each data point according to just its statistical error, and is therefore unable to account for systematic uncertainties on the {\it fit function} itself. Given that NLO $\chi$PT can only be expected to be accurate to $O($5\%) in the 200 - 370 MeV pion-mass range in which the majority of our data lies, the fits over-weight the data in this heavy-mass region resulting in deviations of the fit curve from the light-mass data. In practice the enhanced precision of the near-physical 64I and 48I data partially compensates for the larger number of heavy-mass data points, resulting in only ${\cal O}(1\sigma-2\sigma)$ deviations between these data and the fit curve. However, as the intention of these global fits is only to perform a few-percent mass extrapolation of our near-pristine data, such deviations are unacceptable.%
%
%
While this can be remedied to a certain degree by removing data from the heavy-mass region, there remains pollution from the systematic uncertainty of the fit form. Without going to full NNLO $\chi$PT, one might attempt to reduce this uncertainty by introducing physically motivated `nuisance parameters', perhaps along with Bayesian constraints to confine them within sensible bounds. While this is certainly a valid approach we feel it to be beyond the scope of this work, given that we desire only to perform a small correction to our near-physical data. With this in mind, we instead adopt an alternative approach in which we force the fit curve to pass through our near-physical data by increasing the weight of these data in the $\chi^2$ minimization as follows.

We introduce a measurement-dependent weighting factor $\omega_i$ to the $\chi^2$ determination:
\begin{equation}
\chi^2_J = \sum_i \frac{\omega_i [y_{iJ} - f({\bf x}_{iJ}, {\bf c}_J)]^2 }{\sigma^2_i}\,.
\end{equation}
Note that only the relative values of $\omega_i$ matter as the same parameters that minimize $\chi^2$ will also minimize $r\chi^2$, where $r$ is some common factor. (Of course the algorithm itself has some numerical stopping condition which will need to be adjusted to take into account the change in normalization of $\chi^2$.) In principle one could tune the relative weights based on a combination of the measured statistical error and an estimate of the systematic error of the fit function at each point, but this runs the risk of becoming too complex and arbitrary. Instead, as previously mentioned, we weight the data such that the fit is forced to pass directly through the data points on the 48I and 64I ensembles. To achieve this, we set $\omega_i=\Omega$ for those data, where $\Omega$ is assumed to be large, and $\omega_i=1$ for the remainder. This is performed independently for each superjackknife sample, and does not change the fluctuations on the data {\it between} superjackknife samples. As a result, the statistical error from the overweighted points is unchanged by this procedure. In Appendix~\ref{appendix-weightedfits} we demonstrate that the fit results become independent of $\Omega$ in the limit $\Omega\rightarrow \infty$ and that the procedure has the desired effect of forcing the fit through the physical point data.

For large values of $\Omega$ we must choose small values of the numerical stopping condition on the minimization algorithm, increasing the time to perform the fit and making it more susceptible to finite-precision errors. In the aforementioned appendix we determine that $\Omega=5000$ and a stopping condition of $\delta \chi^2_{\rm min}=1\times 10^{-4}$ is sufficient. 

We emphasize that this procedure is performed separately for each superjackknife sample of our combined data set, such that the error on the fit function evaluated at the parameters associated with the 64I and 48I data is exactly equal to the error on the corresponding data. This can be seen, for example, in Figure~\ref{fig-fpifka2extrap} of Section~\ref{sec-physicalpred}, where we see the $1\sigma$ width of the fit curve exactly aligns with the error bars for the 48I and 64I data.


\FloatBarrier
\section{Fit results and physical predictions}
\label{sec:FitResults}
\begin{table}[tp]
\begin{tabular}{c|l|l|l}
\hline\hline
Ensemble set & \multicolumn{1}{c|}{$m_l$} & \multicolumn{1}{c}{$m_y$} & \multicolumn{1}{|c|}{$\{m_x\}$} \\
\hline
\multirow{6}{*}{\bf 32I}  & \multirow{3}{*}{0.006}        & 0.006 & 0.006, 0.004, 0.002 \\
                      &                               & 0.004 & 0.004, 0.002 \\
                      &                               & 0.002 & 0.002 \\
\cline{2-4} 
     & \multirow{3}{*}{{\bf 0.004}} & 0.006 & 0.006, 0.004, 0.002 \\
                                    &       & {\bf 0.004} & 0.004, {\bf 0.002} \\
                                    &       & {\bf 0.002} & {\bf 0.002} \\     
\hline\hline
\multirow{2}{*}{\bf 24I}  & \multirow{2}{*}{{\bf 0.005}}  	& 0.005       & 0.005, 0.001\\
		      &              			& {\bf 0.001} & {\bf 0.001}\\ 
\hline\hline
\multirow{8}{*}{\bf 32ID} & \multirow{4}{*}{{\bf 0.0042}}    	      & {\bf 0.008}  & 0.008, 0.0042, {\bf 0.001}, {\bf 0.0001} \\
		      &            				      & {\bf 0.0042} & {\bf 0.0042}, {\bf 0.001}, {\bf 0.0001} \\
		      &					              & {\bf 0.001}  & {\bf 0.001}, {\bf 0.0001} \\
		      &               				      & {\bf 0.0001} & {\bf 0.0001} \\
\cline{2-4}
     & \multirow{4}{*}{{\bf 0.001}}	& {\bf 0.008}  & 0.008, 0.0042, {\bf 0.001}, {\bf 0.0001} \\
     &               			& {\bf 0.0042} & {\bf 0.0042}, {\bf 0.001}, {\bf 0.0001} \\
     &               			& {\bf 0.001}  & {\bf 0.001}, {\bf 0.0001} \\
     &               			& {\bf 0.0001} & {\bf 0.0001} \\
\end{tabular}
\caption{The bare light quark masses for the $m_\pi$ and $f_\pi$ data on our older 32I, 24I and 32ID ensembles that we included in our global fits with the 370 MeV pion mass cut. Data in bold are those included in the fits with the lower, 260 MeV cut. Here $m_l$ is the sea light mass, and $m_x$ and $m_y$ are the (partially-quenched) valence masses. The final column gives the full set of available $m_x$ values. Note, each of these points are computed with four different sea strange quark masses that are given in Table~\ref{tab-mhlist_oldens}. \label{tab-mllist_oldens}  }
\end{table}

\begin{table}[tp]
\begin{tabular}{c|l|c|c}
\hline\hline
Ensemble set & $m_h^{\rm sim}$ & \multicolumn{1}{c}{$\{ m_h^{\rm rw} \}$} & \multicolumn{1}{|c}{$\{ m_h^{\rm val} \}$} \\
\hline
32I & 0.03 & 0.029, 0.028, 0.027      & 0.03, 0.025  \\
24I & 0.04 & 0.03775, 0.0355, 0.03325 & 0.04, 0.03\\
32ID & 0.045 & 0.0455, 0.046, 0.0465  & 0.035, 0.045, 0.055
\end{tabular}
\caption{Strange quark masses in the valence and sea sectors on our older 32I, 24I and 32ID ensembles. The second column gives the simulated strange mass, and the third column the subset of reweighted strange masses that are used in our global fits. The final column gives the set of valence strange masses with which we computed the Omega baryon mass, and the kaon mass, decay constant and bag parameters. As discussed in the text, for the 260 MeV pion mass cut, we exclude kaonic data with valence light quark mass $m_x$ if the pion with $m_y=m_x$ is excluded on that ensemble. Similarly, the Omega baryon and the Wilson flow data are excluded if the unitary pion on that ensemble is excluded.\label{tab-mhlist_oldens}}
\end{table}


We performed global fits using the ChPTFV, ChPT and analytic ans\"{a}tze. As discussed in Ref.~\cite{Arthur:2012opa}, we attempt to separate the finite-volume and chiral extrapolation effects by performing the analytic fits to data that is first corrected to the infinite-volume using the ChPTFV fit results. Following Ref.~\cite{Arthur:2012opa}, the ChPTFV and ChPT fits were performed with a 370 MeV pion mass cut on the data (this is set slightly larger than the value used in that paper, as we wish to include in our fit the 32Ifine data with a \mpifour MeV pion). The criteria for excluding the other fitted data are as follows: For $f_\pi$ we exclude the data if the pion mass with the same set of partially-quenched quark masses lies above the cut; for $f_K$ and $m_K$ data points with light valence quark mass $m_x$ and heavy mass $m_y$, we exclude the data if the pion with $m_x=m_y$ on that ensemble is above the pion mass cut; and for $m_\Omega$, $t_0^{1/2}$ and $w_0$ we exclude the data only if the unitary pion on that ensemble is also excluded.

We consider two different pion mass cuts for the analytic fits: the 370 MeV cut used for the ChPTFV and ChPT fits, and a lower, 260 MeV cut. In our previous work we determined that the analytic fits were not able to accurately describe the data over the range from the physical point to the heaviest data, forcing us to use the lower cut. However, in the present analysis the fit predictions are dominated by the near-physical data due to the overweighting procedure, and these data require only a small, percent-scale, chiral extrapolation to correct to the physical light quark mass. This can be seen in Table~\ref{tab-48I64I-datacorrections}, in which we list the sizes of the various corrections required to obtain the physical prediction. We therefore also perform analytic fits with the 370 MeV cut, which includes substantially more data, including a third lattice spacing, that may enable a more precise determination of the dominant $a^2$ scaling behaviour. In practice we find the results to be highly consistent.

Each of the fits with a 370 MeV pion mass cut have 49 free parameters and use 709 data points, giving 660 degrees of freedom; similarly, the analytic fits with the 260 MeV cut have 46 free parameters and use 414 data points, giving 368 degrees of freedom. Note that a substantial amount of the data on the 32ID, 32I and 24I ensembles differ only in their reweighted sea strange quark mass (for which we use four separate values including the simulated value) and are therefore highly correlated. %
The full set of input quark masses for the 32I, 24I and 32ID data that we include in the global fits for each of our two pion mass cuts are summarized in Tables~\ref{tab-mllist_oldens} and~\ref{tab-mhlist_oldens} for convenience. 


The guesses for the parameters in our global fits were input by hand based on a rough order-of-magnitude estimate obtained from previous fits, and within a reasonable basin of attraction we observed no deviations in the fit result (of course wildly different guesses can lead to false minima, but with much much larger $\chi^2$).
 


\setlength{\LTcapwidth}{\textwidth}
\begin{longtable}{c||l||l||l|l|l}
\hline\hline
Quantity & Measured value & Ansatz & $a=0$ & $m_{ud}^{\rm phys}$ & $m_{s}^{\rm phys}$ \\ 
\hline
$f_\pi$(48I) & 0.075799(84) & ChPTFV & -0.0037(73) & -0.00111(30) & 0.00129(30) \\
& & Ana.(370) & -0.0110(67) & -0.00175(20) & -0.00093(44) \\
& & Ana.(260) & -0.0075(80) & -0.00201(24) & -0.00046(33) \\
\hline
$f_\pi$(64I) & 0.055505(95) & ChPTFV & -0.0009(39) & -0.00083(41) & 0.0001(10) \\
& & Ana.(370) & -0.0059(37) & -0.00179(26) & -0.0039(11) \\
& & Ana.(260) & -0.0040(43) & -0.00211(37) & -0.0020(12) \\
\hline
$f_K$(48I) & 0.090396(86) & ChPTFV & -0.0024(58) & -0.00059(14) & -0.00095(68) \\
& & Ana.(370) & -0.0059(54) & -0.00084(10) & -0.00174(73) \\
& & Ana.(260) & -0.0055(62) & -0.00090(12) & -0.00173(75) \\
\hline
$f_K$(64I) & 0.066534(99) & ChPTFV & -0.0009(31) & -0.00047(18) & -0.0061(13) \\
& & Ana.(370) & -0.0032(29) & -0.00085(13) & -0.0074(13) \\
& & Ana.(260) & -0.0029(33) & -0.00093(18) & -0.0073(17) \\
\hline
$f_K/f_\pi$(48I) & 1.1926(14) & ChPTFV & 0.0013(42) & 0.00052(16) & -0.00223(49) \\
& & Ana.(370) & 0.0051(42) & 0.00091(10) & -0.00082(35) \\
& & Ana.(260) & 0.0020(47) & 0.00111(15) & -0.00127(57) \\
\hline
$f_K/f_\pi$(64I) & 1.1987(18) & ChPTFV & 0.0000(23) & 0.00035(23) & -0.00625(89) \\
& & Ana.(370) & 0.0027(23) & 0.00093(13) & -0.00346(68) \\
& & Ana.(260) & 0.0011(25) & 0.00117(22) & -0.0053(13) \\
\hline
$t_0^{1/2}$(48I) & 1.29659(39) & ChPTFV & -0.0276(62) & 0.000122(20) & 0.000204(95) \\
& & Ana.(370) & -0.0260(56) & 0.000120(20) & 0.000176(84) \\
& & Ana.(260) & -0.0259(68) & 0.000140(22) & 0.00023(10) \\
\hline
$t_0^{1/2}$(64I) & 1.74448(98) & ChPTFV & -0.0150(33) & 0.000122(24) & 0.00088(24) \\
& & Ana.(370) & -0.0142(30) & 0.000124(23) & 0.00076(21) \\
& & Ana.(260) & -0.0141(37) & 0.000148(32) & 0.00097(24) \\
\hline
$w_0$(48I) & 1.5013(10) & ChPTFV & 0.0063(59) & 0.000327(40) & 0.00047(20) \\
& & Ana.(370) & 0.0080(54) & 0.000328(41) & 0.00043(19) \\
& & Ana.(260) & 0.0076(66) & 0.000373(48) & 0.00042(18) \\
\hline
$w_0$(64I) & 2.0502(26) & ChPTFV & 0.0034(32) & 0.000322(50) & 0.00199(41) \\
& & Ana.(370) & 0.0043(29) & 0.000335(51) & 0.00183(36) \\
& & Ana.(260) & 0.0041(36) & 0.000388(73) & 0.00179(41) \\

\caption{Data in lattice units on the 48I and 64I ensembles, along with the relative (fractional) correction to the infinite volume limit, in combination with each of the following: the continuum limit, the physical light quark mass and the physical strange mass. The corrections are shown for the ChPTFV fits, the analytic fit with a 260 MeV pion mass cut (labelled `Ana.(260)'), and the analytic fit with a 370 MeV cut (labelled `Ana. (370)'). We include the infinite-volume correction (where applicable) in all of these such that the ChPTFV corrections can be compared directly to those of the analytic fits, where the latter are performed to data that has first been corrected to the infinite volume.\label{tab-48I64I-datacorrections}}
\end{longtable}

The predicted values of the lattice spacings and (unrenormalized) physical quark masses obtained using the ChPTFV ansatz are listed in Table~\ref{tab-finalainvmqbare} alongside the correlated (superjackknife) differences between those and the results for the other ans\"{a}tze. A similar listing of the physical predictions can be found in Table~\ref{tab-predsdiffs}. The corresponding fit parameters for all four ans\"{a}tze are given in Table~\ref{tab-globalfitparams}. For the analytic fit with the 260 MeV cut, the cut excludes the 32Ifine data for which the pion mass is $\mpifour$ MeV, and we are therefore unable to directly obtain the scaling parameters associated with the heavy 32Ifine data; instead we first fit without these data and then determine the remaining unknowns, $Z^{\rm 32Ifine}_{l/h}$ and $R^{\rm 32Ifine}_a$, by including the 32Ifine data while freezing the other fit parameters to those obtained without these data.

\begin{figure}[tp]
\centering
\includegraphics[width=0.5\textwidth]{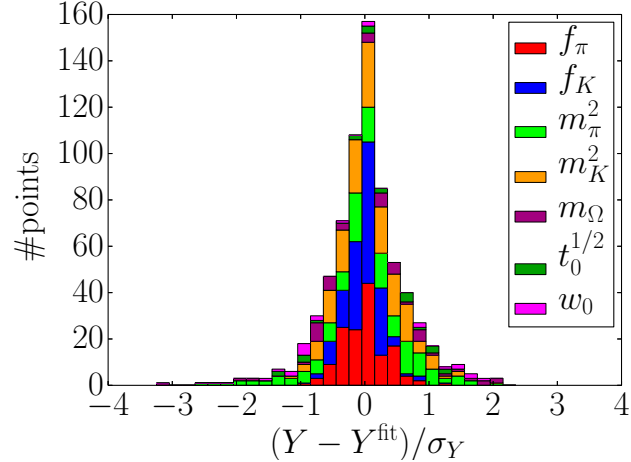}
\caption{A stacked (non-overlapping) histogram of the deviation of the ChPTFV fit curve from our data in units of the statistical error. Different coloured blocks are associated with the different quantities given in the legend. The $3\sigma$ outlier is the $\Omega$ mass on the heavier ($am_l=0.005$) 24I ensemble at the un-reweighted strange mass of 0.04 in lattice units. The jackknife error on this point (not shown) is such that it is consistent with $(y-y_{\rm fit})/\sigma=-2$. \label{fig-chptfv_hist} }
\end{figure}

\begin{figure}[tp]
\centering
\includegraphics[width=0.48\textwidth]{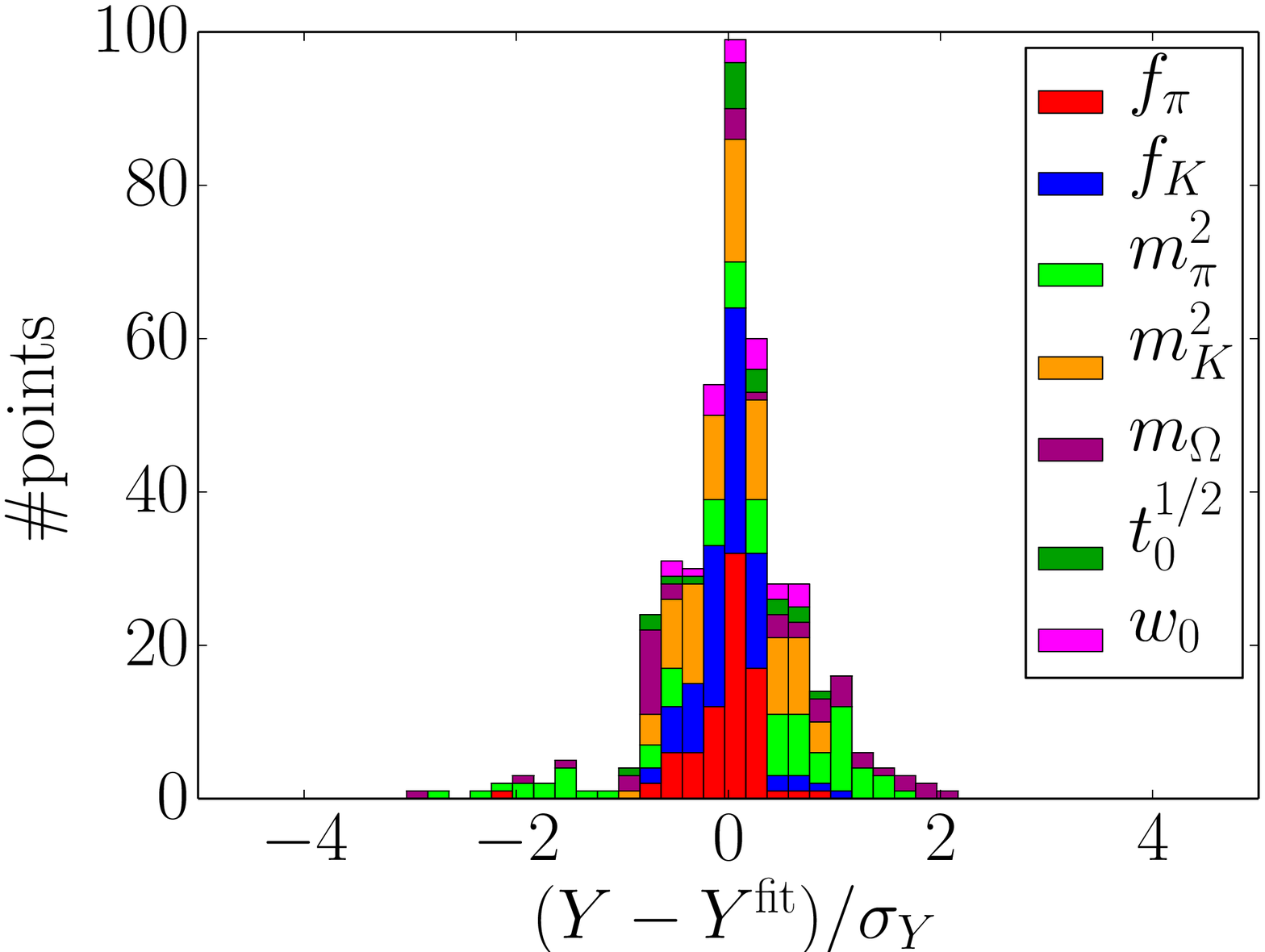}
\includegraphics[width=0.48\textwidth]{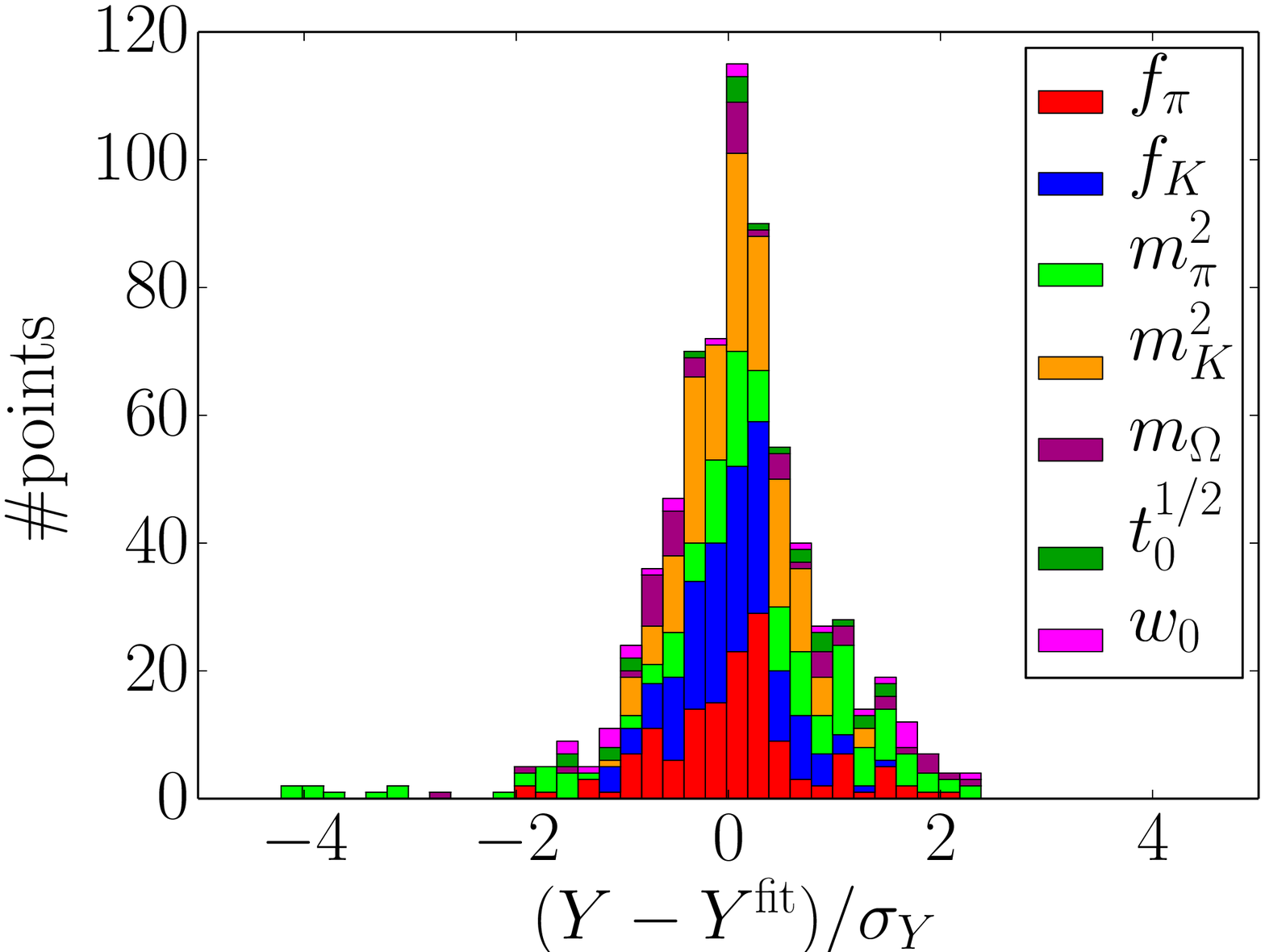}
\caption{A stacked (non-overlapping) histogram of the deviation of the analytic fit curves from our data in units of the statistical error. The left figure is for the 260 MeV pion mass cut, and the right plot for the 370 MeV cut. Different coloured blocks are associated with the different quantities given in the legend. The outliers in the right-hand plot are exclusively from $m_\pi$ on the 32ID ensembles, indicating that the linear curve is deviating from the data due to chiral curvature. \label{fig-ana_hist} }
\end{figure}

\begin{figure}[tp]
\centering
\includegraphics[width=0.48\textwidth]{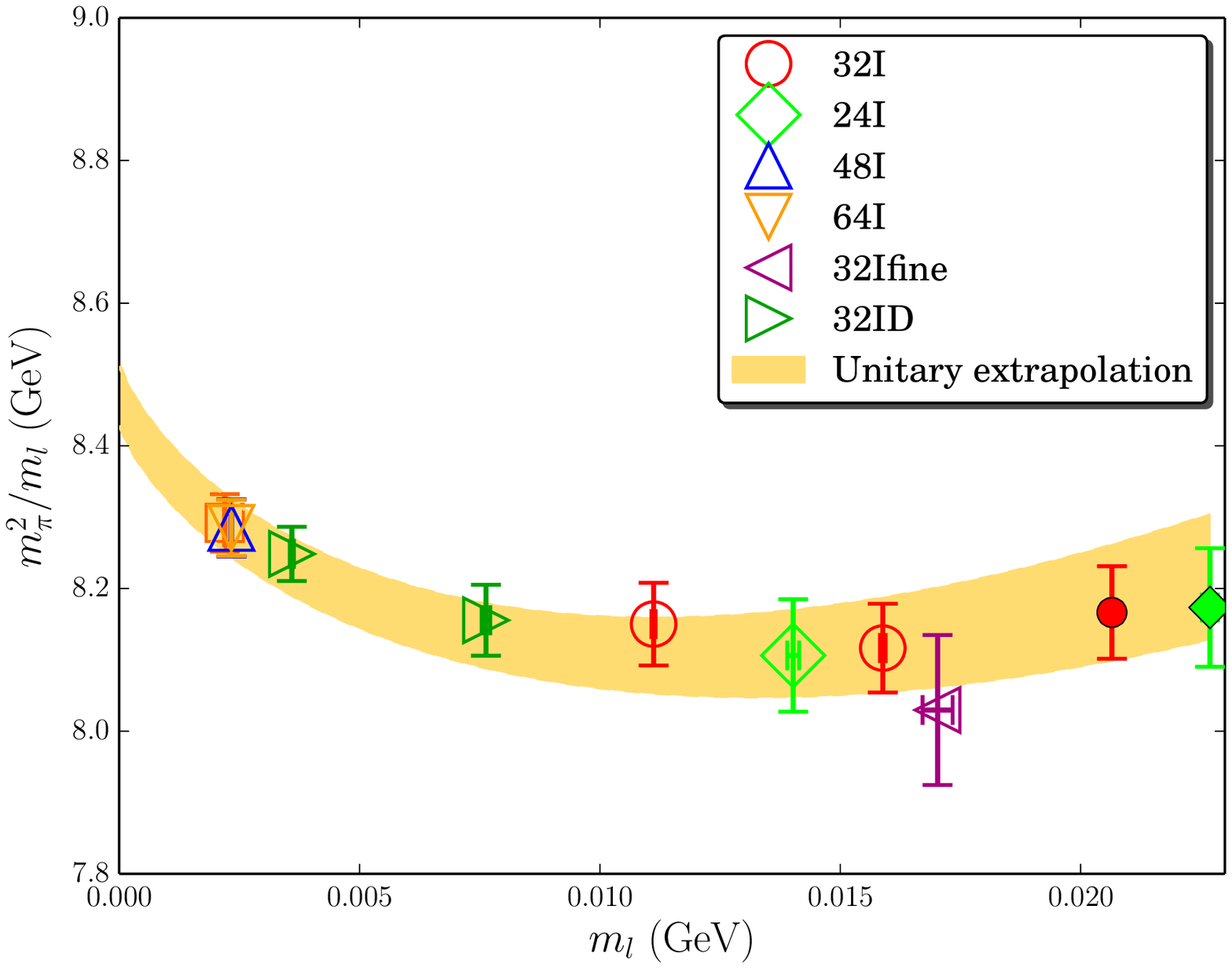}
\includegraphics[width=0.48\textwidth]{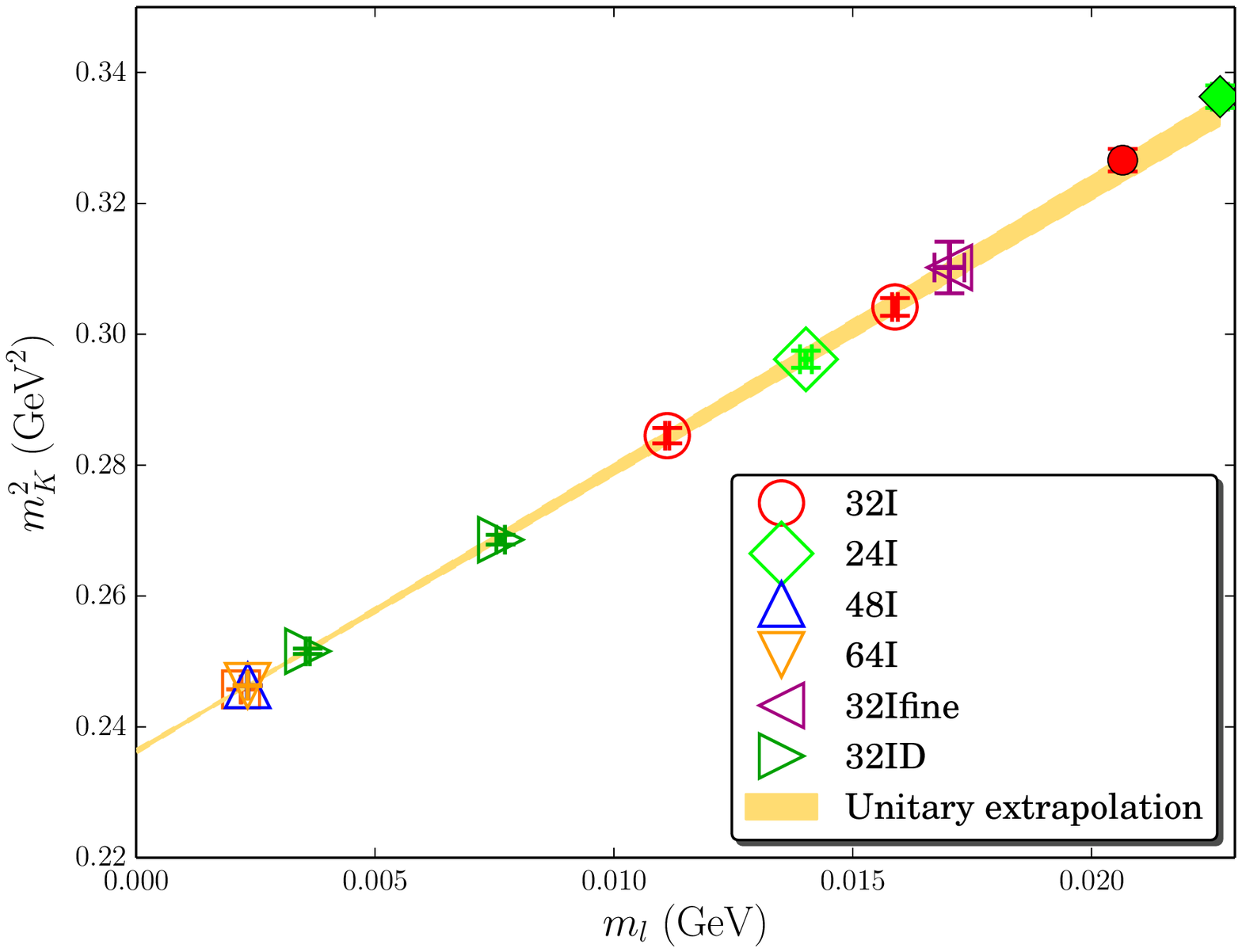}\\
\includegraphics[width=0.48\textwidth]{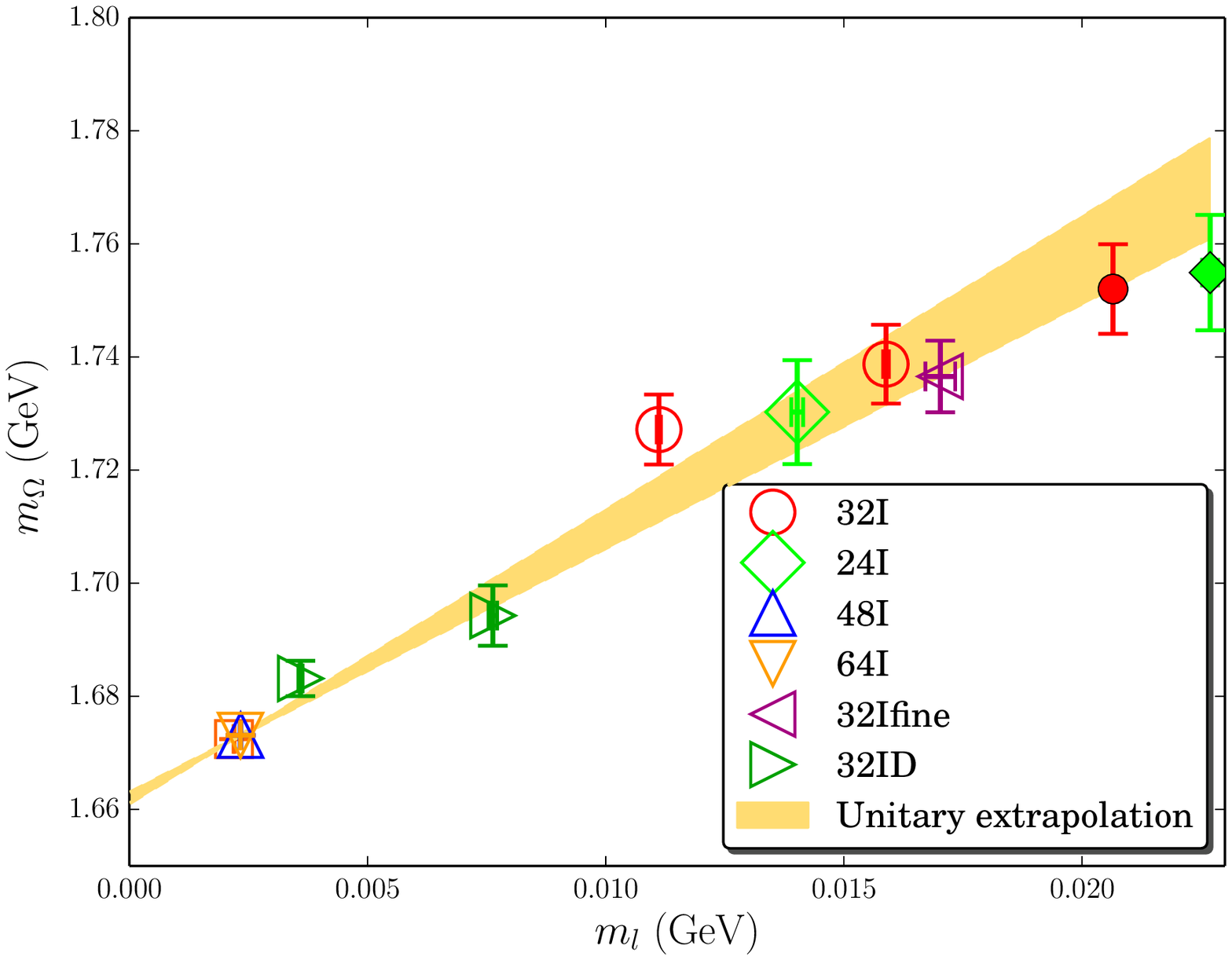}

\caption{$m_\pi^2/m_l$ (upper-left), $m_K^2$ (upper-right) and $m_\Omega$ (lower) unitary data corrected to the physical strange quark mass and the infinite volume limit as a function of the unrenormalized physical quark mass, plotted against the ChPTFV fit curves. Data with hollow symbols are those included in the fit and data with filled symbols are those excluded. The square point is our predicted continuum value. Note the 64I and 48I data lie essentially on top of each other in this figure.\label{fig-mpimkmomegaunitary} }
\end{figure}

In Figure~\ref{fig-mpimkmomegaunitary} we plot the unitary mass dependence of $m_\pi$, $m_K$ and $m_\Omega$, which are used to determine the quark masses and overall lattice scale. In this figure we clearly see that the overweighting procedure forces the curve to pass through the near-physical data as desired, and that this procedure does not introduce any significant tension with the heavier data. In Figure~\ref{fig-chptfv_hist} we plot a histogram of the deviation of the data from the ChPTFV fit, showing excellent general agreement between the fit and the data, and in Figure~\ref{fig-ana_hist} we plot the corresponding histograms for the analytic fits. For the analytic fit with the 370 MeV mass cut we observe ${\cal O}(3-4)\sigma$ deviations of the 32ID pion mass data from the fit curve, which arise because of chiral curvature in the data: the fit is pinned near the physical point by the overweighting procedure and is strongly influenced by the larger volume of data in the heavy mass regime, leading to deviations from the lighter 32ID data that lies between these extremes. Nevertheless, in Tables~\ref{tab-finalainvmqbare},~\ref{tab-predsdiffs} and~\ref{tab-globalfitparams} we generally observe better agreement between the analytic fit with the 370 MeV mass cut and the ChPTFV results than for the lower cut. The total (uncorrelated) $\chi^2/d.o.f.$ are given in Table~\ref{tab-chisq} and are sub-unity for all four ans\"{a}tze.

As previously mentioned, the inclusion of the Wilson flow data in these fits has a significant effect on the precision of the lattice spacings via their influence on the shared $R_a$ parameters. This can be seen in Table~\ref{tab-inclwfloweffect}, in which we show the various scaling parameters, as well as the unrenormalized quark masses and lattice spacings, obtained using the ChPTFV ansatz with and without the Wilson flow data. For the 48I and 64I ensembles, for which the hadronic measurements are very precise, we see only a small improvement in the statistical error. However, for the 32I, 24I and 32Ifine ensembles we observe factors of three or more improvements in precision. The results themselves are very consistent.


\begin{figure}[tp]
\centering
\includegraphics[width=0.5\textwidth]{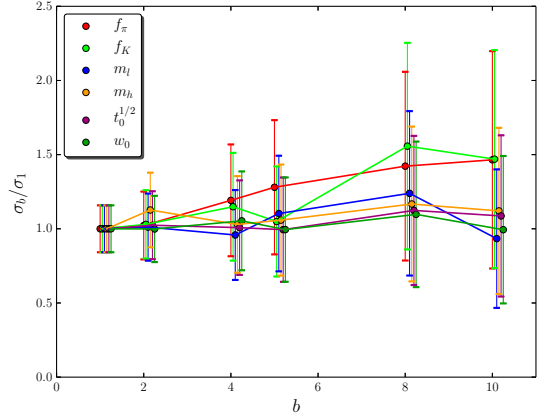}
\caption{The dependence of the error for the predicted physical values, obtained from our global fits with the ChPTFV ansatz, of various quantities as a function of the bin size used for the 64I ensemble. The vertical axis plots the ratio $\sigma_{b}/\sigma_1$ for bin size $b$ along the horizontal axis, where $\sigma$ is the statistical error and the subscript indicates the 64I bin size for which that error was computed. The upper and lower bounds were obtained by varying $\sigma_b$ by $1/\sqrt{N}$, where $N$ is the number of samples. \label{fig-binerrglobal}  }
\end{figure}

In Figure~\ref{fig-binerrglobal} we plot the dependence of our physical predictions on the bin size used for the 64I data. Here we observe no statistically significant dependence on the bin size, further attesting that our chosen bin size of 5 ($5\times 40$ MD time units) is a conservative choice and does not lead to an underestimate in the errors on our physical predictions.

We would like to emphasize that the goal of this analysis is not to extract reliable model parameters but simply to perform a few-percent extrapolation of our pristine near-physical data to the physical point. As we discuss in Section~\ref{sec-chiraldescrspeccal}, we are well aware that NLO ChPT can be expected to fail at the 5\% level in the 200-370 MeV mass range in which the majority of our data lies (and where the fit would be most heavily weighted if we weighted the data by statistical error alone), and we do not want this model failure to unduly influence the quality of our prediction. The overweighting procedure was chosen to ensure that the fits pass through our 48I and 64I data with the heavier data used only to guide the extrapolation. Despite this, we find that the fits are largely insensitive to the pion mass cut and to the fit ansatz such that all of our results agree to a high degree (including their uncorrelated $\chi^2/{\rm d.o.f.}$). In order to gauge the quality of our uncorrelated fits, we present histograms of the deviation of the fit from our data in Figures~\ref{fig-chptfv_hist},~\ref{fig-ana_hist} (and~\ref{fig-bkqqhist} for $B_K$), and we see no spuriously large deviations that cannot be accounted for by higher-order mass dependent terms. Given the high degree of consistency between our results, there is no reason to suggest that any of the fits has converged upon a false minimum. Furthermore, the predictive power of these global fits is highlighted by our numerical discovery of the 3\% shift in lattice spacings between the 48I and 24I ensembles and the smaller 1\% shift between the 64I and 32I ensembles.


\subsection{Systematic error estimation}

In our previous analyses we used the difference between the ChPTFV and ChPT results as a conservative estimate of the higher-order finite-volume errors on our results (recall the ChPTFV formulae incorporate the NLO finite-volume corrections). From a purely $\chi$PT perspective this is a considerable over-estimate of the size of the NNLO and above corrections, which are known to be only a small fraction of the NLO values even at smaller volumes. Our prudence was motivated by Ref.~\cite{Colangelo:2005gd}, in which the authors observed significant deviations between the finite-volume corrections predicted by standard finite-volume chiral perturbation theory and those obtained via a resummed version of the L\"{u}scher formula~\cite{Luscher:1985dn} that relates the finite-volume mass shift of a particle to the infinite-volume Euclidean scattering length of that particle with the pion. Nevertheless, one can conclude from those results that the full finite-volume corrections can be expected to differ from the NLO $\chi$PT predictions by only 30--50\% for the light pions that we are currently using. 

Our present fits are dominated by near-physical data computed on 5.5fm volumes, such that (e.g. in Tables~\ref{tab-finalainvmqbare} and~\ref{tab-predsdiffs}) we observe only very tiny differences between the ChPT and ChPTFV fit results; these differences are typically 10--20\% of the size of the statistical error, and hence have negligible impact upon the total error. Given the small size of these differences and that the true sizes of the higher-order finite-volume effects are expected to be several times smaller, we therefore choose to omit the finite-volume systematic from our error estimate.

The estimate of the chiral extrapolation error is made difficult due to the fact that the global fits combine the chiral and continuum extrapolations together, and in this analysis the latter are larger than the former while being less well determined by the fits (the $a^2$ parameters have typically 50-100\% statistical error). As a result, the established procedure of estimating the chiral error from the difference of the ChPTFV and analytic result with a 260 MeV cut is no longer satisfactory. 

In this analysis we considered analytic fits with both a 260 MeV and a 370 MeV pion mass cut. The latter is clearly applying the linear ansatz outside of its region of applicability, leading to deviations from the 32ID data at the 3-4$\sigma$ level. Despite this there is generally excellent agreement between the continuum predictions of this fit and the ChPTFV. The analytic fit with the 260 MeV mass cut does not suffer from this issue, but at the expense of fitting to a considerably smaller amount of data, including one less lattice spacing. The ChPTFV fit on the other hand is theoretically `clean' in that it is the correct ansatz for the data in the chiral limit, and agrees very well with our data when applied in the 140 to 370 MeV pion mass range. In Table~\ref{tab-48I64I-datacorrections} we see that all four ans\"{a}tze agree at a broad level (given the size of the errors on the $a^2$ terms) as to the size of the continuum extrapolation, and this is by far the dominant correction. The only significant inconsistencies are in the light quark extrapolation, for which the 260 MeV analytic fit gives a larger correction indicating a stronger slope near the physical point. Nevertheless, the differences between the predicted corrections of the ChPTFV and 260 MeV analytic fits are at most on the 0.1$\%$ level.

Given the small size of the observed differences in the corrections to the 48I and 64I data, and our understanding that these are likely a result of deficiencies in the fitting strategies for those ans\"{a}tze, we choose to take the cleaner ChPTFV ansatz, which describes our data very well, as our final result and treat the systematic error associated with the extrapolation to the physical point as negligible.


Finally, we consider the discretization systematic. For Wilson-style fermions the explicit symmetry breaking allows for a dimension-5 clover term of ${\cal O}(a\Lambda_{QCD})$; for domain wall fermions this term is heavily suppressed by the separation of the chiral modes in the fifth dimension, and can be discounted in practice~\cite{Arthur:2012opa}. Our domain wall simulations can be treated as non-perturbatively ${\cal O}(a)$ improved, and further chiral symmetry implies that all terms containing an odd power of the lattice spacing (${\cal O}(a\Lambda_{\rm QCD})$, ${\cal O}(a^3\Lambda^3_{\rm QCD})$, etc) can be neglected; the leading discretization effects therefore enter at the ${\cal O}(a^4\Lambda^4_{\rm QCD})$ level, and
these are of a comparable size~\cite{Arthur:2012opa} to logarithmic corrections to lattice artefacts that are regularly considered negligible. In our previous papers and above (cf. Table ~\ref{tab-48I64I-datacorrections}) we observe that the discretization effects for the coarser 48I ensemble are at the 2\% level, implying a ${\cal O}(0.04\%)$ discretization systematic  that can be neglected. (For our very coarse 32ID ensemble the discretization effects enter at the 5\% level, implying  ${\cal O}(0.25\%)$ discretization errors that can also be discounted.) We could therefore, in principle, obtain precise continuum results from just two lattice spacings, as we have done in previous publications. However, the fits in this document utilize {\it three} widely spaced lattice spacings with the Shamir fermion action. In this document we present several plots overlaying our data with the fitted scaling behavior, from which we observe no evidence of deviations from $a^2$ scaling.


\begin{table}[tp]
\centering
\begin{tabular}{c|c|c|c}
\hline\hline
ChPTFV & ChPT & Analytic (260 MeV) & Analytic (370 MeV)\\
\hline
0.44(13) & 0.44(16) & 0.49(14) & 0.79(18)\\
\end{tabular}
\caption{The $\chi^2/{\rm d.o.f.}$ for each of the four chiral ans\"{a}tze. Here the $\chi^2$ does not include the overweighted data, and the number of degrees of freedom has been correspondingly reduced. For the analytic fits, the pion mass cut is given in parentheses.\label{tab-chisq} }
\end{table}

\begin{table}[tp]
\centering
\begin{tabular}{l||l||r|r|r}
\hline
 & ChPTFV & $\Delta$(ChPT) & $\Delta$ (Analytic [260 MeV]) & $\Delta$ (Analytic [370 MeV])\\
\hline
 $am_l$(32I) & $ 0.000260(13) $  & $ 0.00000152(63) $  & $ -0.0000054(63) $  & $ -0.0000025(58) $  \\
 $am_s$(32I) & $ 0.02477(18) $  & $ 0.000044(15) $  & $ 0.000032(95) $  & $ 0.000072(45) $  \\
 $a^{-1}$(32I) & $ 2.3833(86) $ GeV & $ -0.00234(74) $ GeV & $ -0.0001(51) $ GeV & $ -0.0043(25) $ GeV \\
\hline
 $am_l$(64I) & $ 0.0006203(77) $  & $ 0.00000137(62) $  & $ -0.0000047(60) $  & $ -0.0000031(56) $  \\
 $am_s$(64I) & $ 0.02539(17) $  & $ 0.000039(14) $  & $ 0.000054(88) $  & $ 0.000056(40) $  \\
 $a^{-1}$(64I) & $ 2.3586(70) $ GeV & $ -0.00181(67) $ GeV & $ -0.0021(40) $ GeV & $ -0.0027(19) $ GeV \\
\hline
 $am_l$(24I) & $ -0.001770(79) $  & $ -0.00000048(35) $  & $ -0.0000037(21) $  & $ -0.0000012(20) $  \\
 $am_s$(24I) & $ 0.03224(18) $  & $ 0.0000209(69) $  & $ -0.000054(50) $  & $ 0.000046(18) $  \\
 $a^{-1}$(24I) & $ 1.7848(50) $ GeV & $ -0.00074(21) $ GeV & $ 0.0032(22) $ GeV & $ -0.00194(65) $ GeV \\
\hline
 $am_l$(48I) & $ 0.0006979(81) $  & $ -0.00000049(35) $  & $ -0.0000020(18) $  & $ -0.0000016(19) $  \\
 $am_s$(48I) & $ 0.03580(16) $  & $ 0.0000129(64) $  & $ 0.000015(25) $  & $ 0.000017(13) $  \\
 $a^{-1}$(48I) & $ 1.7295(38) $ GeV & $ -0.00029(16) $ GeV & $ -0.00027(59) $ GeV & $ -0.00042(33) $ GeV \\
\hline
 $am_l$(32ID) & $ -0.000106(17) $  & $ -0.0000069(12) $  & $ -0.000002(13) $  & $ 0.0000004(61) $  \\
 $am_s$(32ID) & $ 0.04625(48) $  & $ -0.000091(27) $  & $ -0.00018(28) $  & $ -0.00016(11) $  \\
 $a^{-1}$(32ID) & $ 1.3784(68) $ GeV & $ 0.00141(37) $ GeV & $ 0.0025(38) $ GeV & $ 0.0020(17) $ GeV \\
\hline
 $am_l$(32Ifine) & $ 0.000058(16) $  & $ 0.0000021(20) $  & $ 0.000024(12) $  & $ 0.0000040(57) $  \\
 $am_s$(32Ifine) & $ 0.01852(30) $  & $ 0.000044(34) $  & $ -0.00019(26) $  & $ 0.00005(10) $  \\
 $a^{-1}$(32Ifine) & $ 3.148(17) $ GeV & $ 0.0003(14) $ GeV & $ 0.0100(99) $ GeV & $ -0.0020(44) $ GeV \\
\end{tabular}
\caption{The unrenormalized physical quark masses in bare lattice units (without $\mres$ included) and the values of the inverse lattice spacing $a^{-1}$ obtained using the ChPTFV ansatz, and the full correlated differences (labelled $\Delta$) between the results obtained using the other ans\"{a}tze and the ChPTFV result. We present analytic fit results obtained using both the 370 MeV and 260 MeV pion mass cut. The latter fit was performed without the 32Ifine data, and a separate fit with fixed parameters was used to obtain the 32Ifine scaling parameters. \label{tab-finalainvmqbare} }
\end{table}

\begin{table}[tp]
\centering
\begin{tabular}{l||l||r|r|r}
\hline
 & ChPTFV & $\Delta$(ChPT) & $\Delta$ (Analytic [260 MeV]) & $\Delta$ (Analytic [370 MeV])\\
\hline
 $f_\pi$ & $ 0.1302(9) $ GeV & $ -0.000375(53) $ GeV & $ -0.00019(45) $ GeV & $ -0.00068(20) $ GeV \\
 $f_K$ & $ 0.1555(8) $ GeV & $ -0.000251(52) $ GeV & $ -0.00035(43) $ GeV & $ -0.00043(17) $ GeV \\
 $f_K/f_\pi$ & $ 1.1945(45) $  & $ 0.00152(12) $  & $ -0.0010(21) $  & $ 0.00297(60) $  \\
 $t_0^{1/2}$ & $ 0.7292(41) $ GeV$^{-1}$ & $ 0.00098(37) $ GeV$^{-1}$ & $ 0.0014(23) $ GeV$^{-1}$ & $ 0.0014(11) $ GeV$^{-1}$ \\
 $w_0$ & $ 0.8742(46) $ GeV$^{-1}$ & $ 0.00114(42) $ GeV$^{-1}$ & $ 0.0013(27) $ GeV$^{-1}$ & $ 0.0016(12) $ GeV$^{-1}$ \\
\end{tabular}
\caption{The physical predictions obtained using the ChPTFV ansatz, and the full correlated differences (labelled $\Delta$) between the results obtained using the other ans\"{a}tze and the ChPTFV result. We present analytic fit results obtained using both the 370 MeV and 260 MeV pion mass cut. The latter fit was performed without the 32Ifine data, and a separate fit with fixed parameters was used to obtain the 32Ifine scaling parameters. \label{tab-predsdiffs} }
\end{table}

\FloatBarrier
\setlength{\LTcapwidth}{\textwidth}

\begin{longtable}{l|rr||l|rr}
\hline
Parameter & ChPT & ChPTFV & Parameter & Analytic (260 MeV) & Analytic (370 MeV)\\
\hline
 \rule{0cm}{0.4cm}$Z_l^{\scriptscriptstyle 24I}$ & $ 0.9727(51) $ & $ 0.9715(54) $ &  & $ 0.9675(70) $ & $ 0.9686(56) $ \\
 $Z_l^{\scriptscriptstyle 48I}$ & $ 0.9727(51) $ & $ 0.9715(54) $ &  & $ 0.9675(70) $ & $ 0.9686(56) $ \\
 $Z_l^{\scriptscriptstyle 32ID}$ & $ 0.9192(67) $ & $ 0.9156(72) $ &  & $ 0.910(13) $ & $ 0.9105(84) $ \\
 $Z_l^{\scriptscriptstyle 32Ifine}$ & $ 1.012(19) $ & $ 1.015(17) $ &  & $ 0.971(19) $ & $ 1.005(15) $ \\
 $Z_h^{\scriptscriptstyle 24I}$ & $ 0.9634(38) $ & $ 0.9628(40) $ &  & $ 0.9637(43) $ & $ 0.9636(36) $ \\
 $Z_h^{\scriptscriptstyle 48I}$ & $ 0.9634(38) $ & $ 0.9628(40) $ &  & $ 0.9637(43) $ & $ 0.9636(36) $ \\
 $Z_h^{\scriptscriptstyle 32ID}$ & $ 0.9159(60) $ & $ 0.9144(63) $ &  & $ 0.9174(82) $ & $ 0.9172(56) $ \\
 $Z_h^{\scriptscriptstyle 32Ifine}$ & $ 1.004(12) $ & $ 1.005(12) $ &  & $ 1.013(16) $ & $ 1.005(12) $ \\
 $R_a^{\scriptscriptstyle 24I}$ & $ 0.7493(22) $ & $ 0.7489(24) $ &  & $ 0.7503(26) $ & $ 0.7494(21) $ \\
 $R_a^{\scriptscriptstyle 48I}$ & $ 0.7263(27) $ & $ 0.7257(28) $ &  & $ 0.7256(29) $ & $ 0.7268(25) $ \\
 $R_a^{\scriptscriptstyle 64I}$ & $ 0.9898(19) $ & $ 0.9896(19) $ &  & $ 0.9888(16) $ & $ 0.9903(18) $ \\
 $R_a^{\scriptscriptstyle 32ID}$ & $ 0.5795(34) $ & $ 0.5783(36) $ &  & $ 0.5794(45) $ & $ 0.5802(33) $ \\
 $R_a^{\scriptscriptstyle 32Ifine}$ & $ 1.3222(44) $ & $ 1.3208(44) $ &  & $ 1.3251(46) $ & $ 1.3224(43) $ \\
\hline
 \rule{0cm}{0.4cm}$B$ (GeV) & $ 4.233(21) $ & $ 4.236(21) $ & \rule{0cm}{0.4cm}$C^{m_\pi}_0$ ($[{\rm GeV}]^2$)& $ 0.00037(15) $ & $ 0.000421(91) $ \\
 $L_8^{(2)}$ & $ 0.000611(41) $ & $ 0.000631(41) $ & $C^{m_\pi}_1$ (GeV)& $ 7.982(80) $ & $ 7.917(51) $ \\
 $L_6^{(2)}$ & $ -0.000145(36) $ & $ -0.000146(36) $ & $C^{m_\pi}_2$ (GeV)& $ 0.190(32) $ & $ 0.219(25) $ \\
 $c_{m_\pi,m_h}$ & $ 6.8(4.1) $ & $ 3.7(4.1) $ & $C^{m_\pi}_3$ (GeV)& $ -0.036(31) $ & $ -0.026(32) $ \\
 $f$ (GeV) & $ 0.12195(94) $ & $ 0.12229(96) $ & $C^{f_\pi}_0$ (GeV)& $ 0.1259(11) $ & $ 0.12593(88) $ \\
 $c_f^{\scriptscriptstyle I}$ ($[{\rm GeV}]^2$)  & $ 0.021(23) $ & $ 0.017(23) $ & $C^{f_\pi,\,\scriptscriptstyle I}_a$ ($[{\rm GeV}]^2$)& $ 0.023(25) $ & $ 0.034(21) $ \\
 $c_f^{\scriptscriptstyle ID}$ ($[{\rm GeV}]^2$) & $ -0.027(30) $ & $ -0.033(30) $ & $C^{f_\pi,\,\scriptscriptstyle ID}_a$ ($[{\rm GeV}]^2$)& $ -0.007(31) $ & $ 0.013(29) $ \\
 $L_5^{(2)}$ & $ 0.000524(78) $ & $ 0.000513(78) $ & $C^{f_\pi}_1$ & $ 1.082(78) $ & $ 0.988(45) $ \\
 $L_4^{(2)}$ & $ -0.000198(64) $ & $ -0.000171(64) $ & $C^{f_\pi}_2$ & $ 0.792(75) $ & $ 0.643(71) $ \\
 $c_{f_\pi,m_h}$ & $ 0.084(46) $ & $ 0.070(46) $ & $C^{f_\pi}_3$ & $ 0.094(54) $ & $ 0.188(46) $ \\
 $m^{(K)}$ ($[{\rm GeV}]^2$) & $ 0.2363(16) $ & $ 0.2363(17) $ & $C^{m_K}_0$ ($[{\rm GeV}]^2$)& $ 0.2363(19) $ & $ 0.2363(15) $ \\
 $\lambda_2$ & $ 0.02825(50) $ & $ 0.02845(50) $ & $C^{m_K}_1$(GeV) & $ 3.782(77) $ & $ 3.828(43) $ \\
 $\lambda_1$ & $ 0.00367(71) $ & $ 0.00371(72) $ & $C^{m_K}_2$ (GeV)& $ 0.54(16) $ & $ 0.478(95) $ \\
 $c_{m_K,m_y}$ (GeV) & $ 3.933(16) $ & $ 3.935(17) $ & $C^{m_K}_3$ (GeV)& $ 3.923(22) $ & $ 3.929(15) $ \\
 $c_{m_K,m_h}$ (GeV) & $ 0.097(86) $ & $ 0.094(86) $ & $C^{m_K}_4$ (GeV)& $ 0.11(15) $ & $ 0.075(83) $ \\
 $f^{(K)}$ (GeV) & $ 0.15123(94) $ & $ 0.15146(97) $ & $C^{f_K}_0$ (GeV)& $ 0.1530(11) $ & $ 0.15304(89) $ \\
 $c_{f^{(K)}}^{\scriptscriptstyle I}$ ($[{\rm GeV}]^2$)& $ 0.012(18) $ & $ 0.010(18) $ & $C^{f_K,\,\scriptscriptstyle I}_a$ ($[{\rm GeV}]^2$)& $ 0.017(19) $ & $ 0.018(17) $ \\
 $c_{f^{(K)}}^{\scriptscriptstyle ID}$ ($[{\rm GeV}]^2$)& $ -0.020(27) $ & $ -0.024(27) $ & $C^{f_K,\,\scriptscriptstyle ID}_a$ ($[{\rm GeV}]^2$)& $ -0.006(28) $ & $ -0.001(26) $ \\
 $\lambda_4$ & $ 0.00620(38) $ & $ 0.00594(39) $ & $C^{f_K}_1$ & $ 0.343(78) $ & $ 0.361(34) $ \\
 $\lambda_3$ & $ -0.00383(79) $ & $ -0.00335(80) $ & $C^{f_K}_2$ & $ 0.653(86) $ & $ 0.573(69) $ \\
 $c_{f_K,m_y}$ & $ 0.2952(51) $ & $ 0.2959(51) $ & $C^{f_K}_3$ & $ 0.3047(60) $ & $ 0.2991(51) $ \\
 $c_{f_K,m_h}$ & $ 0.074(45) $ & $ 0.080(46) $ & $C^{f_K}_4$ & $ 0.113(62) $ & $ 0.124(46) $ \\
 $m^{(\Omega)}$ (GeV)& $ 1.6618(30) $ & $ 1.6620(33) $ & $C^{m_\Omega}_0$ (GeV)& $ 1.6612(43) $ & $ 1.6618(27) $ \\
 $c_{m_\Omega,m_l}$ & $ 4.86(42) $ & $ 4.75(43) $ & $C^{m_\Omega}_1$ & $ 5.14(75) $ & $ 4.89(44) $ \\
 $c_{m_\Omega,m_v}$ & $ 5.565(44) $ & $ 5.583(46) $ & $C^{m_\Omega}_2$ & $ 5.582(63) $ & $ 5.553(42) $ \\
 $c_{m_\Omega,m_h}$ & $ 1.39(45) $ & $ 1.60(47) $ & $C^{m_\Omega}_3$ & $ 1.35(74) $ & $ 1.27(47) $ \\
 $c_{\sqrt{t_0},0}$ ($[{\rm GeV}]^{-1}$)& $ 0.7317(39) $ & $ 0.7307(42) $ & $c_{\sqrt{t_0},0}$ ($[{\rm GeV}]^{-1}$)& $ 0.7323(49) $ & $ 0.7320(37) $ \\
 $c^I_{\sqrt{t_0},a} $ ($[{\rm GeV}]^2$)& $ 0.081(18) $ & $ 0.085(19) $ & $c^I_{\sqrt{t_0},a}$ ($[{\rm GeV}]^2$)& $ 0.079(21) $ & $ 0.080(18) $ \\
 $c^{ID}_{\sqrt{t_0},a}$ ($[{\rm GeV}]^2$)& $ 0.037(13) $ & $ 0.042(14) $ & $c^{ID}_{\sqrt{t_0},a}$ ($[{\rm GeV}]^2$)& $ 0.035(18) $ & $ 0.035(13) $ \\
 $c_{\sqrt{t_0},l}$ ($[{\rm GeV}]^{-2}$)& $ -0.655(81) $ & $ -0.660(81) $ & $c_{\sqrt{t_0},l}$ ($[{\rm GeV}]^{-2}$)& $-0.747(84) $ & $ -0.640(80) $ \\
 $c_{\sqrt{t_0},h}$ ($[{\rm GeV}]^{-2}$)& $ -0.221(43) $ & $ -0.227(43) $ & $c_{\sqrt{t_0},h}$ ($[{\rm GeV}]^{-2}$)& $-0.262(38) $ & $ -0.205(44) $ \\
 $c_{w_0,0}$ ($[{\rm GeV}]^{-1}$)& $ 0.8798(46) $ & $ 0.8787(48) $ & $c_{w_0,0}$ ($[{\rm GeV}]^{-1}$)& $ 0.8805(58) $ & $ 0.8803(43) $ \\
 $c^I_{w_0,a}$ ($[{\rm GeV}]^2$)& $ -0.022(16) $ & $ -0.019(17) $ & $c^I_{w_0,a}$ ($[{\rm GeV}]^2$)& $ -0.022(19) $ & $ -0.024(16) $ \\
 $c^{ID}_{w_0,a}$ ($[{\rm GeV}]^2$)& $ 0.018(12) $ & $ 0.023(13) $ & $c^{ID}_{w_0,a}$ ($[{\rm GeV}]^2$)& $ 0.018(17) $ & $ 0.016(12) $ \\
 $c_{w_0,l}$ ($[{\rm GeV}]^{-2}$)& $ -2.05(12) $ & $ -2.06(12) $ & $c_{w_0,l}$ ($[{\rm GeV}]^{-2}$)& $ -2.30(14) $ & $ -2.03(12) $ \\
 $c_{w_0,h}$ ($[{\rm GeV}]^{-2}$)& $ -0.597(64) $ & $ -0.602(64) $ & $c_{w_0,h}$ ($[{\rm GeV}]^{-2}$)& $ -0.567(67) $ & $ -0.580(65) $ \\
\caption{The fit parameters of each of our chiral ans\"{a}tze. The parameters are given in physical units and with the heavy quark mass expansion point adjusted to the physical strange quark mass {\it a posteriori}. Analytic fit results are presented with a 370 MeV and 260 MeV pion mass cut. The latter was performed without the 32Ifine data, and a separate fit with fixed parameters was used to obtain the 32Ifine scaling parameters. For the ChPTFV and ChPT fits we use a chiral scale of $1.0$ GeV. The fit formulae to which these parameters correspond can be found in Refs.~\cite{Aoki:2010dy,Arthur:2012opa}. \label{tab-globalfitparams} }
\end{longtable}

\newpage

\begin{longtable}[p]{c|ll}
\hline\hline
 & With W.flow & Without W.flow\\
\hline
 $am_l$(32I) & $ 0.000260(13) $  & $ 0.000262(15) $  \\
 $am_s$(32I) & $ 0.02477(18) $  & $ 0.02483(27) $  \\
 $a^{-1}$(32I) & $ 2.3833(86) $ GeV & $ 2.3726(181) $ GeV \\
\hline
 $Z_l$(64I) & $ 1.0(0) $  & $ 1.0(0) $  \\
 $Z_h$(64I) & $ 1.0(0) $  & $ 1.0(0) $  \\
 $R_a$(64I) & $ 0.9896(19) $  & $ 0.9953(60) $  \\
 $am_l$(64I) & $ 0.0006203(77) $  & $ 0.0006175(84) $  \\
 $am_s$(64I) & $ 0.02539(17) $  & $ 0.02531(19) $  \\
 $a^{-1}$(64I) & $ 2.3586(70) $ GeV & $ 2.3615(80) $ GeV \\
\hline
 $Z_l$(24I) & $ 0.9715(54) $  & $ 0.9702(56) $  \\
 $Z_h$(24I) & $ 0.9628(40) $  & $ 0.9612(43) $  \\
 $R_a$(24I) & $ 0.7489(24) $  & $ 0.7494(42) $  \\
 $am_l$(24I) & $ -0.001770(79) $  & $ -0.001767(78) $  \\
 $am_s$(24I) & $ 0.03224(18) $  & $ 0.03236(32) $  \\
 $a^{-1}$(24I) & $ 1.7848(50) $ GeV & $ 1.7779(132) $ GeV \\
\hline
 $Z_l$(48I) & $ 0.9715(54) $  & $ 0.9702(56) $  \\
 $Z_h$(48I) & $ 0.9628(40) $  & $ 0.9612(43) $  \\
 $R_a$(48I) & $ 0.7257(28) $  & $ 0.7291(55) $  \\
 $am_l$(48I) & $ 0.0006979(81) $  & $ 0.0006971(85) $  \\
 $am_s$(48I) & $ 0.03580(16) $  & $ 0.03577(18) $  \\
 $a^{-1}$(48I) & $ 1.7295(38) $ GeV & $ 1.7299(40) $ GeV \\
\hline
 $Z_l$(32ID) & $ 0.9156(72) $  & $ 0.9122(79) $  \\
 $Z_h$(32ID) & $ 0.9144(63) $  & $ 0.9107(70) $  \\
 $R_a$(32ID) & $ 0.5783(36) $  & $ 0.5791(52) $  \\
 $am_l$(32ID) & $ -0.000106(17) $  & $ -0.000099(18) $  \\
 $am_s$(32ID) & $ 0.04625(48) $  & $ 0.04649(53) $  \\
 $a^{-1}$(32ID) & $ 1.3784(68) $ GeV & $ 1.3741(75) $ GeV \\
\hline
 $Z_l$(32Ifine) & $ 1.015(17) $  & $ 0.998(30) $  \\
 $Z_h$(32Ifine) & $ 1.005(12) $  & $ 0.989(21) $  \\
 $R_a$(32Ifine) & $ 1.3208(44) $  & $ 1.308(16) $  \\
 $am_l$(32Ifine) & $ 0.000058(16) $  & $ 0.000078(30) $  \\
 $am_s$(32Ifine) & $ 0.01852(30) $  & $ 0.01907(68) $  \\
 $a^{-1}$(32Ifine) & $ 3.148(17) $ GeV & $ 3.104(45) $ GeV \\
\caption{A comparison of the scaling parameters and the predictions for the lattice spacings and unrenormalized quark masses obtained by fitting using the ChPTFV ansatz with and without the Wilson flow data.\label{tab-inclwfloweffect}}
\end{longtable}

\FloatBarrier

\subsection{Physical predictions}
\label{sec-physicalpred}

In this section we present our predictions.

\subsubsection{$\chi$PT parameters}

The LO and NLO SU(2) partially-quenched $\chi$PT low-energy constants are given in Table~\ref{tab-globalfitparams}. These can be combined into the standard SU(2) $\chi$PT LECs, $\bar l_3$ and $\bar l_4$, giving
\begin{equation}\begin{array}{lcr}
\bar l_3 = 2.73(13) & {\rm and} & \bar l_4 = 4.113(59)\,.
\end{array}\end{equation}
We can also compute the ratio of the decay constant to the LO SU(2) $\chi$PT parameter $f$, for which we obtain:
\begin{equation}
F_\pi/F = 1.0645(15)\,.
\end{equation}
The errors on the above are statistical only; we make no attempt to estimate the systematic errors on these numbers due to higher-order effects or indeed the reliability of $\chi$PT in general. These issues will be investigated in a forthcoming publication.

\subsubsection{Lattice spacings}

For the lattice spacings we obtain the following values:
\begin{equation}
\begin{array}{ll}
a^{-1}_{\rm 32I}      & = 2.3833(86)\ {\rm GeV}\,, \\
a^{-1}_{\rm 64I}      & = 2.3586(70)\ {\rm GeV}\,, \\
a^{-1}_{\rm 24I}      & = 1.7848(50)\ {\rm GeV}\,, \\
a^{-1}_{\rm 48I}      & = 1.7295(38)\ {\rm GeV}\,, \\
a^{-1}_{\rm 32Ifine}  & = 3.148(17)\ {\rm GeV}\,, \\
a^{-1}_{\rm 32ID}     & = 1.3784(68)\ {\rm GeV}\,, \\
\end{array}
\end{equation}
where we quote the statistical error in parentheses. Our previous values~\cite{Arthur:2012opa} for the lattice spacings of the 32I, 24I and 32ID ensembles are as follows:
\begin{equation}\begin{array}{ll}
a^{-1}_{\rm 32I}      & = 2.310(37)(17)(9)\ {\rm GeV}\,, \\
a^{-1}_{\rm 24I}      & = 1.747(31)(24)(4)\ {\rm GeV}\,, \\
a^{-1}_{\rm 32ID}     & = 1.3709(84)(56)(3)\ {\rm GeV}\,,
\end{array}\end{equation}
where the errors are statistical, chiral and finite-volume. We observe a $1.8\sigma$ tension between the new and old values of the 32I lattice spacing, which appears to arise from the introduction of the physical point data; if we look at Figure~\ref{fig-mpimkmomegaunitary} we see that the physical point data appears to favor a stronger light quark mass slope than one would obtain from the heavier data. Nevertheless there do not seem to be any clear discrepancies, except for those that might be attributed to statistical effects. Other than this, our new results are consistent with these values, and are significantly more precise due to the inclusion of the Wilson flow data.

\subsubsection{Decay constants}

\begin{figure}[tp]
\centering
\includegraphics[width=0.48\textwidth]{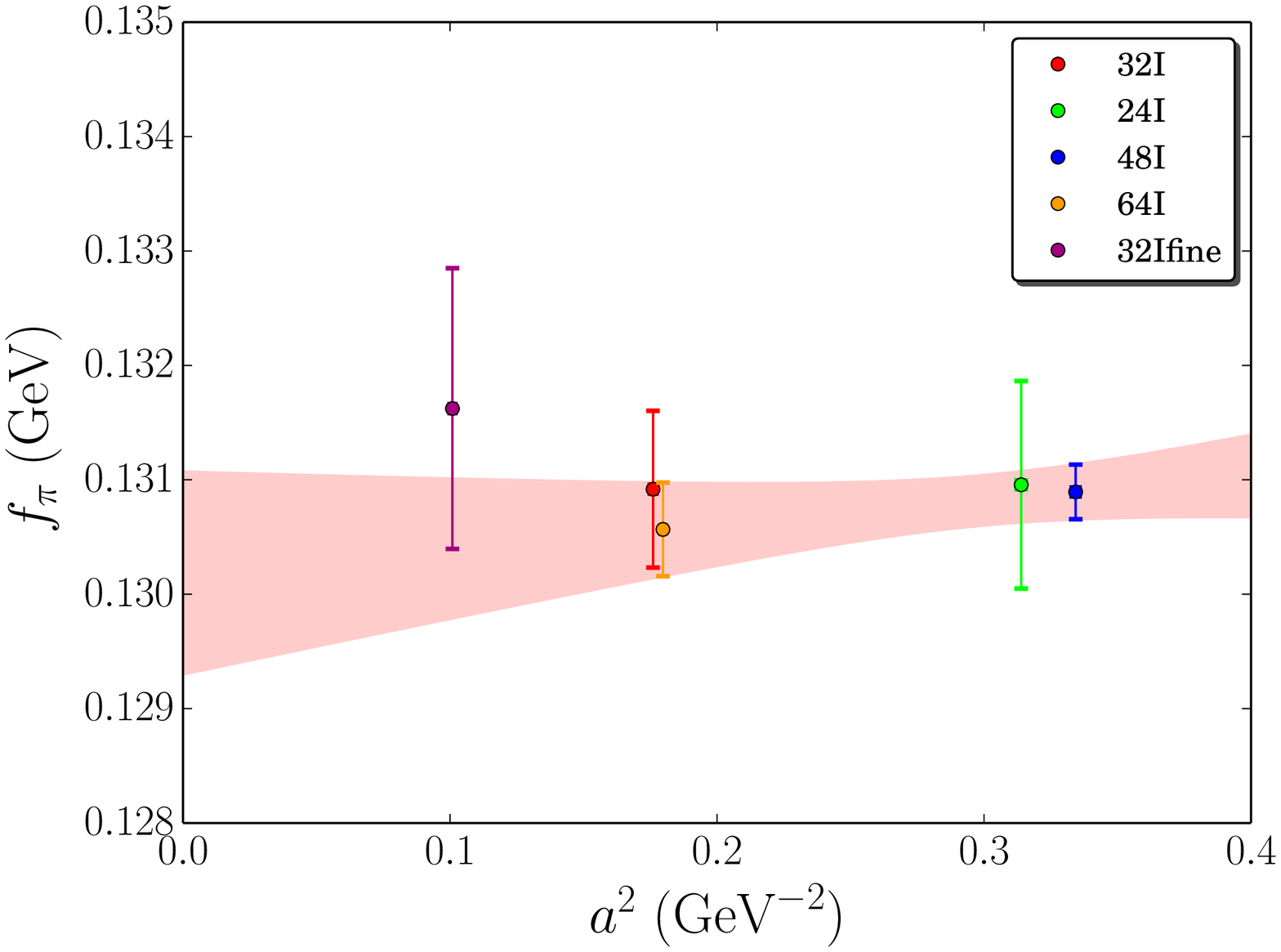}
\includegraphics[width=0.48\textwidth]{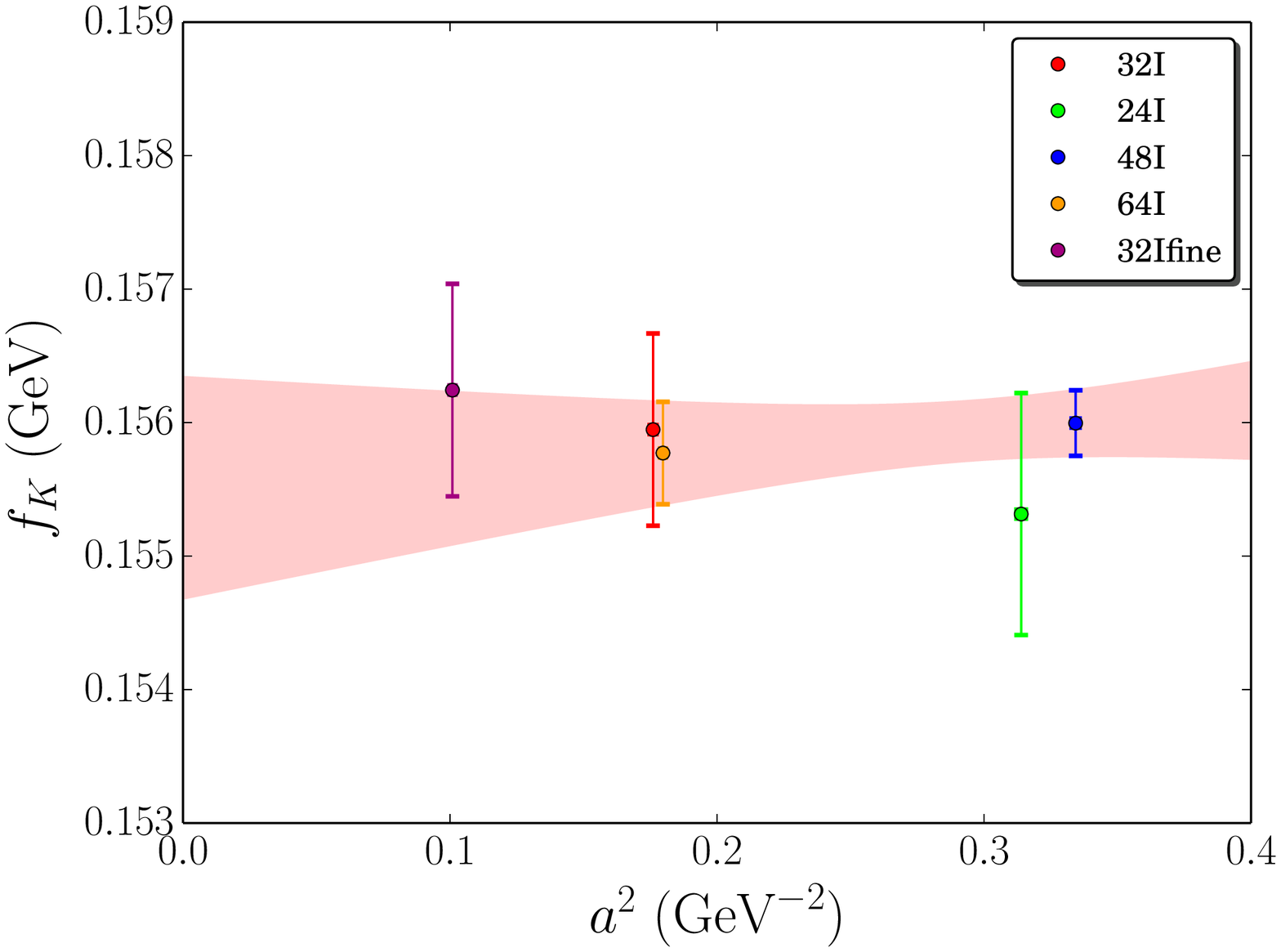}
\caption{$f_\pi$ (left) and $f_K$ (right) data corrected to the physical up/down and strange quark masses and the infinite-volume as a function of the square of the lattice spacing. The curve shows the continuum extrapolation for the Iwasaki action with the ChPTFV ansatz. Here we have not shown the 32ID data point as it has a different gauge action.\label{fig-fpifka2extrap} }
\end{figure}

\begin{figure}[tp]
\centering
\includegraphics[width=0.48\textwidth]{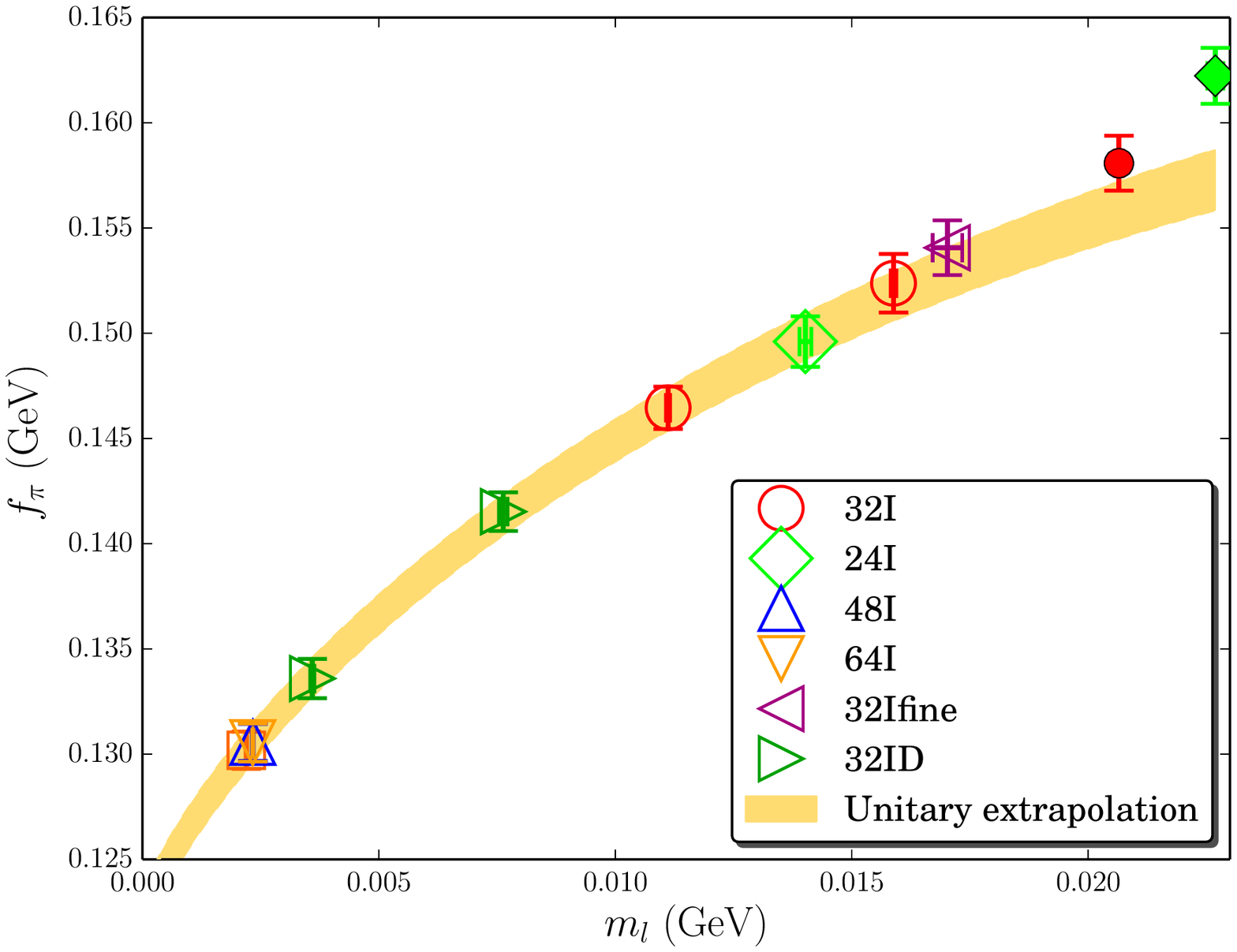}
\includegraphics[width=0.48\textwidth]{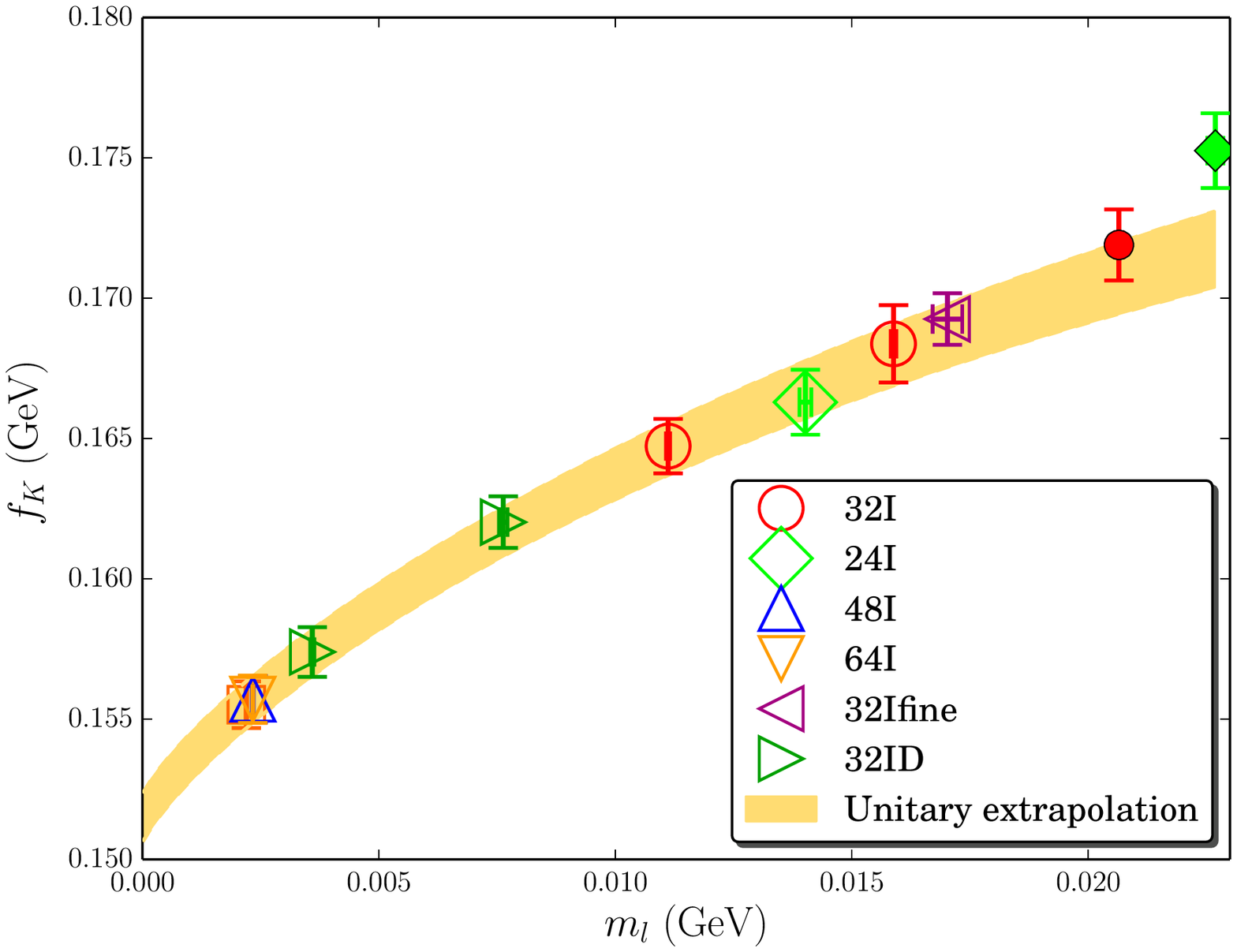}
\caption{$f_\pi$ (left) and $f_K$ (right) unitary data corrected to the physical strange quark mass and the continuum and infinite-volume limits as a function of the unrenormalized physical quark mass, plotted against the ChPTFV fit curves. Data with hollow symbols are those included in the fit and data with filled symbols are those excluded. The square point is our predicted continuum value. Note the 64I and 48I data lie essentially on top of each other in this figure.\label{fig-fpifkunitary} }
\end{figure}

In Table~\ref{tab-predsdiffs} we list the predicted values of $f_\pi$, $f_K$ and $f_K/f_\pi$ obtained using the ChPTFV ansatz, as well as the differences between those results and those of the other ans\"{a}tze. As we now have data at several lattice spacings, we can examine the scaling of both $f_\pi$ and $f_K$ in order to ensure that their dependence on the lattice spacing can be described by a quadratic form. In Figure~\ref{fig-fpifka2extrap} we plot the data, corrected to the physical quark masses and the infinite volume using the ChPTFV fit, as a function of the lattice spacing. In addition we show the scaling curve for the Iwasaki ensembles. We observe excellent consistency between the data and the fit curve for both quantities. In Figure~\ref{fig-fpifkunitary} we show the chiral extrapolation in the continuum/infinite-volume limits with the ChPTFV ansatz, again showing excellent agreement between the data and the fit.

We obtain the following physical predictions:
\begin{equation}
\begin{array}{ll}
f_\pi & = 0.13019(89)\ {\rm GeV}\,,\\
f_K   & = 0.15551(83)\ {\rm GeV}\,,\\
f_K/f_\pi & = 1.1945(45)\,,
\end{array}
\end{equation}
where, as above, the statistical errors are given in parentheses. Previously~\cite{Arthur:2012opa} we obtained 
\begin{equation}
\begin{array}{ll}
f_\pi & = 0.1271(27)(9)(25)\ {\rm GeV}\,,\\
f_K   & = 0.1524(30)(7)(15)\ {\rm GeV}\,,\\
f_K/f_\pi & = 1.1991(116)(69)(116)\,.
\end{array}
\end{equation}
Here we see that the inclusion of the 48I and 64I data, giving statistically precise data at simulated masses very near the physical quark masses, has led to a highly significant improvement in our results. 

In our first global fit analysis~\cite{Aoki:2010dy}, performed only to the 32I and 24I ensembles over a (unitary) pion mass range of 290--420 MeV, we obtained a value for $f_\pi$ from our NLO $\chi$PT fit that was 6.6\% (9 MeV) lower than the experimental value. We concluded that this discrepancy was due to systematic errors in the chiral extrapolation, and introduced the analytic fits as a means of estimating this systematic. When we included the 32ID ensembles into the global fit~\cite{Arthur:2012opa} we observed a marked improvement in the results for the decay constants and a corresponding reduction in the size of the chiral systematic (as estimated by taking the difference between the ChPTFV and analytic fit results). 

Now, with the inclusion of the 48I and 64I data we have essentially eliminated the chiral extrapolation error, and have obtained values for both decay constants that are in excellent agreement with the Particle Data Group (PDG) values~\cite{PDG}, $f_{\pi^-}=0.1304(2)$ GeV and $f_{K^-}=0.1562(7)$ GeV. Here, $f_{\pi^-}$ is determined experimentally using the measured branching fraction and pion lifetime, with $|V_{ud}|$ computed very precisely via nuclear $\beta$ decay, such that the error is dominated by higher order terms in the
decay width formula. On the other hand, the value for $f_{K^-}$ requires $|V_{us}|$ as input, which, for the quoted result, is computed using $|V_{us}|f_+(0)$
determined via semileptonic kaon decays and lattice input for $f_+(0)$. The consistency of our $f_K$ with the PDG value could therefore be taken as both representing the consistency of experiment with the
Standard Model, and the quality of the lattice QCD determinations of both the kaon semileptonic form factor and our determination of the kaon decay constant. 

\subsubsection{Wilson flow scales}

\begin{figure}[tp]
\centering
\includegraphics[width=0.48\textwidth]{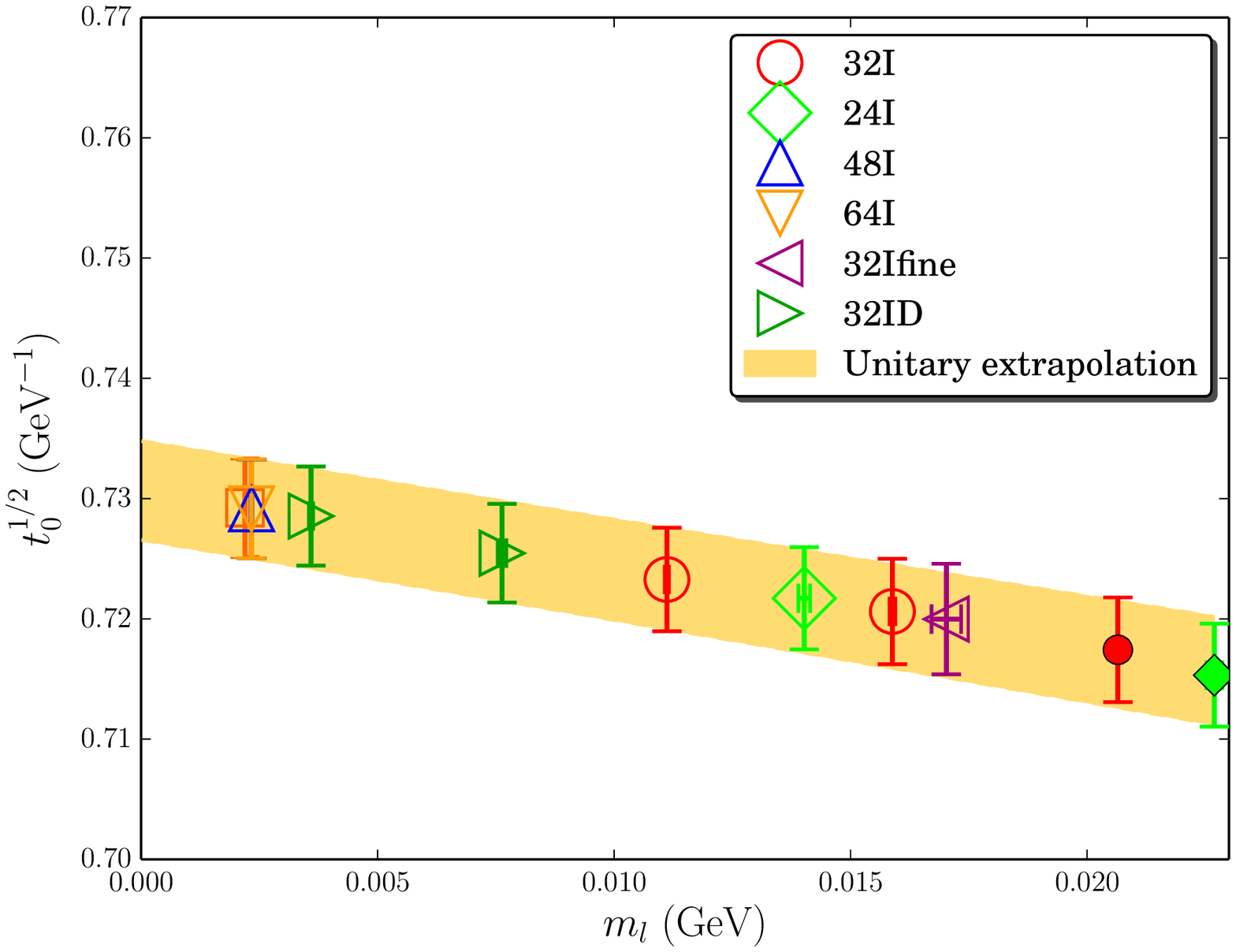}
\includegraphics[width=0.48\textwidth]{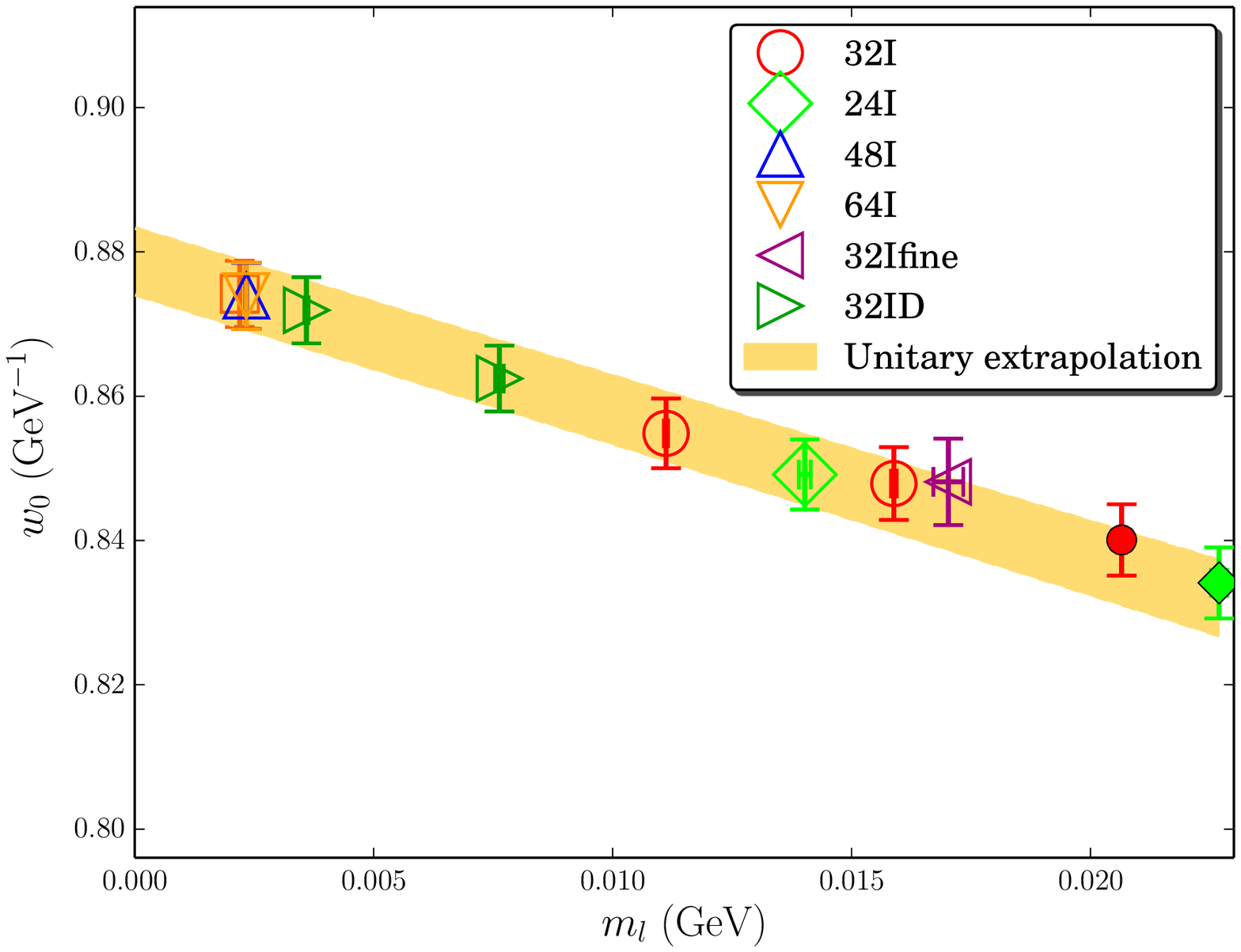}
\caption{$t_0^{1/2}$ (left) and $w_0$ (right) unitary data corrected to the physical strange sea quark mass and the continuum limit as a function of the unrenormalized physical quark mass, plotted against the ChPTFV fit curves. Data with hollow symbols are those included in the fit and data with filled symbols are those excluded. The square point is our predicted continuum value. Note the 64I and 48I data lie essentially on top of each other in this figure.\label{fig-wflowunitary} }
\end{figure}

\begin{figure}[tp]
\centering
\includegraphics[width=0.48\textwidth]{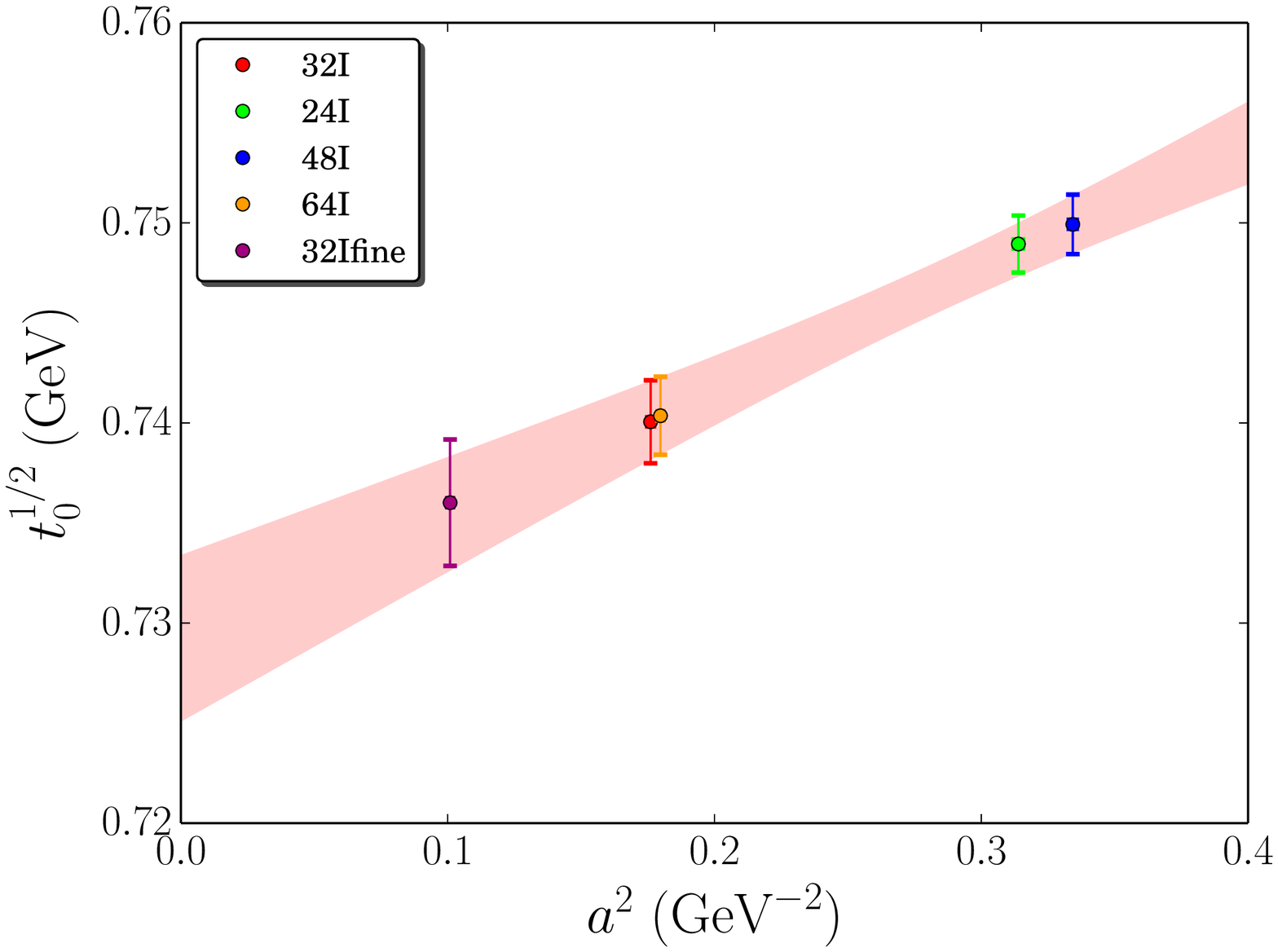}
\includegraphics[width=0.48\textwidth]{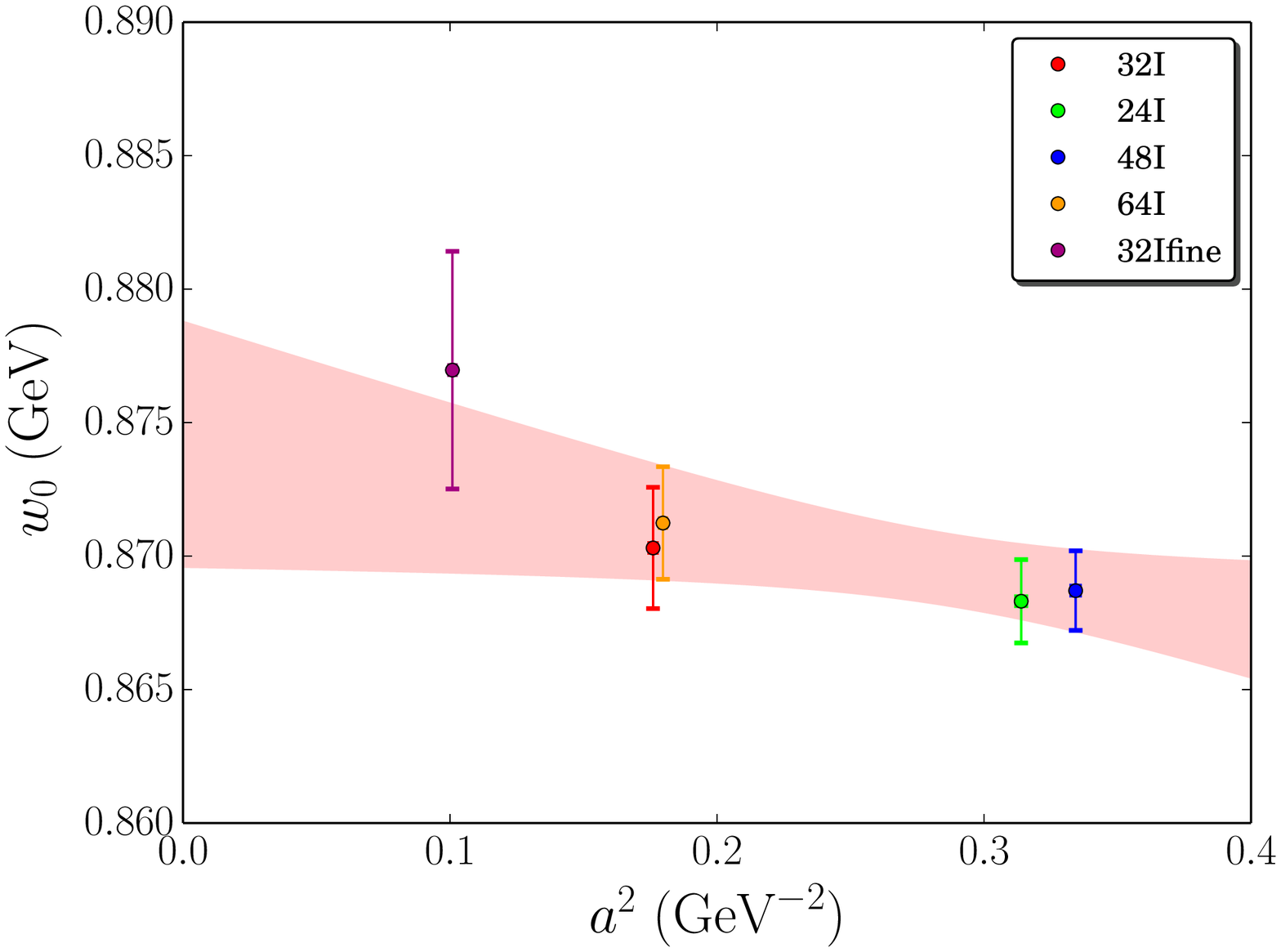}
\caption{$t_0^{1/2}$ (left) and $w_0$ (right) data corrected to the physical up/down and strange sea quark masses as a function of the square of the lattice spacing. The curve shows the continuum extrapolation for the Iwasaki action with the ChPTFV ansatz. Here we have not shown the 32ID data point as it has a different gauge action. \label{fig-wflowa2extrap} }
\end{figure}

In Table~\ref{tab-predsdiffs} we list the predicted values of the Wilson flow scales, $t_0^{1/2}$ and $w_0$, in the continuum limit. The unitary mass dependencies are plotted in Figure~\ref{fig-wflowunitary} and the $a^2$ dependencies in Figure~\ref{fig-wflowa2extrap}. For our final results, we obtain the following continuum predictions:
\begin{equation}
\begin{array}{ll}
t_0^{1/2} & = 0.7292(41)\ {\rm GeV}^{-1}\,,\\
w_0   & = 0.8742(46)\ {\rm GeV}^{-1}\,,\\
\end{array}
\end{equation}
where the statistical error is quoted in parentheses.

The above values can be compared to the following results obtained using 2+1f 2HEX-smeared Wilson fermions~\cite{Borsanyi:2012zs}: 
\begin{equation}
\begin{array}{lll}
t_0^{1/2} & = 0.1465(25)\ {\rm fm} &= 0.7425(127)\ {\rm GeV}^{-1}\,, \\
w_0 &= 0.1755(18)\ {\rm fm}\ &= 0.8894(91)\ {\rm GeV}^{-1}\,,
\end{array}\end{equation}
where we have combined the statistical and systematic errors in quadrature. We find excellent agreement between these and our results.

\FloatBarrier
\subsubsection{Unrenormalized physical quark masses}
\label{sec-unrenquarkmasses}

The quark masses in bare lattice units on the 32I reference ensemble are given in Table~\ref{tab-finalainvmqbare}. In physical units, and including the residual mass, the unrenormalized physical quark masses are given in Table~\ref{tab-mqunrenorm}. Combining these results we obtain the following:
\begin{equation}\begin{array}{rl}
m_{ud}^{\rm unrenorm.} &= 2.198(11)\ {\rm MeV}\,, \\
m_s^{\rm unrenorm.} &= 60.62(24)\ {\rm MeV}\,,\label{eqn-masses-unrenorm}
\end{array}\end{equation}
where the errors are statistical.

\begin{table}[tp]
\begin{tabular}{l|ll}
\hline\hline
Ansatz & $m_{ud}^{\rm unrenorm.}$ (MeV) & $m_s^{\rm unrenorm}$ (MeV)\\
\hline
ChPTFV & 2.198(11)  &  60.62(24) \\
ChPT   & 2.199(10)  &  60.67(22) \\
analytic (260 MeV) & 2.185(16) & 60.70(27) \\
analytic (370 MeV) & 2.188(13) & 60.69(21) \\
\end{tabular}
\caption{Unrenormalized physical quark masses. For the analytic fits, the corresponding pion mass cut is given in parentheses.\label{tab-mqunrenorm}}
\end{table}

\FloatBarrier
\subsection{Renormalized physical quark masses and the chiral condensate}
\label{sec-renormquarkmasses}

The quark masses presented above are defined in the bare lattice normalization of the 32I reference ensemble. On each of the 32I and 24I ensembles independently, we calculate the non-perturbative renormalization factors that are necessary to convert quark masses in the corresponding bare normalization into a variant of the Rome-Southampton RI-MOM scheme~\cite{Martinelli:1994ty} that can be related to $\MSbar$ via perturbation theory. The procedure applied below is identical to that used in Refs.~\cite{Aoki:2010dy} and~\cite{Arthur:2012opa}, and the determination of the renormalization coefficients is documented in Appendix~\ref{appendix-npr}; below we provide only a brief outline.

We compute amputated, projected bilinear vertex functions,
\begin{equation}
\Lambda_{\cal O}(q^2) = {\rm tr}\left[ \Pi_{\cal O}(q^2) \Gamma^{(s)}_{\cal O} \right]\,,
\end{equation}
where ${\cal O}$ is an operator, $\Pi$ are the matrix-valued amputated vertex functions and $\Gamma^{(s)}$ are projection operators, for which the superscript $s$ indexes the particular renormalization scheme (where applicable). We use the `symmetric' RI-MOM schemes, defined by the following condition on the incoming and outgoing quark momenta, $p_{\rm in}$ and $p_{\rm out}$ respectively: $p_{\rm in}^2 = p_{\rm out}^2 = q^2\equiv (p_{\rm in}- p_{\rm out})^2$.

We define renormalization factors by matching to the tree-level amplitude at the scale $\mu^2=q^2$:
\begin{equation}
\frac{Z_O}{Z_q^{n/2}}(\mu,a) \times \Lambda^{bare}_O(\mu,a) =  \Lambda_O^{tree} \,.
\end{equation}
In order to cancel the factors of the quark field renormalization in the denominator, we use
\begin{equation}
Z_m^{(s)}(\mu,a) = \frac{\bar \Lambda_S(\mu,a)} {Z_V \times \bar \Lambda_V^{(s)}(\mu,a)}\,,
\end{equation}
where $\bar \Lambda_{\cal O} \equiv \Lambda^{bare}_{\cal O} \times (\Lambda_{\cal O}^{tree})^{-1}$, $S$ and $V$ are the scalar and vector operators repectively, and $Z_V$ is the vector-current renormalization computed using hadronic variables via the procedure given in Section~\ref{sec-zvdeterm}. We use two different choices of projection operator for the vector vertex, formed from the quantities $\slashed q q^\mu/q^2$ and $\gamma^\mu$; these define the SMOM and SMOM$_{\gamma^\mu}$ schemes respectively. More details on the projection operators and the numerical determination of these quantities can be found in Appendix~\ref{appendix-npr}.

We now describe the procedure by which we obtain the renormalized quark masses given the renormalization factors. In Section~\ref{sec-unrenquarkmasses} we present quark masses normalized according to the bare lattice units of the 32I reference ensemble. For any other ensemble $e$, the quark masses in the associated bare normalization can be obtained simply by dividing the values of $m_{ud}$ and $m_s$ given in Eq.~\eqref{eqn-masses-unrenorm} by $Z_l^e$ and $Z_h^e$ respectively. For each ensemble, the masses renormalized in the RI-SMOM schemes can therefore be computed as
\begin{equation}
(m_f^{\rm SMOM*})^e = (Z_m^{\rm SMOM*})^e m_f^{\rm unrenorm.} / Z_f^e\,,
\end{equation}
where $f\in\{l,h\}$. These measurements contain finite lattice spacing errors associated with the vertex functions
used in the conversion to $\MSbar$. %
%
%
In order to convert our continuum quark masses to the RI-SMOM scheme, and thence
to $\MSbar$, we linearly extrapolate the ratio
\begin{equation}
Z_{mf}^e = Z_m^e/Z_f^e
\end{equation}
in $a^2$ to the continuum. This extrapolation is performed using only two lattice spacings, potentially introducing additional systematic effects. In practice we find that the linear continuum fit results in a 4\% shift in the central values from those computed on our finest ensemble (32I). The good chiral symmetry of the action heavily suppresses ${\cal O}(a^3)$ terms in the Symanzik effective theory and higher order corrections enter only at the ${\cal O}(a^4)$ level. This suggests systematic effects on the order of $(4\%)^2 \sim 0.16\%$, which we treat as negligible. Applied to the quark masses, the products
\begin{equation}
m_f^{\rm SMOM*} = (Z_{mf}^{\rm SMOM*})^{contm.} m_f^{\rm unrenorm.}\,,
\end{equation}
are then free from ${\cal O}(a^2)$ scaling errors and have negligible higher order discretization systematics.

Fixing the renormalization coefficients to a particular scale requires the input of the lattice spacings from the main analysis in order to convert the lattice momenta to physical units; for this we used only the central values of the ChPTFV fits. In order to account for the effect of the statistical and systematic uncertainties on the lattice spacings, we repeated the determination of the renormalization coefficients using two different values of the lattice spacings that differed slightly in value, and from these we estimated the slope of the renormalization coefficients with respect to the input lattice spacing. For each chiral ansatz, we then used the slope to shift the central values of the renormalization coefficients to the lattice spacings determined via that ansatz, and also to inflate the statistical errors of the superjackknife distribution to reflect the uncertainty on those values. The continuum extrapolations of $Z_{ml}$ and $Z_{mh}$ were performed independently for each ansatz, enabling us to determine the full effect of the systematic errors in the final step. The values of $Z_m$ and $Z_{mf}$ thus determined are given in Table~\ref{eqn-zmf}.

Applying the renormalization factors to the masses from the previous section, we obtain the values given in Table~\ref{tab-smommasses}. Converting to the $\MSbar$ scheme and including the additional systematic errors associated with the perturbative matching, we find
\begin{equation}\begin{array}{rl}
m_{ud}(\msbar\,,3.0\,{\rm GeV}) &= 2.997(36)(33)\ {\rm MeV}\,, \\
m_s(\msbar\,,3.0\,{\rm GeV}) &= 81.64(77)(88)\ {\rm MeV}\,.
\label{eq:ren quark mass}
\end{array}\end{equation}
where the errors are statistical and from the perturbative truncation respectively. In the RGI scheme, these correspond to
\begin{equation}\begin{array}{rl}
\hat m_{ud} &= 8.62(10)(9)\ {\rm MeV}\,,\\
\hat m_s &= 235.0(22)(25)\ {\rm MeV}\,.
\end{array}\end{equation}
The quark mass ratio is
\begin{equation}
m_s/m_{ud} = 27.34(21)\,,
\label{eq:quark mass ratio}
\end{equation}
for which there is no systematic error associated with the perturbative matching as it cancels in the ratio.

For comparison, in our previous work~\cite{Arthur:2012opa} we obtained
\begin{equation}\begin{array}{rl}
m_{ud}(\msbar\,,3.0\,{\rm GeV}) &= 3.05(8)(6)(1)(2)\ {\rm MeV}\,, \\
m_s(\msbar\,,3.0\,{\rm GeV}) &= 83.5(1.7)(0.8)(0.4)(0.7)\ {\rm MeV}\,.
\end{array}\end{equation}
and
\begin{equation}
m_s/m_{ud} = 27.36(39)(31)(22)\,,
\label{eq:massratio}
\end{equation}
for which the errors are statistical, chiral and finite-volume. Our new results are highly consistent with these values and again show a substantial improvement in the systematic error as a result of including the near-physical data.

We can also compute the chiral condensate,
\begin{equation}
\Sigma = -\langle \bar u u\rangle_{m_u,m_d\rightarrow 0} = B F^2 = B f^2/2\,,
\end{equation}
by combining the leading-order SU(2) $\chi$PT parameters from Table~\ref{tab-globalfitparams}. Like the quark masses, this quantity must be renormalized. Again we first convert to our intermediate SMOM schemes and subsequently perturbatively convert each to $\MSbar$, using the difference as an estimate of the perturbative truncation systematic. The appropriate renormalization factor can be determined by noting that the leading-order $\chi$PT formula for the pion mass must be renormalization-scheme independent:
\begin{equation}
(m_\pi^2)_{\rm LO} = 2 B^{\rm unrenorm} m_{ud}^{\rm unrenorm} = 2 B^{\rm SMOM*} m_{ud}^{\rm SMOM*} = 2 B^{\rm SMOM*}(Z_{ml}^{\rm SMOM*})^{contm.} m_{ud}^{\rm unrenorm.}.
\end{equation}
This suggests that
\begin{equation}
B^{\rm SMOM*} = B^{\rm unrenorm.}/(Z_{ml}^{\rm SMOM*})^{contm.}\,.
\end{equation}
The subsequent conversion to the $\MSbar$ scheme at 3 GeV can be performed by further dividing by the appropriate scheme change factor.

It is customary to quote the dimension-one quantity $(\Sigma)^{1/3}$. We obtain
\begin{equation}\begin{array}{rl}
\Sigma^{1/3}({\rm SMOM},3.0\ {\rm GeV}) &= 0.2837(19)\ {\rm GeV} \\
\Sigma^{1/3}({\rm SMOM}_{\gamma^\mu},3.0\ {\rm GeV}) &= 0.2791(19)\ {\rm GeV}\,,
\end{array}\end{equation}
which, after converting to $\MSbar$ and combining, gives
\begin{equation}
\Sigma^{1/3}(\msbar,3.0\ {\rm GeV}) = 0.2853(20)(10)\ {\rm GeV}\,,
\end{equation}
where the errors are statistical and from the perturbative matching respectively.

\begin{table}[tp]
\begin{tabular}{lll|lll}
\hline\hline
Scheme & Lattice & Ansatz & $Z_m$ & $Z_{ml}$ & $Z_{mh}$\\
\hline
SMOM & 24I & ChPTFV & 1.4386(12) & 1.4808(82) & 1.4942(63) \\
SMOM & 24I & ChPT & 1.4385(12) & 1.4788(79) & 1.4932(60) \\
SMOM & 24I & analytic (260 MeV) & 1.4390(12) & 1.4874(108) & 1.4931(68) \\
SMOM & 24I & analytic (370 MeV) & 1.4383(12) & 1.4849(86) & 1.4927(57) \\
\hline
SMOM & 32I & ChPTFV & 1.4396(37) & 1.4396(37) & 1.4396(37) \\
SMOM & 32I & ChPT & 1.4393(37) & 1.4393(37) & 1.4393(37) \\
SMOM & 32I & analytic (260 MeV) & 1.4396(37) & 1.4396(37) & 1.4396(37) \\
SMOM & 32I & analytic (370 MeV) & 1.4391(37) & 1.4391(37) & 1.4391(37) \\
\hline
SMOM & cont. & ChPTFV & - & 1.3870(122) & 1.3699(100) \\
SMOM & cont. & ChPT & - & 1.3888(120) & 1.3704(100) \\
SMOM & cont. & analytic (260 MeV) & - & 1.3780(145) & 1.3706(103) \\
SMOM & cont. & analytic (370 MeV) & - & 1.3805(128) & 1.3705(99) \\
\hline\hline
SMOM$_{\gamma^\mu}$ & 24I & ChPTFV & 1.5235(13) & 1.5682(87) & 1.5824(67) \\
SMOM$_{\gamma^\mu}$ & 24I & ChPT & 1.5234(13) & 1.5661(83) & 1.5813(64) \\
SMOM$_{\gamma^\mu}$ & 24I & analytic (260 MeV) & 1.5240(13) & 1.5752(115) & 1.5813(72) \\
SMOM$_{\gamma^\mu}$ & 24I & analytic (370 MeV) & 1.5232(13) & 1.5725(91) & 1.5808(60) \\
\hline
SMOM$_{\gamma^\mu}$ & 32I & ChPTFV & 1.5192(39) & 1.5192(39) & 1.5192(39) \\
SMOM$_{\gamma^\mu}$ & 32I & ChPT & 1.5189(39) & 1.5189(39) & 1.5189(39) \\
SMOM$_{\gamma^\mu}$ & 32I & analytic (260 MeV) & 1.5192(39) & 1.5192(39) & 1.5192(39) \\
SMOM$_{\gamma^\mu}$ & 32I & analytic (370 MeV) & 1.5186(39) & 1.5186(39) & 1.5186(39) \\
\hline
SMOM$_{\gamma^\mu}$ & cont. & ChPTFV & - & 1.4567(126) & 1.4386(103) \\
SMOM$_{\gamma^\mu}$ & cont. & ChPT & - & 1.4585(125) & 1.4389(103) \\
SMOM$_{\gamma^\mu}$ & cont. & analytic (260 MeV) & - & 1.4470(150) & 1.4392(106) \\
SMOM$_{\gamma^\mu}$ & cont. & analytic (370 MeV) & - & 1.4496(134) & 1.4390(102) \\
\end{tabular}
\caption{The non-perturbative renormalization factors calculated at $\mu=3.0$ GeV that are used to convert bare quark masses ($Z_m$) and quark masses in the normalization of the 32I reference ensemble ($Z_{ml}$,$Z_{mh}$). Values are given on the 32I and 24I ensembles and in the continuum limit for the latter quantity.\label{eqn-zmf}}
\end{table}

\begin{table}[tp]
\begin{tabular}{ll|ll}
\hline\hline
Scheme & Ansatz & $m_{u/d}$ (GeV) & $m_s$ (GeV)\\
\hline
SMOM & ChPTFV & 0.003049(37) & 0.08305(80)\\
SMOM & ChPT & 0.003055(36) & 0.08314(79)\\
SMOM & analytic (260 MeV) & 0.003011(50) & 0.08319(87)\\
SMOM & analytic (370 MeV) & 0.003021(42) & 0.08317(76)\\
\hline
SMOM$_{\gamma^\mu}$ & ChPTFV & 0.003202(38) & 0.08721(83)\\
SMOM$_{\gamma^\mu}$ & ChPT & 0.003208(37) & 0.08730(81)\\
SMOM$_{\gamma^\mu}$ & analytic (260 MeV) & 0.003162(51) & 0.08735(89)\\
SMOM$_{\gamma^\mu}$ & analytic (370 MeV) & 0.003172(43) & 0.08733(79)\\
\end{tabular}
\caption{The physical quark masses renormalized at $\mu=3.0$ GeV in the two intermediate RI-SMOM schemes for each of the chiral ans\"{a}tze. The quoted errors are statistical only.\label{tab-smommasses}}
\end{table}

\FloatBarrier

\subsection{Neutral kaon mixing parameter, $B_K$}

The neutral kaon mixing parameter is renormalization scheme dependent, and as such the fits must be performed using renormalized data. As this introduces additional systematic errors, we follow our established procedure of performing these fits separately from the main global fit analysis. Below we first summarize our non-perturbative renormalization procedure for $B_K$ and then present the results of the chiral/continuum fit and finally our physical predictions.

\subsubsection{Renormalization of $B_K$}

In this section we provide a brief outline of the procedure for determining the renormalization coefficients; for more details we refer the reader to Appendix~\ref{appendix-npr} and Refs.~\cite{Aoki:2010pe} and~\cite{Arthur:2012opa}.

As with the quark mass renormalization, we make use of `symmetric' regularization-invariant momentum schemes (RI-SMOM for short), defined by the condition $\mu^2 = p_1^2 = p_2^2 = q^2 \equiv (p_1-p_2)^2$, where $p_1$ and $p_2$ are the momenta of the incoming and outgoing quarks: $d(p_1)\bar s(-p_2)\rightarrow \bar d(-p_1)s(p_2)$. We compute the amputated and projected Green's function of the relevant four-quark operator, ${\cal O}_{LL}$, describing the $K-\bar K$ mixing, normalized by the square of the average between the vector and axial bilinear:
\be
\label{eq:Zbk}
Z^{(s_1,s_2)}_{B_K}(\mu,a) \times \frac {\bar\Lambda^{(s_1)}_{VV+AA}(\mu,a)}{{\bar\Lambda^{(s_2)}_{AV}(\mu,a)}^2}
= 1\,,
\ee
where
\be
\Lambda_{AV} = \frac{1}{2} (\Lambda_V+\Lambda_A)\,,
\ee
and $\bar \Lambda_{\cal O} \equiv \Lambda^{bare}_{\cal O} \times (\Lambda_{\cal O}^{tree})^{-1}$ for the operator ${\cal O}$, as before.

Note that the quark wave function renormalization factor cancels in the ratio. In Appendix~\ref{appendix-npr} we show that the difference between $\Lambda_V$ and $\Lambda_A$ at $3$ GeV is numerically negligible, and therefore the above choice of normalization is irrelevant. The superscript $(s_i)$ refers to choice of projector (cf.~\cite{Arthur:2012opa}): either $\gamma_\mu$ or $\s{q}$. The choices $s_1=s_2=\gamma^\mu$ and $s_1=s_2=\slashed q$ define the so-called SMOM$(\gamma^\mu,\gamma^\mu)$ and SMOM$(\slashed q,\slashed q)$ schemes respectively.

We perform the full analysis separately for each scheme and use the difference to estimate the systematic error associated with the $\MSbar$ matching. While treating the two schemes in an equal fashion is the most rigorous estimate we can make with the current data, we have indications that this might overestimate the error on the SMOM$(\slashed q,\slashed q)$ result: A preliminary study~\cite{JuliensLat2014Proceedings} of step scaling to higher momentum scales suggests that the scale evolution in this scheme agrees with perturbation theory over the full range of scales, whereas the SMOM$(\gamma^\mu,\gamma^\mu)$ scheme evolves into better agreement as the scale is raised. The perturbative truncation error is therefore greater for the SMOM$(\gamma^\mu,\gamma^\mu)$ scheme than for the SMOM$(\slashed q,\slashed q)$ scheme. The complete study of the evolution to higher energy scales requires careful treatment of the charm threshold, and is the subject of further work by RBC and UKQCD. These observations are consistent with our earlier results at lower scales, and the better agreement with the perturbative scale evolution for the SMOM$(\slashed q,\slashed q)$ scheme was the reason we have, in this work and previously, taken our central values for $B_K$ from this scheme~\cite{Aoki:2010pe}.

We compute $Z_{B_K}$ on each ensemble at a number of $q^2$, and interpolate to a chosen high momentum scale at which the matching to $\MSbar$ can be performed. We choose to perform the matching at 3.0 GeV as before. The values of the renormalization coefficients at the various lattice momenta and further details of the analysis are given in Appendix~\ref{appendix-npr}. 

All matrix elements included in the global fit must be renormalized to a common scale of $3.0$ GeV in order that the global fit can extrapolate these to a shared, universal continuum limit. As described in Ref.~\cite{Arthur:2012opa}, due to the coarseness of the 32ID ensemble we are unable to renormalize directly at 3 GeV without introducing potentially sizeable lattice artifacts. Instead we renormalize with a lower momentum scale of $\mu_0 = 1.4363$ GeV, and apply the continuum non-perturbative running $\sigma^{(s_1,s_2)}_{B_K}(\mu,\mu_0)$, extracted from the 32I and 24I lattices (and extrapolated to the continuum), to convert this value to $\mu=3 {\rm GeV}$. More details of this conversion are given in Appendix~\ref{appendix-npr}.

Determining the lattice momentum corresponding to the 3 GeV match point requires the input of the lattice spacings determined in the previous sections. The effects of the uncertainties on the lattice spacings are incorporated by shifting the central values and inflating the errors according to the lattice spacings determined via each of the chiral ans\"{a}tze, using the procedure outlined in the Section~\ref{sec-renormquarkmasses}. The resulting values of $Z_{B_K}$ are given in Table~\ref{tab-zbkaerr}.

\begin{table}
\centering
\begin{tabular}{ll|cccc}
\hline\hline
Scheme & Lattice & ChPTFV & ChPT & Analytic (260 MeV) & Analytic (370 MeV) \\
\hline
\multirow{6}{*}{$(\slashed q,\slashed q)$} & 32I& 0.9787(3)& 0.9787(3)& 0.9787(3)& 0.9786(3)\\
 & 24I& 0.9568(3)& 0.9568(3)& 0.9570(3)& 0.9568(3)\\
 & 48I& 0.9545(1)& 0.9544(1)& 0.9544(1)& 0.9544(1)\\
 & 64I& 0.9782(2)& 0.9781(2)& 0.9781(2)& 0.9781(2)\\
 & 32Ifine& 0.9995(4)& 0.9995(4)& 0.9998(5)& 0.9995(4)\\
 & 32ID& 0.9284(45)& 0.9286(45)& 0.9276(45)& 0.9289(45)\\
 \hline
\multirow{6}{*}{$(\gamma^\mu,\gamma^\mu)$} & 32I& 0.9409(2)& 0.9408(2)& 0.9409(2)& 0.9408(2)\\
& 24I& 0.9161(5)& 0.9161(5)& 0.9162(5)& 0.9160(5)\\
& 48I& 0.9140(1)& 0.9140(1)& 0.9140(1)& 0.9140(1)\\
& 64I& 0.9411(1)& 0.9410(1)& 0.9410(1)& 0.9410(1)\\
& 32Ifine& 0.9617(3)& 0.9617(2)& 0.9619(3)& 0.9617(2)\\
& 32ID& 0.8824(25)& 0.8824(25)& 0.8824(26)& 0.8824(25)\\
\end{tabular}
\caption{$Z_{B_K}$ at 3 GeV in the two intermediate schemes, with the central values shifted and errors inflated to account for the different values of the lattice spacings obtained via each chiral ansatz. \label{tab-zbkaerr}}
\end{table}

\subsubsection{Chiral/continuum fit to $B_K$}

\begin{figure}[tp]
\includegraphics[width=0.48\textwidth]{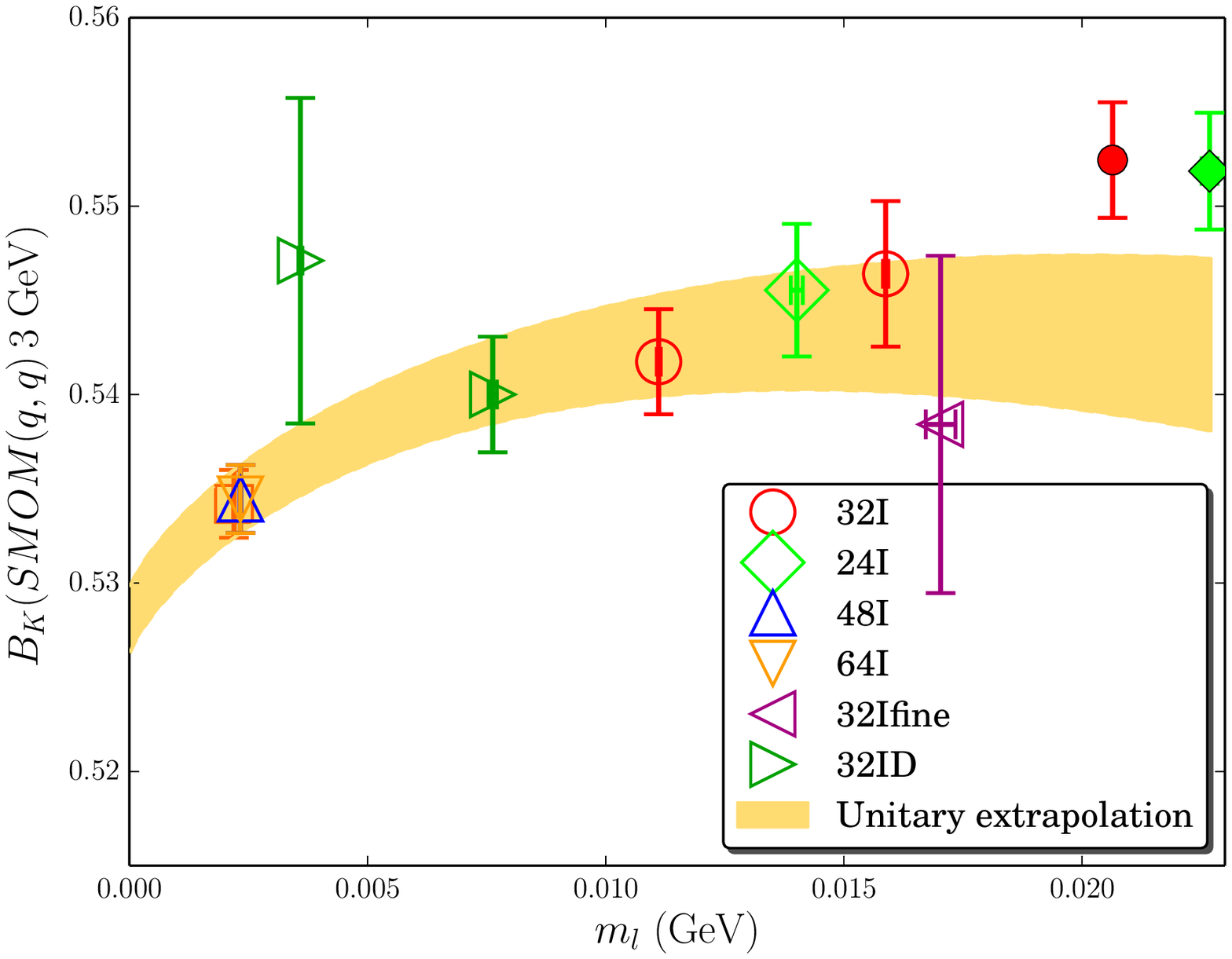}
\includegraphics[width=0.46\textwidth]{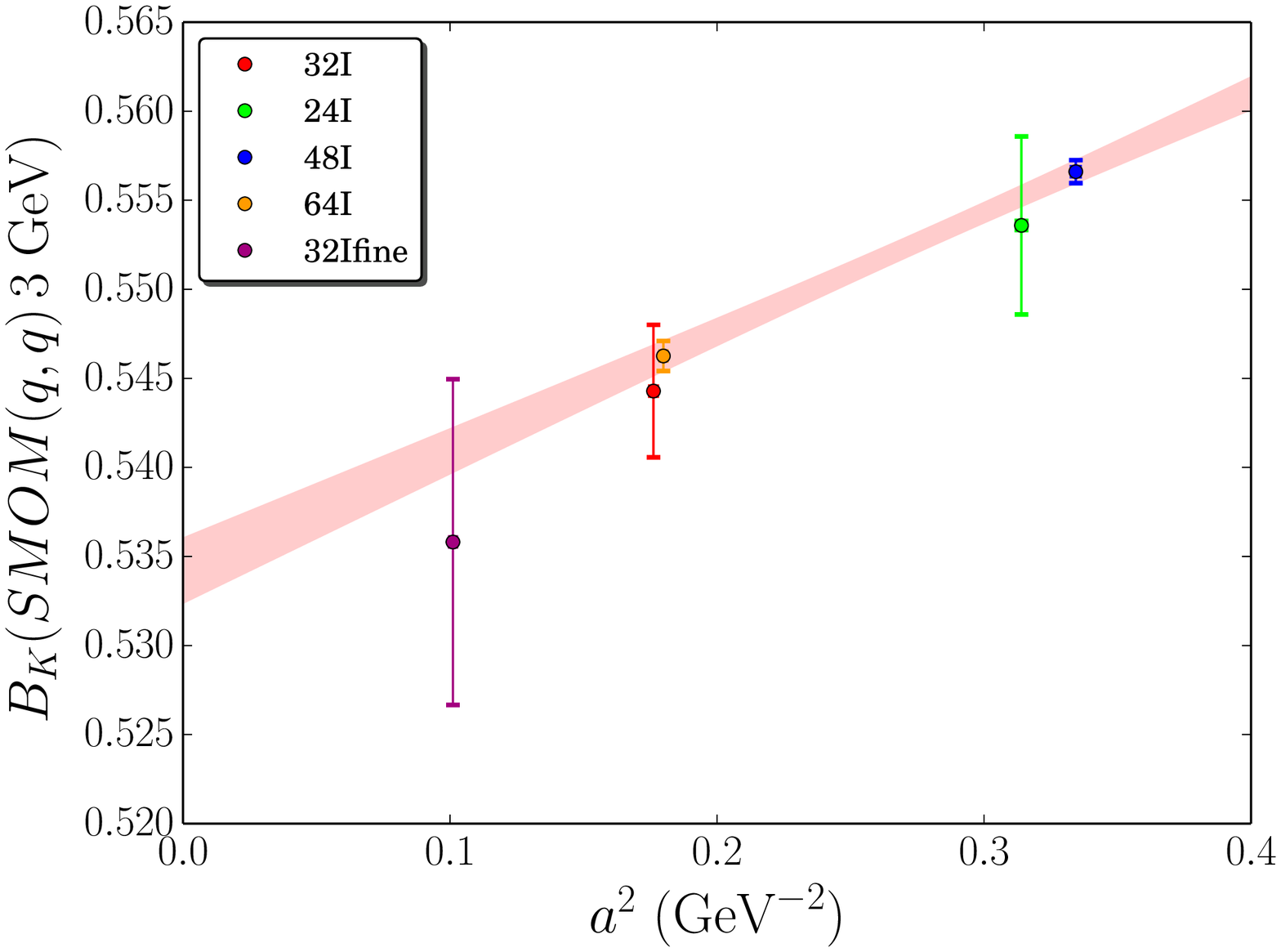}
\caption{The left figure shows the unitary light quark mass dependence of $B_K$ in the SMOM$(\slashed q,\slashed q)$ at 3 GeV. The quark masses are in physical units and in the native normalization of the 32I reference ensemble. Data with hollow symbols are those included in the fit and data with filled symbols are those excluded. The right figure shows the lattice spacing dependence of those data. Here we have not included the 32ID ensemble as it lies on a different scaling trajectory.\label{fig-bkchiral+a2}}
\end{figure}

\begin{figure}[tp]
\includegraphics[width=0.48\textwidth]{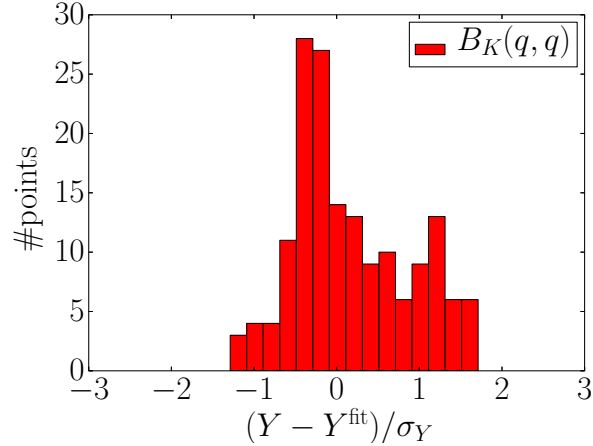}
\caption{A histogram of the deviation of the ChPTFV fit curve from our data in units of the statistical error for the $(\slashed q,\slashed q)$ intermediate scheme. \label{fig-bkqqhist}}
\end{figure}

As above, we describe the chiral dependence using chiral perturbation theory, with and without finite-volume corrections, as well as a linear ansatz with a 260 MeV and 370 MeV pion mass cut. The chiral/continuum fit forms can be found in Ref.~\cite{Aoki:2010pe}. As before, we use separate parameters to describe the lattice spacing dependence of the Iwasaki and Iwasaki+DSDR actions. The fit parameters can be found in Table~\ref{tab-fitpar3gev}, and in Figure~\ref{fig-bkchiral+a2} we show examples of the unitary and continuum extrapolations. In Figure~\ref{fig-bkqqhist}, in which we plot a histogram of the statistical deviations of the data from the ChPTFV fit curve, we see excellent consistency between the data and the fit. The total $\chi^2/{\rm d.o.f.}$ for each of the four ans\"{a}tze are given in Table~\ref{tab-bkchisq}.

The fits to $B_K$ with a 370 MeV pion mass cut have 7 free parameters (the remainder having been determined in our earlier fits, above) and use 163 data points, giving 156 degrees of freedom; for the 260 MeV cut have 7 parameters and 90 data points, giving 83 degrees of freedom.


\begin{table}[tp]
\centering
\begin{tabular}{c|c|c|c|c}
\hline\hline
Scheme & ChPTFV & ChPT & Analytic (260 MeV) & Analytic (370 MeV)\\
\hline
$(\slashed q,\slashed q)$ & 0.55(38) & 0.70(42) & 0.46(35) & 0.51(34) \\
$(\gamma^\mu,\gamma^\mu)$ & 0.62(43) & 0.78(46) & 0.52(40) & 0.58(39) \\
\end{tabular}
\caption{The $\chi^2/{\rm d.o.f.}$ for each of the four chiral ans\"{a}tze and the two intermediate renormalization schemes. Here the $\chi^2$ does not include the overweighted data, and the number of degrees of freedom has been correspondingly reduced. For the analytic fits, the pion mass cut is given in parentheses.\label{tab-bkchisq} }
\end{table}

\begin{table}[tp]
\centering
\begin{tabular}{l|r|r||l|r|r}
\hline\hline
Parameter & ChPT & ChPTFV & Parameter & Analytic (260 MeV) & Analytic (370 MeV)\\
\hline
 \rule{0cm}{0.4cm}$B_K^0$ & $ 0.5280(16) $ & $ 0.5278(16) $ & \rule{0cm}{0.4cm}$C_0^{B_K}$ & $ 0.5316(28) $ & 0.5322(17) \\
 $c_{B_K,a}^{\scriptscriptstyle I}$ & $ 0.125(12) $ & $ 0.128(12) $ & $C_a^{B_K,\,\scriptscriptstyle I}$ & $ 0.145(19) $ & 0.129(12) \\
 $c_{B_K,a}^{\scriptscriptstyle ID}$ & $ 0.148(15) $ & $ 0.153(15) $ & $C_a^{B_K,\,\scriptscriptstyle ID}$ & $ 0.201(33) $ & 0.164(15) \\
 $c_{B_K,m_x}$ & $ 0.00492(64) $ & $ 0.00420(64) $ & $C_1^{B_K}$ & $ -1.0(1.1) $ & 0.37(19) \\
 $c_{B_K,m_l}$ & $ -0.00809(94) $ & $ -0.00728(95) $ & $C_2^{B_K}$ & $ 0.58(68) $ & 0.38(28) \\
 $c_{B_K,m_y}$ & $ 1.316(32) $ & $ 1.324(32) $ & $C_3^{B_K}$ & $ 1.547(96) $ & 1.331(32) \\
 $c_{B_K,m_h}$ & $ -0.13(18) $ & $ -0.06(18) $ & $C_4^{B_K}$ & $ 0.50(55) $ & 0.07(18) \\ 
\end{tabular}
\caption{The $B_K$ fit parameters for each of our chiral ans\"{a}tze in the ${\rm SMOM}(\slashed{q},\slashed{q})$ scheme at 3.0 GeV. The parameters are given in physical units and with the heavy quark mass expansion point adjusted to the physical strange quark mass. For the ChPT and ChPTFV ansatz\"{e} the chiral scale $\Lambda_\chi$ has been adjusted to 1 GeV. \label{tab-fitpar3gev} }
\end{table}

\begin{table}
\centering
\begin{tabular}{l||l||r|r|r}
\hline
 & ChPTFV & $\Delta$ ChPT & $\Delta$ Analytic (260 MeV) & $\Delta$ Analytic (370 MeV)\\
\hline
 $B_K(\slashed q,\slashed q)$ & $ 0.5341(18) $ & $ 0.00020(11) $ & $ -0.0035(25) $ & $ -0.00029(21) $ \\
 $B_K(\gamma^\mu,\gamma^\mu)$ & $ 0.5166(18) $ & $ 0.00027(12) $ & $ -0.0037(24) $ & $ -0.00029(21) $ \\
\hline
 $B_K(\msbar\ {\rm via}\ \slashed q,\slashed q)$ & $ 0.5293(17) $ & $ 0.00020(11) $ & $ -0.0035(24) $ & $ -0.00029(21) $ \\
 $B_K(\msbar\ {\rm via}\ \gamma^\mu,\gamma^\mu)$ & $ 0.5187(18) $ & $ 0.00027(12) $ & $ -0.0037(24) $ & $ -0.00029(21) $ \\
\end{tabular}
\caption{The physical predictions for $B_K$ in the two intermediate schemes and in the $\MSbar$ scheme (via the intermediate schemes) obtained using the ChPTFV ansatz, and the full correlated differences (labelled $\Delta$) between the results obtained using the other ans\"{a}tze and the ChPTFV result. Analytic fit differences are presented with a 370 MeV and 260 MeV pion mass cut.\label{tab-predsdiffsbk} }
\end{table}


\subsubsection{Predicted values}



In Table~\ref{tab-predsdiffsbk} we list the continuum predictions for $B_K$, renormalized in each of the two intermediate schemes, that we obtained using the ChPTFV ansatz, as well as the sizes of the differences between those and the other chiral ans\"{a}tze. In contrast to the other quantities, for $B_K$ we observe that the differences between the ChPTFV and analytic ans\"{a}tze are of the same order as the statistical error, although those differences are poorly resolved. Nevertheless, we choose to continue to neglect the chiral systematic error for the following reasons: We previously chose to treat the chiral extrapolation error as small not just because the differences between the analytic and ChPTFV forms are small, but because we have good evidence to believe that the ChPTFV fits are correctly capturing this behavior in addition to their strong theoretical motivation. This was not the case in our former works where we were extrapolating from heavier masses. There the analytic fits were motivated by the apparent linearity in the available data with full knowledge that they do not correctly describe any underlying chiral curvature and are therefore not applicable over large mass ranges. Given that both fit forms were deficient in different ways, we conservatively took their full difference as an estimate of the error. On the other hand, in our new analysis we have a large amount of data in the light mass regime and the fits are forced to pass through data essentially at the physical point. As a result there is no longer any reason to distrust the ChPTFV results, especially given that they are only being used to perform a 4 MeV extrapolation in the pion mass. On the other hand there is now good evidence of chiral curvature in our results and therefore good reason to discount the analytic results. In fact, it is a testament of the robustness of our procedure that, despite this deficiency, the results obtained using these two ans\"{a}tze differ only at the fraction-of-a-percent level.
 
We use the SMOM$(\slashed{q},\slashed{q})$ result for our central value, giving us a final continuum result in a non-perturbative MOM scheme with 0.3\% total error after all sources of error are accounted for:
%
\begin{equation}
B_K(\slashed{q},\slashed{q},3 \mathrm{GeV}) = 0.5341(18)\,.
\end{equation}
%

This final prediction, and the result in the SMOM$(\gamma^\mu,\gamma^\mu)$ scheme, can be converted into the $\MSbar$ scheme using the following one-loop matching coefficients~\cite{Aoki:2010pe}:
\begin{equation}\begin{array}{rl}
C(\slashed q,\slashed q \rightarrow \msbar) = 0.99113,\,\,\,   &  C(\gamma^\mu,\gamma^\mu \rightarrow \msbar) = 1.00408\,,
\end{array}\end{equation}
using $\alpha_s(3\ {\rm GeV}) = 0.24544$. The resulting $\MSbar$ values are also listed in Table~\ref{tab-predsdiffsbk}.

For the reasons discussed above, we use the value obtained via the SMOM$(\slashed{q},\slashed{q})$ scheme for our final $\MSbar$ result. The matching introduces a perturbative truncation error, which we estimate by taking the full difference between the results obtained using the two RI-SMOM intermediate schemes. We obtain: 
\begin{equation}
B_K(\msbar, 3\ {\rm GeV}) = 0.5293(17)(106)\,,
\end{equation}
where the errors are statistical and from the perturbative matching to $\MSbar$ respectively. 

In the renormalization group invariant (RGI) scheme, the above corresponds to 
\begin{equation}
\hat B_K = 0.7499(24)(150)\,.
\end{equation}

Previously~\cite{Arthur:2012opa} we obtained:
\begin{equation}
B_K(\msbar, 3\ {\rm GeV}) = 0.535(8)(7)(3)(11)\,,
\end{equation}
for which the errors are statistical, chiral, finite-volume and from the perturbative matching respectively. Comparing with the above, we see excellent agreement. Our new result offers a considerable improvement in the statistical error, but the truncation effects are the same as we have not changed the scale, and dominate the final error.

\FloatBarrier
\section{Conclusions}
\label{sec:Conclusions}
Combining decades of theoretical, algorithmic and computational advances, we are finally able to perform $2+1$ flavor simulations with an essentially chiral action directly at the physical masses of the up, down and strange quarks in isospin symmetric QCD with both fine lattice spacings and large physical volumes. In this paper we report on two such ensembles; a $48^3\times 96\times 24$ (48I) ensemble and a $64^3\times 128\times 12$ (64I) ensemble, both using \mobius domain wall fermions.  The inverse lattice spacings are $a^{-1} = \ainvtwo$ GeV and $\ainvthree$ GeV, respectively, and these ensembles have $m_\pi L = \mpiLtwo$ and $\mpiLthree$. We make use of the \mobius kernel with parameters chosen such that the \mobius and Shamir (traditional domain wall) kernels are identical, but the approximation to the sign-function of the four-dimensional effective action is improved in the former, resulting in a smaller residual chiral symmetry breaking for the same computational cost.

The simulated pion masses are \mpitwo$ $ and \mpithree$ $ MeV for the 48I and 64I ensembles respectively. These are slightly above the physical value, requiring a small extrapolation that we performed by combining these ensembles with several of our older Shamir domain wall ensembles in a simultaneous chiral/continuum `global fit', specifically the $24^3\times 64\times 16$ (24I) and $32^3\times 64\times 16$ (32I) ensembles with the Iwasaki gauge action at $\beta=2.13$ and $2.2$ respectively, and the $32^3\times 64\times 32$ (32ID) ensemble with the Iwasaki+DSDR gauge action at $\beta=1.75$. We also include a new $32^3\times 64\times 12$ (32Ifine) Shamir domain wall ensemble with the Iwasaki gauge action at $\beta = 2.37$, corresponding to $a^{-1} = \ainvfour$ GeV, and a heavier \mpifour$ $ MeV pion mass; this enables us to examine the scaling behaviour of our data in the 1.75-3.15 GeV range of inverse lattice spacings to look for deviations from the leading $a^2$ scaling behavior. These ensembles give us access to a wide range of unitary and partially-quenched data ranging from the physical point up to the imposed 370 MeV pion-mass cut. As we use the same kernel for our \mobius and Shamir simulations, we are able to describe all of these ensembles using the same continuum scaling curve, apart from the 32ID ensemble which has a different gauge action.

The global fits are performed using the techniques developed in Refs.~\cite{Aoki:2010dy} and~\cite{Arthur:2012opa}. We fit to the following quantities: $m_\pi$, $m_K$, $f_\pi$, $f_K$, $m_\Omega$ and the Wilson flow scales $w_0$ and $t_0^{1/2}$. A separate fit is performed to the neutral kaon mixing parameter, $B_K$. To describe the mass dependence of these quantities we use NLO partially-quenched chiral perturbation theory with and without finite-volume corrections (referred to as the `ChPTFV' and `ChPT' ans\"{a}tze) and also a linear `analytic' ansatz.

Despite the significantly improved precision of the 48I and 64I data, we found that the fits missed these data by 1-2$\sigma$; this is an artifact of the large number of data points in the heavy-mass regime where $\chi$PT is only reliable to ${\cal O}(5\%)$. We resolve this issue by over-weighting the 48I and 64I data in order that the fit is forced to pass through these points. %
%
%
We emphasize that, while these global fits combine a large amount of data from various sources, the overweighting procedure guarantees that the predictions (and their statistical errors) are dominated by the near-physical data. A simpler procedure in which we simply treated the quark mass mistuning as an additional systematic error, would also obtain a similar statistical precision; the global fits essentially just remove these systematic effects.


The 48I and 64I ensembles each have the same gauge coupling as the corresponding 24I and 32I ensembles, but with smaller residual chiral symmetry breaking (significantly so for the former).  We found that the differences in the fermion action between these two pairs of ensembles, each evaluated at the same gauge coupling, resulted in a 3.2(2)\% difference between the 48I and 24I lattice scales, and a 1.1(2)\% difference between that of the 64I and 32I ensembles. In Appendix~\ref{appendix-mresadep} we show that this can be understood as an unexpectedly large effect of the changes in $L_s$ and the \mobius scale parameter $\alpha$ which distinguish these ensembles, and provide added numerical evidence that these effects are accurately described by such shifts in the lattice scales.

We showed that due to the dominance of the 48I and 64I data, which were measured with near-physical pion masses on large, 5.5fm boxes, the systematic errors associated with the chiral extrapolation and finite-volume can be neglected. The errors on our final results, which we take from the ChPTFV fits, are dominated by statistics, and are themselves very small. For the pion and kaon decay constants we obtain $f_\pi = \fpicont$ MeV and $f_K = \fkcont$ MeV; for the average up/down quark mass and strange quark mass in the $\MSbar$-scheme at 3 GeV, $\mudMSbar$ and $\msMSbar$ MeV; the neutral kaon mixing parameter $B_K$ in the RGI scheme, $\BKRGI$ and the $\MSbar$-scheme at 3 GeV, $\BKMSbar$; and the Wilson flow scales $t_0^{1/2} = \sqrttzero\ {\rm GeV}^{-1}$ and $w_0 = \wzero\ {\rm GeV}^{-1}$. In Table~\ref{tab-conclusions-results} we compare our numbers to the $N_f=2+1$ results compiled by the Flavor Lattice Averaging Group (FLAG) in their Review of Lattice Results~\cite{Aoki:2013ldr}. 

\begin{table}[!ht] 
\centering
\begin{tabular}{c|c|cl}
Quantity & This Work & FLAG Average\\
\hline 
$f_\pi$ & $130.19 \pm 0.89$ MeV & $130.2 \pm 1.4$ MeV &\cite{Follana:2007uv,Bazavov:2010hj,Arthur:2012opa}\\
$f_K$ & $155.51 \pm 0.83$ MeV   & $156.3 \pm 0.9$ MeV &\cite{Follana:2007uv,Bazavov:2010hj,Arthur:2012opa}  \\
$f_K/f_\pi$ & $1.1945 \pm 0.0045 $ & $1.194 \pm 0.005$&\cite{Follana:2007uv,Bazavov:2010hj,Arthur:2012opa,Durr:2011ap} \\
$m_u = m_d(\msbar,3 \; {\rm GeV})$ & $2.997 \pm 0.036 \pm 0.033$ MeV \\
$m_s(\msbar,3 \; {\rm GeV})$ & $81.64 \pm 0.77 \pm 0.88$ MeV \\
$m_s/m_u = m_s/m_d$ & $27.34 \pm 0.21$ & $ 27.46 \pm 0.15$ & \cite{Bazavov:2009fk,Bazavov:2010yq,Durr:2010vn,Durr:2010aw,Arthur:2012opa}\\
$m_u = m_d(\msbar,2 \; {\rm GeV})$ & $3.315 \pm 0.040 \pm 0.036$ MeV & $3.42 \pm 0.06$ MeV & \cite{Bazavov:2010yq,Durr:2010vn,Durr:2010aw,Arthur:2012opa}\\
$m_s(\msbar,2 \; {\rm GeV)}$ & $90.29 \pm 0.85 \pm 0.97$ MeV & $ 93.8 \pm 1.5$ MeV &\cite{Bazavov:2009fk,Durr:2010vn,Durr:2010aw,Arthur:2012opa} \\
$t_0^{1/2}$ & $0.7292 \pm 0.0041$ GeV$^{-1}$ \\
$w_0$ & $0.8742 \pm 0.0046$ GeV$^{-1}$ \\
$B_K({\rm SMOM}(\slashed q,\slashed q),3 \; {\rm GeV})$ & $0.5341 \pm 0.0018$  \\
$B_K(\msbar,3 \; {\rm GeV})$ & $0.5293 \pm 0.0017 \pm 0.0106$  \\
$\hat{B}_K$ & $0.7499 \pm 0.0024 \pm 0.0150$ & $0.7661 \pm 0.0099$ &\cite{Durr:2011ap,Laiho:2011np,Arthur:2012opa,Bae:2013lja}\\
$F_\pi/F$ & $ 1.0645 \pm 0.0015 $ & $1.0624 \pm 0.0021$ & \cite{Bazavov:2009fk,Borsanyi:2012zv,Beane:2011zm} \\
$\left[\Sigma(\msbar,3 \; {\rm GeV})\right]^{1/3}$ & $285.3 \pm 2.0 \pm 1.0$ MeV \\
$\left[\Sigma(\msbar,2 \; {\rm GeV})\right]^{1/3}$ & $275.9 \pm 1.9 \pm 1.0$ MeV & $ 271 \pm 15$ MeV & \cite{Bazavov:2010yq,Aoki:2010dy,Borsanyi:2012zv} \\
$\overline{l}_3 $ & $2.73 \pm 0.13 $ & $3.05 \pm 0.99$ & \cite{Arthur:2012opa,Bazavov:2010yq,Borsanyi:2012zv,Beane:2011zm} \\
$\overline{l}_4 $ & $4.113 \pm 0.059 $ & $4.02 \pm 0.28$ & \cite{Arthur:2012opa,Bazavov:2010yq,Borsanyi:2012zv,Beane:2011zm} \\
\end{tabular}
\caption{Summary of results from the simulations reported here.
The first error is the statistical error, which for most quantities
is much larger than any systematic error we can measure or estimate.
The exception is for the quantities in $\MSbar$ and $\hat{B}_K$.
For these quantities, the second error is the systematic error on
the renormalization, which is dominated by the perturbative matching
between the continuum RI-MOM scheme and the continuum $\MSbar$
scheme.  Comparison of our results to the averages compiled by the
Flavor Lattice Averaging Group~\cite{Aoki:2013ldr} for $N_f=2+1$
flavor isospin symmetric QCD.  Note that for $\hat B_K$, a direct
comparison of the perturbative error is not possible since we use
a different, and we believe more robust, method to estimate it.
This perturbative error is common to our calculation and to the
calculations dominating the FLAG average.  In the rightmost column
we provide the references to the original work that entered the
quoted FLAG-averages.  Light quark masses and the chiral condensate are given in the $\MSbar$ scheme,
evaluated at 2 GeV. Results from this work have been run down from
3 GeV to 2 GeV using the running factor 1.106 from the FLAG
review~\cite{Aoki:2013ldr} and do not include the FLAG-estimated
systematic error due to the omission of the charm sea quark.  
\label{tab-conclusions-results}
}
\end{table}

Our results for the light and strange quark masses, obtained in Section~\ref{sec-renormquarkmasses}, are renormalized in the $\MSbar$ scheme at 3 GeV. The only remaining uncertainties on these quantities are statistical and perturbative matching errors, roughly 1\% each. The renormalization and running of the quark masses were computed nonperturbatively, details of which can be found in Appendix~\ref{appendix-npr}. The masses are quite consistent with our previous determinations, but show significant improvement due to the inclusion of the physical point ensembles. Our masses agree with the FLAG averages, but have errors that are both smaller than those of the average as well as those of any of the individual results used therein~\cite{Bazavov:2009fk,Bazavov:2010yq,Durr:2010vn,Durr:2010aw,Arthur:2012opa}. The ratio of strange to light quark masses, shown in Eq.~\eqref{eq:massratio}, is also consistent with the FLAG average~\cite{Aoki:2013ldr}, but here the error is slightly larger since systematic errors mostly cancel, though it is as small as any individual result used in the average~\cite{Bazavov:2009fk,Bazavov:2010yq,Durr:2010vn,Durr:2010aw,Arthur:2012opa}.

The FLAG average for the standard model kaon bag parameter
is largely dominated by the Budapest-Marseille-Wuppertal collaboration (BMWc) result~\cite{Durr:2011ap},
$\hat B_K=0.7727(81)_{\rm stat}(34)_{\rm sys}(77)_{\rm PT}$,
where the errors are statistical, systematic and from perturbation theory, respectively.
We would like to stress the difficulties one encounters in reliably assessing truncation errors, a point also emphasized by BMWc~\cite{Durr:2011ap}. Among other checks, the BWMc showed that the NLO-perturbative and their non-perturbative running in
the RI-MOM scheme agree between 1.8 and 3.5 GeV within statistical errors (of 2\%), and quote 1\% for the error due to perturbation theory, 2\% being the size of the NLO term in the perturbative expansion. We proceed differently, by evaluating the difference between two different intermediate SMOM schemes, and estimate an error of $2\%$. We believe our procedure is more robust than those that have fed into the FLAG average, since multiple intermediate schemes were used to assess the truncation error. This error can certainly be reduced further in the future by performing the matching to $\overline{\rm MS}$ 
at higher scale or by computing the matching coefficient at NNLO.
We want to emphasize that the errors quoted are different because 
the subjective procedures to estimate these errors are different. For completeness, we also compare the non-perturbative scale evolution to the NLO running 
between $2$ and $3$ GeV. We find a deviation of around $1.5\%$ for the RI-SMOM$(\gamma_\mu,\gamma_\mu)$ and for the RI-MOM schemes,
and of $0.5\%$ for the RI-SMOM$(\s{q},\s{q})$ scheme.

It is useful to compare our results with Ref.~\cite{Durr:2011ap} in the intermediate MOM schemes
(before converting to $\overline{\rm MS}$) as these numbers are purely non-perturbative:
\begin{eqnarray}
B_K^{\rm RI} (3.5\, \GeV)  &=& 0.5308(56)_{\rm stat} (23)_{\rm sys}  \qquad {\rm BMWc} \; \\
B_K^{(\s{q},\s{q})}(3\,\GeV) &=& 0.5341(18)_{\rm stat}  \qquad \mbox{ this work}\,,
\end{eqnarray}
where we neglect the various sources of systematic errors in our result since they are considerably smaller
than the statistical error. These results are in different non-perturbative schemes and at different scales, and are therefore not directly comparable. However, we
can compare their relative total errors: our result and that of BMWc have a $0.3\%$ and a 1.1\% relative error, respectively. We emphasize that in terms of objective statistical errors, and every systematic effect for 
which there is a theoretical framework for estimation (e.g. discretization, mass extrapolation, and finite volume),
our new result is more precise than those entering the FLAG average. This is reflected in the 0.3\% total 
relative error on results in a non-perturbatively defined $\s{q}$ RI scheme. Our assessment
of the (subjective) perturbative systematic uncertainty on the conversion
to $\MSbar$ is more pessimistic than that of FLAG and BMWc, but
we believe that it is better founded on the evidence of multiple intermediate schemes.

Predictions of $\hat B_K$ in lattice QCD have now reached a level of precision where other
ingredients in its utilization for SM-tests are limiting progress (e.g.
our knowledge on $|V_{cb}|$).

The results for the kaon and pion decay constant and their ratio 
are compatible with the FLAG average and amongst the most precise 
$N_f=2+1$ predictions that have been made.  Our results will certainly
allow for further constraining CKM-unitarity tests~\cite{Aoki:2013ldr}.

The most significant remaining differences between our simulations and the
physical world are isospin breaking and EM effects and the effect of quenching the
charm quark. 

Including isospin breaking effects requires using non-degenerate masses for the up and down quarks. This is possible within the domain wall fermion framework with current technology, for example using the rational quotient action or the one-flavor action developed by TWQCD~\cite{Chen:2014hyy}. However, these techniques are computationally demanding, and the effects in question are expected to be similar in size to the electromagnetic effects, hence there is limited value in considering these in isolation. 

The RBC and UKQCD collaborations have performed exploratory calculations using QCD
domain wall configurations with quenched electromagnetic interactions~\cite{Blum:2007cy,Blum:2010ym} and have performed unquenched simulations using reweighting techniques~\cite{Ishikawa:2012ix}. There is increasing effort in the lattice community to control these effects, from more precise electro-quenched calculations~\cite{Tantalo:2013maa,Borsanyi:2013lga} (i.e. with EM included only in the valence sector) up to full QCD+QED
simulations~\cite{Borsanyi:2014jba}. Adding QED to lattice simulations is challenging for many reasons. Firstly, adding a coupling constant to the theory, especially in the context of non-degenerate light quarks, considerably increases the cost of the simulations, particularly when using a chiral action close to the physical point. Secondly, the absence of mass gap in QED implies finite-size effects with power-law dependence on the lattice spatial extent, which are potentially large compared to the QED contributions~\cite{Borsanyi:2014jba,Davoudi:2014qua}. Finally, it is still not clear how to
define quantities such as decay constants in QCD+QED, because the matrix elements are infrared divergent and gauge dependent~\cite{Sachrajda:2014}. Because of these issues, the addition of isospin-breaking effects and electromagnetism remains an important and challenging topic for our future calculations.

Dynamical charm effects are expected to be small for the majority of the quantities studied in this paper, but for quantities such as the $K_L-K_S$ mass difference and $K\rightarrow\pi\pi$ amplitudes they can have significant contributions. This is therefore the most promising avenue for RBC and UKQCD to take, allowing us to address these systematic errors on our flagship calculations. The biggest hurdle for including the charm is the requirement of simulating with finer lattice spacings, which tends to incur freezing of topology as well as requiring large computing power to obtain sufficiently large physical volumes. RBC and UKQCD have developed the `dislocation enhancing determinant' (DED) method~\cite{McGlynn:2013ava} to overcome the effects of the topology freezing, and have already commenced large-scale physical simulations with dynamical charm.

\section*{Acknowledgments}
The generation of the $48^3 \times 96$ and $64^3 \times 128$ \mobius DWF+Iwasaki ensembles was performed using the IBM Blue Gene/Q (BG/Q) ``Mira'' machines at the Argonne Leadership Class Facility (ALCF) provided under the Incite Program of the US DOE, on the STFC funded ``DiRAC'' BG/Q system in the Advanced Computing Facility at the University of Edinburgh, and on the BG/Q machines at Brookhaven National Laboratory (BNL). The BG/Q computers of the RIKEN-BNL Research Center were used to generate the $32^3 \times 64$ fine 3.14 GeV ensemble. The DiRAC equipment was funded by BIS National E-infrastructure capital grants ST/K000411/1, STFC capital grant ST/H008845/1, and STFC DiRAC Operations grants ST/K005804/1 and ST/K005790/1. DiRAC is part of the National E-Infrastructure. Most of the measurements were also performed on the DiRAC and Mira machines, with the remainder performed using the BG/Q computers at the BNL.

The research leading to these results has also received funding from the
European Research Council under the European Community's Seventh
Framework Programme (FP7/2007-2013) ERC grant agreement \#279757.

The software used includes the CPS QCD code (\url{http://qcdoc.phys.columbia.edu/cps.html}), supported in part by the USDOE SciDAC program; and the BAGEL\\ (\url{http://www2.ph.ed.ac.uk/~paboyle/bagel/Bagel.html}) assembler kernel generator for high-performance optimized kernels and fermion solvers~\cite{Boyle:2009vp}. The gauge fixing for the 48I ensemble was performed using the CUTH cluster at Columbia University using the ``GLU'' (Gauge Link Utility) codebase (\url{https://github.com/RJhudspith/GLU}).

T.B is supported by U.S. DOE grant \#DE-FG02-92ER41989.
G.M, N.H.C, R.D.M, and D.J.M are supported in part by U.S. DOE grant \#DE-SC0011941.
A.S, C.J, T.I and C.L are supported in part by US DOE Contract \#AC-02-98CH10886(BNL).
T.I is also supported by Grants-in-Aid for Scientific Research \#26400261.
C.T.S, T.J and A.P acknowledge the STFC grants ST/J000396/1 and ST/L000296/1.
P.A.B, R.D.K, N.G and J.F acknowledge support from STFC grants ST/L000458/1 and ST/J000329/1.
N.G also acknowledges support from STFC under the grant ST/J000434/1 and from the EU Grant Agreement \#238353 (ITN STRONGnet).
C.K is supported by a RIKEN foreign postdoctoral research (FPR) grant.
R.J.H is supported by the Natural Sciences and Engineering Research Council of Canada.

\appendix
\section{Conserved currents of the M\"{o}bius domain wall action}
\label{appendix-mobiusconservedcurrents}

The connection of the \mobius formulation to overlap fermions can be made at the propagator level and
with the familiar DWF physical fields $q_L$ and $q_R$.
In the following subsection we repeat known but important results
connecting the surface-to-surface and surface-to-bulk propagators 
of the M\"{o}bius domain wall action (in our conventions) with the
four dimensional overlap propagator. These results are then used to
establish a practical implementation of the conserved axial and vector
currents for the M\"{o}bius case.

\subsection{Domain wall and overlap propagators, and contact terms}

The approximate overlap operator can be written in terms of our
four dimensional Schur complement matrices as
\begin{eqnarray}
D_{ov} &=& S_\chi(m=1)^{-1} S_\chi(m) .
\end{eqnarray}
Observe that if we solve the following 5-D system of equations,
\begin{eqnarray}
D_\chi^5(m=1)^{-1} D_\chi^5(m) \phi &=& 
\left(
\begin{array}{c}
q\\
0\\
\vdots\\
0
\end{array}
\right),
\end{eqnarray}
and substitute the UDL decomposition, this yields
\begin{eqnarray}
D_S^{-1}(m=1) D_S(m) L(m) \phi &=&
L(m=1) \left(
\begin{array}{c}
q\\
0\\
\vdots\\
0
\end{array}
\right).
\end{eqnarray}
Since $\left( L(m) (q,0,\ldots,0)^T\right)_1 = q $ and
$(L(m)\phi)_1 = \phi_1$, the topmost row of our 5-D system of equations gives the overlap propagator:
\begin{equation}
S_\chi(m=1)^{-1} S_\chi(m) = \left(D^5_\chi(m=1)^{-1}D^5_\chi(m)\right)_{11}.
\end{equation}
This approximate overlap operator can however be expressed
in terms of the $\bar{\psi}$ basis fields, and 
\begin{eqnarray}
D_{ov} &=& S_\chi(m=1)^{-1} S_\chi(m)  \\
      &=& \left[{\cal P}^{-1}{\cal P} D^5_\chi(m=1)^{-1} Q_-^{-1} 
	\gamma_5 \gamma_5
   Q_- D^5_\chi(m) {\cal P}^{-1}{\cal P}\right]_{11}\\
      &=& \left[{\cal P}^{-1} D^5_{GDW}(m=1)^{-1}  D^5_{GDW}(m) {\cal P}\right]_{11}\,.
\end{eqnarray}
The cancellation the Pauli-Villars term can be expressed 
in terms of \emph{unmodified} generalized domain wall matrix $D^5_{GDW}$. 
The overlap contact term can be subtracted from the overlap propagator. Here we define
\begin{eqnarray}
\tilde{D}_{ov}^{-1}&=&\frac{1}{1-m}\left[{D}_{ov}^{-1} - 1\right]\\
 &=&\frac{1}{1-m} 
\left[{\cal P}^{-1} D^5_{GDW}(m)^{-1}  D^5_{GDW}(m=1) {\cal P} - 1\right]_{11}\\
 &=&\frac{1}{1-m} 
\left\{ {\cal P}^{-1} D^5_{GDW}(m)^{-1}  
\left[ D^5_{GDW}(m=1)  -        D^5_{GDW}(m) \right]{\cal P}
\right\}_{11}\,.
\end{eqnarray}

Now, the difference 
$\left[ D^5_{GDW}(m=1)  -        D^5_{GDW}(m) \right]_{ij} = 
(1-m)\left[ P_- \delta_{i,L_s}\delta_{j1} 
+ P_+ \delta_{i,1}\delta_{j,L_s}\right]$.
This relation is simpler to interpret in our convention than with the
convention from Ref.~\cite{Brower:2012vk}: the mass term is applied to our five dimensional
surface fields without field rotation. With this,
\begin{eqnarray}
\tilde{D}_{ov}^{-1}&=&
\left\{ {\cal P}^{-1} D^5_{GDW}(m)^{-1}  R_5 {\cal P}
\right\}_{11}\,.
\end{eqnarray}
This is just the normal valence propagator of the physical DWF fields 
$q = ({\cal P}^{-1} \psi)_1$
and $\bar{q} = (\bar{\psi} R_5 {\cal P})_1$.
We see that the usual domain wall valence propagator has always contained both the contact term
subtraction and the appropriate multiplicative renormalization of the overlap
fermion propagator. As a result, the issues of lattice artifacts in NPR
raised in Ref.~\cite{Maillart:2008pv} have never been present in domain valence analyses.
This was guaranteed to be the case because Shamir's 5-D construction is designed to
exactly suppress chiral symmetry breaking in the limit of infinite $L_s$, including
any contact term. 

For later use, we may also consider the propagator into the bulk from a surface field $q$ for \mobius fermions,
\begin{eqnarray}
\langle Q_s \bar q \rangle &=& 
\left[ 
{\cal P}^{-1} D^5_{GDW}(m)^{-1} R_5 {\cal P} 
\right]_{s1}
\\
&=&
\frac{1}{1-m} 
\left\{
{\cal P}^{-1} D^5_{GDW}(m)^{-1} D^5_{GDW}(1) {\cal P} - \ident
\right\}_{s1}
\\
&=&
\frac{1}{1-m} 
\left\{
D^5_{\chi}(m)^{-1} D^5_{\chi}(1) - \ident
\right\}_{s1}\\
&=&
\frac{1}{1-m} 
\left\{
L^{-1}(m) D^{-1}(m) D(1) L(1) -\ident
\right\}_{s1} \\
&=&
\frac{1}{1-m} 
\left\{
L^{-1}(m) 
\left(
\begin{array}{c|c}
S_\chi^{-1}(m) S_\chi(1) & 0 \\
\hline
0 & \ident
\end{array}
\right)
L(1) -\ident
\right\}_{s1}.
\end{eqnarray}
Now, 
\begin{equation}
\begin{array}{ccc}
L(m) = 
\left(
\begin{array}{c|c}
1 & 0 \\
\hline
\begin{array}{c}
-T^{-(L_s-1)} (P_+-m P_-)\\
\vdots \\
-T^{-1} (P_+-m P_-)
\end{array} & \ident
\end{array}
\right)
&\quad;\quad&
L(m)^{-1} = 
\left(
\begin{array}{c|c}
1 & 0 \\
\hline
\begin{array}{c}
T^{-(L_s-1)} (P_+-m P_-)\\
\vdots\\
T^{-1} (P_+-m P_-)
\end{array} & \ident
\end{array}
\right)
\end{array}
\end{equation}
and so we have,
\begin{eqnarray}
\langle Q_s \bar q \rangle &=&
\frac{1}{1-m} 
\left(
\begin{array}{c|c}
D_{ov}^{-1}(m)  - \ident  & 0 \\
\hline
\begin{array}{c}
T^{-(L_s-1))} [ (P_+ - m P_-)D_{ov}^{-1}(m) - \gamma_5 ]\\
\vdots\\
 T^{-1} [ (P_+ - m P_-)D_{ov}^{-1}(m) - \gamma_5 ]
\end{array}
& 0
\end{array}
\right)_{s1}
\\
&=& 
\label{bulkpropagatormob}
\left(
\begin{array}{c|c}
\left[ P_+ + P_-  T^{-L_s}\right]   & 0 \\
\hline
\begin{array}{c}
T{-(L_s-1)}\\
\vdots\\
T^{-1} \\
\end{array}
& 0
\end{array}
\right)_{s1} \left[ 1+ T_1^{-1}\cdots T_{L_s}^{-1}\right]^{-1} D_{ov}^{-1}(m).
\end{eqnarray}
Finally, applying the permutation matrix, we have the five dimensional propagator from a physical field,
\begin{eqnarray}
G_q  = {\cal P} \langle Q_s \bar q\rangle
&=& [P_+  +P_- T^{-1}] 
\left(
\begin{array}{c}
T^{-(L_s-1)}   \\
T^{-(L_s-2)}  \\
\vdots\\
T^{-1} \\
1
\end{array}
\right)
[1+T^{-L_s}]^{-1} D_{ov}^{-1}.
\label{BulkPropagator}
\end{eqnarray} 

The connection between domain wall systems and the overlap, well established in the literature
and reproduced in this section, is useful in understanding the relation of domain 
wall fermions to their 4-D effective action. 

\subsection{Conserved vector and axial currents}

The standard derivation of lattice Ward identities proceeds
as follows. A change of variables of the fermion fields $\psi$ and $\bar \psi$ 
at a single site $y$ is performed:
\begin{equation}
     \psi^\prime_y = \psi_y - i \alpha \psi_y 
\quad\quad ; \quad\quad
\bar \psi^\prime_y = \bar \psi_y + i \bar \psi_y \alpha\,.
\end{equation}
Under the path integral, the Jacobian is unity, 
and the partition function is left invariant:
\begin{eqnarray}
Z^\prime
&=&
  \int d\bar\psi d\psi e^{-S[\bar\psi,\psi]} 
    \left\{1 - i \alpha \left[ \frac{\delta S}{\delta \psi_y} \psi_y
                     - \bar \psi_y \frac{\delta S}{\delta \bar\psi_y} 
	\right] \right\}
= Z.
\end{eqnarray}
Hence,
\begin{equation}
 \langle \frac{\delta S}{\delta \psi_y}\psi_y - \bar \psi_y \frac{\delta S}{\delta \bar\psi_y} 
\rangle = 0.
\end{equation}
The Wilson action gives eight terms from varying $\bar{\psi}_y$ and
eight terms from varying $\psi_y$ due to the 4-D hopping stencil:
\begin{eqnarray}
\bar\psi \delta_y (D_W) \psi = \Delta^-_\mu J^W_\mu (y)
&=&
\sum_\mu \left[
\begin{array}{c}
-\bar\psi_y \frac{1-\gamma_\mu}{2} U_\mu(y) \psi_{y+\hat\mu}
+\bar\psi_{y-\hat\mu} \frac{1-\gamma_\mu}{2} U_\mu(y-\hat\mu) \psi_{y}\\
-
\bar\psi_{y} \frac{1+\gamma_\mu}{2} U_\mu^\dagger(y-\hat\mu) \psi_{y-\hat\mu}
+\bar\psi_{y+\hat\mu} \frac{1+\gamma_\mu}{2} U_\mu(y)^\dagger  \psi_y
\end{array}
\right]
\label{WilsonDivergence}\\
&=&
\Delta^-_\mu \left[
 \bar\psi_y \frac{1-\gamma_\mu}{2} U_\mu(y) \psi_{y+\hat\mu}
-\bar\psi_{y+\hat\mu} U^\dagger_\mu(y) \frac{1+\gamma_\mu}{2} \psi_y
\right]= 0,
\end{eqnarray}
where $\Delta^-_\mu$ is the backwards discretized derivative.

An equivalent alternate approach may be taken, however, and this is a better
way to approach non-local actions such as the chiral fermions. 
Gauge symmetry leaves the action invariant at $O(\alpha)$ under the simultaneous active substitution,
for a fixed site $y$ of
\begin{equation}
U_\mu(y) \to (1+i\alpha) U_\mu(y) 
\quad\quad ; \quad\quad
U_\mu(y-\hat\mu) \to  U_\mu(y-\hat\mu) (1-i\alpha) 
\end{equation}
and
\begin{equation}
     \psi_y \to (1+ i \alpha )\psi_y 
\quad\quad ; \quad\quad
\bar \psi_y \to \bar  \psi_y (1 - i \alpha)\,.
.\end{equation}

A change of variables on the fermion fields at site $y$
may be performed simultaneously to absorb the phase on the fermions:
\begin{equation}
\psi^{\prime}_y =  (1+ i \alpha) \psi_y 
\quad\quad;\quad\quad
\bar \psi^{\prime}_y = \bar \psi_y (1 - i \alpha).
\end{equation}
Under the path integral, the Jacobian is again unity, and the phase associated with the fermion is absorbed.
We can now view the change in action as being associated with the \emph{unabsorbed}
phases on the eight gauge links connected to site $y$: 
\begin{equation}
Z^\prime = Z =
 \int d\bar\psi^\prime d\psi^\prime e^{-S[\bar\psi^\prime,\psi^\prime,U]}
    \left\{1 + i \alpha \sum_\mu \left[\frac{\delta S}{\delta U_\mu(y)^{ij}} U_\mu(y)^{ij}
    - \frac{\delta S}{\delta U_\mu(y-\mu)^{ij}} U_\mu(y-\mu)^{ij}
	\right] \right\}.
\end{equation}
For a gauge invariant Lagrangian we can \emph{always} 
use a picture where the same change in action, and same current conservation law 
may be arrived at by differentiating with respect to the eight links connected to a site:
\begin{equation}
\langle \sum_\mu \left[
 \frac{\delta S}{\delta U_\mu(y)^{ij}} U_\mu(y)^{ij}
    -                 
 \frac{\delta S}{\delta U_\mu(y-\mu)^{ij}} U_\mu(y-\mu)^{ij}
	\right]\rangle = 0.
\end{equation}
This arises because the phase freedom of fermions and of gauge fields are necessarily coupled and inseparable in a gauge theory.
For the nearest-neighbor Wilson action, this generates the same eight terms entering $\Delta^-_\mu J_\mu =0$.

In the case of non-local actions, the Dirac matrix, whatever its form, can be viewed as a sum of gauge covariant paths. 
When generating a current conservation law from $U(1)$ rotation of the fermion field
at site $y$, we sum over all fields $\bar \psi(x)$ and $\psi(x)$ connecting through the Dirac matrix $D(x,y)$ to the fixed site
$\psi(y)$ and $\bar\psi(y)$. The following sum is always constrained to be zero for all $y$, and is identical to that 
found by Kikukawa and Yamada~\cite{Kikukawa:1998py}:
\begin{equation}
\sum_x \bar{\psi}_x D(x,y) \psi_y - \bar{\psi}_y D(y,x) \psi_x = 0.
\label{sum_df}
\end{equation}
The partitioning of this sum of terms, into a paired \emph{discrete divergence operator} and \emph{current} is not obvious,
and it is cumbersome to generate Kikukawa and Yamada's \emph{non-local kernel}.

It is instructive to consider what happens if we derive the same sum of terms by differentiating with respect to the
8 links connected to site $y$.
\begin{equation}
\langle
\sum_\mu \left[
\begin{array}{c}
 \frac{\delta S}{\delta U_\mu(y)^{ij}} U_\mu(y)^{ij} 
    -                 
 \frac{\delta S}{\delta U_\mu(y-\mu)^{ij}} U_\mu(y-\mu)^{ij}
\end{array}
	\right]
\label{sum_du}
\rangle = 0
\end{equation}
The structure of Eq.~\eqref{sum_du} \emph{always} lends itself interpretation as a backwards finite difference.
For a non-local action, the differentiation Eq.~\eqref{sum_du} appears to generate a lot more terms than the fermion field differentiation
Eq.~\eqref{sum_df}. The reason is clear: these extra terms are constrained by gauge symmetry to sum to zero, but only after
cancellation between the different terms in Eq.~\eqref{sum_du}. 
Specifically, we consider an action constructed as the product of Wilson matrices:
\begin{equation}
S = \sum_{xyzw} \bar \psi_x D_W(x,y) D_W(y,z)D_W(z,w) \psi(w).
\end{equation}
The link variation approach gives three terms, each of which are conserved under a
nearest-neighbor difference divergence: varying with respect to the 8 links we obtain,
via the product rule,
\begin{equation}
\delta_y (\bar\psi D_W D_W D_W \psi) \psi = \psi \left[ (\delta_y D_W) D_W D_W + D_W (\delta_y D_W)D_W  +D_W D_W(\delta_y D_W) \right] \psi\,.
\end{equation}
Each of these contributions contain a backwards difference operator, and it is trivial
to split this into a divergence and corresponding conserved current using Eq.~\eqref{WilsonDivergence}.

The above comment is generally applicable to any function of the Wilson matrix.
We take this approach to establish the exactly-conserved vector current of an approximate overlap operator, where the
approximation is represented by a rational function. 
We will also establish that matrix elements of this current are identical to those of the Furman and Shamir
approach~\cite{Furman:1994ky} in the case of domain wall fermions.
The Furman and Shamir approach will then be used
to also establish an axial Ward identity for our generalized \mobius domain wall fermions under which an explicitly known defect arises.
This is important in both renormalizing lattice operators and also in determining the most appropriate measure
of residual chiral symmetry breaking in our simulations.
We construct the conserved vector current by determining
the variation in the overlap Dirac operator, $\delta_y D_{ov}$:
\begin{eqnarray}
\delta_y D_{ov} &=& \frac{1-m}{2} \gamma_5 \left\{
\delta_y (\frac{1}{1+T^{-Ls}})  [1-T^{-Ls}]
+\frac{1}{1+T^{-Ls}} \delta_y (1-T^{-Ls})
\right\}\nonumber\\
 &=& \frac{1-m}{2} \gamma_5 \left\{
\delta_y (\frac{1}{1+T^{-Ls}}) 
- 
\frac{1}{1+T^{-Ls}} \delta_y(T^{-L_s}) 
\left( 1 - \frac{T^{-L_s}} {1+T^{-Ls}}\right)
\right\}\nonumber\\
&=& (1-m)\gamma_5 \delta_y \left(\frac{1}{1+T^{-Ls}}\right).
\end{eqnarray}

We can similarly find the variation in $T^{-1}$ induced by a variation in $D_W$, 
where the variation in $D_W$ is just the backwards divergence of the standard Wilson conserved 
current operator. Denoting,
\begin{eqnarray}
T^{-1} &=& -(\tilde{Q}_-)^{-1}\tilde{Q}_+\nonumber\\
\tilde{Q}_- &=&  D_+^s P_- - D_- P_+ = D_- \gamma_5 Q_-  \nonumber\\
\tilde{Q}_+ &=&  D_+^s P_+ - D_- P_- = D_- \gamma_5 Q_+  ,
\end{eqnarray}
we see that 
\begin{eqnarray}
\delta_y(T^{-1})  &=& -\tilde{Q}_-^{-1}\left\{ -\delta_y(\tilde{Q}_-) \tilde{Q}_-^{-1}\tilde{Q}_+ +  \delta_y(\tilde{Q}_+) \right\}\nonumber\\
 &=&  -\tilde{Q}_-^{-1}\left\{ \delta_y(\tilde{Q}_-) T^{-1} + \delta_y(\tilde{Q}_+) \right\} \nonumber\\
 &=&  -\tilde{Q}_-^{-1} \delta_y(D_W) \left\{  (b P_-  + c P_+ ) T^{-1} + b P_+ +c P_- \right\}. 
\end{eqnarray}
Since
\begin{eqnarray}
\begin{array}{ccccccc}
\tilde{Q}_- P_- &=& (1+b D_W)P_- 
&;&
\tilde{Q}_+ P_- &=& (c D_W-1)P_- \nonumber\\
\tilde{Q}_- P_+ &=& (c D_W-1)P_+ 
&;&
\tilde{Q}_+ P_+ &=& (1+b D_W)P_+ ,
\end{array}
\end{eqnarray}
we may re-express the identity 
\begin{eqnarray}
 (b+c)(P_+ + P_-) &=&c\tilde{Q}_- P_-  - b\tilde{Q}_+ P_-  +  c\tilde{Q}_+ P_+ - b\tilde{Q}_- P_+ \\
\tilde{Q}_-^{-1}(P_+ + P_-) &=& \frac{\tilde{Q}_-^{-1}}{b+c} \left[ \tilde{Q}_+ (c P_+ - b P_- ) + \tilde{Q}_- (c P_- - b P_+) \right],
\end{eqnarray}
and this lets us find a symmetrical form:
\begin{eqnarray}
(b+c)\delta_y(T^{-1}) &=& 
\left[ b [P_+ - T^{-1}P_-] + c [T^{-1} P_+ - P_- ]  \right]
\delta_y(D_W) 
\left[  b [ P_+ +  P_- T^{-1}] + c [ P_+ T^{-1} + P_- ]\right]. \label{deltaTinv}\nonumber
\end{eqnarray}
We may now look at the variation of the term $T^{-L_s}$
\begin{equation}
\delta_y(T^{-L_s}) =
\sum\limits_{s=1}^{L_s}
T^{-(s-1)} 
\left[ 
\begin{array}{c} 
b [P_+ - T^{-1}P_-] \\+ c [T^{-1} P_+ - P_- ]
\end{array}  \right]
\delta_y(D_W) 
\left[
\begin{array}{c} 
  b [ P_+ +  P_- T^{-1}] \\+ c [ P_+ T^{-1} + P_- ]
\end{array}
\right]
T^{- (L_s-s)}. 
\end{equation}
Compiling these results, we find
\begin{equation}
\delta_y D_{ov} =  -\frac{1-m}{b+c} \gamma_5 \frac{1}{1+T^{-Ls}}
\left( \sum\limits_{s=1}^{L_s}
T^{-(s-1)} \delta_y(T^{-1}) T^{- (L_s-s)} 
\right)
\frac{1}{1+T^{-Ls}}.
\end{equation}

The terms may be expanded until insertions of the
the backwards divergence of the Wilson current are reached (Eq.~\eqref{WilsonDivergence}).
Gauge symmetry then implies the conservation of the obvious current and 
the vector Ward identities can be constructed. For example, we may take
as source $\eta^{j j^\prime\alpha\alpha^\prime} (z) = \delta_{jj\prime}\delta_{\alpha\alpha^\prime} \delta^4(z-x)$
and a two-point function of the conserved current may be constructed as
\begin{equation}
\label{MobiusMatrixElement}
\begin{array}{c}
\Delta^-_\mu \langle \bar \psi \gamma_\nu \psi(x)| {\cal V}_\mu(y) \rangle =
{\rm Tr} \gamma_\nu \gamma_5 \eta^\dagger D_{ov}^{-\dagger} \gamma [1+T^{-L_s}]^{-1}
\left\{\sum\limits_{s=0}^{L_s-1}
T^{-s} \delta_y(T^{-1}) T^{-(L_s-1-s)}
\right\}
[1+T^{-L_s}]^{-1} D_{ov}^{-1}\eta
\end{array}.
\end{equation}

Note that when $c=0$, the insertion of Eq.~\eqref{deltaTinv} contains only terms such as 
\begin{equation}
[ P_-  T^{-1} + P_+ ],
\end{equation} 
which are also present in the surface to bulk propagator Eq.~\eqref{BulkPropagator}.
As one would expect, when we take $b$ and $c$ to represent domain wall fermions, 
the two-point function of our exactly conserved vector current - derived from the four dimensional effective 
action - exactly matches the matrix element of the vector current
constructed by Furman and Shamir~\cite{Furman:1994ky}, Eq.~(2.21), from a five dimensional interpretation
of the action.

Since the Furman and Shamir current was easily constructed from the five dimensional propagator 
Eq.~\eqref{BulkPropagator}, one might hope to do the same in the generalized approach to domain wall fermions.
To play a similar trick for the $c$ term, we would need to generate the terms
\begin{equation}
P_- [1+T^{-L_s}]^{-1} D_{ov}^{-1},
\end{equation} and
\begin{equation}
 P_+  T_1^{-1}[1+T^{-L_s}]^{-1} D_{ov}^{-1}.
\end{equation}
These are not manifestly present in Eq.~\eqref{bulkpropagatormob}.
However, the presence of the contact term on the $s=0$ slice can be removed after a propagator
calculation. We define this slice as 
\begin{equation}
S(x) = \langle Q_0 \bar q\rangle = \frac{1}{1-m}\left( D_{ov}^{-1}(m)  - \ident\right).
\end{equation}
In a practical calculation, the source vector $\eta$ may be used to eliminate the contact term by forming
\begin{equation}
(1-m)S(x)\eta + \eta = D_{ov}^{-1}(m) \eta =  [1+T^{-L_s}] [1+T^{-L_s}]^{-1} D_{ov}^{-1} \eta.
\end{equation}
By applying $P_+$ and $P_-$ we find we have the following set of vectors
\begin{equation}
\left(
\begin{array}{c}
P_+ \\
P_- T^{-L_s}
P_+[1+T^{-L_s}]\\
P_-[1+T^{-L_s}]
\end{array}
\right)[1+T^{-L_s}]^{-1} D_{ov}^{-1},
\end{equation}
and we may eliminate to form a $L_s+1$ vectors from a 4-D source $\eta$
\begin{equation}
T(s) = 
\left(
\begin{array}{c}
1   \\
T^{-1}\\
\vdots\\
T^{-L_s}
\end{array}
\right) \left[ 1+ T_1^{-1}\cdots T_{L_s}^{-1}\right]^{-1} D_{ov}^{-1}(m)\eta.
\end{equation}
This may be used to construct
\begin{equation}
\left[  b [ P_+ +  P_- T^{-1}] + c [ P_+ T^{-1} + P_- ]\right] T^{s},
\end{equation}
for $s\in \{ 0 \ldots L_s-1 \} $, and by contracting these vectors through the Wilson conserved
current the matrix element, Eq.~\eqref{MobiusMatrixElement}, can be formed in a \emph{very} similar manner to the
standard DWF conserved vector current. When $c = 0$ the matrix element reduces to being \emph{identical}
to that for the Furman and Shamir vector current.

A flavor non-singlet axial current, almost conserved under a backwards difference operator, can now also be constructed following Furman and Shamir.
We associate a fermion field rotation 
\begin{equation}
\psi(x,s) \to \left\{ 
\begin{array}{ccc}
e^{i \alpha \Gamma(s)} \psi(x,s) & ; & x=x_0 \\
 \psi(x,s) &;& x\ne x_0
\end{array}
\right.,
\end{equation}
where
\begin{equation}
\Gamma(s) \to \left\{ 
\begin{array}{ccc}
-1 &;& 0\le s < L_s/2\\
1 &;& L_s/2 \le s
\end{array}
\right. .
\end{equation}
We acquire a related (almost-) conserved axial current, whose pseudoscalar matrix element is
\begin{equation}
\begin{array}{c}
\Delta^-_\mu \langle \bar \psi \gamma_5 \psi(x)| {\cal A}_\mu(y) \rangle =\\
{\rm Tr}  [\eta^\dagger \tilde D_{ov}^{-\dagger} \gamma_5 ]
[1+T^{-L_s}]^{-1}
\left\{
\sum\limits_{s=0}^{L_s-1}T^{-s} \Gamma(s) \delta_y(T^{-1}) T^{-(L_s-1-s)}
\right\}
[1+T^{-L_s}]^{-1}
 D_{ov}^{-1} \eta
\end{array}\,.
\end{equation}
The exact vector current conservation induces the same $J_{5q}$ midpoint density defect that
arose for DWF, and the Ward identity is 
\begin{equation}
\label{mobpcac}
\Delta^-_\mu \langle \bar \psi \gamma_5 \psi(x)| {\cal A}_\mu(y) \rangle = \langle \bar \psi \gamma_5 \psi(x)|2m P(y) +2J_{5q}(y) \rangle .
\end{equation}
This allows us to retain the usual definition of the residual mass in the case of M\"{o}bius domain wall fermions.
We emphasize that the definition,
$$m_{res} =\left. \frac{\langle \pi(\vec{p}=0) | J_{5q}\rangle}
{ \langle \pi(\vec{p}=0) | P \rangle}\right|_{m=-m_{res}},$$
via the zero-momentum pion matrix element of $J_{5q}$ is particularly important, because
then our PCAC relation, $$\langle \pi(\vec{p}=0) |2m P +2J_{5q} \rangle = 0,$$  
guarantees that the low momentum lattice pions are massless. This is the appropriate measure of chiral
symmetry breaking for the analysis of the chiral expansion.

Section~\ref{sec-decayconstmeas} discusses methods of using the vector and axial ward identities to 
measure the renormalization of the local vector and axial currents, and their use in our analysis.

\section{Deriving dimensionless global fit forms}
\label{appendix-dimlessforms}

In this section we briefly describe how to obtain the appropriate dimensionless global fit function describing the lattice data for a quantity $Q$ of mass dimension $D$ on a general ensemble $e$. The procedure is as follows:
\begin{enumerate}
 \item Write down the fit formula for $Q$ in {\it physical units} on the reference ensemble, including an $a^2$ term. For example, a linear ansatz might have the following form: $$Q = c_{Q,0}(1 + c_{Q,a}a_r^2) + c_{Q,m_l} \tilde m^r_l + c_{Q,m_h} \tilde m^r_h\,,$$ where we have assumed that there are no partially-quenched data points for simplicity. Here the superscript $r$ on the quark masses indicates that they are in the normalization of the reference ensemble.
 \item To derive the fit form for $Q$ on ensemble $e$, first replace $a_r$ with the lattice spacing, $a_e$, appropriate for that ensemble, then rewrite $a_e$ as $a_e = a_r/R_a^e$:
$$Q = c_{Q,0}(1 + c_{Q,a}a_r^2(R_a^e)^{-2} ) + c_{Q,m_l} \tilde m^r_l + c_{Q,m_h} \tilde m^r_h\,.$$
 \item Multiply by $a_r^D$ and redefine the fit parameters in terms of dimensionless quantities (denoted with a prime superscript): $$a_r^D Q = c'_{Q,0}(1 + c'_{Q,a}(R_a^e)^{-2}) + c'_{Q,m_l} (a_r\tilde m^r_l) + c'_{Q,m_h} (a_r\tilde m^r_h)\,.$$
 \item Using $a_r = R_a^e a_e$, rewrite the function in terms of the lattice spacing on the ensemble $e$: $$(R_a^e)^D (a_e^D Q) = c'_{Q,0}(1 + c'_{Q,a}(R_a^e)^{-2}) + c'_{Q,m_l} R_a^e(a_e\tilde m^r_l) + c'_{Q,m_h} R_a^e(a_e\tilde m^r_h)\,.$$
 \item Finally, use $\tilde m^r = Z_l^e \tilde m^e$ to move the quark masses into the native normalization of ensemble $e$, and divide by $(R_a^e)^D$: $$ (a_e^D Q) = (R_a^e)^{-D} c'_{Q,0}(1 + c'_{Q,a}(R_a^e)^{-2}) + c'_{Q,m_l} (R_a^e)^{1-D} Z_l^e(a_e\tilde m^e_l) + c'_{Q,m_h} (R_a^e)^{1-D} Z_h^e(a_e\tilde m^e_h)\,.$$
\end{enumerate}
This fit function now describes the data in lattice units for the ensemble $e$.
\section{Dependence of the lattice spacing on the fermion action}
\label{appendix-mresadep}

In Sec.~\ref{sec:CombinedChiralFits} we described that, contrary to our
expectations, combining the 24I and 48I ensembles into a single global 
fit required that two lattice spacings, differing by 3.2(2)\%, be used for 
these two, nominally similar ensembles.  (Similar but smaller discrepancies
between the lattice spacings for the 32I and 64I ensembles were also
found.)  In this appendix we will discuss this phenomenon in greater
detail and describe additional measurements that we performed in order to verify that
this assignment of different lattice spacings is correct.  For clarity we will 
focus on the 24I and 48I ensembles, since the explanation for both cases 
is the same. For the 24I ensemble set we consider only the ensemble with the lighter input quark 
mass of $m_f=0.005$

The 24I and 48I ensembles are very similar.  Each uses the same Iwasaki 
gauge action with the same value of $\beta=2.13$.  They differ in the fermion 
formulation used (Shamir and \mobius respectively), the total light quark 
mass ($m_f+\mres= (5.0 +3.154(15))\times 10^{-3} = 8.154(15) \times10^{-3}$ 
and $m_f+\mres= (7.8 +6.102(40))\times 10^{-4} = 13.999(40) \times10^{-4}$, 
respectively) and the degree of residual chiral symmetry breaking, which is
suggested by the differences in the values of the residual quark masses 
just quoted.  For a comparison of the $m_f=0.004$ 32I and 64I ensembles, 
the corresponding numbers are $m_f+\mres=(4.0+0.6664(76))\times10^{-3}  
=  4.6664(76)\times 10^{-3}$ and $m_f+\mres=(6.78 + 3.116(23))\times 10^{-4}
= 9.896(23)\times 10^{-4}$ respectively.  

If we were to describe the low energy Green's functions computed on the 
24I and 48I ensembles as corresponding to separate Symanzik effective 
theories, these two effective theories would be essentially identical, except 
for differences in their low energy constants of order $(ma)^n$.  For example, 
in a theory with chiral fermions the dimension-4 $(F^{\mu\nu})^2$ term, 
closely related to the lattice scale, would have coefficients which differed by 
terms of order $(ma)^2$, terms much too small to be relevant here.  Of 
course, had such a term been important, our global fitting procedure would 
have included its effects by describing both the 24I and 48I ensembles with 
a single Symanzik effective theory, with a single lattice spacing, whose 
mass-dependent coefficients were represented by explicit mass-dependent 
terms in the fit.  In this framework both the 24I and 48I ensembles would be 
described by the same lattice spacing $a$ and the same value of $R_a$.

It may be useful to briefly review the meaning of the lattice spacing $a$ as it is generally defined 
in field theory and specifically defined in the calculation presented here.  Perhaps the simplest 
way to define the cut-off scale is by specifying the value of a ``physical'' quantity, such as the Wilson
flow or three-gluon coupling, at a sufficiently short flow time or large gluon momentum that the process 
can be understood in perturbation theory.  Theories with identical lattice actions but with different quark 
masses will give the same value for the lattice scale up to terms of order $(ma)^2$ if we introduce the lattice
scale $a$ as the natural lower/upper limit on the flow times or momentum scales that are available for such a
short-distance definition.  From this perspective, such mass dependent effects are much too small to result in
the 3\% discrepancy we find.  In our actual approach, we define the lattice spacing through the mass of the
$\Omega^-$.  This requires our global fitting procedure and an explicit extrapolation to a specific value of
input quark masses, specifically those which give physical values for $m_\pi/m_\Omega$ and $m_K/m_\Omega$, in 
order that such a low-energy definition of the lattice scale be well defined.  Necessarily, in this approach
the 24I and 48I ensembles are assigned a common lattice spacing and their different input quark masses are 
completely accounted for in the global fitting procedure (up to negligible systematic effects). For our low-energy definition of the lattice spacing, it 
is not possible to interpret the 3\% difference in $a$ between the 24I and 48I ensembles as resulting from their 
different input masses.


Instead, the change in the lattice spacing between the 24I and 48I ensembles must be attributed to some other change in the lattice action. We are left to conclude that this effect must be a result of the change in fermion formulation. As discussed in Section~\ref{sec:SimulationDetails}, we can consider this change as being accomplished in two steps: we first change $L_s$ from 16 to 48 using the Shamir formulation, and then change from the Shamir ($L_s=48$, $b+c=1$ to the \mobius ($L_s=24$, $b+c=2$) formulation at fixed $L_s(b+c)$.  Since all 4-dimensional Green's functions related by this final change are expected to agree at the $0.1\%$ level, the Shamir to \mobius change is inconsistent with a 3\% change in the lattice spacing, which would naturally result in a 3\% change in such Green's functions.  (For example, a change in the Omega mass of 3\% would result in at least a 3\% change in the Omega propagator.)

Thus, we expect that this 3\% change in lattice spacing would have been observed even if we had continued to use the Shamir action and simply increased $L_s$ from 16 to 48.  While this is a surprisingly large effect for such a change in $L_s$, we believe that it is a plausible explanation.  The effect of the smaller $L_s=16$ value is usually characterized by the value of $\mres a = 3.154(15)\times 10^{-3}$, which is substantially less than 3\%.  However, considerable effort has been devoted to reducing the size of $\mres$, including a careful choice for the domain wall parameter $M_5$ and the choice of the Iwasaki gauge action.  It is possible that, while these choices have significantly reduced $\mres$, they have not correspondingly reduced the size of other $L_s$-dependent effects.   

For example, the value of the lattice spacing, which is determined by the strength of QCD interactions at the scale of $\Lambda_{\mathrm{QCD}}$, is a strong function of the anti-screening produced by QCD vacuum polarization.  The quarks act to reduce this anti-screening, and the Pauli-Villars determinant was originally included in the domain wall fermion action~\cite{Shamir:1993zy} to regulate what would have been a divergent contribution to QCD vacuum polarization coming from the increasing number of fermion species as $L_s\to\infty$.  While, as can be seen by the relation with overlap fermions discussed in Sec.~\ref{sec:SimulationDetails}, these effects have a well defined $L_s\to\infty$ limit, we cannot rule out the possibility that they appear at the 3\% level for $\beta=2.13$ and $L_s=16$.  Instead, we interpret this large shift in $a$ as providing new information about the potential effects of finite $L_s$, and a warning that simple estimates can occasionally be misleadingly low.   In this spirit, we should recognize that the earlier arguments about the insensitivity of the coefficients of the $O(a^2)$ Symanzik correction terms to our change in action may underestimate these effects.  Of course, in this case, if our few tenths of a percent estimate were to become even a 5\% effect, it would not interfere with our current continuum extrapolations.

Since the conclusion, implied by our global fits, that the lattice spacing did indeed change by 3.2\% and 1\% when going from the 24I to 48I and 32I to 64I ensembles respectively, was a surprise, it was important to test this hypothesis.  For that purpose, we generated two additional MDWF+I ensembles with input parameters set equal to those of the lightest 24I and 32I ensembles (i.e. those with $am_l=0.005$ and $am_l=0.004$, respectively), but using the \mobius parameters and $L_s$ values that were used for the 48I and 64I ensembles respectively. We compensated for the reduction in the residual mass by increasing the input bare quark mass in order that the total quark masses remain equal to the 24I and 32I values. If the observed differences in the lattice spacings can indeed be attributed to the change in $L_s$ (that which would have been required if the new ensembles were generated with the Shamir action), then the lattice scales for these new ensembles should match those determined for the 48I and 64I ensembles.

We refer to these new ensembles as the `24Itest' and `32Itest' ensembles. They were generated with \mobius domain wall fermions and the Iwasaki gauge action at $\beta=2.13$ and 2.25 respectively, and with lattice sizes of $24^3\times 64\times 24$ and $32^3\times 64\times 12$. Both ensembles use \mobius parameters of $\alpha=b+c=2$ and $b-c=1$, making them equivalent to Shamir domain wall ensembles with $L_s=48$ and 24 respectively. On the 24Itest ensemble, we measured the residual mass and Wilson flow scales on configurations in the range 120 to 550; the residual mass was measured every 40 configurations, and the Wilson flow scales every 10, and we binned the latter over four successive measurements. Similarly, for the 32Itest ensemble, we performed measurements in the configuration range 200 to 610, measuring the residual mass every 20 and the Wilson flow scales every 10, binning the latter over two successive measurements.

The values of the average plaquette, residual mass, total quark mass and Wilson flow scales are listed in Table~\ref{appendix-ashifttestmeas}. From the table we can immediately observe that, while the total quark masses of the 24Itest and 24I ensembles are closely matched, there are clear differences in the average plaquette and Wilson flow scales; smaller differences are also observable between the 32Itest and 32I measurements. The differences in the Wilson flow scales are $\sim 3\%$ between the 24I and 24Itest ensembles and $~\sim 1\%$ between the 32I and 32Itest, which are very similar to the differences in lattice scales observed between the 24I/48I and 32I/64I ensembles respectively.

We cannot directly compare the computed values on the test ensembles in Table~\ref{appendix-ashifttestmeas} with the corresponding 48I and 64I values, due to the measurements being performed with different quark masses. For a definitive test, we instead include the test ensembles in the global fits. For each ensemble there are associated three free parameters: the scaling parameters $Z_l$, $Z_h$ and $R_a$. The observed differences in the fermion action appear to result in negligible changes to $Z_l$ and $Z_h$, hence we are able to fix those values to those of the 24I/48I (for the 24Itest) and 32I/64I (for the 32Itest ensemble); this leaves only $R_a$ as a free parameter for each ensemble. In Table~\ref{appendix-ravals} we list the values of $R_a$ that we obtain, alongside the corresponding values for the 24I, 32I, 48I and 64I ensembles. We observe excellent agreement between $R_a$ on the 24Itest ensemble and that on the 48I, and similarly between the 32Itest and 64I. This offers clear evidence that the change in $L_s$ is responsible for the observed differences in lattice spacing.  It provides further confidence in our global fitting procedure, which was sufficiently reliable to produce strong evidence for this effect even though it was not expected in advance.

Note, in this explanation we continue to assume the near equality of the \mobius and Shamir 4-D theories for fixed $L_s(b+c)$, and to view the difference in $a$ as what would have been observed had we used only the Shamir action, increasing $L_s$ from 16 to 48 (for 24I/48I) and 16 to 24 (for 32I/64I).  While we believe that this assumption has a strong theoretical justification, the numerical experiment just described does not provide direct evidence for its validity.

\begin{table}[tp]
\begin{tabular}{l|ll|ll}
\hline\hline
Quantity &  24I (0.005) & 24Itest  & 32I (0.004) & 32Itest \\
\hline
$\langle P \rangle$ & 0.588053(4) & 0.587035(6)              & 0.615587(3) & 0.615318(8) \\
\hline
$m_l$               & 0.005        & 0.00746                 & 0.004         & 0.00437    \\
$m_h$               & 0.04         & 0.04246                 & 0.03          & 0.03037     \\
$m_{\rm res}$       & 0.003154(15) & 0.000666(25)            & 0.0006697(34) & 0.000306(9)\\
\hline
$m_l+m_{\rm res}$   & 0.008154(15) & 0.008126(25)            & 0.0046697(34) & 0.004676(9) \\
$m_h+m_{\rm res}$   & 0.043154(15) & 0.043126(25)            & 0.0306697(34) & 0.030676(9) \\
\hline
$t_0^{1/2}$         & 1.3163(6)  & 1.2766(19)          & 1.7422(11) & 1.7226(24)\\
$w_0$               & 1.4911(15) & 1.4485(46)          & 2.0124(26) & 1.9937(57)
\end{tabular}

\caption{Comparison of various quantities in lattice units between the test ensembles and the original ensembles. For the 24I and 32I ensembles we quote values for the residual mass computed at unitary light quark masses (not extrapolated to the chiral limit). These and the average plaquette values were determined in Ref.~\cite{Aoki:2010dy}. The Wilson flow scales on these ensembles are discussed in Appendix~\ref{appendix-new32I24I32IDmeas}. For comparison, the residual masses for the 48I and 64I ensembles are 0.000610(4) and 0.000312(2) respectively.\label{appendix-ashifttestmeas} }

\end{table}

\begin{table}[tp]
\begin{tabular}{c|c|c|c|c|c}
\hline\hline
\multicolumn{3}{c|}{$\beta = 2.13$} & \multicolumn{3}{c}{$\beta = 2.25$} \\
\hline
24I & 48I & 24Itest                 & 32I & 64I & 32Itest \\
\hline
0.7491(23) & 0.7259(27) & 0.7243(28) & 1.0(0) & 0.9897(19) & 0.9877(19)
\end{tabular}
\caption{The values of the lattice spacing ratio $R_a^i = a^{\rm 32I}/a^i$ for ensembles $i$ with $\beta=2.13$ and $\beta=2.25$, including the two test ensembles. \label{appendix-ravals} }
\end{table}

%
%

\section{Weighted fits}
\label{appendix-weightedfits}

We define a weighted $\chi^2$ as
\begin{equation}
\chi^2 = \sum_i \frac{\omega_i [y_i - f({\bf x}_i, {\bf c})]^2 }{\sigma^2_i}\,,
\end{equation}
where $i$ indexes the measurements, $y_i$ and $\sigma_i$ are the measured value and statistical error, ${\bf x}_i$ the associated coordinates, and ${\bf c}$ the set of parameters of the fit function $f$. The quantities $\omega_i$ are set to a value $\Omega$ for some subset of the data, where $\Omega$ is assumed to be large, and to unity for all other data. We demonstrate below that the dependence on $\Omega$ vanishes in the limit $\Omega\rightarrow\infty$ and that this limit is sensible. 

The minimum of $\chi^2$ satisfies
\begin{equation}\begin{array}{rl}
\displaystyle \frac{\partial \chi^2 }{\partial c_\kappa } 
&\displaystyle = \sum_i \frac{\omega_i}{\sigma^2_i}  \frac{\partial\Delta^2_i(\vec x_i, \vec c) }{ \partial c_\kappa }\\
& = 0\,.
\end{array}\end{equation}
Writing out the derivative explicitly and dividing both sides by $\Omega$ gives the following expression:
\begin{equation}
\sum_i \frac{\omega_i}{\Omega}\frac{\Delta_i(\vec x_i, \vec c)}{\sigma^2_i}\cdot  \frac{\partial f(\vec x_i, \vec c) }{ \partial c_\kappa} = 0\,.\label{eqn-chisqmincond}
\end{equation}
If we na\"{i}vely take the $\Omega\rightarrow\infty$ limit of this equation, it appears that all of the data with $\omega_i=1$ drop out entirely and hence do not contribute to the fit. This is certainly true in those cases in which the number of data points with weight $\omega_i=\Omega$ is sufficient to determine the full set of parameters $\vec c$. However when there are fewer points, there is no solution that satisfies Eq.~\eqref{eqn-chisqmincond} in the $\Omega\rightarrow \infty$ limit. We argue that if one first determines the solution for {\it finite} $\Omega$, either analytically or numerically, then afterwards take the limit $\Omega\rightarrow \infty$, the solution remains valid and in fact depends on the data with $\omega_i=1$. The resolution of this apparent paradox is that when the overweighted points are insufficient to determine the parameters, the fit has (almost-)unconstrained directions with infinitesimally small curvature arising from the vanishing unweighted data, and hence there is a well defined minimum.

\subsubsection{Simple example}

It is straightforward to demonstrate the behavior discussed above via a simple example in which we are attempting to determine the parameters of the function
\begin{equation}
f(x) = a + bx 
\end{equation}
by minimizing 
\begin{equation}
\chi^2 = \sum_{i=0}^{N-1}\left( r_i - f(x_i) \right)^2 w_i = 0\,,\label{eqn-chisqex1}
\end{equation}
where $r_i$ are a series of $N$ data points with coordinates $x_i$ and unit variances for simplicity.

Let us first consider a scenario in which we have three data points ($N=3$), two of which are overweighted: $w_0=w_1=\Omega$, and the third is assigned $w_2=1$. Here the overweighted data points are sufficient to determine both parameters and the result of solving for the minimum of Eq.~\eqref{eqn-chisqex1} in the limit of large $\Omega$, and the result of solving at finite $\Omega$ and taking the limit afterwards, are identical:
\begin{equation}\begin{array}{ccc}
a = (r_1 x_0 - r_0 x_1)/(x_0 - x_1)  & {\rm and} &  b=(r_0 - r_1)/(x_0 - x_1)\,.\label{eqn-chsqexsol}
\end{array}\end{equation}
Notice that this result does not contain the unit-weight data point, $r_2$.

Let us now consider just two data points ($N=2$), and take $w_0 = \Omega$ and $w_1=1$ such that the number of overweighted data points is no longer sufficient to determine both parameters. The equations for the minimum of $\chi^2$ are:
\begin{equation}\begin{array}{lc}
\displaystyle \frac{\partial \chi^2 }{\partial a } = -2 \Omega\left( r_0 - f(x_0) \right) -2 \left(r_1 - f(x_1)\right) = 0 & {\rm and} \\  \displaystyle \frac{\partial \chi^2 }{\partial b } = -2 \Omega\left( r_0 - f(x_0) \right)x_0 -2 \left(r_1 - f(x_1)\right)x_1 = 0\,.
\end{array}\end{equation}
Taking the large $\Omega$ limit gives
\begin{equation}\begin{array}{ccc}
-2 \Omega\left( r_0 - f(x_0) \right) = 0 & {\rm and} &  -2 \Omega\left( r_0 - f(x_0) \right)x_0 = 0\,.
\end{array}\end{equation}
These are identical up to a trivial normalization, hence we have two unknowns and only one equation; no unique solution can be found. (Note that the fact that the equations are the same will not be true in a general case with multiple over-constrained data points; there one would instead find expressions that cannot be simultaneously satisfied.) On the other hand we \textit{can} solve for the minimum at finite $\Omega$; the solutions are identical to those given in Equation~\eqref{eqn-chsqexsol}, and are \textit{independent of $\Omega$}, allowing us to take the large $\Omega$ limit \textit{a posteriori} without issue.

\begin{figure}[tp]
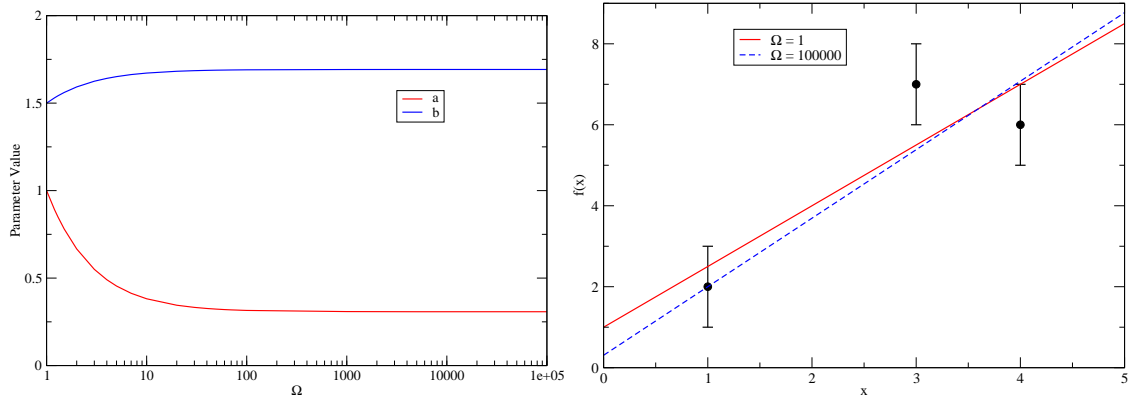

\centering
\includegraphics*[width=0.45\textwidth]{fig/wlim.eps}
\includegraphics*[width=0.45\textwidth]{fig/wcurves.eps}
\caption{(Left) The fit parameters of the function $f(x) = a+bx$ determined from arbitrarily chosen data points, $r_0(x_0=1)=2$,  $r_1(x_1=3)=7$ and $r_2(x_2=4)=6$, plotted against the weighting $\Omega$ of the first point. (Right) The fit curves with $\Omega=1$ (red full line) and $\Omega=100000$ (dashed blue) overlaying the data.\label{fig-fitex1}}
\end{figure}

Finally we consider one further example, again with three data points but this time with only one over-weighted: $N=3$,  $w_0=\Omega$ and $w_1=w_2=1$. Here, as above, the number of overweighted points is insufficient to determine both parameters, but all three points together are more than enough to constrain the parameters (with one degree of freedom). We might therefore expect that the solutions at finite $\Omega$ would be $\Omega$-dependent unlike in the previous example. Indeed this is the case, but it is straightforward to show that the solutions are finite in the limit $\Omega\rightarrow\infty$ and furthermore that they are functions of \textit{all three data points} in this limit. The expressions are somewhat lengthy and we have not reproduced them here, but we have plotted the $\Omega$ dependence of the solutions for a particular set of data points and parameters in Figure~\ref{fig-fitex1}. In the figure we also plot the function before and after the weighting, demonstrating that it does indeed pass through the over-weighted data point.

\subsubsection{Determination of the optimal $\Omega$ value in the global fits}

\begin{figure}[tp]
\centering
\includegraphics*[width=0.45\textwidth]{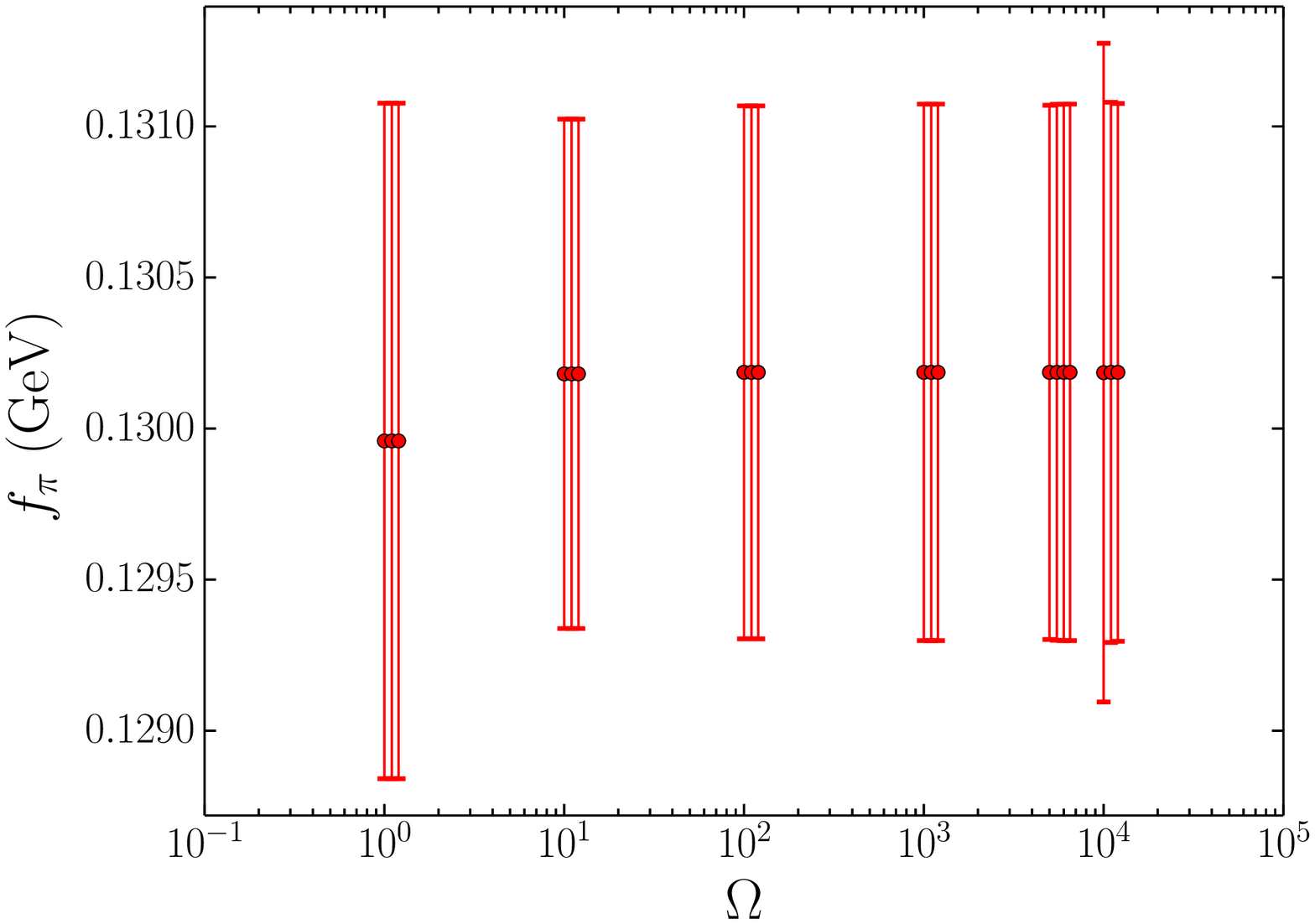}
\includegraphics*[width=0.45\textwidth]{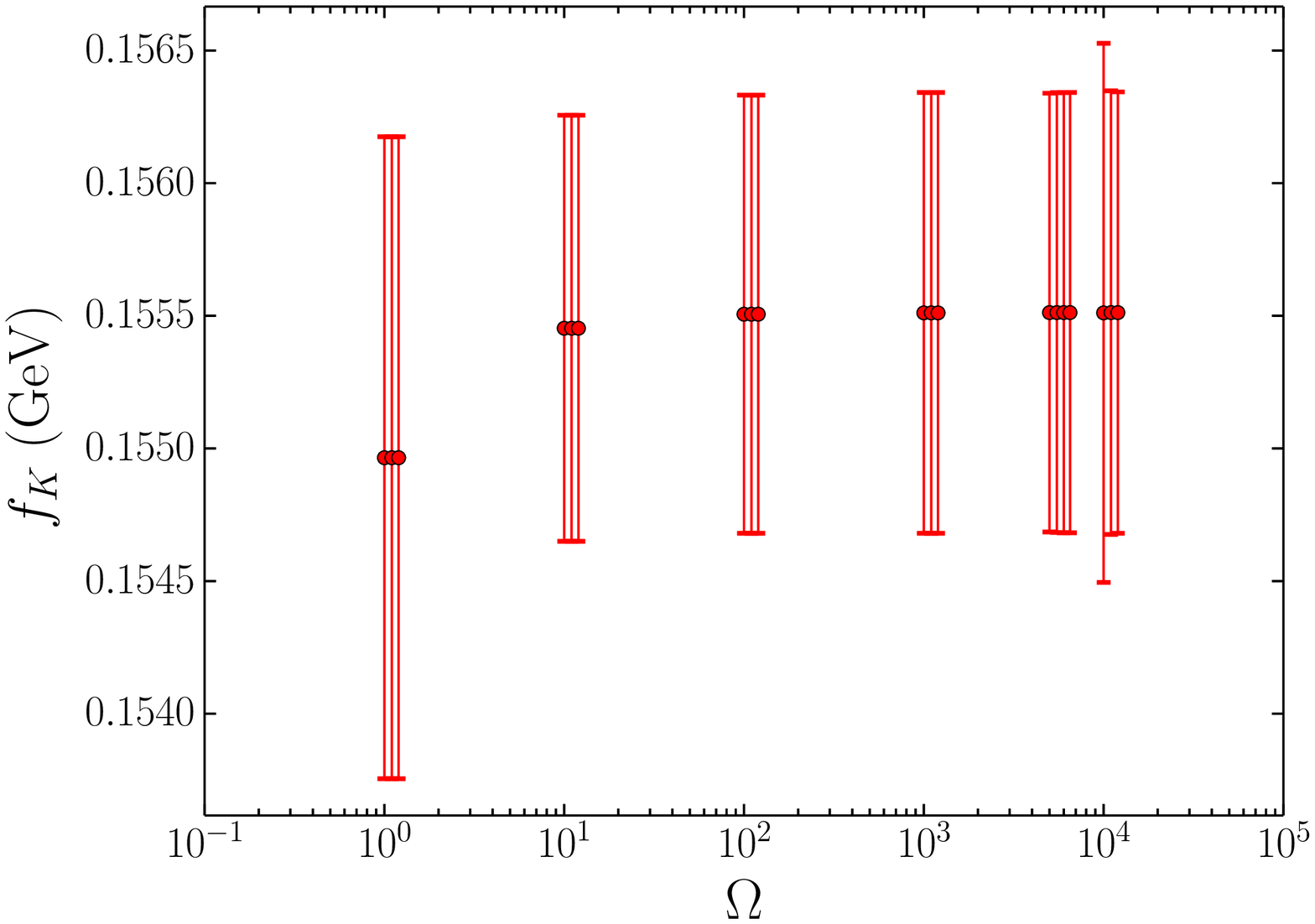}\\
\includegraphics*[width=0.45\textwidth]{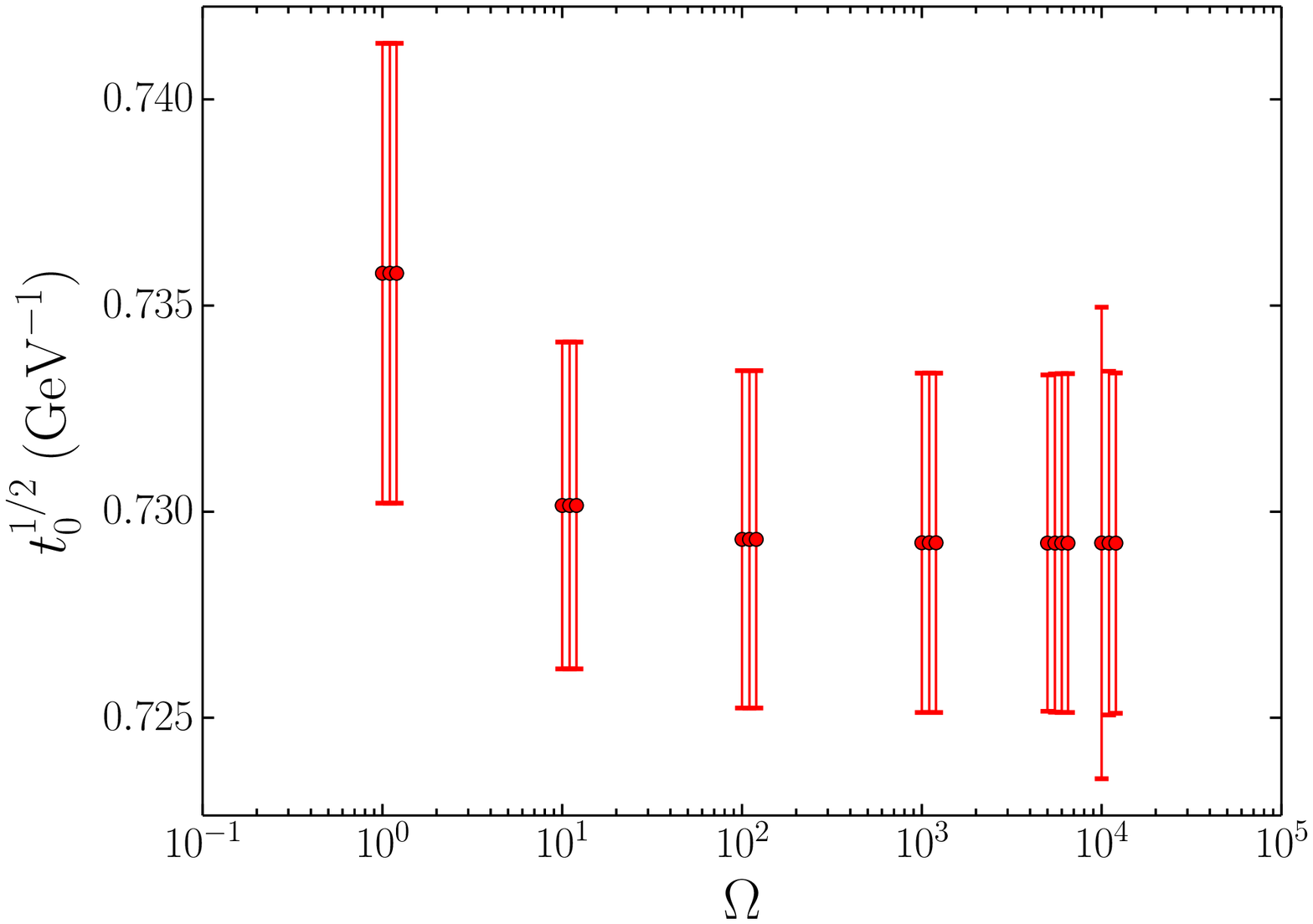}
\includegraphics*[width=0.45\textwidth]{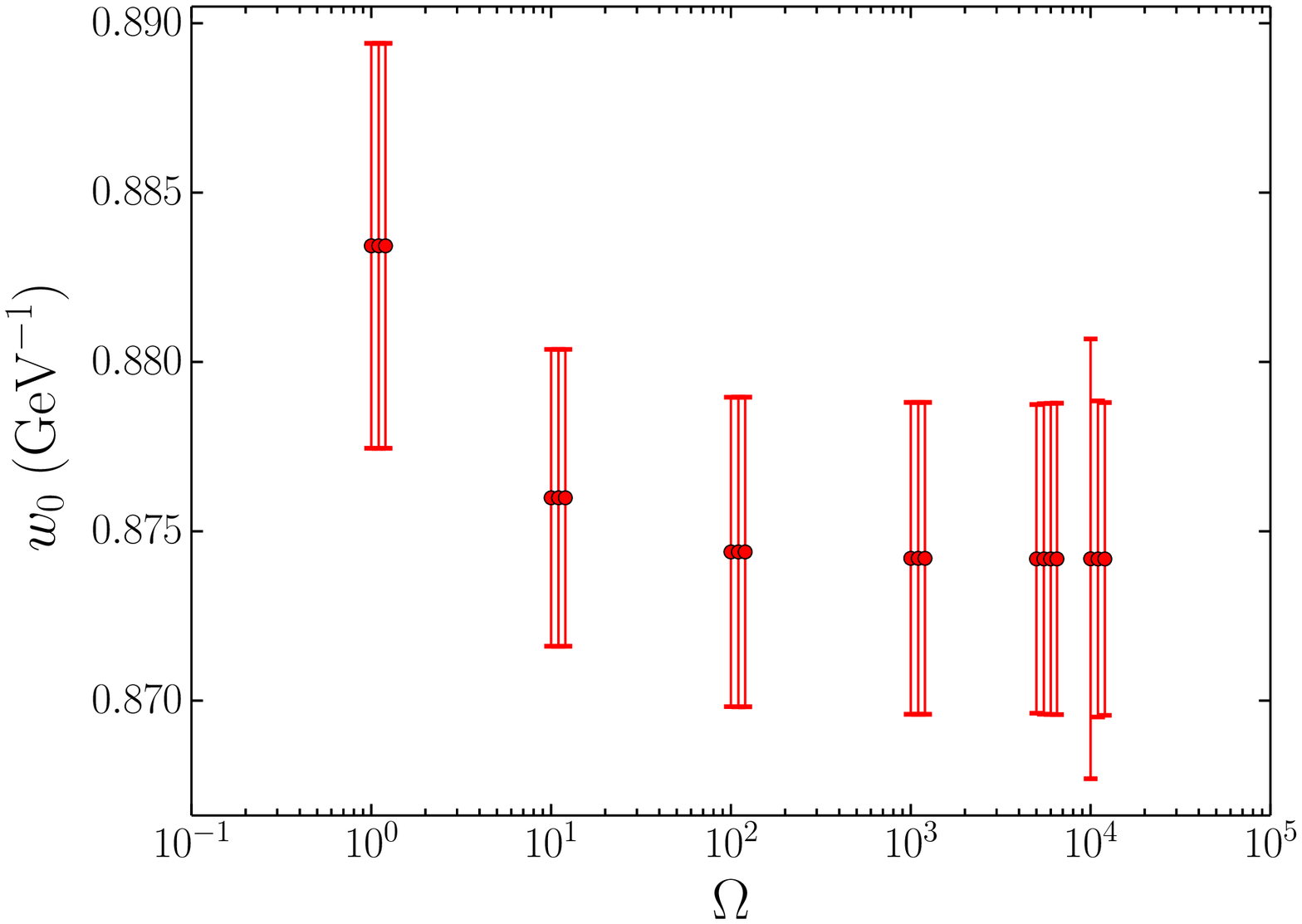}\\
\caption{Plots of the predicted continuum value for $f_\pi$ (upper-left), $f_K$ (upper-right), $t_0^{1/2}$ (lower-left) and $w_0$ (lower-right) as a function of the weight $\Omega$ applied to the physical point ensembles in the fit. Fits were performed with $\Omega =1,10,100,1000,5000,10000$ and 50000. We also considered three different values of the stopping condition $\delta \chi^2_{\rm min}$: $1\times 10^{-3}$, $1\times 10^{-4}$ and $1\times 10^{-5}$. For the point at $\Omega = 5000$ we also considered a fourth value, $\delta \chi^2_{\rm min}=1\times 10^{-6}$, and we only consider two values for $\Omega = 50000$ where the errors are clearly less well controlled.  For each choice of $\Omega$, the results for each value of $\delta \chi^2_{\rm min}$ have been offset for clarity, with the largest value the left-most point of each cluster, with the largest error. \label{omega_dep_fpi_fk}  }
\end{figure}

It remains to demonstrate the limiting behaviour in the more complex environment of the global fits. As the minimization is performed numerically via the Marquardt-Levenberg algorithm we must be careful in our choice of algorithmic parameters; the algorithm terminates when the change in $\chi^2$ under a shift of the fit parameters is less than some chosen value, $\delta \chi^2_{\rm min}$. As we increase $\Omega$ at fixed $\delta \chi^2_{\rm min}$, the relative effects of fluctuations in the unweighted data are reduced and the fit becomes more tolerant to increasingly large deviations of the fit from the unit-weight data. This manifests as an increase in the jackknife statistical error of our predictions. We must therefore choose a value of $\delta \chi^2_{\rm min}$ that is small enough to properly take into account the constraints from the unit-weight data. The choice is limited by the increased time for the fit to reach its minimum coupled with the inevitable limits of finite precision. For fixed $\delta \chi^2_{\rm min}$, the time to perform the fit also naturally increases with $\Omega$ due to the increase in the overall scale of the fluctuations. We must therefore determine an optimal value for $\Omega$ that is large enough that our predictions are no longer noticeably dependent on its value while small enough for the fits to complete in a reasonable time and to be unaffected by finite precision errors.

In Figure~\ref{omega_dep_fpi_fk} we show examples of the $\Omega$ dependence on the predicted values of $f_\pi$, $f_K$, $w_0$ and $t_0^{1/2}$. The plots also show the result of reducing the stopping condition $\delta \chi^2_{\rm min}$ by several orders of magnitude. We observe percent-scale shifts in the central values of these quantities from the unweighted fit results, and we clearly see the behavior flattens out at around $\Omega = 1000$. We choose $\Omega = 5000$ as a value large enough to be well within the flat region while small enough to avoid the difficulties discussed above. For the chosen value of $\Omega$ we observed no significant dependence of the results on $\delta \chi^2_{\rm min}$, but to be conservative we chose $1\times 10^{-4}$ as our final value. Note that we observed stronger dependence of our results on $\delta \chi^2_{\rm min}$ for some alternate choices of guess parameters, but with tighter stopping conditions the results stabilized and agreed with those presented in this document. To be certain, all fits presented within the body of this work were repeated with tighter stopping conditions, and no significant changes from the given values were observed.
\section{Additional measurements on the 32I, 24I and 32ID ensembles}
\label{appendix-new32I24I32IDmeas}

In this work we include additional data for the 24I and 32I ensembles, specifically measurements of the Wilson flow scales, $t_0^{1/2}$ and $w_0$, and also an improved measurement of the vector current renormalization coefficient that we use to normalize our decay constants. To remind the reader, these ensembles have lattice volumes of $24^3\times 64\times 16$ and $32^3\times 64\times 16$, and use the Shamir domain wall fermion action with the Iwasaki gauge action at bare couplings $\beta=2.13$ and $2.25$ respectively, and were originally described in Refs.~\cite{Allton:2008pn} and~\cite{Aoki:2010dy}. We also perform measurements of the Wilson flow scales on the 32ID ensemble, which has a lattice volume of $32^3\times 64\times 32$, Shamir domain wall fermions with the Iwasaki+DSDR gauge action at $\beta = 1.75$, and was described in Ref.~\cite{Arthur:2012opa}. 

\subsection{Wilson flow scales}

The procedure for determining the Wilson flow scales is described in Section~\ref{sec-wflowmeas}. We have three 32I ensembles with bare light quark masses of $am_l = 0.004$, 0.006 and 0.008, upon which we perform measurements using 300, 312 and 252 configurations respectively (separated by 10 MD time units) following our earlier analyses. The measurements are binned over four successive configurations to take account of autocorrelations. For the 24I ensemble set, we have two ensembles with $am_l = 0.005$ and 0.01, and we measure on 202 and 178 configurations respectively (separated by 40 MD time units) and use a bin size of 2. Finally, for the 32ID ensemble set we have two ensembles with $am_l=0.001$ and $0.0042$, and we measure on 180 and 148 configurations respectively (with 8 MD time units separation) and bin over 4 configurations. Note that the results for the 32I $m_l=0.008$ ensemble and the 24I $m_l=0.01$ ensemble are not included in the global fits due to the pion mass cut, but we include the results here for completeness.

On all three ensembles we use reweighting in the sea strange quark mass to constrain the mass dependence. The number of reweighting steps and the mass ranges used are given in the aforementioned papers. For the results presented in this section, we list only the simulated value and the closest reweighted value to the physical strange quark mass. The simulated strange quark masses are 0.03, 0.04 and 0.045 for the 32I, 24I and 32ID ensembles respectively, and the physical strange masses are as follows: $(am_s)^{\rm 32I} = 0.0248(2)$, $(am_s)^{\rm 24I} = 0.0322(2)$ and $(am_s)^{\rm 32ID} = 0.0462(5)$. 

The values we obtain are given in Table~\ref{tab-wflowoldens}.

\begin{table}[t]
\centering
\begin{tabular}{ll|ll}
\hline\hline
Ens. set & $(am_l, am_s)$ & $t_0^{1/2}/a$ & $w_0/a$ \\
\hline
32I      & (0.004, 0.03)  & 1.7422(11) & 2.0124(26)\\
         & (0.004, 0.025) & 1.7510(14) & 2.0310(34)\\
32I      & (0.006, 0.03)  & 1.7362(9) & 1.9963(19)\\
         & (0.006, 0.025) & 1.7439(15) & 2.0136(34)\\
32I      & (0.008, 0.03)  & 1.7286(11) & 1.9793(24) \\
         & (0.008, 0.025) & 1.7359(12) & 1.9913(24) \\
\hline
24I      & (0.005, 0.04)     & 1.3163(6)   & 1.4911(15) \\
         & (0.005, 0.03225)  & 1.3237(12)  & 1.5071(22) \\
24I      & (0.01, 0.04)      & 1.3050(7)   & 1.4653(14) \\
         & (0.01, 0.03225)   & 1.3126(12)  & 1.4808(30) \\
\hline
32ID     & (0.001, 0.045)     & 1.0268(3) & 1.2178(7) \\
         & (0.001, 0.04625)   & 1.0262(3) & 1.2088(10) \\
32ID     & (0.0042, 0.045)     & 1.0225(3) & 1.2042(7) \\
         & (0.0042, 0.04625)   & 1.0220(3) & 1.2031(8)     
\end{tabular}
\caption{The Wilson flow scales in lattice units on the 32I, 24I and 32ID ensembles at the simulated strange quark mass and the reweighted mass closest to the physical value. The quark masses are given in bare lattice units.\label{tab-wflowoldens} }
\end{table}

\subsection{Vector current renormalization}

In Section~\ref{sec-decayconstmeas} we describe how the renormalization coefficient relating the domain wall local axial current to the physically-normalized Symanzik current can be determined via the quantity $Z_V/Z_{\cal V}$, which relates the local vector current $V_\mu$ to the conserved 5D current ${\cal V}_\mu$. This quantity is used to renormalize the decay constants. In our earlier works~\cite{Aoki:2010dy,Arthur:2012opa} we obtained $Z_V$ by fitting directly to the ratio of two-point functions,
\begin{equation}
\frac{Z_V}{Z_{\cal V}} = \frac{ \sum_{i=1}^3 \sum_{\vec x} \langle {\cal V}_i^a(\vec x,t) V_i(\vec 0,0)\rangle }{ \sum_{i=1}^3 \sum_{\vec x} \langle V_i^a(\vec x,t) V_i(\vec 0,0)\rangle }\,.
\end{equation}
Since the lightest state that couples to the vector operator is the noisy $\rho$ meson, for this work we instead determine the ratio for the 48I, 64I and 32Ifine ensembles via the three-point function, $\langle\pi | V_\mu | \pi\rangle$, as described in Section~\ref{sec-zvdeterm}; this procedure gives a substantially more precise result than the above. In the global fits we attempt to describe the aforementioned ensembles, along with 32I and 24I ensemble sets, using the same continuum scaling trajectory. In order to guarantee consistent scaling behavior we must therefore recompute $Z_V$ on the 32I and 24I ensemble sets using the new method. This is not necessary for the 32ID ensembles, which are described by a different scaling trajectory.

On the 24I ensemble set we measured on 147 and 153 configurations of the $am_l = 0.005$ and 0.01 ensembles respectively. We also included 85 measurements on the heavier $am_l = 0.02$ ensemble and 105 measurements on the $am_l = 0.03$ ensemble described in Ref.~\cite{Allton:2008pn}. For the 32I ensembles we measure on 135, 152 and 120 configurations of the $am_l = 0.004$, 0.006 and 0.008 ensembles respectively. In Table~\ref{tab-32I24IZV} we list the measured values on each ensemble and extrapolated to the chiral limit.

\begin{table}[tp]
\begin{tabular}{cl|l}
\hline\hline
Ens. set. & $(am_l)$ & $Z_V$\\
\hline
24I & 0.03  & 0.71611(8) \\
    & 0.02  & 0.71498(13) \\
    & 0.01  & 0.71409(20) \\
    & 0.005 & 0.71408(58) \\
    & $-am_{\rm res}$ & 0.71273(26) \\
\hline
32I & 0.008 & 0.74435(42) \\
    & 0.006 & 0.74387(55) \\
    & 0.004 & 0.74470(99) \\
    & $-am_{\rm res}$ & 0.74404(181)
\end{tabular}
\caption{$Z_V$ measured on the 24I and 32I ensembles, and the extrapolated value in the chiral limit.\label{tab-32I24IZV} }
\end{table}

\section{Non-perturbative renormalization}

\label{appendix-npr}

In order to determine the renormalization coefficients for the quark masses and $B_K$, we use what is now the standard framework for our collaboration: the Rome-Southampton non-perturbative renormalization schemes~\cite{Martinelli:1994ty} with momentum sources, twisted boundary conditions and non-exceptional kinematics \cite{Gockeler:1998ye,Aoki:2007xm,Arthur:2010ht,Sturm:2009kb}. 
This setup has already been described in several previous publications~\cite{Arthur:2010ht,Arthur:2011cn,Boyle:2011cc,Arthur:2012opa}, and results in tiny statistical errors, infra-red contamination suppression, 
and consistent removal of $a^2$ discretization effects in the vertex functions.

A key aspect of the RI-MOM approach is that any other, potentially regularization dependent, scheme may be easily converted into the RI-MOM scheme
using momentum-space scattering amplitudes determined (either perturbatively or non-perturbatively) solely within that other scheme. This makes RI schemes a very useful intermediate scheme for converting
between lattice calculations and $\MSbar$.

The amputated vertex functions $\Pi_O$ of the operators of interest $O$ (in this paper $O$ represent flavour non-singlet bilinear and four-quark fermion operators) are computed on Landau gauge-fixed configurations, for which we use the timeslice by timeslice FASD algorithm~\cite{Hudspith:2014oja}). We use non-exceptional `symmetric' momentum configurations, defined by the condition
\be
p_1^2=p_2^2= q^2\,,
\ee
where, for bilinear vertices, $p_1$ and $p_2$ are the incoming and outgoing quark momenta respectively, and for the four-quark vertices used to compute $Z_{B_K}$ the quark momenta are assigned as follows: $d(p_1)\bar s(-p_2)\rightarrow \bar d(-p_1)s(p_2)$. In the above, $q=p_1-p_2$ is the momentum transfer. 

In contrast to the symmetric scheme, the original RI-MOM scheme defined in Ref.~\cite{Martinelli:1994ty}, which we do not include here, corresponds to the zero-momentum transfer kinematics, i.e. $p_1=p_2$, and suffers from enhanced non-perturbative effects at high energies arising from low-momentum loop effects; in particular the effects of the dynamical chiral symmetry breaking are greatly enhanced~\cite{Aoki:2010dy}.

We compute projected, amputated vertex functions of the form
\be
\Lambda^{bare}_O(\mu,a) = {\p}\{\Pi_O(q^2,a)\}_{\mu^2=q^2}\,.
\ee
Precise definitions of the projectors $\p$ depend on the choice of operator, the kinematics, and the choice of scheme. In practice the Green's functions are first computed 
at finite values of the quark mass and then extrapolated to the chiral limit; this quark mass dependence is however very mild for the non-exceptional schemes considered here and we omit it below for the purpose of clarity.

The renormalization factors are defined by imposing
\be
\label{eq:Zdef}
\frac{Z_O}{Z_q^{n/2}}(\mu,a) \times
\Lambda^{bare}_O(\mu,a) =  \Lambda_O^{tree} \,.
\ee
where $Z_q$ is the quark wave function renormalization factor, and $n$ the number of fermion fields in $O$. A second, separate condition is required in order to extract $Z_q$. Note that the right-hand side of the above depends on the choice of projector.

In order to simplify the equations, we introduce the following notation:
\be
\bar\Lambda_O = \Lambda^{bare}_O \times (\Lambda_O^{tree})^{-1} \,.
\ee
%
projection scheme and for each ensemble, as a function of the external momenta.

In this work we are only interested in quantities that renormalize multiplicatively, such that the $Z$-factors and the $\Lambda$s are simply scalars. For a general lattice action with non-zero chiral symmetry breaking, the four-quark operator responsible for $K-\bar K$ mixing in fact mixes with other operators, and $Z_O$ and $\Lambda_O^{bare}$ become matrix-valued~\cite{Arthur:2011cn}. However for our choice of action, the residual chiral symmetry breaking is negligible and only multiplicative renormalization is required.

Once a bare matrix element $\la O\ra^{bare}(a)$ of the operator $O$ has been computed on a lattice with lattice spacing $a$, the $Z$-factor can be used to convert it into the corresponding MOM-scheme:   
\be
\label{eq:OR}
\la O \ra ^{MOM}(\mu,a)  = \left( \frac{Z_O}{Z_q^{n/2}} (\mu,a) \right)^{MOM} \times \la O \ra ^{bare}(a) \;.
\ee
In order to connect the lattice results to phenomenology, they have 
to be matched to a scheme suitable for a continuum computation, 
such as $\MSbar$; this is performed using perturbation theory. The final equation reads:
\bea
\la O \ra^{\msbar}(\mu,a) &=& 
c^{\msbar \leftarrow MOM}(\mu) \times 
\la O \ra ^{MOM}(\mu,a)  
\eea
This quantity has a well-defined continuum limit as any potential divergences are absorbed by the $Z$-factors. 

We remind the reader that the $Z$-factors defined above are scheme dependent. The renormalization scheme is fixed by the choice of projectors and of kinematics; specifically, with the choice of symmetric kinematics given above, it depends on the projector used for the operator $O$ and that used to extract $Z_q$. For both the quark mass renormalization factor, $Z_m$, and the $B_K$ renormalization factor, $Z_{B_K}$ we use two SMOM schemes; for the former these are the RI-SMOM and the RI-SMOM{$_{\gamma_\mu}$~\cite{Sturm:2009kb} schemes, and for the latter the SMOM($\gamma^\mu,\gamma^\mu)$ and SMOM$(\slashed q,\slashed q)$~\cite{Aoki:2010pe} schemes. In the main analysis we use the difference between the $\MSbar$ results computed using these two intermediate schemes as an estimate of the systematic error associated with the truncation of the perturbative series used to compute the SMOM$\rightarrow \MSbar$ matching factors.

\subsubsection{Renormalization of the quark masses}

Our determination of the quark masses from the global fits uses an intermediate scheme that is hadronically defined and explicitly dependent on our choice
of lattice regulator. The renormalization factors from bare masses to this temporary hadronic scheme are denoted $Z_l$ and $Z_h$ for light and strange quarks respectively.
For quark masses we can convert this temporary scheme to an SMOM scheme by determining the SMOM renormalization $Z_m^{\rm SMOM}$ in the usual way and then determining the continuum limit of the
ratio $\frac{Z_m^{RI}}{Z_{l/h}}$, and from there to $\MSbar$ in the usual way. This is described in more detail in Section~\ref{sec-renormquarkmasses}.

We first introduce the renormalization factor of the flavour non-singlet bilinears. We define $\Lambda_S$ and $\Lambda_P$, the amputated and projected Green's functions of the scalar and pseudoscalar bilinear operators respectively, as
\begin{align}
\Lambda_S=
\mathrm{tr}\left[\Pi_S\cdot I\right]\,, & \hspace{1cm}
\Lambda_P=
\mathrm{tr}\left[\Pi_P\cdot \gamma_5\right]\,.
\label{eqn-LambdaSPproject}
\end{align}
Similarly, for the local vector and axial currents we define:
\be
\Lambda_{V,A} = 
\mathrm{tr}\left[\Pi_{V_\mu,A_\mu}\cdot {\Gamma}^{(s)}_{V_\mu,A_\mu} \right] \;.
\ee
where $(s)$ denotes the choice of projector. Following Ref.~\cite{Sturm:2009kb}, we define the $\gamma_\mu$ and the $\s{q}$-schemes (or projectors) in the following way:
\begin{equation}\begin{array}{lllclll}
\Gamma^{(\gamma_\mu)}_{V_\mu}   &=& \gamma_\mu\,, & {\rm and} & \Gamma^{(\gamma_\mu)}_{A_\mu}   &=& \gamma_\mu \gamma_5\,,
\end{array}\end{equation}
and
\begin{equation}\begin{array}{lllclll}
\Gamma^{(\s{q})}_{V_\mu}   &=& \s{q}q_\mu/q^2\,, & {\rm and} & \Gamma^{(\s{q})}_{A_\mu}   &=& \s{q}q_\mu \gamma_5/q^2\,.
\end{array}\end{equation}

For completeness, we also renormalize the tensor current. The vertex function is $\Pi_{\sigma_{\mu\nu}}$, 
where
\be
\sigma_{\mu\nu}= \frac{i}{2} \left[\gamma_\mu,\gamma_\nu\right] \,,
\ee
and the amputated and projected vertex are
\be
\Lambda_{T} = 
\mathrm{tr}\left[\Pi_{\sigma_{\mu\nu}}\cdot {\Gamma}^{(s)}_{\sigma_{\mu\nu} } \right] \;.
\ee
For the projectors, we use
\begin{equation}\begin{array}{lllclll}
\Gamma^{(\gamma_\mu)}_{\sigma_{\mu\nu}}  &=& \sigma_{\mu\nu}\,, & {\rm and} & \Gamma^{(\s{q})}_{\sigma_{\mu\nu}}  &=& \sigma_{\nu\rho} q_\rho q_\mu/q^2 \,.
\end{array}\end{equation}

The corresponding renormalization factors $Z_{S,V,T,A,P}/Z_q$ can then obtained by imposing Eq.~\eqref{eq:Zdef} with $n=2$.

To obtain the renormalization factor of the quark mass, $Z_m$, we take the ratio of the vector and scalar bilinears in order to cancel the quark wave-function renormalization: 
\be
\label{eq:Zm}
Z_m^{(s)}(\mu,a) = \frac{\bar \Lambda_S(\mu,a)} {Z_V(a) \times \bar \Lambda_V^{(s)}(\mu,a)}
\,,
\ee
where $Z_V$ is computed hadronically via the procedure given in Section~\ref{sec-zvdeterm}. In the previous equation, we have used the fact that 
$Z_m=1/Z_S=1/Z_P$ in the chiral limit. Similarly, we should expect $Z_A=Z_V$ up to some small corrections arising, for example, from the fact that we work at finite $L_s$,
or due to infrared contaminations. In our estimate of the systematic errors, we have also replaced $\Lambda_S$ by $\Lambda_P$ and $\Lambda_V$ by $\Lambda_A$ in Equation~\eqref{eq:Zm}.

\subsubsection{Renormalization of the kaon bag parameter}

The renormalization factor $Z_{B_K}$ is defined in a similar manner. The amputated Green's function of the relevant four-quark operator ${\cal O}_{VV+AA}$ describing $K-\bar K$ oscillations in the Standard Model is computed numerically with a certain choice of kinematics and projected onto its tree-level value. We normalize by the square of the average between the vector and axial bilinear:
\be
\label{eq:Zbkappendix}
Z^{(s_1,s_2)}_{B_K}(\mu,a) \times \frac {\bar\Lambda^{(s_1)}_{VV+AA}(\mu,a)}{{\bar\Lambda^{(s_2)}_{AV}(\mu,a)}^2}
= 1\,,
\ee
where
\be
\Lambda_{AV} = \frac{1}{2} (\Lambda_V+\Lambda_A)\,,
\ee
such that the quark field renormalization cancels in the ratio. In practice we find that the difference between the vector and axial vertices are very small, hence choosing the average rather than simply $\Lambda_V$ or $\Lambda_A$ in the denominator, has no discernable effect.

In Eq.~\eqref{eq:Zbkappendix}, the superscripts $s_1$ and $s_2$ label the choice of projectors. 
We refer the reader to Refs.~\cite{Aoki:2010pe,Arthur:2012opa} for the details on the implementation, including the explicit 
definitions of projectors. 

\subsection{Numerical details and discussion}

\subsubsection{Quark mass renormalization}

For the quark mass renormalization we require only the values on the 32I and 24I ensembles, which together are sufficient to perform the continuum extrapolation of $Z_m/Z_{l/h}$. Here we discuss an update of the analysis performed in Ref.~\cite{Arthur:2012opa} using the newly-determined lattice spacings and a number of additional data points.

In the Rome-Southampton method, the projected vertex functions are first computed at finite quark mass before being extrapolated to the chiral limit. For each ensemble, we use unitary valence quark masses and extrapolate linearly in the quark mass. In the sea sector, the strange quark mass remains fixed to - or close to - its physical value. Since we do not observe any relevant quark mass dependence in our data, we neglect the systematic error associated with the fact that the sea strange quark mass is not extrapolated to zero.

We use partially-twisted boundary conditions to obtain momenta of the following form:
\bea
p_{in}  &=& \frac{2\pi}{L} (-\tilde m, 0, \tilde m,0)\,,\\
p_{out} &=& \frac{2\pi}{L} (0, \tilde m,  \tilde m,0) \,,
\eea
where $\tilde m$ combines the Fourier mode with the twist angle $\theta$ 
\be
\tilde m = m +\theta/2\,, \quad m\in \mathbb{N}\,.
\ee
The fact that these momenta all point in the same direction up to hypercubic rotations means that they lie upon a common continuum scaling curve (i.e. their $a^2$ dependence is the same), allowing us to unambiguously take the continuum limit.

For the 24I lattice, in addition to the momenta listed in~\cite{Arthur:2012opa}, we have generated additional points closer to the $3$ GeV point at which we ultimately evaluate the Z-factors. More precisely, the twist angle $\theta$ is chosen to be  $n \times 3/16$, with $n=15,16,\ldots 19$. The results can be found in the following section.

\subsubsection{Renormalization of $B_K$}

As $B_K$ is a scheme dependent quantity, we must perform our global fits to renormalized data, and as a result we require values of the renormalization coefficients to be computed on all of the ensembles used in the analysis: the 32I, 24I, 48I, 64I, 32Ifine and 32ID. This differs from the quark mass determination, for which we used a hadronically defined intermediate scheme during the continuum extrapolation and converted to $\MSbar$ {\it a posteriori}. In this appendix we present updated values of the 32I, 24I and 32ID $Z_{B_K}$ results in Ref.~\cite{Arthur:2012opa}, as well as new values for the 48I, 64I and 32Ifine.

For our new ensembles, we have considered only one value of the valence quark mass $m_l^{sea} = m_l^{val}$. Again, due to the modest chiral dependence previously observed for the non-exceptional schemes, we expect the associated systematic error to be negligible compare to the other sources of errors (in particular the perturbative matching).

As the 32ID ensemble is comparatively coarse, we renormalize at a lower scale $\mu_0 \sim1.4 \,\GeV$ and use the non-perturbative continuum step-scaling factor $\sigma^{(s_1,s_2)}_{B_K}(\mu,\mu_0)$ to run to 3 GeV. This procedure is discussed in Ref.~\cite{Arthur:2012opa}. The step-scaling factor is obtained by performing a continuum extrapolation of the ratio 
\begin{equation}
\sigma^{(s_1,s_2)}_{B_K}(\mu,\mu_0;a) = Z^{(s_1,s_2)}_{B_K}(\mu;a)/Z^{(s_1,s_2)}_{B_K}(\mu_0;a)\,,
\end{equation}
computed on the 32I and 24I lattices. 

Since the values of the lattice spacings have been updated, the numbers quoted here differ slightly from our previous work. The strategy is the following: we use the same 32ID lattice renormalization coefficient, $Z_{B_K}^{(s_1,s_2)}(\mu_0,a_{\rm 32ID})$, as used previously, but notice that the corresponding value of $\mu_0$ obtained with the new lattice spacings is $1.4363$ GeV rather than 1.426 GeV. As a result we must recompute the step-scaling factor. The results for $Z^{(s_1,s_2)}_{B_K}$ at $\mu_0$ can be found in Table~\ref{tab:finalZbk} and our 
updated results for $\sigma^{(s_1,s_2)}_{B_K}$ are reported in Table~\ref{tab:sigmabk}. For each scheme, the 32ID renormalization factor evaluated at $\mu=3$ GeV is then simply given by
\be
Z^{(s_1,s_2)}_{B_K}(\mu,a_{\rm 32ID}) = \sigma^{(s_1,s_2)}(\mu,\mu_0) \times Z^{(s_1,s_2)}_{B_K}(\mu_0,a_{\rm 32ID})\,.
\ee 

\subsection{Numerical results}

\subsubsection{Bilinears and quark mass renormalization}

The values for the amputated vertex functions $\bar\Lambda$ (normalized by the tree level value) at finite quark mass and in the chiral limit computed on the 24I ensemble are given in Tables~\ref{tab-billambdag24I} and~\ref{tab-billambdaq24I} for the SMOM$_{\gamma^\mu}$ and SMOM schemes respectively. The corresponding numbers for the 32I ensembles are given in Tables~\ref{tab-billambdag32I} and ~\ref{tab-billambdaq32I}. Recall that we use only one choice of projector for the scalar and pseudoscalar vertices, specifically those given in Eq.~\eqref{eqn-LambdaSPproject}. The results for these vertices computed on the 24I and 32I ensembles are included in Tables~\ref{tab-billambdag24I} and~\ref{tab-billambdag32I} respectively.

\begin{table}[tp]
\begin{tabular}{c|ccccc}
\hline
\multicolumn{6}{c}{$am = 0.01 $}\\
\hline
 $q$/GeV & 2.911997& 2.973955& 3.035912& 3.097870& 3.159827\\
\hline
 S & $1.1492(3)$ & $1.1455(3)$ & $1.1422(3)$ & $1.1390(2)$ & $1.1360(2)$ \\
 V & $1.0530(2)$ & $1.0537(2)$ & $1.0543(2)$ & $1.0550(2)$ & $1.0557(2)$ \\
 T & $1.0225(2)$ & $1.0244(2)$ & $1.0263(2)$ & $1.0281(2)$ & $1.0299(2)$ \\
 A & $1.0527(2)$ & $1.0534(2)$ & $1.0541(2)$ & $1.0548(2)$ & $1.0556(2)$ \\
 P & $1.1520(3)$ & $1.1480(3)$ & $1.1444(3)$ & $1.1409(2)$ & $1.1377(2)$ \\
\hline 
\multicolumn{6}{c}{$am = 0.005 $}\\
\hline
 $q$/GeV & 2.911997& 2.973955& 3.035912& 3.097870& 3.159827\\
\hline
 S & $1.1491(2)$ & $1.1455(2)$ & $1.1421(1)$ & $1.1390(1)$ & $1.1360(1)$ \\
 V & $1.0529(1)$ & $1.0536(1)$ & $1.0542(1)$ & $1.0549(1)$ & $1.0556(1)$ \\
 T & $1.0225(2)$ & $1.0244(2)$ & $1.0262(1)$ & $1.0281(1)$ & $1.0299(1)$ \\
 A & $1.0528(1)$ & $1.0534(1)$ & $1.0541(1)$ & $1.0548(1)$ & $1.0556(1)$ \\
 P & $1.1517(2)$ & $1.1478(2)$ & $1.1441(2)$ & $1.1407(2)$ & $1.1375(2)$ \\
\hline 
\multicolumn{6}{c}{$am = -am_{\rm res}$}\\ 
\hline
 $q$/GeV & 2.911997& 2.973955& 3.035912& 3.097870& 3.159827\\
\hline
 S & $1.1491(7)$ & $1.1455(6)$ & $1.1421(6)$ & $1.1389(5)$ & $1.1359(5)$ \\
 V & $1.0527(5)$ & $1.0534(4)$ & $1.0540(4)$ & $1.0547(4)$ & $1.0555(4)$ \\
 T & $1.0223(5)$ & $1.0243(5)$ & $1.0261(5)$ & $1.0280(5)$ & $1.0299(5)$ \\
 A & $1.0528(4)$ & $1.0535(4)$ & $1.0542(4)$ & $1.0549(4)$ & $1.0556(4)$ \\
 P & $1.1512(7)$ & $1.1473(7)$ & $1.1436(7)$ & $1.1402(6)$ & $1.1370(6)$ 
\end{tabular}
\caption{Projected, amputated vertex functions $\bar\Lambda$ for the vector, axial-vector and tensor operators in the SMOM$_{\gamma^\mu}$ scheme computed on the two 24I ensembles, and in the chiral limit, at scales close to the chosen renormalization scale of 3 GeV. In this table we also include the projected, amputated scalar and pseudoscalar vertices.\label{tab-billambdag24I} }
\end{table}

\begin{table}[tp]
\begin{tabular}{c|ccccc}
\hline
\multicolumn{6}{c}{$am = 0.01 $}\\
\hline
 $q$/GeV & 2.911997& 2.973955& 3.035912& 3.097870& 3.159827\\
\hline
 V & $1.1159(4)$ & $1.1160(3)$ & $1.1163(3)$ & $1.1166(3)$ & $1.1171(3)$ \\
 T & $1.0225(2)$ & $1.0244(2)$ & $1.0263(2)$ & $1.0281(2)$ & $1.0299(2)$ \\
 A & $1.1156(3)$ & $1.1158(3)$ & $1.1160(3)$ & $1.1164(3)$ & $1.1169(3)$ \\
\hline 
\multicolumn{6}{c}{$am = 0.005 $}\\
\hline
 $q$/GeV & 2.911997& 2.973955& 3.035912& 3.097870& 3.159827\\
\hline
 V & $1.1158(3)$ & $1.1159(3)$ & $1.1162(2)$ & $1.1165(2)$ & $1.1169(2)$ \\
 T & $1.0225(2)$ & $1.0244(2)$ & $1.0262(2)$ & $1.0281(1)$ & $1.0299(1)$ \\
 A & $1.1156(3)$ & $1.1158(2)$ & $1.1160(2)$ & $1.1163(2)$ & $1.1167(2)$ \\
\hline 
\multicolumn{6}{c}{$am = -am_{\rm res}$}\\ 
\hline 
 $q$/GeV & 2.911997& 2.973955& 3.035912& 3.097870& 3.159827\\
\hline
 V & $1.1156(9)$ & $1.1158(9)$ & $1.1159(8)$ & $1.1161(8)$ & $1.1164(8)$ \\
 T & $1.0224(5)$ & $1.0243(5)$ & $1.0262(5)$ & $1.0280(5)$ & $1.0299(5)$ \\
 A & $1.1156(9)$ & $1.1158(8)$ & $1.1160(8)$ & $1.1162(7)$ & $1.1165(7)$ 
\end{tabular}
\caption{Projected, amputated vertex functions $\bar\Lambda$ in the SMOM scheme computed on the two 24I ensembles, and in the chiral limit, at scales close to the chosen renormalization scale of 3 GeV.\label{tab-billambdaq24I} }
\end{table}

\begin{table}[tp]
\begin{tabular}{c|ccccccc}
\hline
\multicolumn{8}{c}{$am = 0.008 $}\\
\hline
 $q$/GeV & 1.186382& 1.581953& 2.067155& 2.397900& 2.769988& 3.100733& 3.431478\\
\hline
 S & $1.5760(95)$ & $1.4124(23)$ & $1.2881(7)$ & $1.2346(4)$ & $1.1920(2)$ & $1.1648(2)$ & $1.1446(1)$ \\
 V & $1.0568(13)$ & $1.0425(4)$ & $1.0376(1)$ & $1.0368(1)$ & $1.0374(1)$ & $1.0387(0)$ & $1.0405(0)$ \\
 T & $0.9072(10)$ & $0.9403(3)$ & $0.9668(1)$ & $0.9796(1)$ & $0.9915(1)$ & $1.0005(0)$ & $1.0083(0)$ \\
 A & $1.0357(9)$ & $1.0369(4)$ & $1.0364(1)$ & $1.0362(1)$ & $1.0371(1)$ & $1.0385(0)$ & $1.0404(0)$ \\
 P & $1.8453(92)$ & $1.4853(22)$ & $1.3065(9)$ & $1.2425(5)$ & $1.1956(3)$ & $1.1665(2)$ & $1.1457(2)$ \\
 \hline
\multicolumn{8}{c}{$am = 0.006 $}\\
\hline
 $q$/GeV & 1.186382& 1.581953& 2.067155& 2.397900& 2.769988& 3.100733& 3.431478\\
\hline
 S & $1.5818(54)$ & $1.4178(29)$ & $1.2906(10)$ & $1.2358(6)$ & $1.1930(3)$ & $1.1656(2)$ & $1.1451(1)$ \\
 V & $1.0544(7)$ & $1.0413(4)$ & $1.0376(2)$ & $1.0370(2)$ & $1.0376(1)$ & $1.0388(1)$ & $1.0406(1)$ \\
 T & $0.9081(5)$ & $0.9402(3)$ & $0.9669(3)$ & $0.9798(2)$ & $0.9917(1)$ & $1.0006(1)$ & $1.0084(1)$ \\
 A & $1.0357(8)$ & $1.0374(3)$ & $1.0367(2)$ & $1.0366(2)$ & $1.0374(1)$ & $1.0387(1)$ & $1.0405(1)$ \\
 P & $1.8124(66)$ & $1.4745(23)$ & $1.3048(6)$ & $1.2419(6)$ & $1.1956(3)$ & $1.1669(2)$ & $1.1458(2)$ \\
 \hline
\multicolumn{8}{c}{$am = 0.004 $}\\
\hline
 $q$/GeV & 1.186382& 1.581953& 2.067155& 2.397900& 2.769988& 3.100733& 3.431478\\
\hline
 S & $1.5697(61)$ & $1.4163(24)$ & $1.2915(6)$ & $1.2363(3)$ & $1.1927(2)$ & $1.1653(1)$ & $1.1448(1)$ \\
 V & $1.0542(10)$ & $1.0418(2)$ & $1.0373(2)$ & $1.0368(1)$ & $1.0374(1)$ & $1.0387(1)$ & $1.0405(1)$ \\
 T & $0.9078(8)$ & $0.9404(2)$ & $0.9668(2)$ & $0.9798(1)$ & $0.9917(1)$ & $1.0005(1)$ & $1.0083(1)$ \\
 A & $1.0396(8)$ & $1.0383(2)$ & $1.0366(2)$ & $1.0365(1)$ & $1.0373(1)$ & $1.0386(1)$ & $1.0404(1)$ \\
 P & $1.8346(113)$ & $1.4761(21)$ & $1.3050(9)$ & $1.2414(4)$ & $1.1948(3)$ & $1.1663(1)$ & $1.1455(1)$ \\
\hline
\multicolumn{8}{c}{$am = -am_{\rm res}$}\\
\hline
 $q$/GeV & 1.186382& 1.581953& 2.067155& 2.397900& 2.769988& 3.100733& 3.431478\\
\hline
 S & $1.5605(173)$ & $1.4219(57)$ & $1.2955(14)$ & $1.2384(7)$ & $1.1934(4)$ & $1.1659(3)$ & $1.1453(3)$ \\
 V & $1.0510(25)$ & $1.0413(6)$ & $1.0369(3)$ & $1.0369(3)$ & $1.0375(2)$ & $1.0388(2)$ & $1.0405(1)$ \\
 T & $0.9085(18)$ & $0.9406(6)$ & $0.9668(4)$ & $0.9799(3)$ & $0.9918(2)$ & $1.0006(2)$ & $1.0084(1)$ \\
 A & $1.0438(18)$ & $1.0400(5)$ & $1.0368(4)$ & $1.0368(3)$ & $1.0376(2)$ & $1.0388(1)$ & $1.0405(1)$ \\
 P & $1.7969(254)$ & $1.4637(53)$ & $1.3030(22)$ & $1.2400(11)$ & $1.1940(7)$ & $1.1659(4)$ & $1.1452(3)$ 
 \end{tabular}
 \caption{Projected, amputated vertex functions $\bar\Lambda$ for the vector, axial-vector and tensor operators in the SMOM$_{\gamma^\mu}$ scheme computed on the three 32I ensembles, and in the chiral limit, at scales close to the chosen renormalization scale of 3 GeV. In this table we also include the projected, amputated scalar and pseudoscalar vertices.\label{tab-billambdag32I} }
\end{table} 

\begin{table}[tp]
\begin{tabular}{c|ccccccc}
\hline
\multicolumn{8}{c}{$am = 0.008 $}\\
\hline
 $q$/GeV & 1.186382& 1.581953& 2.067155& 2.397900& 2.769988& 3.100733& 3.431478\\
\hline
 V & $1.1756(30)$ & $1.1387(12)$ & $1.1145(4)$ & $1.1047(2)$ & $1.0979(2)$ & $1.0945(1)$ & $1.0928(1)$ \\
 T & $0.9077(10)$ & $0.9406(3)$ & $0.9668(1)$ & $0.9796(1)$ & $0.9916(1)$ & $1.0005(0)$ & $1.0083(0)$ \\
 A & $1.1659(27)$ & $1.1359(12)$ & $1.1137(4)$ & $1.1043(2)$ & $1.0977(2)$ & $1.0943(1)$ & $1.0927(1)$\\
\hline
\multicolumn{8}{c}{$am = 0.006 $}\\
\hline
 $q$/GeV & 1.186382& 1.581953& 2.067155& 2.397900& 2.769988& 3.100733& 3.431478\\
\hline
 V & $1.1735(15)$ & $1.1367(10)$ & $1.1147(6)$ & $1.1049(4)$ & $1.0981(3)$ & $1.0946(3)$ & $1.0929(3)$ \\
 T & $0.9089(6)$ & $0.9405(3)$ & $0.9669(3)$ & $0.9798(2)$ & $0.9917(1)$ & $1.0006(1)$ & $1.0084(1)$ \\
 A & $1.1661(15)$ & $1.1347(10)$ & $1.1142(6)$ & $1.1046(4)$ & $1.0979(3)$ & $1.0945(3)$ & $1.0928(3)$\\
\hline
\multicolumn{8}{c}{$am = 0.004 $}\\
\hline
 $q$/GeV & 1.186382& 1.581953& 2.067155& 2.397900& 2.769988& 3.100733& 3.431478\\
\hline
 V & $1.1760(19)$ & $1.1377(8)$ & $1.1138(4)$ & $1.1045(2)$ & $1.0979(2)$ & $1.0944(1)$ & $1.0927(1)$ \\
 T & $0.9086(7)$ & $0.9408(2)$ & $0.9667(2)$ & $0.9798(1)$ & $0.9917(1)$ & $1.0005(1)$ & $1.0083(1)$ \\
 A & $1.1713(20)$ & $1.1365(8)$ & $1.1134(4)$ & $1.1043(2)$ & $1.0978(2)$ & $1.0943(1)$ & $1.0926(1)$ \\
\hline
\multicolumn{8}{c}{$am = -am_{\rm res} $}\\
\hline 
 $q$/GeV & 1.186382& 1.581953& 2.067155& 2.397900& 2.769988& 3.100733& 3.431478\\
\hline
 V & $1.1769(54)$ & $1.1368(23)$ & $1.1131(9)$ & $1.1042(5)$ & $1.0979(4)$ & $1.0944(3)$ & $1.0926(3)$ \\
 T & $0.9097(17)$ & $0.9411(6)$ & $0.9667(3)$ & $0.9799(3)$ & $0.9918(2)$ & $1.0006(2)$ & $1.0084(1)$ \\
 A & $1.1777(53)$ & $1.1372(23)$ & $1.1132(9)$ & $1.1043(5)$ & $1.0980(4)$ & $1.0944(3)$ & $1.0926(3)$
\end{tabular}
\caption{Projected, amputated vertex functions $\bar\Lambda$ in the SMOM scheme computed on the three 32I ensembles, and in the chiral limit, at scales close to the chosen renormalization scale of 3 GeV.\label{tab-billambdaq32I} }
\end{table}

\begin{table}[tp]
\begin{tabular}{cc|ccccc}
\hline\hline
Lattice & Scheme & S & V & T & A & P\\
\hline
\multirow{2}{*}{24I} & $\gamma^\mu$ & 1.1441(6) & 1.0536(4) & 1.0251(5) & 1.0538(4) & 1.1457(7)\\
                     & $\slashed q$ &   -       & 1.1158(8) & 1.0251(5) & 1.1159(8) &    -     \\
\hline
\multirow{2}{*}{32I} & $\gamma^\mu$ & 1.1736(4) & 1.0383(2) & 0.9981(2) & 1.0383(2) & 1.1737(5)\\
                     & $\slashed q$ &   -       & 1.0957(3) & 0.9978(2) & 1.0958(3) &    -     \\
\end{tabular}
\caption{The bilinear amputated, projected vertex functions $\bar\Lambda$ interpolated to 3 GeV. Note that these errors do not include the lattice spacing uncertainty. \label{tab-appendixSVTAPrenfacs24I32I} }
\end{table}
In Table~\ref{tab-appendixSVTAPrenfacs24I32I} we present $\bar\Lambda$ interpolated to $3$ GeV using a polynomial ansatz in the momenta. For the 24I lattice, since we have a very fine resolution, we take 
the five momenta quoted in the tables. For the 32I results we use $q\sim 2.77,3.10$ and $3.43\,\GeV$ in the interpolation. 

We show the values of the quark mass renormalization in Table~\ref{tab-appendixZm32I24I}. Using Table~\ref{tab-appendixSVTAPrenfacs24I32I} we can gauge the size of the systematic error on $Z_m$ by comparing the S and P vertices and the A and V vertices. We observe that the differences between the vector and axial vector vertices are very small, and can therefore be neglected. The differences between the scalar and pseudoscalar vertices are slightly larger, but these correspond to only 0.01\% changes if used in the computation of $Z_m$, and can therefore be ignored. As discussed above, the systematic error associated with not taking the chiral extrapolation of the sea strange quark mass can also be ignored. Note that the uncertainties on the lattice spacings are incorporated in these quantities in the main analysis.

\begin{table}[tp]
\begin{tabular}{c|cc}
\hline\hline
                & 24I & 32I\\
\hline
${\gamma_\mu}$ &  $1.523(1)$ & $1.519(4)$  \\
$\s{q}$       &  $1.439(1)$ & $1.440(4)$  \\
\end{tabular}
\caption{Quark mass renormalization factors $Z_m^{(s)}(3\,\GeV,a)$ computed on the 24I and 32I lattices at $3\,\GeV$ in the
two SMOM-schemes. Note that these errors do not include the lattice spacing uncertainty.\label{tab-appendixZm32I24I} }
\end{table}

\subsubsection{Renormalization of $B_K$ }

We quote the results for the projected vertex function
$\bar \Lambda_{VV+AA}$ at finite masses and in the chiral limit
for various momenta on the 32I and 24I ensembles in Tables~\ref{tab:Lbk24IWqg} and~\ref{tab:Lbk32IWqg}. The corresponding values each computed at a single quark mass on the 48I, 64I and 32Ifine ensembles are given in Tables~\ref{tab:Lbk48gq},~\ref{tab:Lbk64qg} and~\ref{tab:Lbk32Ifineqg} respectively.

To obtain the final results we construct the ratio given in Equation~\eqref{eq:Zbk} at finite quark masses for a few momenta surround the desired scale, either $\mu_0= 1.4363\,\GeV$ or $\mu=3\,\GeV$, take
the chiral limit and then perform the interpolation with a polynomial ansatz. Similarly to the quark mass case, the procedure is very robust and does not depend 
on the order we perform these operations, nor on the details of the interpolation. The final results for $Z_{B_K}$ on the various ensembles are given in Table~\ref{tab:finalZbk}, and the continuum step-scaling factors used to run the 32ID renormalization factor to 3 GeV are quoted in Table~\ref{tab:sigmabk}.

As with the quark mass renormalization the only significant source of systematic error on these results arises from the perturbative matching to $\MSbar$, which we estimate using the full difference between our final predictions for $B_K$ determined via the two intermediate SMOM schemes. As above, we incorporate the uncertainties on the lattice spacings into our renormalization factors in the main analysis.

\begin{table}[tp]
\begin{tabular}{c|cccc}
\hline\hline
\multicolumn{5}{c}{$(\gamma^\mu,\gamma^\mu)$ scheme, lowest momenta}\\
\hline
 $q$/GeV & 1.172282& 1.563201& 1.858201& 1.920141\\
\hline
  $am = 0.01 $ & $1.1453(14)$ & $1.1617(9)$ & $1.1702(5)$ & $1.1722(5)$ \\
  $am = 0.005 $ & $1.1458(16)$ & $1.1600(8)$ & $1.1688(4)$ & $1.1708(4)$ \\
  $am =-am_{\rm res} $ & $1.1466(48)$ & $1.1574(26)$ & $1.1665(14)$ & $1.1685(13)$ \\
\hline
\multicolumn{5}{c}{$(\gamma^\mu,\gamma^\mu)$ scheme, highest momenta}\\
\hline  
 $q$/GeV & 2.973122& 3.035062& 3.097002& 3.158942\\
\hline
  $am = 0.01 $ & $1.2116(5)$ & $1.2145(5)$ & $1.2174(5)$ & $1.2205(5)$ \\
  $am = 0.005 $ & $1.2113(3)$ & $1.2142(3)$ & $1.2171(3)$ & $1.2202(3)$ \\
  $am =-am_{\rm res} $ & $1.2108(13)$ & $1.2137(13)$ & $1.2167(14)$ & $1.2197(14)$ \\
\hline\hline
\multicolumn{5}{c}{$(\slashed q,\slashed q)$ scheme, lowest momenta}\\
\hline
 $q$/GeV & 1.172282& 1.563201& 1.858201& 1.920141\\
\hline
  $am = 0.01 $ & $1.3017(25)$ & $1.2921(14)$ & $1.2851(10)$ & $1.2846(10)$ \\
  $am = 0.005 $ & $1.2996(23)$ & $1.2876(15)$ & $1.2825(8)$ & $1.2821(8)$ \\
  $am =-am_{\rm res} $ & $1.2962(64)$ & $1.2803(49)$ & $1.2782(30)$ & $1.2779(27)$\\
\hline
\multicolumn{5}{c}{$(\slashed q,\slashed q)$ scheme, highest momenta}\\
\hline   
 $q$/GeV & 2.973122& 3.035062& 3.097002& 3.158942\\
\hline
  $am = 0.01 $ & $1.3012(7)$ & $1.3037(7)$ & $1.3064(8)$ & $1.3092(8)$ \\
  $am = 0.005 $ & $1.3008(5)$ & $1.3033(5)$ & $1.3059(4)$ & $1.3087(4)$ \\
  $am =-am_{\rm res} $ & $1.3003(18)$ & $1.3027(18)$ & $1.3052(18)$ & $1.3078(18)$
\end{tabular}
\caption{Chiral extrapolation of $\bar \Lambda_{VV+AA}$ in both schemes on the 24I ensemble for the momentum points in the vicinity of the $~1.4$ GeV scale, and those in the vicinity of the 3 GeV matching scale. }
\label{tab:Lbk24IWqg}
\end{table}

\begin{table}[tp]
\begin{tabular}{c|ccccccc}
\hline\hline
\multicolumn{8}{c}{$(\gamma^\mu,\gamma^\mu)$ scheme}\\
\hline
 $q$/GeV & 1.186382& 1.581953& 2.067155& 2.397900& 2.769988& 3.100733& 3.431478\\
\hline
  $am = 0.008 $ & $1.0985(21)$ & $1.1117(11)$ & $1.1240(2)$ & $1.1311(3)$ & $1.1399(2)$ & $1.1483(2)$ & $1.1573(1)$ \\
  $am = 0.006 $ & $1.0991(20)$ & $1.1111(6)$ & $1.1246(5)$ & $1.1318(4)$ & $1.1404(3)$ & $1.1487(3)$ & $1.1577(3)$ \\
  $am = 0.004 $ & $1.1008(16)$ & $1.1120(2)$ & $1.1236(5)$ & $1.1312(4)$ & $1.1401(3)$ & $1.1483(2)$ & $1.1573(2)$ \\
  $am =-am_{\rm res} $ & $1.1034(43)$ & $1.1129(11)$ & $1.1237(13)$ & $1.1316(9)$ & $1.1404(7)$ & $1.1485(6)$ & $1.1574(5)$ \\
\hline
\multicolumn{8}{c}{$(\slashed q,\slashed q)$ scheme}\\
\hline
 $q$/GeV & 1.186382& 1.581953& 2.067155& 2.397900& 2.769988& 3.100733& 3.431478\\
\hline
  $am = 0.008 $ & $1.2473(40)$ & $1.2352(21)$ & $1.2262(5)$ & $1.2233(3)$ & $1.2238(3)$ & $1.2266(2)$ & $1.2316(2)$ \\
  $am = 0.006 $ & $1.2471(25)$ & $1.2334(13)$ & $1.2271(10)$ & $1.2242(7)$ & $1.2243(6)$ & $1.2271(5)$ & $1.2321(5)$ \\
  $am = 0.004 $ & $1.2515(28)$ & $1.2345(13)$ & $1.2253(9)$ & $1.2231(4)$ & $1.2239(4)$ & $1.2266(3)$ & $1.2316(2)$ \\
  $am =-am_{\rm res} $ & $1.2566(77)$ & $1.2341(38)$ & $1.2251(22)$ & $1.2231(12)$ & $1.2241(10)$ & $1.2267(8)$ & $1.2316(7)$ 
\end{tabular}
\caption{Chiral extrapolation of  $\bar \Lambda_{VV+AA}$ in both schemes on the 32I ensemble for all simulated momenta. }
\label{tab:Lbk32IWqg}
\end{table}

\begin{table}[tp]
\begin{tabular}{c|ccccc}
\hline\hline
\multicolumn{6}{c}{$(\gamma^\mu,\gamma^\mu)$ scheme}\\
\hline
$q$/GeV & 2.72125 & 2.88132 & 2.96136 & 3.04139 & 3.20147 \\
\hline
VV+AA & 1.20472(14) & 1.21216(8) & 1.21604(8) & 1.21996(8) & 1.22827(6) \\
V & 1.05201(5) & 1.05371(3) & 1.05463(4) & 1.05557(3) & 1.05753(1) \\
A & 1.05196(3) & 1.05368(2) & 1.05458(4) & 1.05553(3) & 1.05745(4) \\
\hline
\multicolumn{6}{c}{$(\slashed q,\slashed q)$ scheme}\\
\hline
$q$/GeV & 2.72125 & 2.88132 & 2.96136 & 3.04139 & 3.20147 \\
\hline
VV+AA & 1.29658(31) & 1.30250(14) & 1.30598(10) & 1.30955(10) & 1.31773(25) \\
V & 1.11640(15) & 1.11660(7) & 1.11697(4) & 1.11749(5) & 1.11902(15) \\
A & 1.11633(13) & 1.11659(5) & 1.11695(4) & 1.11747(5) & 1.11902(14) 
\end{tabular}
\caption{Vertex functions of the four-quark operators $\bar \Lambda_{VV+AA}$ and the bilinears
 $\bar \Lambda_{V}$ and $\bar \Lambda_{A}$ needed for $Z_{B_K}$, 
computed in both schemes on the 48I ensemble  with $am=0.00078$.
}
\label{tab:Lbk48gq}
\end{table}

\begin{table}[tp]
\begin{tabular}{c|ccccc}
\hline\hline
\multicolumn{6}{c}{$(\gamma^\mu,\gamma^\mu)$ scheme}\\
\hline
$q$/GeV & 2.7823 & 2.94596 & 3.0278 & 3.10963 & 3.27329 \\
\hline
VV+AA & 1.13936(9) & 1.14363(10) & 1.14575(6) & 1.14798(6) & 1.15261(4) \\
V & 1.03721(4) & 1.03783(3) & 1.03818(2) & 1.03859(2) & 1.03949(2) \\
A & 1.03715(2) & 1.03780(3) & 1.03815(2) & 1.03856(2) & 1.03949(1) \\
\hline
\multicolumn{6}{c}{$(\slashed q,\slashed q)$ scheme}\\
\hline
$q$/GeV & 2.7823 & 2.94596 & 3.0278 & 3.10963 & 3.27329 \\
\hline
VV+AA & 1.22136(20) & 1.22299(22) & 1.22387(12) & 1.22501(11) & 1.22760(12) \\
V & 1.09622(9) & 1.09451(11) & 1.09379(6) & 1.09323(5) & 1.09239(6) \\
A & 1.09619(9) & 1.09449(11) & 1.09377(6) & 1.09321(5) & 1.09238(7) 
\end{tabular}
\caption{Vertex functions of the four-quark operators $\bar \Lambda_{VV+AA}$ and the bilinears
 $\bar \Lambda_{V}$ and $\bar \Lambda_{A}$ needed for $Z_{B_K}$, 
computed in both schemes on the 64I ensemble  with $am=0.000678$.
}
\label{tab:Lbk64qg}
\end{table}

\begin{table}[tp]
\begin{tabular}{c|cccccc}
\hline\hline
\multicolumn{7}{c}{$(\gamma^\mu,\gamma^\mu)$ scheme}\\
\hline
$q$/GeV & 2.18326 & 2.61991  & 3.05656  & 3.49322  & 3.92987 & 4.36652  \\
\hline
VV+AA & 1.08793(30) & 1.09602(53) & 1.10356(30) & 1.11137(28) & 1.11909(12) & 1.12751(11) \\
V & 1.03160(12) & 1.03005(15) & 1.02984(8) & 1.03049(8) & 1.03162(5) & 1.03327(4) \\
A & 1.03076(11) & 1.02976(17) & 1.02972(8) & 1.03042(8) & 1.03159(5) & 1.03324(5) \\
\hline
\multicolumn{6}{c}{$(\slashed q,\slashed q)$ scheme}\\
\hline
$q$/GeV & 2.18326 & 2.61991  & 3.05656  & 3.49322  & 3.92987 & 4.36652  \\
\hline
VV+AA & 1.18143(36) & 1.17928(112) & 1.17863(65) & 1.18050(63) & 1.18350(28) & 1.18855(25) \\
V & 1.10353(55) & 1.09220(57) & 1.08465(34) & 1.08013(31) & 1.07727(14) & 1.07605(12) \\
A & 1.10308(55) & 1.09196(57) & 1.08453(34) & 1.08005(31) & 1.07723(14) & 1.07601(12) 
\end{tabular}
\caption{Vertex functions of the four-quark operators $\bar \Lambda_{VV+AA}$ and the bilinears
 $\bar \Lambda_{V}$ and $\bar \Lambda_{A}$ needed for $Z_{B_K}$, 
computed in both schemes on the 32Ifine ensemble  with $am=0.0047$.
}
\label{tab:Lbk32Ifineqg}
\end{table}

\begin{table}[tp]
\centering
\begin{tabular}{c|cccccc}
\hline\hline
\multicolumn{7}{c}{$Z_{B_K}^{(s_1,s_2)}(3\,\GeV,a)$}\\
\hline
                        & 24I          & 32I           &  48I           & 64I  & 32Ifine   & 32ID       \\
\hline
$(\gamma_\mu,\gamma_\mu)$ &  $0.9161(5)$ & $0.9409(2)$   &  $0.91397(3)$  & $0.94106(2)$  & $0.9617(1)$ & - \\
$(\s{q},\s{q})$         &  $0.9568(2)$ & $0.9787(1)$   &   $0.954452(4)$ & $0.978152(2)$ & $0.9995(1)$ & - \\
\hline
\multicolumn{7}{c}{$Z_{B_K}^{(s_1,s_2)}(1.4363\,\GeV,a)$}\\
\hline
$(\gamma_\mu,\gamma_\mu)$ &  0.9546(10)  & 0.9809(93) & - &- & -&0.9210(8)  \\
$(\s{q},\s{q})$         &  1.0488(16)  & 1.0638(20) &- &- & -& 0.9992(11) \\
\end{tabular}
\caption{$B_K$ renormalization factors $Z_{B_K}^{(s_1,s_2)}$ computed on the various ensembles. For the 32I, 24I we quote values at both 1.4363 GeV and 3 GeV, which are used to compute the step-scaling factor. For the coarse 32ID ensemble we only quote the value at the lower scale, and for the 48I, 64I and 32Ifine we do not quote the values at the lower scale as they are not needed for our analysis. These values do not include the effect of the uncertainty on the lattice spacing in their errors.}
\label{tab:finalZbk}
\end{table}

\begin{table}[tp]
\centering
\begin{tabular}{c |c}
\hline\hline
$(\gamma_\mu,\gamma_\mu)$ &  0.9573(21) \\
$(\s{q},\s{q})$         &  0.9103(31)   
\end{tabular}
\caption{Continuum non-perturbative scale evolution $\sigma^{(s_1,s_2)}_{B_K}(\mu,\mu_0)$ extracted from the 24I and 32I lattices in 
two SMOM-schemes. As explained in the text, we choose $\mu_0=1.4363$ GeV and $\mu=3$ GeV. These values do not include the effect of the uncertainty on the lattice spacing in their errors.}
\label{tab:sigmabk}
\end{table}

\section{Random number generator}

\label{appendix-rng}

After all the data presented in this paper was generated, it was found that the U(1) noise generated from the freshly initialized random number generator (RNG) in CPS~\cite{Jung:2014ata} is vulnerable to correlations, such that the expectation value of $ \left\|\sum_{x} \sum_{j=1}^{N} e^{-i\theta(x)_j}\right\|^2/V$ deviates from N.
This correlation is not observed when U(1) noise is replaced with gaussian noise, for which the
accept/reject procedure used in generating the gaussian random numbers appears to eliminate the observed correlation. We also confirmed that the U(1) noise generated from the CPS RNG for the thermalized gauge configurations on our previous ensembles do not show the correlation, due to the de-correlating effect of the gaussian RNG used for the pseudofermion fields. 

To further test the robustness of gaussian random numbers generated from CPS RNG,
we reproduced the 2+1 flavor DWF ensemble used in Ref.~\cite{Blum:2011pu}, with the RNG's replaced with the Mersenne Twister~\cite{Matsumoto:1998}, implemented in C++11. Each $2^4$ hypercube of lattice sites was initialized with randomized seeds. We confirmed that the plaquette agrees to within 1 standard deviation: 0.588064(12) from 8460 MD units compared to 0.588052(9) from the configurations used in Ref.~\cite{Blum:2011pu}. 
All the random numbers generated from CPS RNG for the work presented here were gaussian random numbers. The only exception are the Z(3) random numbers for $Z_3$ box source used for the $\Omega$ baryon in Section~\ref{section:omega}, which was generated independently from the CPS RNG.

\FloatBarrier
\section{Global fit forms}
\label{appendix-fitforms}

The ChPT forms and their associated finite-volume corrections were originally determined in Ref.~\cite{Allton:2008pn} and the analytic forms in Refs.~\cite{Aoki:2010dy,Aoki:2010pe}. We have subsequently~\cite{Aoki:2010dy,Aoki:2010pe,Arthur:2012opa} added additional terms describing the scaling behavior and the dependence of the quantities on the heavy sea and valence quark masses where appropriate. In this analysis we also introduce linear fit forms to describe the Wilson flow scales. For the convenience of the reader we collect these disparate formulae below.

The ChPT forms for the pseudoscalar mass and decay constant are~\cite{Allton:2008pn,Aoki:2010dy,Arthur:2012opa}:
\begin{align}
m_{xy}^2 &= \frac{\chi_x+\chi_y}{2}\left[ 1+ L^{m_\pi}(\chi_x,\chi_y,\chi_l)\right] + c_{m_\pi,m_h} \frac{m_x+m_y}{2}(m_h - m_h^{\rm phys})\,, \\
f_{xy} &= f \left[ 1 + c_f^{\bf A} a^2 + L^{f_\pi}(\chi_x,\chi_y,\chi_l)\right] + c_{f_\pi,m_h}(m_h - m_h^{\rm phys})\,.
\end{align}
Here $m_x$ and $m_y$ are the (partially-quenched) valence light quark masses, $m_l$ is the sea light quark mass and $m_h$ the sea heavy quark mass. The quantity $\chi_x = 2B m_x$, and the superscript ${\bf A}$ above the $a^2$ coefficient denotes the gauge action. We use the following notation for the gauge actions: ${\bf I}$ for the Iwasaki action and ${\bf ID}$ for the Iwasaki+DSDR. The logarithmic terms $L^{m_\pi}$ are defined in Eq. B32 and B33 of Ref.~\cite{Allton:2008pn} for non-degenerate and degenerate valence quark masses, respectively. Similarly, $L^{f_\pi}$ are given in Eqs. B36 and B37 of the same document. For the kaon mass, decay constant and bag parameter we use the following forms~\cite{Allton:2008pn,Aoki:2010dy,Aoki:2010pe,Arthur:2012opa}:
\begin{align}
m_{xy}^2 &= m^{(K)}\left[1 + \frac{\lambda_1\chi_l}{f^2} + \frac{\lambda_2\chi_x}{f^2}\right] + c_{m_K,m_y}(m_y - m_h^{\rm phys}) + c_{m_K,m_h}(m_h - m_h^{\rm phys})\,, \\
f_{xy} &= f^{(K)}\left[1 + c_{f_K,a}^{\bf A} a^2 + \frac{\lambda_3\chi_l}{f^2} + \frac{\lambda_4\chi_x}{f^2} + L^{f_K}(\chi_x,\chi_l)\right] + c_{f_K,m_y}(m_y - m_h^{\rm phys}) + c_{f_K,m_h}(m_h - m_h^{\rm phys})\,, \\
B_{xy} &= B_K^0\left[ 1 + c_{B_K,a}^{\bf A} a^2 + \frac{ c_{B_K,m_l}\chi_l }{f^2} + \frac{ c_{B_K,m_x}\chi_x }{f^2} - \frac{ \chi_l }{32\pi^2 f^2}\log\left(\frac{\chi_x}{\Lambda_\chi^2}\right) \right] +  \nonumber\\
       &\hspace{2cm}c_{B_K,m_y}(m_y - m_h^{\rm phys}) + c_{B_K,m_h}(m_h - m_h^{\rm phys})\,,
\end{align}
where $m_y$ and $m_x$ are the heavy and light valence quark masses, respectively, and $m_l$ and $m_h$ are as above. Here the logarithmic term $L^{f_K}$ is defined in Eq. B47 of Ref.~\cite{Allton:2008pn}. For the Omega baryon mass we use
\begin{align}
m_{vvv} = m^{(\Omega)} + c_{m_\Omega,l}m_l + c_{m_\Omega,v}(m_v - m_h^{\rm phys}) + c_{m_\Omega,v}(m_h - m_h^{\rm phys})\,,
\end{align}
where $m_v$ is the valence heavy quark mass.

The analytic forms for the pseudoscalar mass and decay constant are~\cite{Aoki:2010dy,Arthur:2012opa}
\begin{align}
m_{xy}^2 &= C_0^{m_\pi} + C_1(m_x + m_y)/2 + C_2 m_l + C_3(m_h - m_h^{\rm phys}) \\
f_{xy} & = C_0^{f_\pi}(1 + C_a^{f_\pi,{\bf A}}a^2) + C_1(m_x + m_y)/2 + C_2 m_l + C_3(m_h - m_h^{\rm phys}) 
\end{align}
where again, $m_x$ and $m_y$ are the valence light quark masses, and $m_l$ and $m_h$ are the sea light and heavy quark masses. For the kaon mass, decay constant and bag parameter~\cite{Aoki:2010dy,Aoki:2010pe,Arthur:2012opa},
\begin{align}
m_{xy}^2 &= C_0^{m_K} + C_1^{m_K}m_x+ C_2^{m_K} m_l + C_3^{m_K}(m_y - m_h^{\rm phys}) + C_4^{m_K}(m_h - m_h^{\rm phys})\,,\\
f_{xy} &= C_0^{f_K}(1+C_a^{f_K,{\bf A}}a^2) + C_1^{f_K} m_x+ C_2^{f_K} m_l + C_3^{f_K}(m_y - m_h^{\rm phys}) + C_4^{f_K}(m_h - m_h^{\rm phys})\,,\\
B_{xy} &= C_0^{B_K}(1+C_a^{B_K,{\bf A}}a^2) + C_1^{B_K} m_x+ C_2^{B_K} m_l + C_3^{B_K}(m_y - m_h^{\rm phys}) + C_4^{B_K}(m_h - m_h^{\rm phys})\,,
\end{align}
where, as before, $m_y$ represents the heavy valence quark. Finally the analytic function for the Omega baryon mass is
\begin{align}
m_{vvv} &= C_0^{m_\Omega} + C_1^{m_\Omega} m_l + C_2^{m_\Omega}(m_v - m_h^{\rm phys}) + C_3^{m_\Omega}(m_h - m_h^{\rm phys})\,,
\end{align}
where again $m_v$ is the valence heavy quark mass. In general, the coefficients for these analytic functions are ordered as follows (skipping entries as appropriate): The valence light quark mass dependence; the sea light quark mass dependence; the valence heavy quark mass dependence; and the sea heavy quark mass dependence.

For this analysis we also define the following functions for the Wilson flow scales $t_0^{1/2}$ and $w_0$:
\begin{align}
w_0 &= c_{w_0,0}(1+c_{w_0,a}^{\bf A}a^2) + c_{w_0,l} m_l + c_{w_0,h}(m_h - m_h^{\rm phys})\,,\\
\sqrt{t_0} &= c_{\sqrt{t_0},0}(1+c_{\sqrt{t_0},a}^{\bf A}a^2) + c_{\sqrt{t_0},l} m_l + c_{\sqrt{t_0},h}(m_h - m_h^{\rm phys})\,.
\end{align}
These fit functions are used for both the ChPTFV/ChPT and analytic ans\"{a}tze.

Note that in the expressions above we do not show the $a^2$ coefficient for the pion, kaon and Omega baryon masses as they are fixed to zero by our choice of scaling trajectory (cf. Section V.A of Ref.~\cite{Aoki:2010dy}).

\FloatBarrier
\bibliography{paper}

\end{document}